\documentclass[11pt,a4paper]{article}
\usepackage{geometry}
\usepackage[utf8]{inputenc}
\usepackage[T1]{fontenc}
\usepackage{lmodern}
\usepackage{textcomp}
\usepackage{newunicodechar}
\usepackage{amsmath}
\numberwithin{figure}{section} % Commented out to avoid duplicate figure identifiers
\usepackage{jlcode}
\usepackage{amssymb}
\usepackage{amsthm}
\usepackage{stmaryrd}
\usepackage{graphicx}
\usepackage[hidelinks]{hyperref}
\usepackage{cleveref}
\usepackage{float}
\usepackage{listings}
\usepackage{natbib}
\usepackage{lineno}

\usepackage{tikz}
\usetikzlibrary{tikzmark, arrows.meta, positioning, shapes.geometric}
\usepackage{tikz-cd} % Load this separately as it's a specific package based on TikZ

\usepackage{listings}

\definecolor{jlrule}{rgb}{0.7,0.7,0.7}
\definecolor{jlcomment}{rgb}{0.5,0.5,0.5}
\definecolor{jlstring}{rgb}{0.8,0.2,0.2}
\definecolor{jlkeyword}{rgb}{0.0,0.0,0.8}
\definecolor{jlbackground}{rgb}{1.0,1.0,1.0}
\definecolor{jlop}{rgb}{0.6,0.0,0.6}
\definecolor{jlstrnum}{rgb}{0.0,0.6,0.0}
\definecolor{jlbase}{rgb}{0.0,0.4,0.6}
\definecolor{jltype}{rgb}{0.8,0.4,0.0}
\definecolor{jlmacro}{rgb}{0.6,0.4,0.0}
\definecolor{jlsymbol}{rgb}{0.4,0.4,0.4}
\definecolor{jlfuncall}{rgb}{0.2,0.2,0.8}
\definecolor{jlnumber}{rgb}{0.0,0.6,0.0}
\definecolor{jlfunctions}{rgb}{0.2,0.4,0.8}
\definecolor{jltext}{rgb}{0.0,0.0,0.0}
\definecolor{jlident}{rgb}{0.2,0.2,0.2}
\definecolor{jlerror}{rgb}{0.8,0.0,0.0}
\definecolor{jlwarning}{rgb}{0.8,0.6,0.0}
\definecolor{jlmodule}{rgb}{0.4,0.6,0.8}
\definecolor{jlbuiltins}{rgb}{0.6,0.2,0.8}
\definecolor{jloperator}{rgb}{0.6,0.0,0.6}
\definecolor{jlliteral}{rgb}{0.0,0.6,0.4}
\definecolor{jlspecial}{rgb}{0.8,0.2,0.6}

\lstdefinelanguage{Julia}{
    keywords={abstract,begin,break,case,catch,const,continue,do,else,elseif,end,export,false,for,function,global,if,import,let,local,macro,module,mutable,otherwise,quote,return,struct,switch,true,try,type,using,while},
    keywordstyle=\color{jlkeyword}\bfseries,
    ndkeywords={Int,Float64,String,Vector,Matrix,Array,Dict,Set,Tuple,Int64,Bool,Any,Nothing},
    ndkeywordstyle=\color{jltype}\bfseries,
    identifierstyle=\color{jlident},
    sensitive=true,
    comment=[l]{\#},
    commentstyle=\color{jlcomment}\ttfamily,
    stringstyle=\color{jlstring}\ttfamily,
    morestring=[b]',
    morestring=[b]",
    basicstyle=\ttfamily\footnotesize,
    breaklines=true,
    showstringspaces=false,
    numbers=left,
    numberstyle=\tiny\color{jlrule},
    frame=single,
    rulecolor=\color{jlrule},
    backgroundcolor=\color{jlbackground},
}

\newcommand{\green}[1]{\textcolor{green}{#1}}
\newcommand{\red}[1]{\textcolor{red}{#1}}
\theoremstyle{definition}
\newtheorem{definition}{Definition}[section]

\theoremstyle{remark}
\newcommand{\Aa}{{\blacksquare\square}}
\newcommand{\La}{{\square\blacksquare}}
\newcommand{\Aal}{\textcolor{red}{\blacksquare}\square}
\newcommand{\Aar}{\blacksquare\textcolor{red}{\square}}
\newcommand{\Lal}{\textcolor{red}{\square}\blacksquare}
\newcommand{\Lar}{\square\textcolor{red}{\blacksquare}}
\title{
Macroeconomic Foundation of Monetary Accounting\\by Diagrams of Categorical Universals 
}
\author{Renée Menéndez, Viktor Winschel\footnote{SRH University of Applied Sciences, email: \texttt{viktor.winschel@gmail.com}. 
Many thanks to Dusko Pavlovic, Alexander Kurz, Melanie Swan and Boyan Beronov
for many past and ongoing discussions using category theory for theories and models in economics and econometrics.
}}
\date{\today}
\begin{document}
\maketitle
%--------------------------------
\begin{abstract}
We present a category theoretical formulation of the Monetary Macroeconomic Accounting Theory (MoMaT) of~\cite{MW2025}.
We take macroeconomic (national) accounting systems to be composed from microeconomic double-entry systems with real and monetary units of accounts.
Category theory is the compositional grammar and module system of mathematics which we use to lift micro accounting consistency to the macro level. 
The main function of money in MoMaT is for the repayment of loans and not for the exchange of goods, bridging the desynchronisation of input and output payments of producers.
Accordingly, temporal accounting consistency is at the macroeconomic level.
We show that the accounting for macroeconomies organised by a division of labor can be consistent and stable as a prerequisite for risk and GDP sharing of societies.
We exemplify the theory by five sectoral agents of Labor and Resource owners, a Company as the productive sector, a Capitalist for profits, and a Bank 
as the financial sector providing loans to synchronise the micro and the macro levels of an economy.
The dynamics is described by eight sectoral macroeconomic bookings in each period demonstrating stable convergence of the MoMaT in numerical simulations. 
The categorical program implements a consistent evolution of hierarchical loan repayment contracts by an endofunctor.
The universal constructions of a limit verify all constraints as the sectoral investment and learning function at the macroeconomic level.
The dual colimit computes the aggregated informations at the macro level as usual in the mathematics of transitions from local to global structures.
We use visual diagrams to make complex economic relationships intuitive.
This paper is meant to map economic to categorical concepts to enable interdisciplinary collaboration for digital twins of monetary accounting systems.
\end{abstract}

\newpage
\tableofcontents
\setcounter{tocdepth}{3}
\newpage

%\section*{List of Definitions}
%\input{money.theory.cat.sim.def.tex}

\newpage
\listoffigures

\newpage
%-----------------------------------------------------------------------------------------------------------------------------------------------------------------------------
\newpage
\section{Introduction}
    We present a categorical foundation of the Monetary Macroeconomic Accounting Theory (MoMaT) of \cite{MW2025}.
    It is a theory of the fundamental hierarchical multi-sector-agent macroeconomic process of consumption, production, investment in monetary and real units.
    The dynamics of such a complex system involves different types of concepts at different hierarchical levels: macroeconomic invariances, markets, prices, production, goods, 
    demand of the sectors, therein individual agents' decisions being companies, households, or banks.
    We validate constraints, factorise computations of universals in pullbacks and pushouts over
    accounts in structural functorial constructions of a consistent operational.
    The hierarchical macroeconomic accounting system is to be designed to book into consistent double-entry microeconomic accounting systems.
    For that we categorify the equations and functions of a recursive formulation of MoMaT with functors and natural transformations operating on categories as its state.
    Simulations of the same data are illustrated in three different ways as recursive, categorically typed recursive and categorical programs.
    
    Formal visual diagrams of the structures form an integrated modelling environment for economics.
    Complex economic relationships are more intuitive in formal geometric visualisations.
    They keep local parts and global properties of whole or macroeconomic systems consistent.
    The abstract nonsense attributed to category theory makes sense when used to separate syntax and semantics in designing domain specific programmning languages (DSL) in compilers.
    The DSL of this paper implements macroeconomic sectors and their macroeconomic accounting structures for a faithfully cooperation 
    enabled by the sound consistency of the macroeconomic accounting system, organised and kept consistent 
    by lifted double-entry accounting consistency conditions to also encompase macroeconomic accounting consistency.
    Categorical structures enforce type discipline for the economic flows captured in eight sectoral bookings of basic events of the processes of consumption, production and investment.
    The events take place every period consistently connected by natural transformations as monadic endofunctors as evolution of time 
    over a memory of past legally created and partially fulfilled contractual future obligations as a generalized economic constraint problem and its solution and consistent evolution.
    A monad formalises in functional programming language semantics referentially transparent functions and side effects like changing a variable's assignment in programs like $x=x+1$.
    Side effects and external effects in economics and institutional dynamics where domains and codomains of the functions inolved change, 
    polycentrism of~\cite{Ostrom2005Book} and endogenous econometricians of the Lucas Critique can be modelled by universal constructions.
    Categories offer a mathematical foundation for monetary and real units as well as creation of types as currency unions and exchange rate regimes 
    typable in initial and final categorical algebras and coalgebras, respectively.
    
    An economically important result of the categorification is that sectoral investment learning and aggregation are both universal.
    The macroeconomic accounting system can represent quadruple accounting as in the standard for national accounting systems in~\cite{SNA2009}.
    Quadruple accounting system book parallely in two micro double-entry ($4=2\times 2$) accounting systems in each macroeconomic booking.
    In our approach they are decentralisable by compositionality and parallel and hierarchical combinators.
    Micro consistency is about booking in each micro booking the same one number into the credit and debit sides of two T-accounts of one decentralized and consistent double-entry booking system.
    An algebra of resources can be constructed on the linear, resource aware logic of these one numbers booked in several decentralized places of the system.\footnote{Linear 
    logic is a resource aware logic, meaning that logical resources can only be used once in a proof, where they are used up. 
    For example, in a linear logic of the logical resources $\{A,A\rightarrow B, A\rightarrow C\}$,
    only $B$ or $C$ can be proved using A, since $A$ is used up in one proof being not available anymore for the other.
    To proof both $B$ and $C$ we need to copy $A$ first and with the resources $\{A,A,A\rightarrow B, A\rightarrow C\}$ we can proof $B$ and $C$ by using the modus ponens
    $\{A,A\rightarrow B \vdash B\}$ twice $\{A,A\rightarrow C \vdash C\}$. Quantum computers run on linear logic which makes them ideal to represent property right transfers, 
    secured by physical laws whereas digital computers need redundant blockchains to ensure that costless copy-and-paste operations of digital computers 
    are not misused to fake property right transfers.
    }% end footnote
    It also needs an operator to represent the parallel processes of a decentralized market economy.
    The parallel, sequential and hierarchical combinators in this paper enable us to fully type macroeconomic accounting systems for nations and institutions.
    
    In MoMaT the macroeconomic consistency encompasses a market where the price matches consumers' demands and the planned sales of goods of producers to cover the costs and profits in universals.
    Economic optimisation is lifted to categorical universals and adjunctions.
    Category theory offers computable closure operators over composed macroeconomies for currency areas, closed over debt contracts of institutional subobject structures
    like central banks, governments or generations over time as deontic modalities of legal units of polycentric management.
    Another important closure we also calculate as usual in economics is the closure over endowment, production and consumption of goods, to get what is usually meant by a {\it closed} economy.
    The institutional closures of meta rules like macroeconomic budgets, or liabilities equal assets in debt systems of loans and repayments 
    is complementing the production closure in the categorified MoMaT.
    This opens a path for an economic theory of sharing the wealth and values over universaly factorized production and aggregation, 
    complementing the theory of labor sharing and macroeconomic accounting for hierarchical risk sharing.
    Money is then a medium of labor, risk and GDP sharing.
    The monetary macroeconomic system of loans and repayments is consistently booked over hierarchical structural functorial evolutions.
    Extensions of the theory for more agents, accounts, levels and consistencies are compositional by the categorical semantics of the DSL.
    Universal categorical constructions are adjunct functors and generalise the economic operators of maximisation 
    to higher-order functions, morphisms, functors with natural transformations and universals for institutional dynamics of systemic macroeconomic entities 
    as functorial views on the system at hand.
    
    In mathematics adjunctions are everywhere as in economics.
    Universals formulate systemic rationality formally as validations of economic constraints and implementations of the programs for decentral and hierarchical economic planning and coordination.
    The simulated data shows stable convergence of the monetary systems, it is not a Ponzi scheme, neither in need to be stabilised by governmental debt policies (especially by unsustainable ones).
    Categorical stability proofs and proofs of the universal construction by verifying the universal constructions used in this paper are left for future research. 
    We are focused in this paper to map the economics concepts to the categorical ones.
    The {\it proofs} so far are the numerical simulations, to be taken as constructive witnesses of the soundness of the categorical constructions.
    
    The DSL can be used to program a national accounting system as for the one in MoMaT.
    As an architectual toolbox it improves the one we have developed in industrial projects on 
    liquidity management for holdings of decentralized subsidiaries running windparks in and outside of the EU,
    a platform for smart financial contracts on liquidity exchange among companies in supply chains, 
    an IoT data based simulation of decentralized energy exchange systems of mobile decentralized energy consumption and production, 
    a dialy production decision of a diary holding of 7000 farmers producing 1600 products from 20 Mio liters of milk a day in 9 factories,
    delivering intermediate and final products to 300 logistic hubs and super markets,
    and a decentralized ERP system for digital farmers' outlets or shops with regional logistic hubs for dozens of shops that we equipped with standard hardware.
    The research started in the PhD of the first author on money theory~\cite{Menendez1989} and the follow up in \cite{menendez_theorie_2010}.
    The second author's PhD was on central bank policies in econometric models of exchange rate regimes, see~\cite{Winschel2005}.
    
    In Appendix~(\ref{app:category_theory}) we give the categorical definitions used in this paper.
    The order of sections in this paper corresponds roughly to the sequence of macroeconomic bookings in the recursive program.
    It is a usual presentation of the economic theory and rationals in the macroeconomic economy.
    However, we extend this sequential order by parallel combinators.
    In programming language design theory the tensor product is used for that, in function spaces of monoidal cartesian closed categories.
    We use that to represent the economics of the eight events and their macroeconomic bookings.
    A mathematician might feel more comfortable to read the paper in the reversed order starting from Section~\ref{sec:categoricalstateevolution}
    with the definition of $\mathcal{C}_{Economy}$ the category of the economy.
    There the whole macro thing, the macroeconomy, is defined.
    The endo functor on this category is the evolution in time as a natural transformation.
    The micro foundation of macroeconomics follows the order of the sections 
    and the macroeconomic foundation of microeconomics starts top down from the category of the economy and its endofunctor in Section~\ref{sec:categoricalstateevolution}.
    The circle from micro to macro and from macro to micro forms a {\it hermeneutical} circle, where the parts constitute the whole and can only be understood after understanding the whole.
    Similarly, in architecture it is said that only the city, composed of places, gives to places their functions.
    
    The categorical, process oriented model of consciousness as Memory Evolutive Systems (MES) by~\cite{EV2007} understands life to
    emerge from synchronised, i.e. consistent, time scales that are increasing bottom up to the top level control.
    Cells, tissues and organs self-organise into the identities of cognitive living beings,
    formed by colimits of concept formation in the brains' neural networks.
    Death occurs as a desynchronisation of the hierarchical levels.
    Similarly, we can understand economic institutions as synchronisation devices over hierarchical levels.
    Control is to be synchronised as well over time, levels and subobjects or parts of the economic system management.
    The coregulators in MES and polycentric management are both about decentralized control and our open games in~\cite{GHWZ2018} can capture 
    decentralised control by compositional economic modelling within the functorial and by that compositinal national accounting system of this paper.
    Nations may be taken to fail and emerge from de/synchronisation of micro- and macroeconomic levels.
    We state the basic consistency conditions towards such types of theories and institutions.

% ------------------------------------------------------------------------------------------------------------------------------------------------------------
\section{Macroeconomic Accounting}

\begin{figure}[h!]
    \centering
    \includegraphics[width=1.0\textwidth]{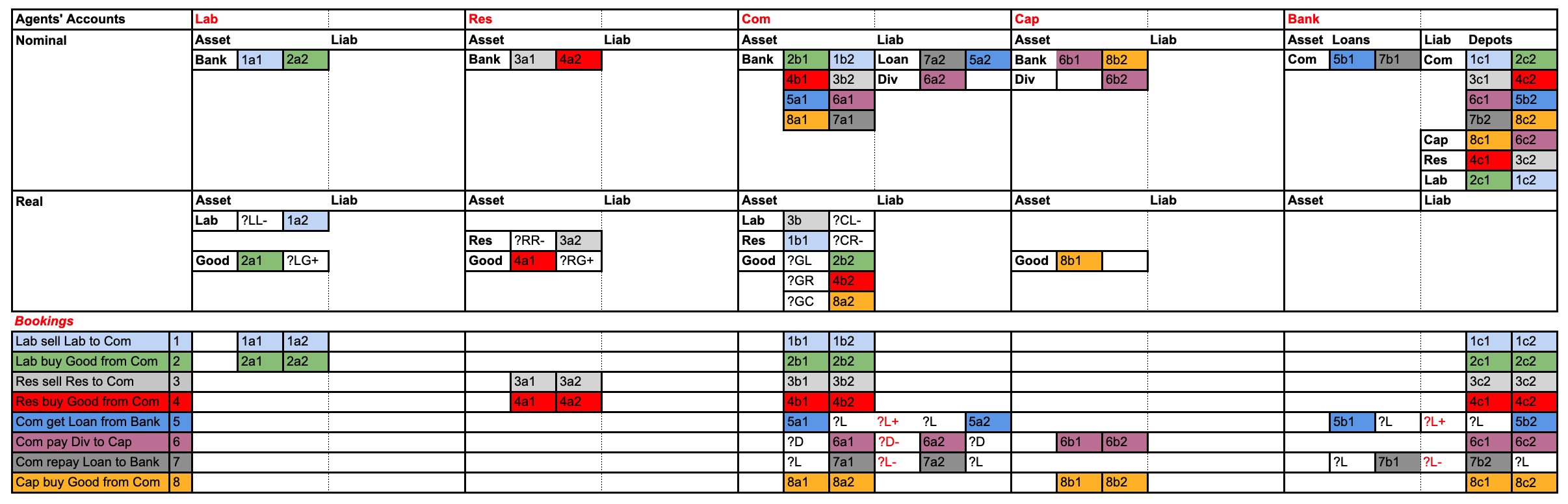}
    \caption{8 Macro Accounting Bookings of 5 Agents in 20 Accounts}
    \label{a}    
\end{figure}

\begin{definition}[Macroeconomic Accounting]
    In the Monetary Macroeconomic Accounting Theory (MoMaT) of~\cite{MW2025})
    money does not primarily function as a medium of exchange between goods,
    but rather as a medium of loan processing in solving the fundamental intertemporal problem of capitalist production.
    This problem is that the producers must pay their suppliers before receiving money from their customers.
    This payment carries the risk of not being paid for unsold products.
    Therefore, for producing, we need to have the institution of a monetary unit for measuring debts, loans and liabilities 
    and a thing called money representing some amounts of the monetary units as a means to repay the loans by transfering the property right on the money thing, 
    see~\cite{MW2025} for the legal principles underlying this view on money.
    We do not consider the problem to circumvent the need to exchange goods in a double coincidence of wants, as the fundamental functionality that money solves.
    The exchange of goods via money encompasses only a sales contract with goods and money transferred intratemporally at the same time settling debts in the good and the money.
    So, even a simple sales contract say a bike for money, creates debts, the seller owes a bike and the buyer some money.
    Risk is involved if a defunct central bank (king or finance minister) issues the money used in the exchange.
    In MoMaT the real factors labor and resources and the real goods can be exchanged via money or bank account transfers between the agents involved, 
    as well as the dividend payments to the dividend owner called capitalist.
    These are processes with money as a means of payment and exchange, but the actually important dynamics of debt creation and repayment are not addressed.
    This is what we focus on in MoMaT, where the intertemporal consistency needs define what a functioning central bank is essentially about.

    To prepare the technical tools to define what makes a functioning central bank and what consistency conditions are needed, where and at what levels, is the subject of this paper.
    In MoMaT, the argument goes that the fundamental intertemporal problem in capitalist economies characterised by a division of labor and risk can and is being solved every day.
    For this, we model bank loans and bank transfers for good production and labor and resources and goods exchange in a macroeconomic accounting system for five sectors also called agents.

    Producers need loans from banks because labour and resource owners want money in order to buy goods for consumption immediately in exchange for their real input factors to production.
    Labour and resource owers do not want to bear any risk of production and sales, if so they play the role of a bank or investor.
    The distinction between risk takers like the company, bank or investor versus labor and resource owners is important and is modelled hierarchically in this paper.
    The company and the investment projects are at the micro level and the bank at the macroeconomic level.
    The bank provides loans to the producer and creates by that consistency conditions and invariances at the macroeconomic level of accounting 
    over the micro level of double-entry accounting systems of the agents involved.
    A hierarchical risk absorption takes place, first at the company, then bank then central bank and government at the macroeconomic level, if the basic MoMaT is extended 
    for a third meso level then of banks and central banks and governments at the then macroeconomic level.
    Once producers have sold their products as goods to consumers, the labor, resource and dividend owners or capitalists, they can repay the loans to the bank.
    We formulate the macroeconomic consistency conditions for the agents involved.
    The banking sector can price risks at the aggregated level and compute from it the interest rate 
    as a risk premium of the insurance of investments, much like mandatory insurance for bank deposits.
    The closures we formulate are categorical constructions over the loan, closing investments by repayments and over endowments, closing production by consumption, 
    which is usually called {\it closed} economy.
    With a categorical closure over real and nominal units of account, we can compose and connect to formulations of circular economies and environmental economics 
    for example of~\cite{georgescu-roegen_entropy_1971},
    with physical invariances like geophysical world macroeconomic resources 
    or Emmy Noether's rescue of Einstein's relativity theory with energy (and value) preservation in an inflating universe by endogenous units of coordinates (and accounts).
    We categorize the nominal and real units, the loan repayments and production consumption closure as the universal investment pullback construction for loan contract verification 
    and the pushouts for the aggregated calculations involved while keeping the program of the macroeconomic accounting system consistent.

    Company owners can be called capitalists, profit owners or dividend owners.
    They consume based on dividend payments rather than like workers and resource owners from their sales.
    There are of course important dogma historical economic, philosophical and economic policy discussions around these loaded concepts.
    However, in this paper we are concerned about money theory providing the tools to be able to inform these debates on complex systems by categorical types.

    In Figure \ref{a} we see the agents, their T-accounts in their double-entry systems and the eight macroeconomic bookings that capture the dynamics within each period.
    The visual representation that has proven useful for a macroeconomic system of microeconomic accountings consists of representing the agents' systems 
    as T-balances with Asset $\Aa$ and Liability$_{\La}$ T-accounts of the micro systems on the left and right sides of the T-balance.
    This allows visual distinction of the formally important in$_{\Aal}$- and outflows$_{\Aar}$ on Asset and out$_{\Lal}$- and inflows$_{\Lar}$
    on Liability account types for a transfer into categorical morphisms.
    In the lower eight lines are the macroeconomic accounting bookings within each period of the five economic agents and their accounts.
    We see question marks for some bookings among the eight which will be answered by understanding the categorical types involved.

    The bookings 1, 2, 3, 4 of the Labor and Resource agents record the sales of their endowments for consumption in each period.
    Resources and Labor are created anew each period and consumption is the purpose of running a labour sharing economy.
    The real inputs and finished, produced goods are also booked for completeness purposes in real accounts, because we can, and should as economists.
    Accountants routinely keep such real accounts as Excel spreadsheets 
    without them being formally integrated in the accounting systems, where accounts are kept in monetary units,
    while the real goods and inputs are kept in hours, kg or other types and units of account.
    In German the data of these spreadsheets are called {\it Mengen-gerüst}, which is litarally translated as something like {\it amounts-constructions}.
    Of course, this is called inventory or warehouse management software and they are more or less integrated with the accounts kept in monetary units in the double-entry accounting systems.
    In this sense we unify such programs or systems in the two functors for real and nominal accounts which in a national accounting setup can categorically become a resource theory.
    The real inputs and finished goods are, and actually intermediate goods could be, also booked for completeness.
    Creation of goods or entities in general are categorically handled in initial and final objects, which we do not do for now in this paper.
    These have the universal property of a morphism pointing from the initial to every object, and a morphism pointing into the final object.
    These two universals allow to model resources creation and deletion or consumption or usage in production and the like, like contract creation and fullfillment.
    A resource theory in this sense seems to be missing for economics and can be developed and properly typed in terms of initial and final objects and their universal morphisms.
    Symmetric monoidal categories formalise multiplicative aspects of linear logic in tensor products of combining resources.
    The question marks in the dividend booking 6 and especially in the loan contract bookings 5,7 will be captured by the interesting mathematical structure of contra variant functors.
    This points into the direction to capture the abstract legal principle of {\it quid pro quo} or {\it tit for tat} in economics by covariant and contra variant functors
    which may include a change of domains, codomains or units of accounts.

    In the upper part of Figure \ref{a} we see the monetary accounts of the agents 
    Labor owner (Lab), Resource owner (Res), Company or producer (Com), Capitalist, dividend or company owner (Cap) and Bank (Bank).
    In the upper part are the nominal accounts arranged as assets and liabilities in a T-balance,
    which contains the T-accounts of the agents' double-entry bookkeeping systems.
    The accounts are arranged as asset and liability accounts in the T-balance on the left and right hand side, respectively. 
    Note that T-accounts have the inflows on the left being an asset or right being a liability account, and outflows accordingly on the right and left for assets and liability accounts.
    The in- and outflow ports of each T-account are labelled by xyz to count all the in- and outflow ports involved in the eight bookings,
    with x denoting the macroeconomic booking number,
    y counts the agents by a,b,c, for 2 as a,b and for 3 agents as a,b,c that form a macroeconomic booking.
    The z groups the T-accounts involved in the bookings by 1,2 for the in- and outflow ports.
    This weired index system counts all flows, it is implicit in the recursive implementation and categories free us elegantly 
    from the need to keep track of consistency by such low level, confusing and error-prone indices.
    The xyz labels in the upper, accounts part of Figure~\ref{a}, index the ports involved and, if projected down to the lower eight lines of the figure, 
    connect the eight macroeconomic bookings to the account ports of in- and outflows.
    The xyz labels in each T-account, vertically stacked in the upper part, are the in- and outflows of one whole period with all eight bookings.
    As a sum the in- and outlows are added and substracted to and from the accounts' states at the end of each period, 
    representing a yearly financial report as a balance sheet of asset and liability accounts.
    These flows change the state of the macroeconomic accounting systems' accounts from one to the next period.
    This is the whole dynamics of the system with the ports as the places where bookings flow values in and out of the double-entry microeconomic systems, the parts of the macroeconomic accounting system.

    With T-accounts, one can avoid the minus $-a$ as a unary negative operator and use it only as a binary $a-b$ subtraction operator in the equity definition of assets $-$ liabilities.
    To avoid unary $-$ and use binary $-$ was in 1452 the state of knowledge, when Paccioli published his history changing book on double entry bookkeeping.
    Negative numbers where only understood by mathematicians, let alone accountants some centuries later.
    The negative numbers in the outflow of assets and liabilities 
    are arithmetically represented in equations like $a-b-c=a+(-b)+(-c)=a-(b+c)$ of $a$ as asset and $b,c$ as liability accounts.
    Without a unary $-$ operator (defined by the middle expression), we can only use the left or right expressions, where the right expression is the equity definition, using a sum 
    over the positive numbers representing liabilities in $(b+c)$.
    T-accounts allow to represent liabilities and outflows (to assets and liabilities) as positive numbers and only use substraction by a binary minus in the top level calculation of equity,
    in Equity = Assets - Liabilities.
    Outflows to liabilities as negative negative assets would have probably cracked the brain of an accountant anno 1452, which is now simply $-(-b)=b$, 
    saying that less liabilities is more equity, a favourable event from an accountant's point of view.
    This technique has made T-accounts into a bookkeeping technology where only binary and not unary $-$ operations are neeeded to represent negative numbers of outflows and liabilities.
    The double-entry bookkeeping system is a revolutionalry protocol that records parallel, decentralized events of exchanges 
    while keeping the whole thing, centrally consistent as its essential characteristic, see~\cite{katis_partita_2008}.
    
    We clarify how this mechanics of consistency keeping through micro bookings where one and the same amount of value as a number is written on the debit and credit sides of two T-accounts,
    can be transferred to national accountings in quadruple entry systems, where one and the same number is written in two double-entry systems' accounts as a flow of value 
    between the debitor and creditor involved in a loan contract.
    We clarify this as the content of the macroeconomic foundation of microeconomics.
    Categorically typed, we extend the consistency maintenance of micro- to macroeconomic bookkeeping as a pullback verified calculation by pushouts.
    However, to begin with, the deterministic version needs to be as equipped with consistency tools as in this paper.
    A stochastic version, for example by Giry~\cite{giry1982categorical} monads, will enable to model more suitable risk estimates at the macroeconomic level of banks.
    It is missing in macroeconomics and to the best of our knowledge a theory of quadruple accounting is not formulated or software systems provided.
    
    A mathematical and computer scientific formulation with implementation details for double entry systems can be found in~\cite{rambaud_algebraic_2010}.
    We have programmed its algebra in Julia with some first categorifications using the categorical library \texttt{Catlab.jl}, see~\cite{patterson2022} for the tools at work.
    The algebra of double-entry bookkeeping is formulated in terms of state machines or automata on vectors of real numbers.
    The state vector with elements being the balances of each T-account are normalized by one account as the consistency condition to sum to zero, with equity taken as an account.
    The asset and liability accounts are represented as signed, i.e. negative and positive states and state transitions as vectors of the same length as the state vector,
    with in- and outflows as positive and negative numbers.
    So, bookings are state transitions as element wise additions of change vectors to consistent state vectors.
    This representation of the state and the state transitions uses unary $-$ in negative numbers,
    while in usual T-account representations numbers are never negative, which is the point of using T-accounts first of all.
    The double-entry algebra and code can be compositionaly, i.e. categorically, integrated into our categorical macroeconomic accounting system,
    using standard theoretical computer scientific formulations of automata, for example in a coalgebraic semantic of~\cite{rutten_universal_2000}.
 
    To recap: Lab has a nominal bank account and real accounts for labor and goods.
    Res has a nominal bank account and real accounts for resources and goods.
    Com has a nominal bank account, a liability account for managing loan repayments, a liability account for dividends, and three real accounts for goods, resources and labor.
    Cap has a nominal bank account, an asset account for dividends, and a real account for goods. 
    Bank has only nominal and no real accounts, one for company loans as an asset account and the bank deposits of agents as liability accounts.
    Only Com and Bank maintain liability accounts in the maintained invariances since we are interested in the closure of the dynamics of loans with investments and repayments 
    as two intertemporal payments of a debt relation over time.
    The real and nominal account product categories are for the tit for tat relations of real and nominal flows.
    The dividend flows can be properly extended to become a tit for tat flow of values in contra and covariant functors of some contract on property rights on the compny.
    However, this part of the contract between the capitalist and the company is not included in the invariances conditions of MoMaT.
    Only the Bank is involved in invariances since we are modelling the loan functionality that synchronises the desynchronised payments of 
    input factors of production and its outputs.
    There are 20 accounts for 5 agents to be managed in the macroeconomic accounting system of debts, investments, loans and repayments.
    Bookings 5, 7 involve only two agents, company and bank, in the investment and repayment bookings.
    The other bookings involve three agents.
    These bookings repeat in each period to form the overall dynamics of the macroeconomic economy which we are categorifying in this paper.

    The producer's liability in the form of dividends can be closed by some form of company ownership that can be formally specified as a contract of the Capitalist type of agent.
    The company pays dividends to the capitalist but this contract does not appear in MoMaT as an invariance of the money and credit system.
    We focus to model the money for loans functionality and study the closure of debt relations in the loan contracts of investments and repayments.
    The contract between capitalist and company can be specified as some form of debt as loan or contribution of the capitalist to the initial endowment of the company.
    But these additional contracts do not change the insights from categorifying the debt relationships 
    between company and bank in the loan contract and its repayment flows.
    Ownership contracts between capitalist and company would be necessary if we wanted to model and program their creation, transfer and deletion.
    But even then they would be additional functors and consistency conditions in natural transformations and another functionality of the macroeconomic accounting system. 
    We opted to not model this complication for now without further insights into the nature of the monetary loan system.
    
    The states in the formulation of MoMaT as recursion on vectors of real numbers as opposed to the categorical formulation 
    also represent the states of information within the system about the system at each point in time.
    Categories contain much more mathematical information than the recursive formulation of numbers in the simply typed vectors.
    The categorical structures of economic accounts (category), flows (functor), parameters (category) and their relationships (morphisms in categories, bookings) such as
    decisions (morphisms and functors), dynamics (functors), consistency conditions (natural transformations)
    and optimality (universality) conditions
    are mathematically much more informative, for researchers, agents in the theory and users of economic programs in the economy, 
    institutional designers or architects of macroeconomic systems.
    
    The categorical formulation made programming in our workflow more into an engineered experience working out the types and specification and the compiler for running the programs.
    The compiling process is actually the non-automated way of doing economics by an economist who is translating the economic argumentation lines into, say, a simply typed recursive program.
    If management is about doing economics then compositional economic modelling is useful for decentralized management.
    This involves a centrally planned versus a decentrally organised market economy or a hierarchically decomposed polycentric management of~\cite{Ostrom2005Book}.
    The pattern is similar to the structure involved in the so called {\it simultaneous engineering} at BMW.
    It denotes the simultaneous production of cars and the production of tools to produce cars. 
    This reflexivity made BMW very adaptive and contributed much to its success in the past.
    
    The workflow of doing economics by extending our MoMaT, for example by a central bank and a government or public sector, 
    is to define their accounts, bookings of basic events and the consistency conditions or invariances involved.
    The purpose of a DSL is to enable its users and modellers to formulate their theories using this high-levelness for economic concepts,
    which the compiler translates into verified, consistent low level implementations like iterations.
    But the categorical types do not only inform compilers what to compute,
    but also real agents in the economy, agents in theory, and economists - they create own functorial views on the economy in its digital twins.
    The categorical approach to mathematics is reminiscent of the Kantian view that {\it we can not grasp the thing itself} and all experience is mediated by our senses.
    The Yoneda lemma as a most categorical essence, states that we do not need to grasp the thing.
    Categorically it is enough to consistently glue different points of view on the thing in order to understand and define it mathematically.
    Combined with the {\it social way} of defining properties of things by its behaviour in its environment, also as a prerequisite to achieve modularity and compositionality,
    we get by category theory a mathematics that feels like a dream of a mathematical social scientist comes true.
    But, the learning curve for category theory is high, even for a mathematical economist, in need of a map from the basic economic concepts into the categorical ones.
    This map is the point of this paper, providing a categorical semantics for macroeconomics.
    This paper has a paedagogical purpose to exemplify category theory to economists and economics to computer scientists.
    We are well aware that both parts are too simple and too lengthy for both adressed groups of researchers, but we hope to provide a starting point for both.
    
    The categorical formulation ensures formal consistency, which in the recursive one is only implicitly encoded via the basic types of numbers and vectors and recursive functions.
    The low levelness of the indexing in Figure~\ref{a} by the names "1a1,1a2,1b1,1b2,...,2c2,...,8c2" of the in- and outflow ports of the macroeconomic accounting system 
    is similary the result of the poorly typed recursive formulation.
    The functions and equations are not typed either, as invariances, decisions, bookings or memory functionality.
    They are functions or variable asserts by the usual syntax in programming languages using $=$ to define calculations, state transitions or function definitions.
    The functions as well are not, but can be annotated with their types as types of domains and codomains combined into the type of a function, like in $f:\mathbb{R}\rightarrow\mathbb{N}$,
    by the type operator $\rightarrow$.
    The bottom line is that the recursive program does not use modern compiler and type defining (by \texttt{struct}) in Julia and type checking (with multiple dispatch) technology 
    like in the categorical programs.
    
    The problem of missing types and units of account has profoundly shocked the second author as a young economist in the first five minutes of lectures in economics, 
    especially being trained in unit discipline for consistency checks in physics at school.
    This is stunning since arithmetics in any exchange in economies involves nominal and real units, like in a price of 5.35~\$/kg.
    Economic theory should be explicit about units of accounts to verify their consistencies.
    This became more a problem then merely a stunning situation during the PhD on exchange rate regimes,
    where units are not only to be there but are to be created for a monetary union by political agents.
    The mathematical illiteracy for representing the type of the economical events and agents in an economic theory on money and exchange rate regimes was the motivation towards this paper.
    What is the type of the creation of the type (or unit of account) of a currency union?
    This obscure sounding question points to a profound field and effort of computer science,
    namely that type specifications are the proper way to think about DSLs, compilers, unit checking, correct programs, in short, the effort to give programs, describing a computation, 
    a semantics and a guaratee of correctness of the program, in the sense of deriving a correct implementation given a specification.
    
    The motivation for verified programs, for example at NASA, is that errors in programs may lead to exploding Space Shuttles and dead pilots.
    The motivation for the same technology in economics is that errors in inconsistent economic theories and their implementations in software 
    may lead to currency crises and devastating results, riots or even death for whole continents.
    This was the content of a discusssion with~\cite{Fischer2003} on the topic of AutoBayes, a Prolog program for generating data analysis programs from the specification of statistical models at NASA.
    What are the types of the units, agents and hierarchical levels and the types of agents creating and deleting types?
    What is the political economics of consistency and transparency with the delicate issue in macroeconomic policies 
    where those agents, entities, groups or decision units, who fail to build consistent systems as part of their job, might not want to be checked for consistency?
    Constitutions and institutions are meta rules for finding good rules or laws in the political process.
    In this sense a type theory and specification, say of the national accounting system, has to be able to represent the constitution of a central bank,
    which is translated into the programs of the central bank.
    A central bank is created together with the type of a monetary unit of account, the currency, while its operational processses create money, i.e. units of this type.
    However, private type creations are also part of our economic reality called a newly created crypto currency with tokens of this type being created operationally, day to day.
    Similarly, a cost center in internal or cost accounting within companies might be usefully conceived as a custom currency as well.
    However, type theory should be a highly welcomed mathematical tool available in a future economics.
\end{definition}
    
\begin{definition}[Accounts]\label{def:macroaccountingaccounts}
A macroeconomic accounting system is a system with double entry accounting systems for each agent.
The 5 agents in MoMaT are Labor (Lab) and Resource Owners (Res), Producer (Com), Profit Owner (Cap) and Bank (Bank).
Figure~\ref{a} shows the T-accounts of these agents and their inflow and outflow ports.
In the lower 8 lines of Figure~\ref{a} are the 8 macroeconomic bookings describing the dynamics within a period.
The account names are typeset in equation~(\ref{eq:macroaccountingaccounts}) as used in categorical formulas.
In equation~(\ref{eq:macroaccountingaccountsascode}) they are typeset as used in the programs, 
as variable names of a real number type $\mathbb{R}$ representing the state of the account in nominal or real units like $[EU], [kg], [h]$.
These numbers are typed by functors, however we have not programmed functionalities for creating custom units of accounts.
They can be integrated by defining functors with unit checks in natural transformations.
The 8 macroeconomic bookings and their executing events happen every period anew.
The sequential order in the run time of the recursive program of these 8 bookings
does not capture the economic logic, i.e. the sequential and parallel order of the economic events and of the bookings is not precisely enough defined in the (sequential) program.
We need parallel combinators which in computer science are taken to be the tensor operation in the categorical foundations of computing theory.
We will see the parallel combinators in diagrams with a categorical semantic which can be used to formalise the economics of parallel resource usages.

An important result of the categorification is to clarify the parallel and sequential execution order of the events and the bookings 
and to use the categorical tools made for that in monetary macroeconomics.
Since we are modelling the hierarchy of the system, we need sequential, parallel and hierarchical combinators.
The order of the 8 bookings can be said for now to be vaguely consistent with a macroeconomic consistent sequential, parallel and hierarchical order.
We see an example of the methodic difficulties of underspecification in booking 2.
In terms of DSL design, the simply typed recursive language is not expressive enough to represent the real world phenomenon of parallel, sequential and hierarchical executions, i.e. the core of economics.
Alternatives can not be properly evaluated, if not represented in theory, i.e. we expect welfare losses from using underspecified economic theories.
The Labor agent in booking 2 consumes even so the resource for production is not yet bought by the company from the resource agent.
What happens during production, can the company use up its stocks of resources in its warehouse, i.e. in the real T-accounts, and when and how are the combined constraints of the agents taken into account?
The production and consumption decisions interact with the amounts of labor and resources at the company in their warehouses and how we define them to be used and to vanish.
Answers to such questions can happen to become quite complicated, as we see them even in the simplest setup of five agents of MoMaT.
The only candidate, it seems to us, for a formal tool to scale to nation wide or macroeconomic systems, is category theory, and we need to start somewhere, 
here from the simplest economic setup of a monetary system in MoMaT.

We see that economics without parallel, sequential and hierarchical combinators, 
and operations for creating and deleting objects, leaves us with an underspecified theory, for programming purposes but also for consistency verifications. 
This is a problem especially since economics is about labour sharing in decentralised, i.e. parallel and sequentially more or less well ordered and to be improved orders, say by assembly lines.
Hierarchical and temporal combinators have to specify desired properties of systems involving as complex situations as the wants of or fairness deliberations for future generations.

The categorical typing results in string and commutative diagrams in the parallel arrangement of the dependencies between state and variables of interest 
as intermediate calculations of decisions up to the state updates as bookings in the T-accounts of all inflows and outflows.
We can think about the 2-dimensional commutative diagrams to represent two sides of an equation which as a consistency condition
characterises macroeconomic invariances or desired, specified properties in general.

The following table shows the types of the T-accounts for each agent together with their in- and outflows in our MoMaT.

{\scriptsize
\begin{equation}\label{eq:macroaccountingaccounts}
\begin{array}{|ll|ll|ll|ll|ll|ll|ll|lllll} \hline
Lab & Type & Res & Type & Com & Type & Com & Type & Cap & Type & Bank & Type & Bank & Type  \\ \hline
Lab^{Bank}_\Aa & [EU] & Res^{Bank}_\Aa & [EU]  & Com^{Bank}_\Aa & [EU] & Com^{Loan}_\La & [EU]    & Cap^{Bank}_\Aa & [EU] & Bank^{Loan}_\Aa & [EU]  & Bank^{Com}_\La  & [EU]    \\
            &      &            &         &&& Com^{Div}_\La  & [EU]                            & Cap^{Div}_\Aa  & [EU]                         &&& Bank^{Cap}_\La & [EU] \\
            &      &            &      &&&&&            &                                                                                      &&& Bank^{Res}_\La  & [EU]  \\
            &      &            &      &            &          &            &                                        &&                        &&& Bank^{Lab}_\La  & [EU]  \\ \hline
Lab^{Lab}_\Aa  & [h]  &            &      & Com^{Lab}_\Aa  & [h]                           &&&&    &&        && \\
            &      & Res^{Res}_\Aa  & [kg] & Com^{Res}_\Aa  & [kg] &&&&&&                                     &&  \\
Lab^{Good}_\Aa & [G]  & Res^{Good}_\Aa & [G]  & Com^{Good}_\Aa & [G]      &&& Cap^{Good}_\Aa & [G]  &&&&                    \\ \hline
\multicolumn{14}{|c|}{}\\ \hline
\multicolumn{4}{|c|}{\text{In- and Outflows for Types}}               & \multicolumn{10}{|c|}{} \\
\multicolumn{2}{|c|}{\text{Asset}}  & \multicolumn{2}{|c|}{\text{Liability}} & \multicolumn{10}{|c|}{} \\
\Aal & \Aar & \Lal & \Lar                                                    & \multicolumn{10}{|c|}{}  \\ \hline
\end{array}
\end{equation}
}
The in- and outflows of the accounts are left$_{\Aal}$ and right$_{\Aar}$ for Asset$_{\Aa}$ and right$_{\Lar}$ and left$_{\Lal}$ for Liability$_{\La}$ accounts.
The company Com produces from 2 real input factors Lab [h] and Res [kg] of the Labor and Resource Owners (Lab, Res)
a real good (Good [G]) with the help of the loan contract between company Com and Bank.
There are 11 nominal (in [EU] monetary units) and 9 real (in [kg, h, G] units) accounts, 14 are Asset and 6 are Liability accounts.    
\hfill $\blacksquare$
\end{definition}

\begin{definition}[Accounts as Code]\label{def:macroaccountingaccountsascode}
    Equation~(\ref{eq:macroaccountingaccountsascode}) shows the accounts written as code with initial and repeated endowments.
    The initial endowments at $t_0=0$ of the company are 110 for Lab and 20 for Res.
    The repeated endowments in each period, $\forall t>t_0$, are 100 for Lab and Res in real units.

{\tiny
\begin{equation}\label{eq:macroaccountingaccountsascode}
    \begin{array}{||l|l||l|l||l|l||l|l||l|l||} \hline
    Lab (A)    & (L) & Res (A) & (L) & Com (A) & (L) & Cap (A) & (L) & Bank (A) & (L) \\ \hline
    AccLabBank && AccResBank && AccComBank & AccComLoan & AccCapBank && AccBankLoan & AccBankCom \\
            &&            &&            & AccComDiv  & AccCapDiv  &&             & AccBankCap \\
            &&            &&            &            &            &&             & AccBankLab \\ \hline
    AccLabLab, 100, \forall t &&            && AccComLab, 110, t_0 &            & AccCapGood &&& \\
            && AccResRes, 100, \forall t  && AccComRes, 20, t_0 &            & AccResRes  &&& \\
    AccLabGood && AccResGood && AccComGood &            & AccCapGood &&& \\ \hline
    \end{array}
\end{equation}
}
\hfill $\blacksquare$
\end{definition}

\begin{definition}[Bookings as Flows]\label{def:macrobookingsflows}
The 8 macroeconomic bookings of the 5 agents in the macroeconomic accounting system define the interactions as channels on which the payments flow in between the 20 T-accounts.
Since we not only manage nominal but also real T-accounts these involve not only payment flows but also flows of real goods, resources and labour,
where the real flows are measured, evaluated and kept consistent with the flows in nominal units.
The 6 bookings 1,2,3,4,6,8 encompass 3 agents while bookings 5,7 only 2 agents.
A quadruple accounting system always books in at least two microeconomic or double-entry accounting systems of the agents.
The real T-accounts are involved in bookings 1,2,3,4,8 but not in bookings 5,7, which are nominal bookings of the loan contract.
Booking 6 is between nominal T-accounts as the flow of dividends.
The 100 new labor and resources every period and the 110 and 20 initial endowments of the company can be, but are not yet modelled elegantly 
as parts of the universal initial and final objects of some category of resources.
We see the problems involved popping up in the simulated data of increasing and zero stated real T-accounts.
This is to be combined into properly booked dynamics of the endowments.
The stability of the system is in need to be checked over different processes and endowments to evaluate how the monetary system copes with them and their shocks.
The 110 and 20 initial real endowments of the company and the 100 renewabled ones at Lab and Res are just one set of parameters 
where stability is demonstrated to be reached but a careful sensitivity analysis like in~\cite{HMSW2019} is needed as well.
Categorically, we can check for the analytic covergences in terms of the atomic analytically factorized and decomposed functions of transformations.
By that nonlinear functions in economic theory formulations can be taken to modern methods of advanced numerics like cohomologies, homotopy or sheaf theoretical methods.

The price functor transforms the units between real and nominal account types.
We handle functorially the types and units of both, state and flow quantities.
Flows translate between units of type $[t]$ and $[t+1]$ of states and verfiy the consistency with the units of flows being $[[t+1]-[t]]$.
Beware that the last expression is a type expression, composed from types of two account states in two subsequent periods by the $-$ as a type operator.
The consistency between in- and outflows and between real and nominal units is represented by functors with natural transformations verified by the pullback constructions.
We distinguish between the channels of the bookings on the one side which we create functorially and the bookings themselves on the other side, 
which transfer information and value via the channels as morphisms in the category of accounts.

This is an important distinction and it is to be separated from the econometric way to understand {\it structurality}.
In {\it structural econometrics} we mean by {\it structural} that decisions are derived as maximisations as opposed to the way macroeconomics has been conducted since Keynes,
by postulating causalities between macroeconomic aggregates like inflation or interest rates without stating how they come about from individual decisions.
This is the program of the microeconomic foundation of macroeconomics.
However, these kind of structural theories are still missing the macroeconomic foundation namely stating whether micro decisions are consistent with the macroeconomic aggregates
which is what we do in this paper in its very basic form by stating how a macroeconomic aggregate is to be represented mathematically, which we answer by colimits.
The decisions in MoMaT are not structural in the decision theoretic sense but we type them by universals which can represent maximizations 
or even more general selection functions, see~\cite{GHWZ2018}.
{\it Structural} in the sense of {\it institutional} are categorically definable as functorial dynamics 
which can represent the change of domains and codomains of transformations. 
We can take home that institutional economics is functorial 
and that universals are suited to represent an {\it institutional structurality} but also the {\it structurality of explicit decision making} 
derived from maximising utilities or preferences of agents subject to constraints.

An interesting feature is that bookings may occur logically simultaneously, i.e. in parallel.
Even more interesting is that they do not need to occur simultaneously in real time.
This even more points to the need for parallel and sequential combinators to check consistency conditions, 
including plain vanila logical modalities of necessity and possibility of truth or should and could as legal deontic modalities.
These technologies are available as coalgebraic constructions.
The simplest algebra is probably formed by addition on natural numbers of type $\mu:\mathbb{N}\times \mathbb{N}\rightarrow \mathbb{N}$, defined by $\mu(n,m)=n + m$.
The simplest coalgebra is probably the observation of the first element of an infinite list of natural numbers of type $\Delta : N^\omega \rightarrow N \otimes N^\omega$ 
with $\Delta(n)=n_0 \otimes(n_1, n_2, n_3, ...)$.
It splits an infinite list of natural numbers $n=(n_0, n_1, n_2, n_3, ...)$ also called stream into a first (or finite first part) of element $n_0$ and the infinite rest of the stream $(n_1, n_2, n_3, ...)$.
We see that coalgebras $X\rightarrow \mathcal{F}(X)$ are dual to algebras $\mathcal{F}(X)\rightarrow X$, arrows reversed, see~\cite{rutten_universal_2000}.
Coalgebras are the mathematics of infinite streams and trees and formalise observable data of unobservable structures 
which in economics are for example underlying the well known Kalman filter, see~\cite{AW2015} for a game theoretical economic application of infinite trees.

In the Lucas Critique the observing econometrician agent and the observed agents, within the theory of the econometrician, and some logically needed statement of the true state of the system 
might be differentiated by the amount of information available to the agents about the system.
These differences are conceptualised by the notion of {\it model communism}: do the observed and the observer share the same information and model (i.e. theory)?
In quantum physics the observer changes the system while in the economics of the Lucas Critique the observed system also changes the observer, the laws of the dynamics are not given 
by the universe (or if prefered by God) as in clasical mechanical or quantum physics.
In physics these kind of discussions are headed under the notion of multiple Lapacian daemons. 

As a starting point of these deliberations we can use the categorification of Tarski's theory, model and homomorphisms 
by Lavwere's conceptualisation thereof by category, functor and natural transformations.
Categories representing system states and theories thereof and functors being views on the system with natural transformations to encode the relation of the observer and the observed.
As programs and formulas, the algorithms used by the observer and the observed might be the same, implementing for example some likelihood functions for classical or Bayesian estimators.
However, even this set up might not be general enough since the relations of the observed and the observer do change dynamically as the core of their interaction, 
the theory does change the system and the system the theories.
Hence, the suitable mathematical generalisation of natural transformations to dynamically and endogenously changing relations and consistency conditions between both has to be found.
The notion of the true state of the economy which includes all agents and their perfect or imperfect information about the system is moreover a case for emergent properties in complex systems,
like formalised in the MES of~\cite{EV2007}.
The discussions in the Dagstuhl seminar~(\cite{Dagstuhl2015}) on the coalgebraic semantics of reflexive (Lucas Critique resolving) economics can be now framed within an extended MoMaT.
The operations of the observer and the observed might be algorithmically the same, namely the available AI and econometric tech for both, the observer and the observed,
but economically they are different, being possibly based on different data and priors processed.
We might think about the provision of the observer's or econometrician's information as an institution to coordinate the informations of the observed agents in the economy.
How all these concepts unite and how to categorify them, is an interesting topic for future research.

These theoretical problems do start with the question of what kind of processes are involved in macroeconomies and the simplest thereof are what we model in MoMaT as macroeconomic bookings.
In order to categorify them, we first take a look on the dynamics described by the 8 macroeconomic bookings involved in the following table.

{\tiny
\begin{equation}\label{eq:bookingsall}
\begin{array}{|llllllllll|} \hline
    1.\;\text{Lab sells Lab to Com}    &1a: Lab^{Lab}_{\Aar}  & \xrightarrow{h:EU}    & Lab^{Bank}_{\Aal} &1b: Com^{Bank}_{\Aar}  & \xrightarrow{EU:h}    & Com^{Lab}_{\Aal}  &1c: Bank^{Com}_{\Lal} & \xrightarrow{EU} & Bank^{Lab}_{\Lar}  \\ \hline \\ \hline
    2.\;\text{Lab buys Good from Com}  &2a: Lab^{Good}_{\Aar} & \xleftarrow{G:EU}     & Lab^{Bank}_{\Aar} &2b: Com^{Bank}_{\Aal}  & \xleftarrow{EU:G}     & Com^{Good}_{\Aar} &2c: Bank^{Com}_{\Lar} & \xleftarrow{EU}  & Bank^{Lab}_{\Lal}  \\ \hline \\ \hline
    3.\;\text{Res sells Res to Com}    &3a: Res^{Res}_{\Aar}  & \xrightarrow{kg:EU}   & Res^{Bank}_{\Aal} &3b: Com^{Bank}_{\Aar}  & \xrightarrow{EU:kg}   & Com^{Res}_{\Aal}  &3c: Bank^{Com}_{\Lal} & \xrightarrow{EU} & Bank^{Res}_{\Lar}  \\ \hline \\ \hline
    4.\;\text{Res buys Good from Com}  &4a: Res^{Good}_{\Aar} & \xleftarrow{G:EU}     & Res^{Bank}_{\Aar} &4b: Com^{Bank}_{\Aal}  & \xleftarrow{EU:G}     & Com^{Good}_{\Aar} &4c: Bank^{Com}_{\Lar} & \xleftarrow{EU}  & Bank^{Res}_{\Lal}  \\ \hline \\ \hline
    5.\;\text{Com gets Loan from Bank} &                       &                       &                    &5a: Com^{Bank}_{\Aal}  & \red{EU:Loan}         & Com^{Loan}_{\Lar} &5b: Bank^{Com}_{\Lar} & \red{EU}         & Bank^{Loan}_{\Aal} \\ \hline \\ \hline
    6.\;\text{Com pays Div to Cap}     &6b: Cap^{Div}_{\Aar}  & \xrightarrow{Div:EU}  & Cap^{Bank}_{\Aal} &6a: Com^{Bank}_{\Aar}  & \red{EU:Div}          & Com^{Div}_{\Lal}  &6c: Bank^{Com}_{\Lal} & \xrightarrow{EU} & Bank^{Cap}_{\Lar}  \\ \hline \\ \hline
    7.\;\text{Com repays Loan to Bank} &                       &                       &                    &7a: Com^{Bank}_{\Aar}  & \red{EU:Loan}         & Com^{Loan}_{\Lal} &7b: Bank^{Com}_{\Lal} & \red{EU}         & Bank^{Loan}_{\Aar} \\ \hline \\ \hline
    8.\;\text{Cap buys Good from Com}  &8b: Cap^{Good}_{\Aar} & \xleftarrow{G:EU}     & Cap^{Bank}_{\Aar} &8a: Com^{Bank}_{\Aal}  & \xleftarrow{EU:G}     & Com^{Good}_{\Aal} &8c: Bank^{Com}_{\Lar} & \xleftarrow{EU}  & Bank^{Cap}_{\Lar}  \\ \hline 
\end{array}
\end{equation}
}
The arrow heads indicate inflows while the arrow origins denote outflows in between the T-accounts.
However, this interpretation as flowing from out- to inflows in T-accounts does not always hold.
Some arrows in bookings 5,6,7 are missing and their would-be-captions are shown in red.
In micro booking 5a both T-accounts involved increase as well as both in micro booking 5b.
This ultimately, speaking in high level concepts, reflects the fact that the bank is securitising loans by its own assets, a bancrupt company not repaying the loan implies a loss of the bank.
We do see that in the macroeconomic invariance of equation~(\ref{eq:inv_macro}) formally stated as a zero condition to be fullfilled by a consistent system, i.e. an non-bancrupt bank.
This happens as well in creating loans for banks by the central banks as a prerequisite of a functional and operating bank, 
which clarifies that banks to NOT create money since then, they could never go bancrupt and the investment processes of banks (selecting suitable companies) would not improve 
due to the natural selection of banks implemented in markets for loans.
The universality of the pullback, as to be shown later, representing the {\it learning} of the banking sector, would collapse and the economy would not be adaptive but inherently unstable, 
just as centrally planned economies are.
Their macroeconomy is understood to be unable to properly process available microeconomic informations, like some well-known inabilities of a company or enterpreneur at the microeconomic level,
not transported to the macroeconomic level, where control decisions are located.

The flows and their directions involved in bookings 5,7 operate on a different level compared to the other bookings.
There it involves two temporally opposing micro bookings - the investment inflow to the company and repayment outflow from the company back to the bank.
In micro booking 7a both T-accounts decrease as in 7b when the loan is repaid and the risk of the bank vanishes.
Booking 6a shows dividend debt settlement from company to the capitalist via bank transfers.
Here, unlike in booking 5,7 of investment and repayment, there is no flow back.
It is just the profit which is actually typed by the capitalist agent or role, which we could equally leave in an T-account of the company.
But then, we would need to model the company to be a consumer as well, since the point of profits is to consume.
It is a modelling choice without any more insights into the nature of monetary systems.
The unclear structure of this core functionality of monetary systems are to be resolved once we type the entities involved.
To anticipate, we will find that the functors must really be on another level, i.e. being typed as so called covariant or contra variant functors, pointing in opposite directions,
for a consistent whole of the type of flows of values in real and monetary units, in between the involved agents and their T-accounts.
Especially confusing is that there are real flows which interactions with nominal flows which will admittingly not be fully clarifiable without initial and final objects and some resource theory 
which we will not approach in this paper.
However, we do need anyway to first type the product categories of real and nominal account categories with co- and contra variant functors operating on them, 
done in Section~\ref{sec:categoricaltypes} for properly typing the 8 macroeconomic bookings.
\hfill $\blacksquare$
\end{definition}
            
\begin{definition}[Invariances]\label{def:invariances}
The invariances of the macroeconomic accounting system encompass
the 4 bank accounts of the agents Labor, Resources, Company, Capitalist,
the loan contract of the Company with the Bank and 
one macroeconomic invariance as solvency requirement for the Bank.
The bank invariances $I_{Lab}^{B}$, $I_{Res}^{B}$, $I_{Cap}^{B}$, $I_{Com}^{B}$ 
say that the bank accounts of the agents Lab, Res, Cap, Com are managed by the Bank which must have the same state or balance in the double-entry micro accounting systems of the bank and the agents.
The invariance $I_{Com}^{L}$ says that the loan contract of the company with the bank must be the same in the Bank's and in Company's Loan accounts.
The macroeconomic invariance $I_{Mac}$ says that the bank must always be balanced, because it is insolvent or bankrupt otherwise, which is the core goal of managing liquidity in banking.
In the real world the banks all together balance each day at the interbanking market organised and regulated by the central bank.
This market maintains a macroeconomic accounting systems just as markets and banks do in general.
In MoMaT the macroeconomic invariance $I_{Mac}$ represents this market's or the Bank's functionality.
An extension by including a central bank would bring about cash and money-in-circulation T-accounts by money creation at the central bank,
with appropariate functors and natural transformations representing consistency conditions on the appropriate T-accounts.
A nominal exchange rate regime with central banks has more of this kind of accounts, functors and consistency conditions.
The recursive program calculates the invariances as calculations that must always produce a zero, being by that as a predicate or test that the T-accounts 
are equal invariantely for all periods, bookings and flows.
The categorical formulation uses pullbacks to ensure consistency between accounts and pushouts to synchronize flows between accounts as computations.
\begin{align}
I_{Lab}^{B} &= Lab^{Bank}_{\Aa} - Bank^{Lab}_{\La}  & \text{Labor Bank Invariance} \label{eq:inv_lab} \\
I_{Res}^{B} &= Res^{Bank}_{\Aa} - Bank^{Res}_{\La}  & \text{Resource Bank Invariance} \label{eq:inv_res} \\
I_{Cap}^{B} &= Cap^{Bank}_{\Aa} - Bank^{Cap}_{\La}  & \text{Capitalist Bank Invariance} \label{eq:inv_cap} \\
I_{Com}^{B} &= Com^{Bank}_{\Aa} - Bank^{Com}_{\La}  & \text{Company Bank Invariance} \label{eq:inv_com} \\
I_{Com}^{L} &= Bank^{Loan}_{\Aa} - Com^{Loan}_{\La} & \text{Company Loan Invariance} \label{eq:inv_loan} \\
I_{Mac} &= I_{Lab}^{B} + I_{Res}^{B} + I_{Cap}^{B} + I_{Com}^{B} + I_{Com}^{L} & \text{Macroeconomic Accounting Invariance} \label{eq:inv_macro}
\end{align}

These differences test for identical T-account states in two double-entry bookkeeping systems.
The need for these consistency conditions arises from the booking of one number in two microeconomic accounting systems being (decentral) subparts of the macroeconomic accounting system.
A resource theory based on linear, resource aware logic can be formulated to check the invariances as a proof (like in a logical program in Prolog).
The formulation as differences obscures that the invariances are equality constraints.
The liabilities we are interested in, exist between bank and company, as loans that resolve the desynchronization of payments for input factors and goods.
This is the core of MoMaT where money exist to solve this desynchronization.
The bank invariances arise as the tit for tat content of the contract of the agents having a bank deposit.
This is the core of the usual approach in money theory to resolve the double concidence of wants of barter exchange without money.
It is a rather simple functionality of the banking systems as opposed to the desynchronization problem and both need to be modelled at the same time,
which we see in the macroeconomic invariance $I_{Mac}$ of the bank.
The bank is balanced, if the sum of the bank balances of the agents' deposits and the loan account are balanced
into the macroeconomic invariance $I_{Mac}$ of the bank, the only inhabitant of the {\it macroeconomic} level in MoMaT.
This closes the relation that for each debtor there must be a creditor, the agents in any loan contract.
All these invariances guarantee a sound and faithful monetary macroeconomic accounting with loans and repayments over time and bank deposits for payments of real goods, labor and resources exchanges.
We have the contra and covariant loan and bank deposit contracts represented in equation~(\ref{eq:inv_loan}) 
as $I_{Com}^{L} = Bank^{Loan}_{\Aa} - Com^{Loan}_{\La}$ where the $-$ obscures the contra variance 
which could be highligted by $I_{Com}^{L} = Com^{Loan}_{\La} - Bank^{Loan}_{\Aa}$.

\hfill $\blacksquare$
\end{definition}

\begin{definition}[Parameters]\label{def:parameters}
We use the following 22 parameters $\in\mathbb{R}^{22}$
to specify the economic behaviours, decisions and contracts as the money theory of MoMaT (actually $\tau\in \mathbb{N}$).

    {\scriptsize
    \begin{equation}
    \begin{array}{lllll}
    \text{Parameter} & \text{Value}  & \text{Description} \\ \hline
    \textbf{Investment}\\
    \tau &= 10  & \text{investment length} \\
    \lambda &= 0.2  & \text{investment to labor vs resources} \\
    \sigma_a &= 20.0  & \text{learner sigmoid constant} \\
    \sigma_b &= 480.0  & \text{learner sigmoid nominator} \\
    \sigma_c &= 200.0  & \text{learner sigmoid \texttt{DemandSurplus} factor} \\
    \textbf{Consumption}\\
    \rho_r &= 0.8  & \text{resource's part to consume} \\
    \rho_l &= 0.95  & \text{labor's part to consume} \\
    \rho_c &= 0.6  & \text{capitalist's part to consume} \\
    \textbf{Profit}\\
    \mu &= 0.5  & \text{markUp} \\
    \omega &= 0.5  & \text{windfall profit} \\
    \textbf{Dividends}\\
    \delta_c &= 0.15  & \text{dividend for companies} \\
    \delta_b &= 0.4  & \text{dividend for banks} \\
    \\
    \end{array}
    \quad
    \begin{array}{lllll}
    \text{Parameter} & \text{Value}  & \text{Description} \\ \hline
    \textbf{Prices}\\
    p_r &= 25.0  & \text{price of resources} \\
    p_l &= 12.0  & \text{price of labor} \\
    p_0 &= 30.0  & \text{initial price of good} \\
    \textbf{Production}\\
    \gamma &= 0.75  & \text{elasticity of substitution} \\
    \alpha &= 0.42  & \text{factor} \\
    \textbf{Decay}\\
    \beta_l &= 0.95  & \text{labor's good} \\
    \beta_r &= 0.7  & \text{resource's good} \\
    \beta_c &= 0.6  & \text{rapitalist's good} \\
    \textbf{New Resources} \\
    \nu_l &= 100.0  & \text{new labor} \\
    \nu_r &= 100.0  & \text{new resources}\\
    \textbf{Memory} \\
    \mathcal{H}_{wages} &= \mathbb{R}^{\tau=10}_{\geq 0} & \text{wages to pay by contracts} \\
    \mathcal{H}_{repays} &= \mathbb{R}^{\tau=10}_{\geq 0} & \text{repays by loan contracts} \\
    \end{array}
    \end{equation}
    }
\hfill $\blacksquare$
\end{definition}

%\clearpage
\subsection{Economic Decisions}
\begin{definition}[Decisions]\label{def:decisions}
We manage a memory that stores and remembers future payment obligations from past contracts on labor and loans.
The parameter $\tau=10$ defines the length of the memory.
It contains $\tau$ real numbers for 10 wage and repayment payments as obligations from labor and loans contracts,
given by equation~(\ref{eq:dec:memory}) as the result of past investment decisions.
Labor and loan contracts have very different risks or termination conditions and we opt for the simplest case.
We take the 2$\times$10 payments as those to be paid from both contracts for 10 periods.
The memory remembers these payments that are to be done in each of the following 10 periods.\footnote{The memory is used to calculate the sum of the last agreed $\tau=10$ payments.
    We implement this as a FILO (First In Last Out) stack.
    The last entries are forgotten at the end of the period and the numbers calculated in the current period from the investment decision
    and how this affects future wage payments and loan repayments are written to memory.
    The implementation in Julia is a simple concatenation with vectors and elements therein.
    \texttt{1:end-1} takes the elements from the first to the second-to-last element of vector \texttt{hist},
    adds \texttt{newelem} before all these other elements via the operator \texttt{;} and from that forms the new vector 
    with the vector generator [] representing the new returned memory.
    The implementation of the memory function \texttt{p2H} is \texttt{p2H(hist::Vector{Float64}, newelem::Float64) = [newelem; hist[1:end-1]]}, in equation~(\ref{eq:dec:memory:update}).
    This is the memorize function of the memory.
    The recall reads the state \texttt{wagehist} and \texttt{repayhist} and adds them by \texttt{sum} as an implementation of equation~(\ref{eq:dec:memory}).
    In the recursive program in Appendix~\ref{app:code_appendix_recursive} in lines 66-68 the global \texttt{push\_to\_history} 
    and the local \texttt{p2H} in line 74, implement the memory function.
    } % footnote end
The memory starts initially with no obligations, represented by 10 zeros in the vectors which reduce to 9 in the first period until in the 10th period the memory is full.

\begin{equation}\label{eq:dec:memory}
\begin{array}{llll}
WagesPayment  &=  \Sigma_{i=1}^{\tau=10} \mathcal{H}_{wage}[i] \\
RepaysPayment &= \Sigma_{i=1}^{\tau=10} \mathcal{H}_{repay}[i] 
\end{array}
\end{equation}

The economic decision on {\bf consumption} is given by equation~(\ref{eq:dec:demand}).
Consumption of labour, resource agents and capitalist are assumed to be simple fractions of the state of their bank accounts.
These are then the amounts each consumer decides to spend on goods meassured in monetary units and together they sum to the total \textbf{Demand $D$} of the economy.

\begin{equation}\label{eq:dec:demand}
\begin{array}{lclll}
    C_{Lab} &=& \rho_l \;\; Lab^{Bank} \\
    C_{Res} &=& \rho_r \;\; Res^{Bank} \\
    C_{Cap} &=& \rho_c \;\; Cap^{Bank} \\
    Demand = D &=& C_{Lab} + C_{Res} + C_{Cap} 
\end{array}
\end{equation}

There are three parameters more that govern consumption, $\beta_l$, $\beta_r$, $\beta_c$.
They specify the decay of goods as a percentage of outflows as a kind of warehouse technology.
The economic meaning is that hording and not consuming costs.
The parameters could be explicitely derived as a universal by utility maximisation, for example,
rationalising the behavioural decisions as an economic tradeoff between consuming today versus hording and consuming tomorrow.

The producer wants to pay the wages, repay the loans and make a profit, calculated as a markup on the costs 
which we call the \textbf{DemandPlan $D^*$} of the company.
It is important to note that this is a decision about future events.
The company only plans to realise the profits discribed by a markup $\mu$ over the costs.

\begin{equation}\label{eq:dec:demandplan}
\begin{array}{lclll}
D^* &= WagesPayment + RepaysPayment + Profit^* \\
    &= (WagesPayment + RepaysPayment) * (1 + \mu)
\end{array}
\end{equation}

We could excplicitely take the future into account in a maximisation of goals or utilities as a behavoural model.
We could use structural econom(etr)ic theories like in~(\cite{WK2010}) 
or categorical and compositional game theory by~(\cite{GHWZ2018})
with a teleological causality from the (contra variant) future while observing the (covariant) past.
The decision theories can be all united in the categorical universal constructions as the top level of our constructions.
Universals unite and generalise optimality conditions and may be so called reduced form theories, like in equation~(\ref{eq:dec:demandplan}),
structural theories of intertemporal maximisations,
strategic open games, 
or in the most general way as an adjunction of functors.

The market matches \textbf{Demand $D$} of consumers and \textbf{DemandPlan $D^*$} of the producing company in \textbf{DemandSurplus}.
This happens by the price adjustment behaviour specified in equation~(\ref{eq:dec:goodprice}) also influencing the investment decisions in equation~(\ref{eq:dec:investment}).

\begin{equation}\label{eq:dec:demandsurplus}
DemandSurplus = D - D^*
\end{equation}

The \textbf{GoodProduction} follows a standard Cobb-Douglas production function with input factors of labor and resources in equation~(\ref{eq:dec:production}).
The real accounts $Com^{Lab}$ and $Com^{Res}$ represent the state of the warehouse of the company.
We denote the unit of the good and its accounts to be [G].
Type G is not formaly defined in our categorical programs but goods are typed and integrated into the real and nominal verification by functors and natural transformations.
The custom units are so to say empty such that the types of transformed units are different but look the same.

\begin{equation}\label{eq:dec:production}
GoodProduction = 1 + \alpha \;\; (Com^{Lab})^{\gamma} \;\; (Com^{Res})^{1 - \gamma}
\end{equation}

The good production and demand surplus together drive the \textbf{GoodPrice} in equation~(\ref{eq:dec:goodprice}),
as a fraction of \texttt{DemandPlan $D^*$} of \texttt{GoodProduction} starting with 30.0 and a additional price tag if the demand is larger than the planned demand of the company.
This expression denotes a mix of aspects and types with the initial state of the price being a parameter \texttt{InitialGoodPrice} 
used in the decision function which is a program of an \texttt{if\_then\_else} condition.
This defines a locally linear function with kinks by inequalities.

\begin{equation}\label{eq:dec:goodprice}
\begin{array}{lclll}
GoodPrice &=& \texttt{30.0 in } t_0 \\
            &+& \frac{DemandPlan}{GoodProduction} \\
            &+& \texttt{DemandSurplus * $\omega$ If DemandSurplus > 0.0} \\
\end{array}
\end{equation}

The \textbf{investment} decision is the most important one in the monetary economy.
It happens after all other decisions are made, the markets have equilibrated and the company's success is determined to inform the investment decision.
It is a sigmoid activation of the $DemandSurplus$ in equation~(\ref{eq:dec:investment}).
It represents the learning of the banking sector to invest in companies and to share the investment 
among resources (\ref{eq:dec:investmentres}) and labour (\ref{eq:dec:investmentlab}).
The repayment scheme of 10 equal repayments over 10 years is given in equation~(\ref{eq:dec:investment:scheme}).
The investments are saved in the memory as a FILO stack in equation~(\ref{eq:dec:memory:update}).
In general, if the repayment scheme were not specified to be 10 equal fractions of the investment, but any general function of time or other variables,
the memory would be an element of a functional space or category.

\begin{equation}\label{eq:dec:investment}
Investment = \sigma_a + \frac{\sigma_b}{1 + e^{-\frac{DemandSurplus}{\sigma_c}}}
\end{equation}

\begin{equation}\label{eq:dec:investmentres}
InvestmentRes = Investment * (1 - \lambda)
\end{equation}

\begin{equation}\label{eq:dec:investmentlab}
InvestmentLab = Investment * \lambda
\end{equation}

\begin{equation}\label{eq:dec:investment:scheme}
Repayment = Investment / \tau
\end{equation}

The investment decision \textbf{updates the memories} of wage payments and repayments, which also contain the other payment obligations of the past,
contracted among company, labor agent and bank.
Since every number remembered in the memory is there for $\tau=10$ periods, each number appears 10 times in the sums of repayments over the memory in equation~(\ref{eq:dec:memory}).
The FILO stack is a simple way to implement the remembering of the 10 constant payment schema of the labour and loan contracts.

\begin{equation}\label{eq:dec:memory:update}
\begin{array}{lcll}
\mathcal{H}_{wage} &=& \texttt{p2H}(\mathcal{H}_{wage}, InvestmentLab) \\
\mathcal{H}_{repay} &=& \texttt{p2H}(\mathcal{H}_{repay}, Repayment) 
\end{array}
\end{equation}

The \textbf{dividend} decision is programmed in the bookings as the state updates of the T-accounts.
The dividend depends on the actual realisation of the turnover and realised profit of the company calculated in $Diff$.
The calculation is again a non-linear, kinked function arising from the inequality checks for positive profits and bank account state of the company, 
as a mix of concerns and aspects.

\begin{equation}\label{eq:dec:dividend}
\begin{array}{lcll}
Diff &=& InAccComBank - outAccComBank \\
DividendDecision &=& Diff * \delta_C \texttt{ If } Diff > 0.0 \\
                 &+& \texttt{AccComBank * $\delta_B$ If AccComBank > 0.0} \\
                 &+& \texttt{0.0 If AccComBank < 0.0}
\end{array}
\end{equation}
Diff like some of the other formulations of the theory mixes aspects of behavioural assumptions, checks of conditionals in programs and mathematical expressions.
This is all to be properly sorted out by the types of the parts and the whole.
Also the missing of explicit sequential, parallel and hierarchical combinators of composable parts of the systems
makes economic interpretation complicated and programming difficult.
The same for popping up endowments and used up production inputs without initial and final objects in a category of resources.
The contractual content of the memory vectors is of concern for legal tech by modal logic where coalgebras for infinite time series and stochastics are needed anyway.
This then allows to include the dividend, loan and bank contracts as timed expressions as musts and cans which can be mapped into programs and formulas
for unambiguous and transparent legal obligations and options implemented by logical programs or theorem provers.
String diagrams and commutative diagrams will be helpful to disentangle these concerns and aspects of programs and mathematical types
by making the sequential, parallel and hierarchical structures explicit.
Modal logic for theory formulation in logic programs (like Prolog) based on coalgebraic semantics is for further research.

\hfill $\blacksquare$
\end{definition}    

%\clearpage
\subsection{Recursive Program}\label{sec:implementations}
We have implemented three programs in the Julia language, a simply typed recursive, a categorically typed recursive and a categorical program.
The listing are in Appendix~\ref{app:code_appendix}.
The recursive program in Appendix~\ref{app:code_appendix_recursive} takes the account states 
as the important state of vectors of real numbers beside memory, parameters and initial states and endowments.
The time series are calculated as an iteration or recursion using initial values, parameters and memory to update the states of the T-accounts.
The categorically typed recursive program in Appendix~\ref{app:code_appendix_categorically_typed} 
implements categorical structures as record types, called \texttt{struct} in Julia, with fields for the actual categorical data.
The categorical types as \texttt{struct} are wrapped around the numbers calculated in an iteration or recursion.
The categorically typed program implements a minimal categorical state of parameters, flows and accounts as categories and adds 
by generators the theory of five agents, parameters, initial states, accounts, bookings and consistency conditions.
These generators are the helper functions as they are called in the categorically typed program.
This implementation, even so it does not implement the actual evolution of time as natural transformations (even so the functions are defined),
makes the advantage of categorical modelling more explicit than the third, categorical program.
We see the abstract syntax of the compiler at work as the categorical \texttt{struct} being defined and used as a custom type by the multiple dispatch system of Julia.
We see the endogenous economist or econometrician at work in the helper functions, 
which allow an agent in the theory or us, as observers, to construct the theories or econometric models represented in categories, functors and natural transformations.
The agent in the theory could write this paper as an explanation of the results of the simulated data.
Finally, the categorical program is in Appendix~\ref{app:code_appendix_categorical}.
Time is an evolution represented by applying the endo functor to the category of the economy $\mathcal{C}_{Economy}$, the whole macroeconomic object.

In Figure~\ref{p} we see the time series plots of the simulation of the recursive program.
The recursive formulation is computationally effective because it delivers the correct results.
We see this in the invariance variables, which are (red) zero for all calculations in the periods.
The period grows as a time index, also the resources and labor and the goods produced.
This reminds us that the deletion and creation of resources in set theory is functorial in the domains and codomains of functions.
Since category theory is in need of dependent types (which is implemented in the categorical data types), we can use them for structural theories of networks.
The economics of double entry bookkeeping is about networks and keeping decentralised exchange in a consistent accounting and booking system.
We extend these consistency needs from decentralised needs in microeconomic accounting to hierarchical needs in macroeconomic accounting.

\begin{definition}[Recursive State]\label{def:state:rec}
The state of the recursive formulation is given by

\begin{equation}\label{eq:state:rec}
    S^{rec} = (\theta::\mathbb{R}^{22}, \mathcal{H}_{wage}::\mathbb{R}^{10}, \mathcal{H}_{repay}::\mathbb{R}^{10}, \texttt{Acc}::\mathbb{R}^{20})
\end{equation}
where $S^{rec}$ is a composite record type of vectors of parameters $\theta::\mathbb{R}^{22}$,
of wage history $\mathcal{H}_{wage}$,
of repayment history $\mathcal{H}_{repay}$,
and of 20 states of accounts $\texttt{Acc}$, each a real (positive or zero, non-negative) number.

In the recursive program bookings are state transitions of the changing state as balances of the accounts.
Parameters and initial values are constant parts of the state.
Beware, the categorical state will be the whole mathematical description of the theory and economy, not just the balances of the T-accounts.
Accordingly, all parts of the economy and theories can be changed, functorially, we can elegantly model institutional changes as part of the decisions of the agents involved.

The company has an initial endowment of resources and labor in its real accounts $Com^{Lab}_{\Aa}=110 [h]$, $Com^{Res}_{\Aa} = 20 [kg]$.
The memory of wages and payments $\mathcal{H}_{wage}$ and $\mathcal{H}_{repay}$ are used and updated at the beginning and end of each period.
The memory is part of the state of the system and the state of the system is part of the state of the recursive program.
The resources and labor endowments of labor and resource agents pop up each period in the real accounts as described by two parameters $\nu_l = 100 [h], \nu_r = 100 [kg]$.

The new real resources endowments every period are a main dynamic driving force.
They are transferred by their owners into the units of the bank account for buying goods.
The company buys labor and resources to produce the goods according to a production function.
The company pays dividends and repays loans while the profit, labor and resource owners consume.
This in turn gives revenues to the company and the bank accompanies all these events by granting loans, making bank transfers and receiving loan repayments.
This is what happens economically and what is translated into the recursive program.
The state is used to iterate on using and updating the memory, calculating banking, loan and macroeconomic invariances, 
the behavioral decisions and the actual bookings in the accounts as state transitions.
The recursive formulation uses what we call simply typed numbers and untyped equations for the invariances, memory usage and update, decisions and bookings.

Categorifying the recursive formulation means to type parameters, state and lift structural morphisms over operational bookings.
The equations in the Julia code uncover the mathematical need for a type discipline
which we take to the mathematical extreme abstraction of the categorical data types of categories, functors, natural transformations and calculations as pullbacks and pushouts.
In the categorical formulation we also use some helper functions to define, update and apply the categorical data structures and constructions
for the concrete agents, accounts and bookings.
In the categorical formulation the mathematical formulas are programs.

In the recursive formulation, the state is a vector of real numbers (actually, $\tau$ is of type natural number).
The recursive state $S^{rec}$ is defined in equation~(\ref{eq:state:rec}).
The categorical state $S^{cat}$ encompasses categories, functors, natural transformations in equation~(\ref{eq:state:cat}).
The recursive state $S^{rec}$ can be derived from the categorical state $S^{rec}=\mathcal{F}(S^{cat})$ by a functor.
The helper functions for managing the categorical state are mathematically formulated and allow an endogenisation of the economist as modeler
or the econometrician as an estimator of parameters of empirical theories on data of the model.
In Figure~(\ref{p}) we see the time series plots of the recursive formulation of the five agent theory of MoMaT.

\begin{figure}[t!]
    \centering
    \includegraphics[width=0.8\textwidth]{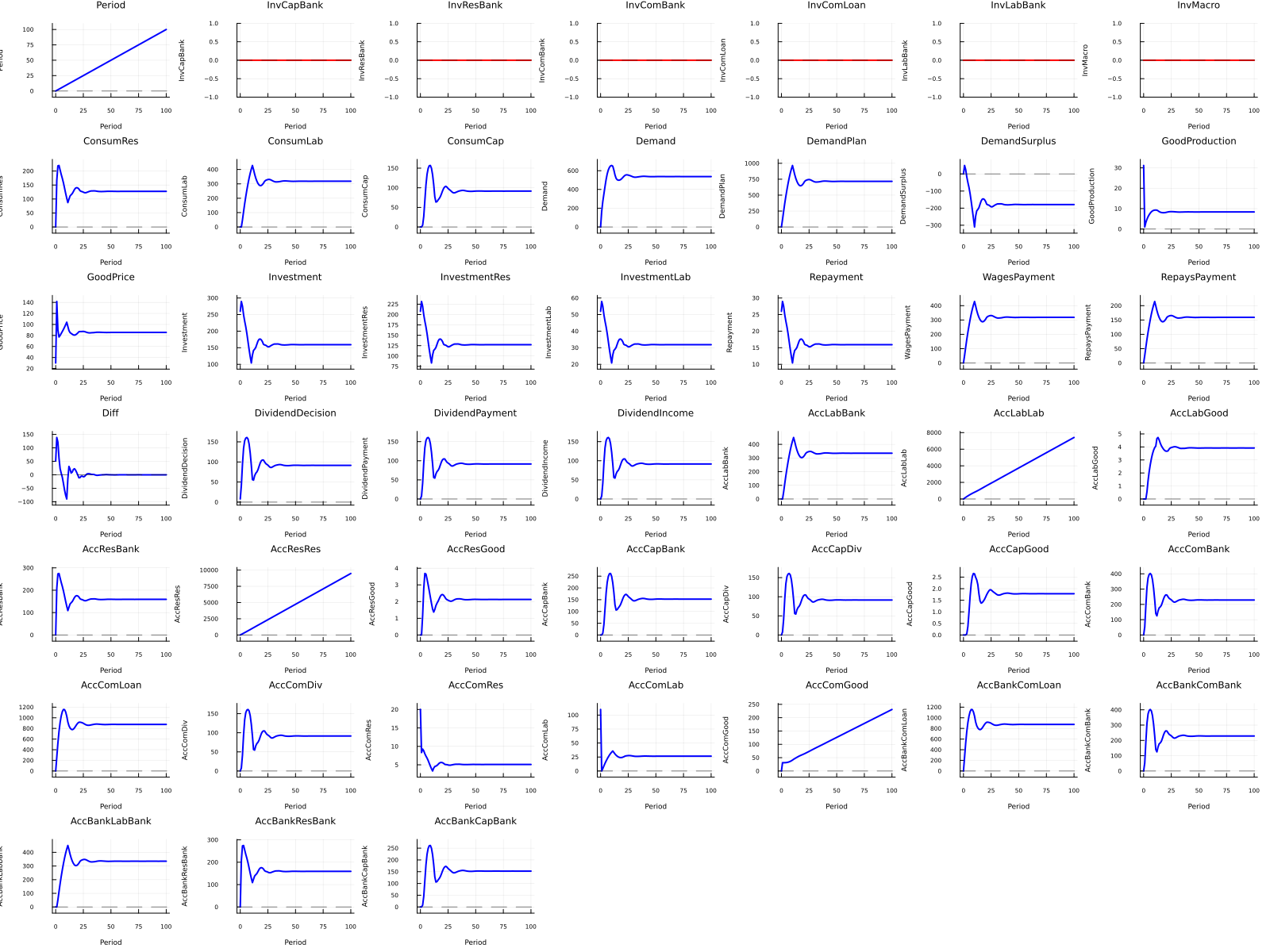}
    \caption{Time series plots of variables in the recursive simulation}
    \label{p}
\end{figure}

\hfill $\blacksquare$
\end{definition}

% --------------------------------------------------------------------------------------------------------------------------------------------------------------
\section{Categories for Macroeconomics}

We have seen a simply typed recursive formulation and program of the macroeconomic theory of MoMaT.
The formulation has difficulties in the mathematical and software specification.
We reformulate the key economic issues from the point of view of the categorication to do.

The categorical implementation forms a domain-specific language (DSL) for monetary macroeconomic accounting. 
This is best seen in the categorically typed recursive program.
The data structures are defined, wrapped arround the calculated numbers and the creation of the categorical structures are explicit in the helper functions.
However, the natural transformation functions are not used, i.e. the iteration is not a categorical application of natural transformations but simply an iteration.
The implementations represent the theory as a category with objects as accounts,
morphisms as monetary flows between the accounts,
pullbacks and pushouts for consistency verification and calculations,
functors for different views on the system,
an endofunctor as top level functor on the category of the economy as dynamics or evolution.

While the recursive implementation is a rather simple program of the dynamics, the categorical version provides a more abstract representation 
that enforces the mathematical structure of the theory. 
All three implementations produce the same simulated data.
The categorical formulation offers several advantages.
Mathematical rigor through categorical structure.
Clear separation of abstract theory and concrete computation.
Multiple interpretations via different functors.
Natural representation of temporal evolution and of type checks.
Built-in consistency checks through categorical constraints.
A compiler from a high-level language (DSL) to Julia code.
The DSL allows to define a concrete syntax for different uses in science, engineering, or media.
Proofability of code which is important for security checks and financial regulation and compliance.
A separation of aspects in operational morphisms with the category as the state and functorial morphisms as the structural dynamics, possibly including a change in the context or institution.
The theory is the code, making the working of economists explicit as typing and formulating the constraints and consistency conditions.
Category theory connects economics to other sciences via the categorical interface.

\textbf{The Fundamental Economic Problem:}
The MoMaT addresses the core challenge of capitalist production: temporal desynchronisation between production and consumption. 
Producers must pay their suppliers (workers and resource owners) before receiving payments from customers, creating a fundamental financing gap. 
This problem cannot be solved through simple exchange of paper money for solving intratemporally the double coincidence of wants.
If it is extended to intermediate goods and companies we get the same intertemporal problem of MoMaT.
To solve it, a sophisticated hierarchical credit system that enables intertemporal synchronisation is needed.
We use money in MoMaT also as a neutral medium of exchange but the crucial role of money is in credit creation and debt repayment as the essential monetary structure in MoMaT.
Therein we argue that credit creation and repayment through the banks' investments and repayments and not profit growth and money circulation 
are the essential features of the macroeconomic evolution that enables complex production in decentralised market economies organised by the division of labour.
We will see that this financial technology can be stable, consistently typed, composable, implementable and verifiable.

\textbf{The Five-Agent Economic Model:}
The closed economy with five distinct agent and sector types describes specific economic roles with suitable account structures.
The Labor agent are workers who provide labor services and consume goods. 
They seek immediate payments for their services and avoid production risk.
The Resource owners supply raw materials and resources who, like the workers, demand immediate payments and avoid production risk.
The Company produces and has a fundamental financing problem. 
It must pay suppliers before receiving customer payments, requiring loans from the banking system.
The Capitalist receives dividends and consumes only based on these profits rather than by owning productive resources. 
He represents ownership claims on productive assets.
The Bank is the financial intermediary that provides loans to bridge the temporal gap between production costs and sales revenues and also processes payments.
The Bank performs the system critical function of learning to aggregate risk and temporal coordination, the oppposite of producing bad banks,
booked or more like burried in black wholes of interbanking double-entry systems, central banks, public sector's and national accounting systems.
The Bank is the only agent that can create loans, the other agents are only able to use money and the loans.
The money is created implicitely in the MoMaT so far.
In a next paper the central bank is to be typed as the entity which initiates the only booking of macroeconomic accounting that involves only one single double-entry micro system. 
By this booking the central bank has the authority to {\it create} money in its cash asset T-account with the opposite booking in the circulating money liabilitites T-account, 
as discussed in~\cite{MW2025}.
It will be a universal booking of the central banks that is uniquely authorised to create money and threfore will be part of the learning pullback of the banking sector.
For consistency some more accounts are needed for the contracts of banks with central banks but the investment functionality will stay at the banking level, searching and selecting investments.
The creation of money is cental bank's authority, to declare papers with registered IDs to be the legally accepted means 
of debt repayments in loan contracts, of debts in exchange contracts of goods for money or in the payments of the liabilities of the company to the capitalist.

\textbf{The Eight Macroeconomic Bookings:}
The dynamics of the system are determined by eight fundamental events and macroeconomic bookings that represent all monetary and real flows in the system. 
All bookings are processed with bank intermediation for the payments of the dividends, production factors and goods.
The details of the legal principles of abstraction and separation to legally type the events are quite intricate in their relation to 
the issues of interest and pitfalls of the theory of money, as discussed in~\cite{MW2025}.
The legal principles may even involve intergenerational aspects like protecting the natural environment of future generations in §20a of the German constitution.
Institutional goals or agents are usually modelled in economics with a planning horizon without known end, i.e. by infinite horizon models, 
for which coalgebraic semantics have been developed in computer science.
The legal principles help us to disentangle the issues involved in contracts, like quid pro quo, or that opposing contracts of the 
original overall contract is settled by two mutual contracts on exchanging the promissed goods or money among the exchanging agents.
The details are not of interest here, but that macroeconomic bookings in different double-entry systems result from the fundamental activities as the eight bookings:
1. {Labor sells Labor to Company}: Workers provide labor services in exchange for wages.
2. {Labor buys Goods from Company}: Workers consume goods using their wages.
3. {Resource owners sell Resources to Company}: Resource suppliers provide inputs for production.
4. {Resource owners buy Goods from Company}: Resource suppliers consume goods from selling resources.
5. {Company gets Investment as a Loan from Bank}: The crucial credit creation that enables production and risk sharing.
6. {Company pays Dividends to Capitalist}: Profit distribution to shareholders.
7. {Company Repays Loan to Bank}: Credit contract deletion by repayments of the loans.
8. {Capitalist buys Goods from Company}: Dividend recipient consumes goods.
Bookings 5 and 7 (loan creation and repayment) are the purely monetary transactions that solve the fundamental temporal problem. 
These involve only nominal accounts and represent the core of the credit system.
Dividend payments is the other pure monetary transaction that does not involve repayments.

\textbf{The Puzzle of Marx:}
The categorical framework sheds light on Marx's puzzle of profit realisation:
How can profits be realized when workers don't have enough money to buy the products they produce?
The light emerges through {debt spirals} rather than money circulation and profit dynamics.
The question is actually why does it look like money circulates while the crucial dynamics is in the creation and repayment of debts.
1. {Credit Creation}: Banks create purchasing power by extending credit to companies.
2. {Production Financing}: Companies use credit to pay workers and resource owners before sales.
3. {Consumption Enabling}: Workers and resource owners can consume using their wages and revenues from selling the resources.
4. {Profit Realisation}: Sales revenue enables loan repayment and profit distribution.
5. {Temporal Coordination}: The credit system bridges the gap between production costs and sales revenue.
This process creates spirals of debt and repayment that enable complex production without requiring workers to have sufficient money. 
The banking system performs the crucial function of risk aggregation and temporal synchronisation.

\textbf{Categorical Implementation:}
The categorical ansatz provides several advantages over the recursive formulation.
{Type Safety:} Categorical structures ensure proper typing like nominal vs. real accounts, asset vs. liability positions, and flow vs. stock variables.
{Compositional Structure:} Categories, functors, and natural transformations make parallel, sequential and hierarchical relationships in economic decisions, institutions and accounting explicit.
{Universal Constructions:} Pullbacks verify loan contracts and investment decisions, while pushouts aggregate consistency conditions across multiple double-entry systems.
{Diagrammatic Reasoning:} String diagrams and commutative diagrams provide visual representations of complex economic relationships, 
making the parallel, sequential and hierarchical structures of economic processes explicit.
The categorical state encompasses layers of embeded categories in the top level category of the economy and its endofunctor of time evolution.
There are accounts, flows, parameters, two functors of price and flow transformations, and natural transformations for the functors.
They ensure consistency across temporal, hierarchical, parallel sequential dynamics, nominal, real or any scientific units.

\textbf{Hierarchical Risk Management:}
The model demonstrates hierarchical risk management through the banking system:
1. {Micro Level:} Individual agents maintain double-entry accounting consistency.
2. {Macro Level:} Banks aggregate individual risks and provide system-wide liquidity.
3. {Temporal Risk:} Credit creation enables production despite the timing mismatches of costs and revenues.
4. {Systemic Stability:} Proper categorical typing ensures overall system consistency.
This hierarchy can be extended to include central banks as "lenders of first liquidity" or government spending financed by taxation.

\textbf{Implementation:}
The economic theory has been implemented as a recursive simulation which has to be typed for a categorical simulation.
{22 Parameters}: Rules for agent behavior and system dynamics.
{8 Bookings}: State transitions between periods to become the evolution by a functorial composition over time.
{5 Invariance Conditions}: Macro accounting consistency conditions constitute the contracts of the financial sector.
{Investment Function}: Bank learning and verification process to become a universal construction to represent the generalized optimisation behaviour of the sector.
{Categorical Structures}: Mathematical foundation for programming and proofs of systemic properties.
We want to use category theory for typing a simulation, providing both theoretical insights and practical computational methods for monetary economics.
MoMaT demonstrates a hierarchical risk management system through the banking system at the macroeconomic level.
The mathematical formulation and the specification of the program are the same.
The program in the high level operations of the DSL in terms of categories of banking is a type system, 
specifying what is to be compiled as the macroeconomic accounting system.

%\clearpage
\subsection{Categories as Economic Structures}
\textbf{Visualising the Macro of Economics:}
This section provides the basic categorical infrastructure needed to categorify the macroeconomics of the eight bookings.
We model a macroeconomic or global structure and accordingly can use the geometric and visual tools of category theory first developed in algebraic geometry.
We think about the theory as a stack of 3D layers of the category of parameters $C_{pars}$, flows $C_{flow}$ and accounts $C_{accs}$.

\begin{tikzpicture}[x={(1cm,0cm)}, y={(0.5cm,0.5cm)}, z={(0cm,1cm)}]
\node (e) at (3.8,0,6.5) {};% Parameters layer (top)
\draw[fill=yellow!20,opacity=0.5] (0,4,4) -- (4,4,4) -- (6,2,4) -- (2,2,4) -- cycle;
\node at (3.8,3.0,4.2) {$\blacksquare C_{pars}$}; % Flows layer (middle) 
\draw[fill=blue!20,opacity=0.5] (0,4,2) -- (4,4,2) -- (6,2,2) -- (2,2,2) -- cycle;
\node at (3.8,3.0,2.2) {$\blacksquare C_{flow}$}; % Accounts layer (bottom)
\draw[fill=green!20,opacity=0.5] (0,4,0) -- (4,4,0) -- (6,2,0) -- (2,2,0) -- cycle;
\node at (3.8,3.0,0.2) {$\blacksquare C_{accs}$}; % Vertical connectors between layers
\draw[dashed] (0,4,0) -- (0,4,4);
\draw[dashed] (4,4,0) -- (4,4,4);
\draw[dashed] (6,2,0) -- (6,2,4);
\draw[dashed] (2,2,0) -- (2,2,4); % Functor arrows
\draw[->,thick] (3.4,3.0,4.2) -- (3.4,3.0,2.4) node[pos=0.7,left=0.05cm] {$\mathcal{F}_{p2f}$};
\draw[->,thick] (3.4,3.0,2.4) -- (3.4,3.0,0.4) node[pos=0.7,left=0.05cm] {$\mathcal{F}_{f2a}$};
\draw[->,thick] (1.8,3.0,4.0) -- (1.8,3.0,0.0) node[midway,right=0.12cm] {$\mathcal{F}_{p2a}$};
\end{tikzpicture}

Since we are not dealing with a $41\times 100$ matrix of real numbers of 41 variables representing the state of a macroeconomic economy over 100 periods,
but with a mathematically complete description, we need appropriate tools for dimension reduction and composition.
Functors can also used for 2D slices of the 3D layers of categories of geometric macroeconomic structures and to rotate layers around corners in triangles, for example.

\begin{tikzcd}[row sep=4em, column sep=4em, baseline=(current  bounding  box.center)]
C_{pars} \arrow[d, "\mathcal{F}_{p2f}"'] \arrow[dd, bend right=50, "\mathcal{F}_{p2a}"'] & \\
C_{flow} \arrow[d, "\mathcal{F}_{f2a}"] & \\
C_{accs} &
\end{tikzcd}
\begin{tikzcd}[row sep=4em, column sep=4em, baseline=(current  bounding  box.center)]
C_{pars} \arrow[rd, "\mathcal{F}_{p2a}"'] 
            \arrow[r, "\mathcal{F}_{p2f}"] 
& C_{flow} \arrow[d, "\mathcal{F}_{f2a}"] \\
& C_{accs}
\end{tikzcd}

Natural transformations between functors ensure consistency.
The visualisation shows that parameters influence flows 
which in turn affect account balances, with natural transformations ensuring consistency between different paths between the system and its parts.
The types, functors and universal constructions decompose and compose the system and its parts, zoom in and out as different views in the system, see \cite{willems_behavioral_2007}.
The visualisation provides a comprehensive view of the categorical state, i.e. categories, functors, natural transformations and memory state while executing the program.
The categorical ansatz is the IME's (Integrated Modelling Environment) of mathematics with a cockpit's view on a multi-layered visualisation of the multi-sector-agent system.
The purpose of an IME in economics is to observe and analyze foundational hierarchical structures and institutions that compose into governances of macroeconomic behaviour, 
to design institutions, i.e. meta-rules.
Category theory provides tools into the mathematical foundations of this kind of structural economic dynamics. 
With the visual tools of category theory organigrams of organisations can become the program of the digital twin.
The accounts can be visualized as categorical networks of graphs where vertices represent economic objects and
edges represent morphisms, functors or natural transformations and universal constructions as generalisation of economic maximisation operators of economic theory.
The simulation updates the categories, functors and natural transformations, rather then reals in vectors.
The category is the state and the formulas are the program.
We see during simulations how accounts balance, update and how categorical structures change, visualise, witness and calculate constructively the informations needed.

We will take accounts to be the objects in the fundamental categorical structure where the category is the state.
There are different ways to define and implement the mathematical structures of MoMaT.
In the categorically typed recursive program we take the state to be represented by three categories of parameters, flows and accounts and the memory of the contracts
with functors and natural transformations for keeping always all units and accounts consistent.
The category in Section~\ref{sec:categoricalstateevolution} is finally used for the categorical program which is to be discussed in the next sections.
For now, we start with a category of accounts.

\begin{definition}[Category as State]
    The economic category $\mathcal{C}_{Economy}$ consists of:
    \begin{itemize}
        \item Objects $A\in \text{Ob}(\mathcal{C}_{Economy})$ - economic accounts with current value, as amounts of monetary, real or other units of account
        \item Morphisms $f, id_A\in \text{Mor}(\mathcal{C}_{Economy})$ - economic transactions with metadata
        \item Composition: $f: A \rightarrow B, g: B \rightarrow C \Rightarrow g \circ f: A \rightarrow C$
        \item Identity morphisms: $\text{id}_A: A \rightarrow A$ for each object $A$
    \end{itemize}

    Some of the objects in \textbf{periods 0, 1, 2} that represent the account balances:
    \begin{align}
    \text{Ob}(\mathcal{C}_0) &= \{Lab^{Bank}: 0.0, Res^{Bank}: 0.0, Com^{Bank}: 0.0, \ldots\}\nonumber\\
    \text{Ob}(\mathcal{C}_1) &= \{Lab^{Bank}: 0.0, Res^{Bank}: 208.0, Com^{Bank}: 52.0, \ldots\}\nonumber\\
    \text{Ob}(\mathcal{C}_2) &= \{Lab^{Bank}: 52.0, Res^{Bank}: 273.19, Com^{Bank}: 190.50, \ldots\}\nonumber
    \end{align}        
    
    \begin{align}
    Lab^{Bank}: \mathbb{R}_{\geq 0} &\rightarrow \mathcal{C}_{\text{accounts}}\nonumber\\
    Res^{Bank}: \mathbb{R}_{\geq 0} &\rightarrow \mathcal{C}_{\text{accounts}}\nonumber\\
    Com^{Bank}: \mathbb{R}_{\geq 0} &\rightarrow \mathcal{C}_{\text{accounts}}\nonumber\\
    Cap^{Bank}: \mathbb{R}_{\geq 0} &\rightarrow \mathcal{C}_{\text{accounts}}\nonumber\\
    Bank^{Loan}: \mathbb{R}_{\geq 0} &\rightarrow \mathcal{C}_{\text{accounts}}\nonumber
    \end{align}
    \hfill$\blacksquare$
\end{definition}

\subsection{Functors as Value Transformations}

\begin{definition}[Functor as Value Transformation]
Functors as maps between categories are for transforming values between categories as representations or variables of interests.
In MoMaT we have the price functor as a functor between the real and nomimal categories of accounts.
The second functor is between the states and the flow category, also transforming units and points of view.
This ensures the stock flow consistency.
Both are accompanied by two natural transformations to keep the functors consistent among each other.
We take a look at the price functors.
The price functor transforms values expressed in nominal units to values expressed in real units.

\begin{align}
\text{Object mapping} &:& F(A) \in \text{Ob}(\mathcal{D}) && \forall A \in \text{Ob}(\mathcal{C})\nonumber\\
\text{Morphism mapping} &:& F(f: A \rightarrow B) &=& F(f): F(A) \rightarrow F(B)\nonumber\\
\text{Composition preservation} &:& F(g \circ f) &=& F(g) \circ F(f)\nonumber\\
\text{Identity preservation} &:& F(\text{id}_A) &=& \text{id}_{F(A)} \forall A \in \text{Ob}(\mathcal{C})\nonumber
\end{align}
\begin{align}
\mathcal{F}_{\text{price}}            &:\mathcal{C}_{\text{nominal}} \rightarrow  \mathcal{C}_{\text{real}}\nonumber\\
\mathcal{F}_{\text{price}}(GoodPrice) &: \mathbb{R}_{> 0} \rightarrow \mathbb{R}_{> 0}\nonumber\\
\mathcal{F}_{\text{price}}(LaborPrice) &: \mathbb{R}_{> 0} \rightarrow  \mathbb{R}_{> 0}\nonumber\\
\mathcal{F}_{\text{price}}(ResourcePrice) &: \mathbb{R}_{> 0} \rightarrow  \mathbb{R}_{> 0}\nonumber
\end{align}
\begin{align}
\mathcal{F}_{\text{price}}(GoodPrice_0) &= 30.0 \rightarrow 30.0\nonumber\\
\mathcal{F}_{\text{price}}(GoodPrice_1) &= 141.7 \rightarrow 141.7\nonumber\\
\mathcal{F}_{\text{price}}(GoodPrice_2) &= 89.94 \rightarrow 89.94\nonumber
\end{align}

Although in economics types are not used in theory as functors, like here where the reals are mapped to reals, the mathematical structure is nevertheless there, the functors.
We can define any currency or real or custom unit of account.
We have already noted the different units like $[EU]$ for monetary units or $[kg]$ for resources, $[h]$ for labor, $[G]$ for goods.
The definition of custom types is not programmed but could be integrated into the functorial set up.
However, important types are implicitly defined in the recursive code and are defined as the second functor for the flows between the accounts.
The account types have units of account and their time differences in the flows as in equation~(\ref{eq:bookingsall}) accordingly as well.
Their category $\mathcal{C}_{flow}$ has objects of type $flow::[[t+1]-[t]]$.
The minus $-$ is in between types is a type operator or functor.

\hfill$\blacksquare$
\end{definition}

%-------------------------------------------------------------------------------------------------------------------------------------------------------------------------
%\clearpage
\subsection{Natural Transformations as Evolution}
    Natural transformations model temporal evolution while preserving categorical structure.

\begin{definition}[Natural Transformation as Evolution]\label{def:nattransev}
    A natural transformation $\eta: F \Rightarrow G$ assigns to each object $A$ a morphism $\eta_A: F(A) \rightarrow G(A)$ such 
    that for every morphism $f: A \rightarrow B, \eta_B \circ F(f) = G(f) \circ \eta_A$
    The typed natural transformation has the following components.
    \begin{align}
        \eta_{\text{time}}&: F_t \Rightarrow F_{t+1}\nonumber\\
        \eta_{Lab^{Bank}}&: F_t(\mathbb{R}_{\geq 0}) \rightarrow F_{t+1}(\mathbb{R}_{\geq 0})\nonumber\\
        \eta_{Res^{Bank}}&: F_t(\mathbb{R}_{\geq 0}) \rightarrow F_{t+1}(\mathbb{R}_{\geq 0})\nonumber\\
        \eta_{Com^{Bank}}&: F_t(\mathbb{R}_{\geq 0}) \rightarrow F_{t+1}(\mathbb{R}_{\geq 0})\nonumber
    \end{align}
    The temporal evolution natural transformation $\eta_{\text{time}}$ preserves account relationships.
    \begin{align}
        \eta_{\text{time}}(Lab^{Bank})&: 0.0 \rightarrow 0.0 \rightarrow 52.0\nonumber\\
        \eta_{\text{time}}(Res^{Bank})&: 0.0 \rightarrow 208.0 \rightarrow 273.19\nonumber\\
        \eta_{\text{time}}(Com^{Bank})&: 0.0 \rightarrow 52.0 \rightarrow 190.50\nonumber
    \end{align}
\hfill$\blacksquare$
\end{definition}

\begin{figure}[htbp]
\centering
\tikzset{
    stringdiagram/.style={
        baseline=(current bounding box.center),
        %every node/.style={draw=none},
        every path/.style={thick}
    }
}
\begin{tikzpicture}[stringdiagram]
    \begin{scope}[shift={(0,0)}]
        % Left diagram: F_t composition
        \node at (1.5,3.5) {$\mathcal{C}_{\text{acc}}$};
        \node[draw, rectangle, minimum width=1.2cm, minimum height=0.6cm] (Ft) at (1.0,2.5) {$F_t$};
        \node at (1.5,1.5) {$\mathcal{C}_{\text{flow}}$};
        \node[draw, rectangle, minimum width=1.2cm, minimum height=0.6cm] (eta1) at (1.0,0.5) {$\eta_{\text{time}}$};
        \node at (1.5,-0.5) {$\mathcal{C}_{\text{flow}}$};
        \node at (9.0,1.5) {$=$};
        \node at (10.5,3.5) {$\mathcal{C}_{\text{acc}}$};
        \node[draw, rectangle, minimum width=1.2cm, minimum height=0.6cm] (Ft1) at (10.0,1.5) {$F_{t+1}$};
        \node at (10.5,-0.5) {$\mathcal{C}_{\text{flow}}$};
        \draw (1.0,-1) -- (eta1);
        \draw (eta1) -- (1.0,1.5);
        \draw (1.0,1.5) -- (Ft);
        \draw (Ft) -- (1.0,4);
        \draw (10.0,-1) -- (Ft1);
        \draw (Ft1) -- (10.0,4);
        \node[right] at (2.2,3.5) {$\{Lab^{Bank}: 0.0, Res^{Bank}: 0.0\}$};
        \node[right] at (2.2,1.5) {$\{FlowLab: 0.0, FlowRes: 0.0\}$};
        \node[right] at (2.2,-0.5) {$\{FlowLab: 52.0, FlowRes: 273.2\}$};
        \node[right] at (11.2,3.5) {$\{Lab^{Bank}: 0.0, Res^{Bank}: 0.0\}$};
        \node[right] at (11.2,-0.5) {$\{FlowLab: 52.0, FlowRes: 273.2\}$};
    \end{scope}
\end{tikzpicture}
\caption{String Diagram: Natural Transformation $\eta_{\text{time}}: F_t \Rightarrow F_{t+1}$ showing functor composition equivalence for temporal evolution}
\label{m}
\end{figure}
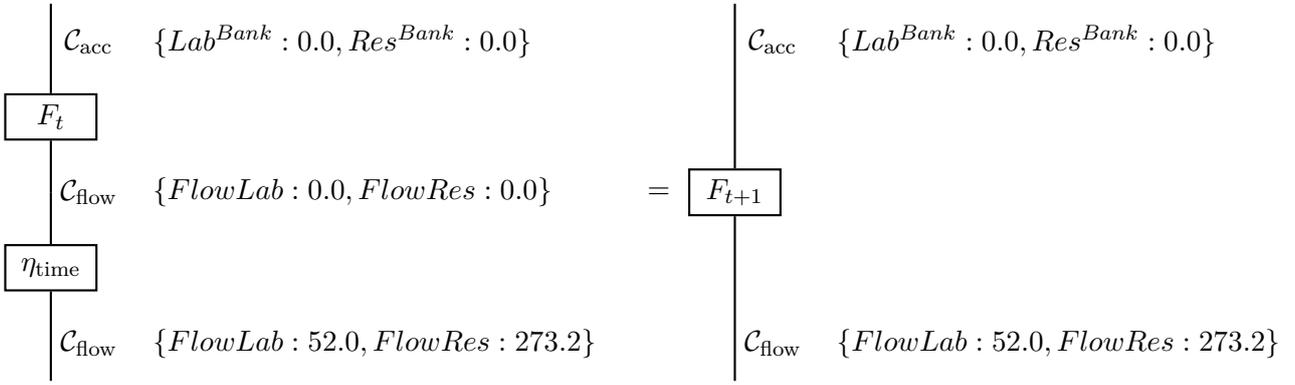

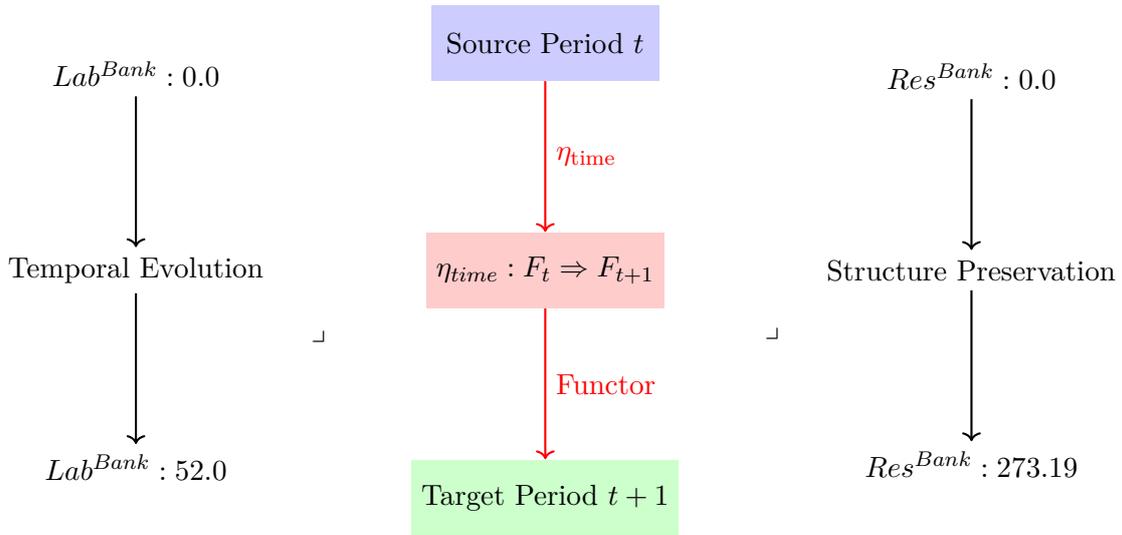
\begin{figure}[htbp]
\centering
\begin{tikzpicture}[node distance=2cm]
\node[rectangle, fill=blue!20, minimum width=3cm, minimum height=1cm] (source) {Source Period $t$};
\node[rectangle, fill=red!20, minimum width=3cm, minimum height=1cm, below=of source] (nattrans) {$\eta_{time}: F_t \Rightarrow F_{t+1}$};
\node[left=of nattrans] (eta1) {Temporal Evolution};
\node[right=of nattrans] (eta2) {Structure Preservation};
\node[rectangle, fill=green!20, minimum width=3cm, minimum height=1cm, below=of nattrans] (target) {Target Period $t+1$};
\node[below=of eta1] (lab2) {$Lab^{Bank}: 52.0$};
\node[below=of eta2] (res2) {$Res^{Bank}: 273.19$};
\node[above=of eta1] (lab0) {$Lab^{Bank}: 0.0$};
\node[above=of eta2] (res0) {$Res^{Bank}: 0.0$};
\draw[->, thick, red] (source) -- (nattrans) node[midway,right] {$\eta_{\text{time}}$};
\draw[->, thick, red] (nattrans) -- (target) node[midway,right] {Functor};
\draw[->, thick] (lab0) -- (eta1);
\draw[->, thick] (res0) -- (eta2);
\draw[->, thick] (eta1) -- (lab2);
\draw[->, thick] (eta2) -- (res2);
\node[below right=0.5cm of eta1] {$\lrcorner$};
\node[below left=0.5cm of eta2] {$\lrcorner$};
\end{tikzpicture}
\caption{Commutative Diagram: Natural Transformation Pullback Flow}
\label{n}
\end{figure}

\begin{figure}[htbp]
    \centering
    {\footnotesize
    \begin{tikzcd}[column sep=8em, row sep=7em, arrows={-stealth}] 
        {\begin{matrix} \green{D} \\ \text{Period Evolution} \\ \text{(0 $\to$ 1 $\to$ 2)} \end{matrix}} \\
        & {\begin{matrix} \red{P} \\ \text{Natural Transformation} \\ \eta_{time} \end{matrix}} & {\begin{matrix} B \\ \text{Target Period} \\ \text{(t+1)} \end{matrix}} \\
        & {\begin{matrix} A \\ \text{Source Period} \\ \text{(t)} \end{matrix}} & {\begin{matrix} \mathcal{C}_{acc} \\ \text{Account Category} \\ \mathbb{R}_{\geq 0} \end{matrix}} \\
        \arrow["{\begin{matrix} \exists ! \; u \\ Lab^{Bank}: 000.0 \to 000.0 \to 052.0 \\ Res^{Bank}: 000.0 \to 208.0 \to 273.2 \\ Com^{Bank}: 000.0 \to 052.0 \to 190.5 \end{matrix}}"{right=1.5cm}, pos=0.1, color={rgb,255:red,92;green,214;blue,92}, dashed, from=1-1, to=2-2]
        \arrow["{\begin{matrix} d_A = p_A \circ u \\ \text{Source Period} \\ \text{Evolution Mapping} \end{matrix}}"'{left=0.1cm}, pos=0.3, color={rgb,255:red,92;green,214;blue,92}, from=1-1, to=3-2]
        \arrow["{\begin{matrix} d_B = p_B \circ u \\ \text{Target Period} \\ \text{Evolution Mapping} \end{matrix}}"{right=1.5cm}, pos=0.7, color={rgb,255:red,92;green,214;blue,92}, from=1-1, to=2-3]
        \arrow["{\begin{matrix} p_A \\ \text{Source Period} \\ \text{Projection} \end{matrix}}"'{left=0.1cm}, pos=0.2, color={rgb,255:red,214;green,92;blue,92}, from=2-2, to=3-2]
        \arrow["{\begin{matrix} p_B \\ \text{Target Period} \\ \text{Projection} \end{matrix}}"{below=0.2cm}, pos=0.3, color={rgb,255:red,214;green,92;blue,92}, from=2-2, to=2-3]
        \arrow["{\begin{matrix} f \\ \text{Account Category} \\ \text{Preservation} \end{matrix}}"'{above=0.5cm}, from=3-2, to=3-3]
        \arrow["{\begin{matrix} g \\ \text{Temporal} \\ \text{Consistency} \end{matrix}}"{right=0.4cm}, pos=0.6, from=2-3, to=3-3]
    \end{tikzcd}
    }
    \vspace{-2.5cm}
    \caption{Natural Transformation Pullback: Temporal evolution validation across periods 0, 1, 2. 
    The pullback ensures that account transformations preserve categorical structure while evolving through time.}
    \label{o}
\end{figure}
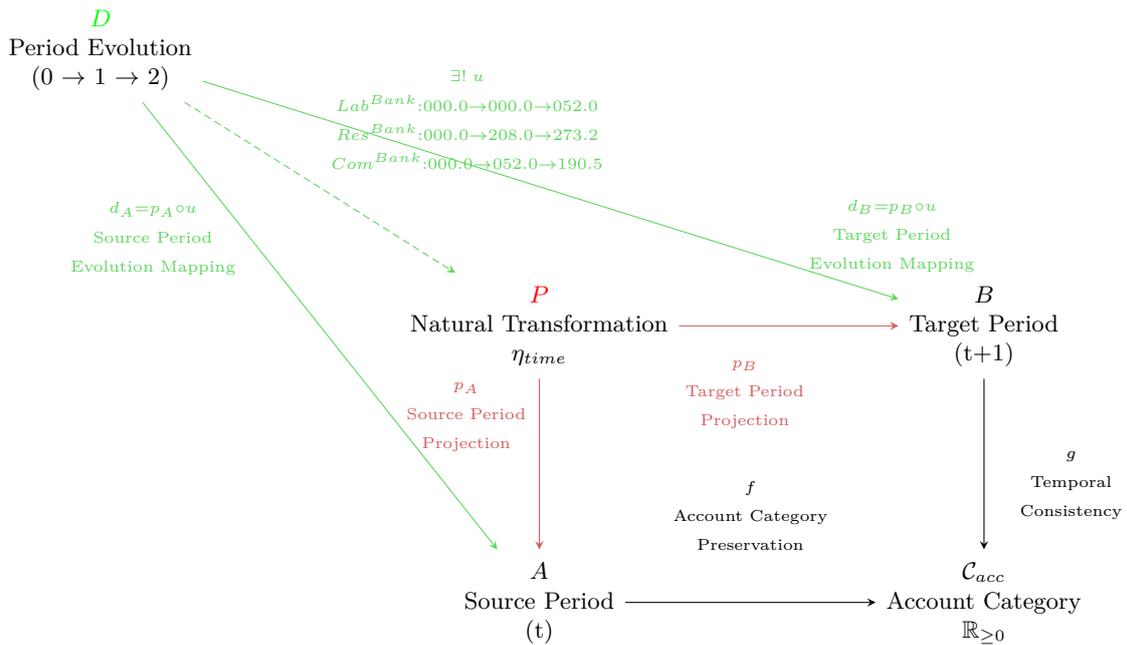

\textbf{Economics: Period Evolution Through Natural Transformations:}
The evolution in definition~\ref{def:nattransev} is visualised in time in the string diagram~\ref{m}.
In the commutative diagram~\ref{n} we see the natural transformations at work in time with the pullback denoted by the corners $\lrcorner$.
The consistency condition in Figure~\ref{o} is more informative 
than the zeros of equations~(\ref{eq:inv_lab})-(\ref{eq:inv_macro}).

The natural transformation pullback captures the fundamental economic principle of temporal consistency across accounting periods. 
The typed morphism $F_{\text{Evol}}: \mathcal{C}_t \rightarrow \mathcal{C}_{t+1}$ ensures that account balances evolve coherently through natural transformations 
that preserve the categorical structure of the economy.  

Economically, this means that the transition from period 0 through period 1 to period 2 maintains accounting identities allowing for consistent booking and investment. 
The pullback constraint validates that each period's state is reachable from the previous period through valid economic transactions. 
For instance, $Lab^{Bank}$ evolving from $0.0 \to 0.0 \to 52.0$ reflects labor owner's accumulation of bank deposits through wage payments, 
while $Res^{Bank}$ growth from $0.0 \to 208.0 \to 273.19$ shows resource owner's increasing wealth through sales to the company.

The commutative diagram's \ref{n} three-layer structure shows the categorical nature of time: the purple layer represents the source period's economic state, 
the red layer embodies the transformation mechanism of the natural transformation, and the green layer shows the target period's evolved state. 
The pullback corners $\lrcorner$ note that this evolution respects categorical limits.
Economically this means no units of accounts are created or destroyed arbitrarily, and all flows among the double-entry systems are consistent across temporal boundaries
which is what macroeconomic consistency is about.

%-------------------------------------------------------------------------------------------------------------------------------------------------------------------------
\section{Evolution}
\subsection{Eight Macroeconomic Bookings}

\begin{figure}[htbp]
\centering
\tikzset{
    stringdiagram/.style={
        baseline=(current bounding box.center),
        every node/.style={draw=none},
        every path/.style={thick}
    }
}
\begin{tikzpicture}[stringdiagram]
    \begin{scope}[shift={(0,0)}]
        \node at (1.6,3.5) {$\mathcal{C}_{\text{Real}}$};
        \node[draw, rectangle, minimum width=1.2cm, minimum height=0.6cm] (real) at (1.0,2.5) {Real};
        \node at (1.6,1.5) {$\mathcal{C}_{acc}$};
        \node[draw, rectangle, minimum width=1.2cm, minimum height=0.6cm] (pushout1) at (1.0,0.5) {Pushout};
        \node at (1.6,-0.5) {$\mathcal{C}_{flow}$};
        \node at (5.5,1.5) {$+$};
        \node at (8.6,3.5) {$\mathcal{C}_{Nom}$};
        \node[draw, rectangle, minimum width=1.2cm, minimum height=0.6cm] (nominal) at (8.0,2.5) {Nominal};
        \node at (8.5,1.5) {$\mathcal{C}_{acc}$};
        \node[draw, rectangle, minimum width=1.2cm, minimum height=0.6cm] (pushout2) at (8.0,0.5) {Pushout};
        \node at (8.6,-0.5) {$\mathcal{C}_{flow}$};
        \node at (12.5,1.5) {$=$};
        \node at (15.0,3.5) {$\mathcal{C}_{\text{Booking}}$};
        \node[draw, rectangle, minimum width=1.2cm, minimum height=0.6cm] (universal) at (14.0,1.5) {$\exists ! u$};
        \node at (15.0,-0.5) {$\mathcal{C}_{\text{Conserve}}$};
        \draw (1.0,-1) -- (pushout1);
        \draw (pushout1) -- (1.0,1.5);
        \draw (1.0,1.5) -- (real);
        \draw (real) -- (1.0,4);
        \draw (8.0,-1) -- (pushout2);
        \draw (pushout2) -- (8.0,1.5);
        \draw (8.0,1.5) -- (nominal);
        \draw (nominal) -- (8.0,4);
        \draw (14.0,-1) -- (universal);
        \draw (universal) -- (14.0,4);
        \node[right] at (2.2,3.5) {$\{Lab \to Com: 52.0\}$};
        \node[right] at (2.2,1.5) {$\{Real\}$};
        \node[right] at (2.2,-0.5) {$\{Flow: 52.0\}$};
        \node[right] at (9.2,3.5) {$\{Bank \to Lab: 52.0\}$};
        \node[right] at (9.2,1.5) {$\{Nominal\}$};
        \node[right] at (9.2,-0.5) {$\{Flow: 52.0\}$};
        \node[right] at (15.2,4.0) {$\{8 Bookings\}$};
        \node[right] at (14.8,0.0) {$\{\Sigma \Delta Accounts = 0\}$};
    \end{scope}
\end{tikzpicture}
\caption{String Diagram: 8 Bookings Pushout $Real + Nominal \rightarrow Conservation$ showing universal aggregation property}
\label{b}
\end{figure}
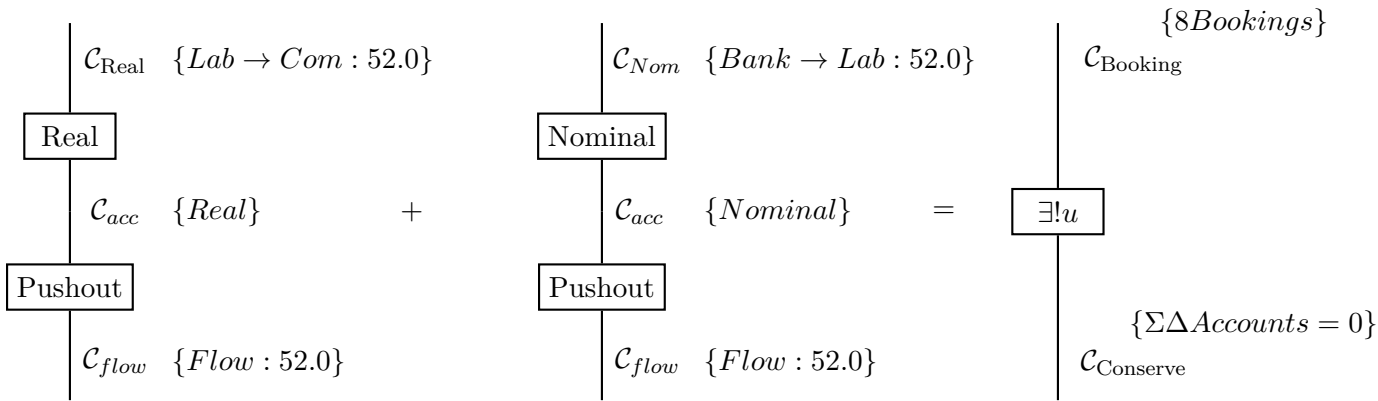

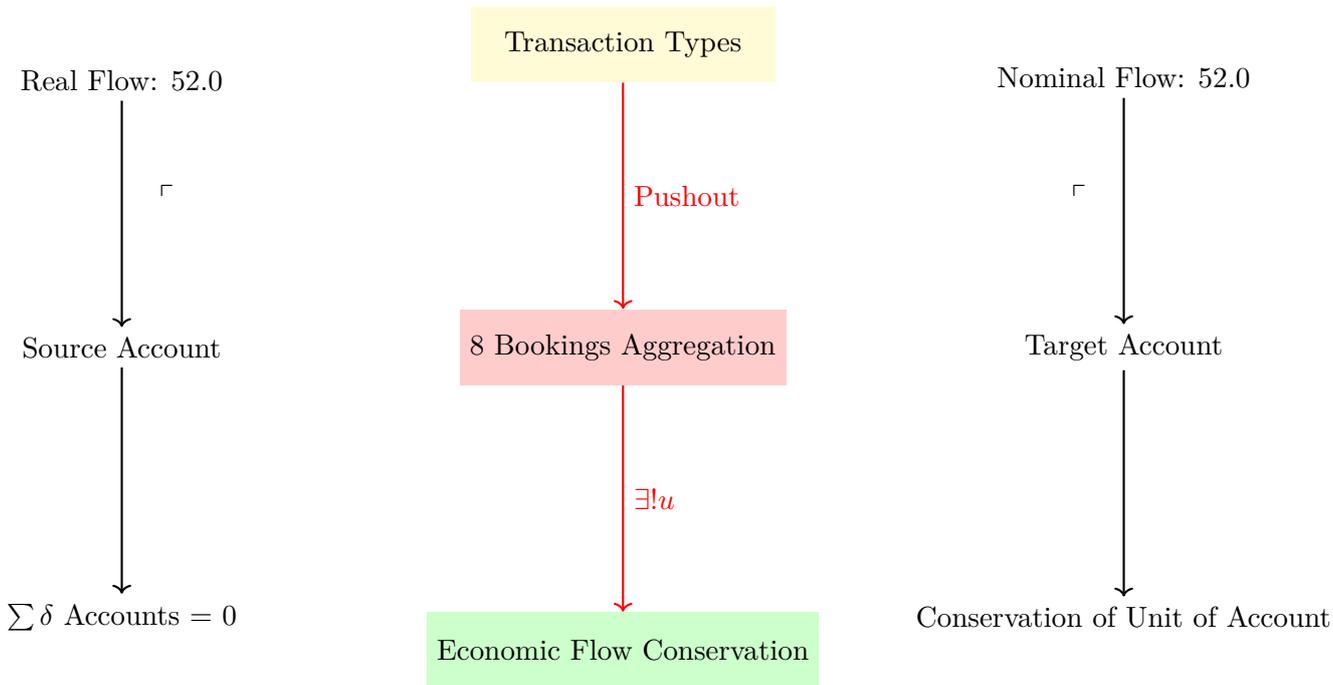
\begin{figure}[htbp]
\centering
\begin{tikzpicture}[node distance=3cm]
\node[rectangle, fill=yellow!20, minimum width=4cm, minimum height=1cm] (transtype) {Transaction Types};
\node[rectangle, fill=red!20, minimum width=4cm, minimum height=1cm, below=of transtype] (aggregation) {8 Bookings Aggregation};
\node[left=of aggregation] (source) {Source Account};
\node[right=of aggregation] (target) {Target Account};
\node[above=of source] (real) {Real Flow: 52.0};
\node[above=of target] (nominal) {Nominal Flow: 52.0};
\node[rectangle, fill=green!20, minimum width=4cm, minimum height=1cm, below=of aggregation] (econflow) {Economic Flow Conservation};
\node[below=of source] (balance) {$\sum \delta$ Accounts = 0};
\node[below=of target] (money) {Conservation of Unit of Account};
\draw[->, thick, red] (transtype) -- (aggregation) node[midway,right] {Pushout};
\draw[->, thick, red] (aggregation) -- (econflow) node[midway,right] {$\exists ! u$};
\draw[->, thick] (real) -- (source);
\draw[->, thick] (nominal) -- (target);
\draw[->, thick] (source) -- (balance);
\draw[->, thick] (target) -- (money);
\node at (-6,-2) {$\ulcorner$};
\node at (6,-2) {$\ulcorner$};
\end{tikzpicture}
\caption{Commutative Diagram: 8 Bookings Pushout Flow}
\label{c}
\end{figure}

\begin{figure}[htbp]
    \centering
    {\footnotesize
    \begin{tikzcd}[column sep=8em, row sep=7em, arrows={-stealth}]
        {\begin{matrix} C \\ \text{Transaction Type} \\ \text{(Real/Nominal)} \end{matrix}} & {\begin{matrix} B \\ \text{Target Agent} \\ \text{(Com, Lab, ...)} \end{matrix}} \\
        {\begin{matrix} A \\ \text{Source Agent} \\ \text{(Lab, Res, ...)} \end{matrix}} & {\begin{matrix} \red{P} \\ \text{Booking Aggregation} \\ \text{(8 Bookings)} \end{matrix}} & \\
        && {\begin{matrix} \green{D} \\ \text{Economic Flow} \\ \text{(Balanced Accounts)} \end{matrix}} & \\
        \arrow["{\begin{matrix} f \\ \text{Real Flow} \\ Lab\to Com: 52.0 \end{matrix}}"{right}, pos=0.2, from=1-1, to=2-1]
        \arrow["{\begin{matrix} g \\ \text{Nominal Flow} \\ Com\to Lab: 52.0 \end{matrix}}"{above=0.3cm}, pos=0.5, from=1-1, to=1-2]
        \arrow["{\begin{matrix} p_A \\ \text{Source Agent} \\ \text{Aggregation} \end{matrix}}"{above=0.4cm}, pos=0.6, color={rgb,255:red,214;green,92;blue,92}, from=2-1, to=2-2]
        \arrow["{\begin{matrix} p_B \\ \text{Target Agent} \\ \text{Aggregation} \end{matrix}}"{right=0.2cm}, pos=0.7, color={rgb,255:red,214;green,92;blue,92}, from=1-2, to=2-2]
        \arrow["{\begin{matrix} d_A \\ \text{Source Flow} \\ \text{Conservation} \end{matrix}}"{left=1.0cm}, pos=0.6, color={rgb,255:red,92;green,214;blue,92}, from=2-1, to=3-3]
        \arrow["{\begin{matrix} d_B \\ \text{Target Flow} \\ \text{Conservation} \end{matrix}}"{right=0.6cm}, pos=0.9, color={rgb,255:red,92;green,214;blue,92}, from=1-2, to=3-3]
        \arrow["{\begin{matrix} \exists ! \; u \\ \text{Conservation} \\ \text{$\Sigma\Delta$ Accounts} = 0 \end{matrix}}"{description}, pos=0.2, color={rgb,255:red,92;green,214;blue,92}, dashed, from=2-2, to=3-3]
    \end{tikzcd}
    }
    \vspace{-2.5cm}
    \caption{8 Bookings Pushout: Flow aggregation of real flows and nominal flows into balanced economic flows. The pushout ensures conservation of accounts across all transactions.}
    \label{d}
\end{figure}
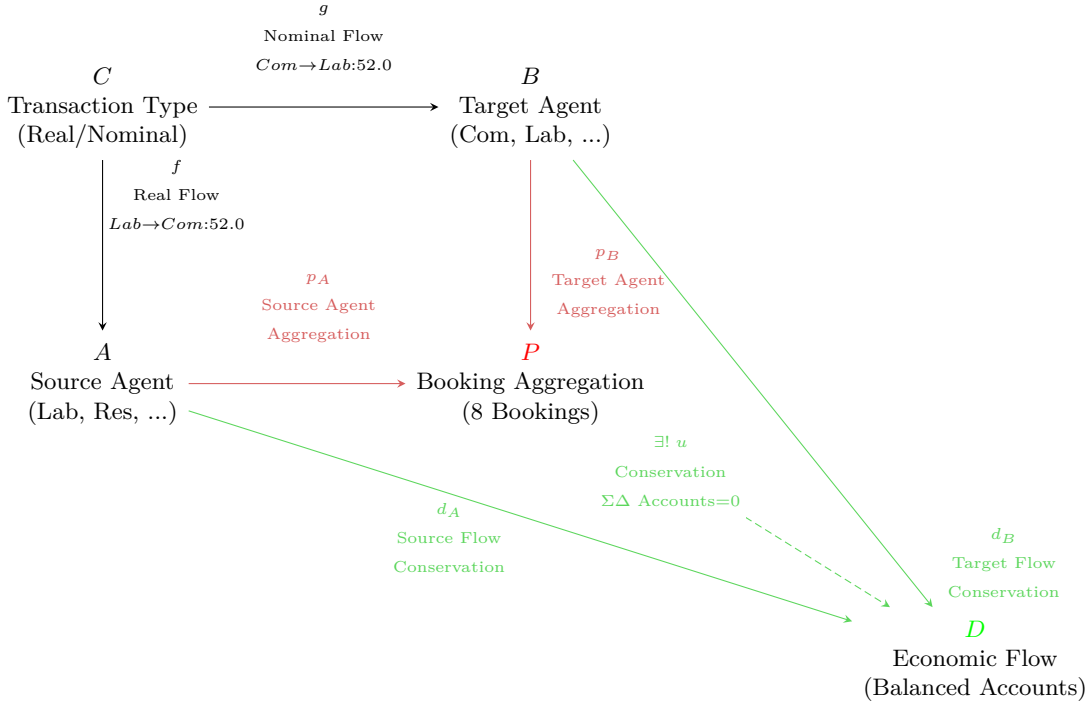
    
\textbf{Economics of Transaction Aggregation and Flow Conservation:}
The 8 bookings pushout embodies the fundamental principle of double-entry accounting lifted to categorical level. 
Each booking represents a simultaneous exchange where real flows and nominal flows move in different directions, creating the pushout structure 
that aggregates individual transactions into the economy's global flow pattern.
The fundamental economic flows in equation~(\ref{eq:bookingsall}) become morphisms in categorically typed constructions.

Economically, this captures how elementary flows compose into complex economic interactions. 
The pushout universal property ensures that the aggregation preserves the conservation laws: when labor sells services to the company (Booking 1), real labor flows from Labor to Company 
while nominal euros flow from Company to Labor, mediated by the Bank. 
The commutative diagram \ref{c} reveals that the Real Flow (52.0) and Nominal Flow (52.0) are equal and opposite, maintaining the zero-sum property of monetary consistency.

The categorical structure shows that Source Agents and Target Agents are connected through Transaction Types, which then pushout to Economic Flow Conservation. 
This means that all individual bilateral exchanges aggregate coherently into a global pattern that respects both microeconomic transaction constraints and macroeconomic conservation laws. 
The pushout corners in the commutative diagram mark the points where individual transaction morphisms compose into the universal aggregate morphism, 
ensuring that the sum of all real flows equals the sum of all nominal flows with opposite signs in Figure~\ref{d}.
Figure~\ref{e} shows the natural transformation sequence from abstract validation to concrete booking instances (Booking 1, Booking 3, Booking 5) 
with the pullback's universal property ensuring transaction validity.

\begin{figure}[htbp]
\centering
\tikzset{
stringdiagram/.style={
    baseline=(current bounding box.center),
    every node/.style={draw=none},
    every path/.style={thick}
}
}
\begin{tikzpicture}[stringdiagram]
\begin{scope}[shift={(0,0)}]
\node at (1.5,4.5) {$\mathcal{C}_{\text{Validation}}$};
\node[draw, rectangle, minimum width=2.0cm, minimum height=0.8cm, fill=green!20] (validation) at (1.0,3.5) {Balance Check};
\node at (0.0,2.5) {$\mathcal{C}_{\text{Transaction}}$};
\node[draw, rectangle, minimum width=2.0cm, minimum height=0.8cm, fill=red!20] (transaction) at (1.0,1.5) {Typed Morphism};
\node at (0.0,0.5) {$\mathcal{C}_{acc}$};
\node[draw, rectangle, minimum width=2.0cm, minimum height=0.8cm, fill=blue!20] (accounts) at (1.0,-0.5) {Source/Target};
\node at (6.5,4.5) {$\mathcal{C}_{\text{Concrete}}$};
\node[draw, rectangle, minimum width=2.0cm, minimum height=0.8cm, fill=green!20] (book1) at (6.0,3.5) {$0.0 \to 52.0$};
\node at (5.0,2.5) {$\mathcal{C}_{\text{Morphisms}}$};
\node[draw, rectangle, minimum width=2.0cm, minimum height=0.8cm, fill=red!20] (morph1) at (6.0,1.5) {$p_A \circ u$};
\node at (5.0,0.5) {$\mathcal{C}_{\text{Types}}$};
\node[draw, rectangle, minimum width=2.0cm, minimum height=0.8cm, fill=blue!20] (type1) at (6.0,-0.5) {$\mathbb{R}_{\geq 0}$};
\node[draw, rectangle, minimum width=2.0cm, minimum height=0.8cm, fill=green!20] (book3) at (11.0,3.5) {$208.0 \to 231.59$};
\node[draw, rectangle, minimum width=2.0cm, minimum height=0.8cm, fill=red!20] (morph3) at (11.0,1.5) {$p_B \circ u$};
\node[draw, rectangle, minimum width=2.0cm, minimum height=0.8cm, fill=blue!20] (type3) at (11.0,-0.5) {Balance};
\node[draw, rectangle, minimum width=2.0cm, minimum height=0.8cm, fill=green!20] (book5) at (16.0,3.5) {$260.0 \to 289.49$};
\node[draw, rectangle, minimum width=2.0cm, minimum height=0.8cm, fill=red!20] (morph5) at (16.0,1.5) {$\exists ! u$};
\node[draw, rectangle, minimum width=2.0cm, minimum height=0.8cm, fill=blue!20] (type5) at (16.0,-0.5) {Consistency};
\draw[->, thick] (validation) -- (transaction) node[midway,right=0.0cm] {$\pi_1$};
\draw[->, thick] (transaction) -- (accounts) node[midway,right=0.0cm] {$\pi_2$};
\draw[->, thick] (book1) -- (morph1) node[midway,right=0.0cm] {$\phi_1$};
\draw[->, thick] (morph1) -- (type1) node[midway,right=0.0cm] {$\phi_2$};
\draw[->, thick] (book3) -- (morph3) node[midway,right=0.0cm] {$\psi_1$};
\draw[->, thick] (morph3) -- (type3) node[midway,right=0.0cm] {$\psi_2$};
\draw[->, thick] (book5) -- (morph5) node[midway,right=0.0cm] {$\zeta_1$};
\draw[->, thick] (morph5) -- (type5) node[midway,right=0.0cm] {$\zeta_2$};
\draw[->, thick, red] (validation) -- (book1) node[midway,above] {$\eta_{\text{val1}}$};
\draw[->, thick, red] (transaction) -- (morph1) node[midway,above] {$\eta_{\text{trans1}}$};
\draw[->, thick, red] (accounts) -- (type1) node[midway,above] {$\eta_{\text{acc1}}$};
\draw[->, thick, red] (book1) -- (book3) node[midway,above] {$\eta_{\text{val3}}$};
\draw[->, thick, red] (morph1) -- (morph3) node[midway,above] {$\eta_{\text{trans3}}$};
\draw[->, thick, red] (type1) -- (type3) node[midway,above] {$\eta_{\text{acc3}}$};
\draw[->, thick, red] (book3) -- (book5) node[midway,above] {$\eta_{\text{val5}}$};
\draw[->, thick, red] (morph3) -- (morph5) node[midway,above] {$\eta_{\text{trans5}}$};
\draw[->, thick, red] (type3) -- (type5) node[midway,above] {$\eta_{\text{acc5}}$};
\node at (0.3,2.9) {$\lrcorner$};
\node at (5.3,2.9) {$\lrcorner$};
\node at (10.3,2.9) {$\lrcorner$};
\node at (15.3,2.9) {$\lrcorner$};
\node at (3.5,3.5) {$\otimes$};
\node at (3.5,1.5) {$\otimes$};
\node at (3.5,-0.5) {$\otimes$};
\node at (8.5,3.5) {$\otimes$};
\node at (8.5,1.5) {$\otimes$};
\node at (8.5,-0.5) {$\otimes$};
\node at (13.5,3.5) {$\otimes$};
\node at (13.5,1.5) {$\otimes$};
\node at (13.5,-0.5) {$\otimes$};
\draw[->, thick, purple, dashed] (validation) to[bend left=25] (book5) node at (8.5,6.0) {$\exists ! u$};
\end{scope}
\end{tikzpicture}
\caption{Commutative Diagram: Booking Validation Pullback Structure}
\label{e}
\end{figure}
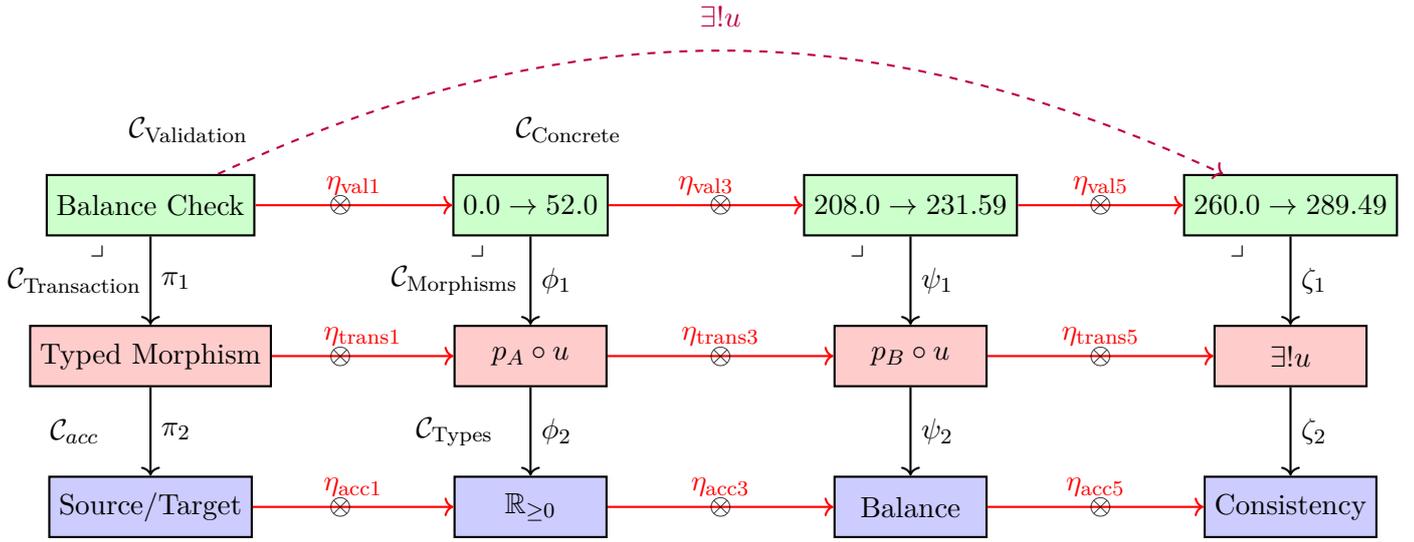

\begin{figure}[htbp]
\centering
\tikzset{
stringdiagram/.style={
baseline=(current bounding box.center),
every node/.style={draw=none},
every path/.style={thick}
}
}
\begin{tikzpicture}[stringdiagram]
\begin{scope}[shift={(0,0)}]
\node at (1.5,4.5) {$\mathcal{C}_{\text{Validation}}$};
\node[draw, rectangle, minimum width=1.8cm, minimum height=0.8cm, fill=green!20] (validation) at (1.0,3.5) {Validation};
\node at (0.0,2.5) {$\mathcal{C}_{\text{Morphism}}$};
\node[draw, rectangle, minimum width=1.8cm, minimum height=0.8cm, fill=red!20] (morphism) at (1.0,1.5) {Morphism};
\node at (0.0,0.5) {$\mathcal{C}_{\text{Type}}$};
\node[draw, rectangle, minimum width=1.8cm, minimum height=0.8cm, fill=blue!20] (acctype) at (1.0,-0.5) {$\mathbb{R}_{\geq 0}$};
\node at (8.5,4.5) {$\mathcal{C}_{\text{Bookings}}$};
\node[draw, rectangle, minimum width=1.8cm, minimum height=0.8cm, fill=green!20] (check1) at (8.0,3.5) {$0.0 \to 52.0$};
\node at (7.0,2.5) {$\mathcal{C}_{\text{Domains}}$};
\node[draw, rectangle, minimum width=1.8cm, minimum height=0.8cm, fill=red!20] (domain) at (8.0,1.5) {Domain};
\node at (7.0,0.5) {$\mathcal{C}_{\text{Accounts}}$};
\node[draw, rectangle, minimum width=1.8cm, minimum height=0.8cm, fill=blue!20] (source) at (8.0,-0.5) {Source};
\node[draw, rectangle, minimum width=1.8cm, minimum height=0.8cm, fill=green!20] (check3) at (15.0,3.5) {$208.0 \to 231.59$};
\node[draw, rectangle, minimum width=1.8cm, minimum height=0.8cm, fill=red!20] (codomain) at (15.0,1.5) {Codomain};
\node[draw, rectangle, minimum width=1.8cm, minimum height=0.8cm, fill=blue!20] (target) at (15.0,-0.5) {Target};
\draw[->, thick] (validation) -- (morphism) node[midway,right=0.0cm] {$\pi_1$};
\draw[->, thick] (morphism) -- (acctype) node[midway,right=0.0cm] {$\pi_2$};
\draw[->, thick] (check1) -- (domain) node[midway,right=0.0cm] {$\phi_1$};
\draw[->, thick] (domain) -- (source) node[midway,right=0.0cm] {$\phi_2$};
\draw[->, thick] (check3) -- (codomain) node[midway,right=0.5cm] {$\psi_1$};
\draw[->, thick] (codomain) -- (target) node[midway,right=0.5cm] {$\psi_2$};
\draw[->, thick, red] (validation) -- (check1) node[midway,above] {$\eta_{\text{book1}}$} node[midway] {};
\draw[->, thick, red] (morphism) -- (domain) node[midway,above] {$\eta_{\text{dom}}$} node[midway] {};
\draw[->, thick, red] (acctype) -- (source) node[midway,above] {$\eta_{\text{src}}$} node[midway] {};
\draw[->, thick, red] (check1) -- (check3) node[midway,above] {$\eta_{\text{book3}}$};
\draw[->, thick, red] (domain) -- (codomain) node[midway,above] {$\eta_{\text{cod}}$};
\draw[->, thick, red] (source) -- (target) node[midway,above] {$\eta_{\text{tgt}}$};
\node at (0.3,2.9) {$\lrcorner$};
\node at (7.3,2.9) {$\lrcorner$};
\node at (14.3,2.9) {$\lrcorner$};
\node at (4.5,3.5) {$\otimes$};
\node at (4.5,1.5) {$\otimes$};
\node at (4.5,-0.5) {$\otimes$};
\node at (11.5,3.5) {$\otimes$};
\node at (11.5,1.5) {$\otimes$};
\node at (11.5,-0.5) {$\otimes$};
\draw[->, thick, purple, dashed] (validation) to[bend left=20] (check3) node at (8.0,5.5) {$\exists ! u$};
\end{scope}
\end{tikzpicture}
\caption{Commutative Diagram: Booking Validation Flow}
\label{f}
\end{figure}

\begin{figure}[htbp]
\centering
\begin{tikzpicture}[node distance=2cm]
\node[rectangle, fill=green!20, minimum width=4cm, minimum height=1cm] (validation) {Booking Validation};
\node[rectangle, fill=red!20, minimum width=4cm, minimum height=1cm, below=of validation] (morphism) {Typed Morphism Validation};
\node[left=of morphism] (domain) {Domain Checks};
\node[right=of morphism] (codomain) {Codomain Checks};
\node[above=of domain] (check1) {Booking 1: 0.0 $\to$ 52.0};
\node[above=of codomain] (check3) {Booking 3: 208.0 $\to$ 231.59};
\node[rectangle, fill=blue!20, minimum width=4cm, minimum height=1cm, below=of morphism] (acctype) {Account Type $\mathbb{R}\geq 0$};
\node[below=of domain] (source) {Source Accounts};
\node[below=of codomain] (target) {Target Accounts};
\draw[->, thick, red] (validation.south) -- (morphism.north) node[midway,right] {$\exists ! u$};
\draw[->, thick, red] (morphism.south) -- (acctype.north) node[midway,right] {Pullback};
\draw[->, thick] (check1.south) -- (domain.north);
\draw[->, thick] (check3.south) -- (codomain.north);
\draw[->, thick] (domain.south) -- (source.north);
\draw[->, thick] (codomain.south) -- (target.north);
\node at (-4,-1) {$\lrcorner$};
\node at (4,-1) {$\lrcorner$};
\end{tikzpicture}
\caption{Commutative Diagram: Booking Validation Pullback Flow}
\label{g}
\end{figure}

\begin{figure}[htbp]
    \centering
    {\footnotesize
    \begin{tikzcd}[column sep=8em, row sep=7em, arrows={-stealth}]
    {\begin{matrix} \green{D} \\ \text{Booking Validation} \\ \text{(Balance Check)} \end{matrix}} \\
    & {\begin{matrix} \red{P} \\ \text{Transaction} \\ \text{(Typed Morphism)} \end{matrix}} & {\begin{matrix} B \\ \text{Target Account} \\ \text{(Com, Lab)} \end{matrix}} \\
    & {\begin{matrix} A \\ \text{Source Account} \\ \text{(Bank, Res)} \end{matrix}} & {\begin{matrix} C \\ \text{Account Type} \\ \mathcal{R}_{\geq 0} \end{matrix}} \\
    \arrow["{\begin{matrix} \exists ! \; u \\ Booking 1: 000.0 \to 052.0 \\ Booking 3: 208.0 \to 231.59 \\ Booking 5: 260.0 \to 289.49 \end{matrix}}", pos=0.9, color={rgb,255:red,92;green,214;blue,92}, dashed, from=1-1, to=2-2]
    \arrow["{\begin{matrix} d_A = p_A \circ u \\ \text{Source Account} \\ \text{Validation} \end{matrix}}"'{left=0.1cm}, pos=0.3, color={rgb,255:red,92;green,214;blue,92}, from=1-1, to=3-2]
    \arrow["{\begin{matrix} d_B = p_B \circ u \\ \text{Target Account} \\ \text{Validation} \end{matrix}}"{right=1.5cm}, pos=0.7, color={rgb,255:red,92;green,214;blue,92}, from=1-1, to=2-3]
    \arrow["{\begin{matrix} p_A \\ \text{Source Account} \\ \text{Projection} \end{matrix}}"'{left=0.1cm}, pos=0.3, color={rgb,255:red,214;green,92;blue,92}, from=2-2, to=3-2]
    \arrow["{\begin{matrix} p_B \\ \text{Target Account} \\ \text{Projection} \end{matrix}}"{below=0.2cm}, pos=0.3, color={rgb,255:red,214;green,92;blue,92}, from=2-2, to=2-3]
    \arrow["{\begin{matrix} f \\ \text{Account Type} \\ \text{Preservation} \end{matrix}}"'{above=0.4cm}, from=3-2, to=3-3]
    \arrow["{\begin{matrix} g \\ \text{Balance} \\ \text{Consistency} \end{matrix}}"{right=0.5cm}, pos=0.7, from=2-3, to=3-3]
    \end{tikzcd}
    }
    \vspace{-2.5cm}
    \caption{Booking Validation Pullback: Each of the 8 bookings is validated as a typed morphism with domain and codomain checks. The pullback ensures transaction validity before execution.}
\label{h}
\end{figure}
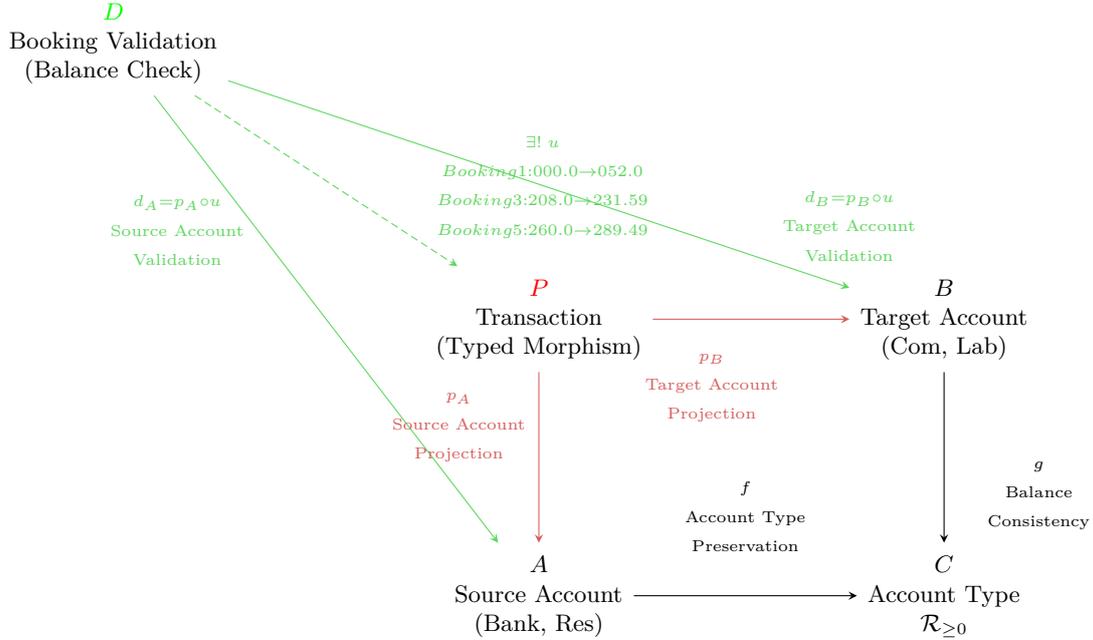
    
\textbf{Economics of Transaction Validation and Type Safety:}
The booking validation pullback ensures economic transactions respect fundamental constraint structures before execution. 
The typed morphism $\mathcal{F}_{\text{Validation}}: \text{SourceAccounts} \times \text{TargetAccounts} \rightarrow \{0,1\}$ acts as a categorical gatekeeper, 
verifying that proposed transactions satisfy accounting rules and economic logic.

Economically, this represents the critical validation layer that prevents impossible transactions: accounts cannot go negative (unless explicitly allowing credit), 
funds cannot be created ex nihilo, and transaction types must match account capabilities. 
The pullback structure validates that Booking 1 (0.0 $\to$ 52.0) and Booking 3 (208.0 $\to$ 231.59) represent feasible state transitions from their source accounts to target accounts.

The three-layer architecture of the diagram~\ref{g} shows validation as a categorical limit: the yellow layer represents proposed transactions, 
the blue layer embodies the validation mechanism (typed morphism validation), and the green layer confirms the transaction as either valid (1) or invalid (0). 
The pullback corners ensure that validation respects both source account constraints and target account capacities. 
This categorical approach to transaction validation provides mathematical foundations for economic transaction processing systems, ensuring 
that only economically feasible operations propagate through the system.
Figure~(\ref{f}) shows the natural transformation between abstract validation structure and concrete booking instances, with pullback universal property ensuring transaction validity.
% --------------------------------------------------------------------------------------------------------------------------------------------------------------
%\clearpage
\subsubsection{Categorically Typed Recursive Program}

\textbf{Formal Operations on Economic Categories:}
We define now the creation of the accounting system itself, mathematically and as a program.
By that we can obey to the Lucas Critique to endogeneize the observing econometrician and theory constructor.
This equips the agent in the theory with the same abilities as the observing econometrician to learn from data.
The econometrician can provide, as we do in this paper as observing economists, implementations and theories of a monetary macroeconomic accounting system.
This can be done categorically by taking the operations to be the endogenous creation and change of the categorical state of the monetary accounting system.
As an econometric generalisation of the pullback of the investment learning of the banking sector can be some portfolio selection,
other sectoral financial instruments,  market making and clearing structures of the bank sector or the central bank. 
The endogenous creation of the categorical state of the monetary accounting system is done by the helper functions.

Mathematical formulas for the construction of an economic category is a first step to endogeneize the econometric observer, the economist and econometrician,
maintaining an economic theory about data sharing or not in the memory of the economy as in a Bayesian game of incomplete information.
Helper functions are the ability of the observer and observed likewise to construct perspectives on or transformations of the system functorially.
This can be used with universal initial and final objects functionality for creating, deleting and copying structures.
Functors and categorical morphisms allow us to distinguish between structures and operations, observers' and observeds' views in hierarchical orders.
We can use them for any functorial change of the domains or codomains that describe the state and institutional structures of the economy.
We can extend MoMaT by categorifying structural macroeconometric theories like in~\cite{WK2010} by open games of~\cite{GHWZ2018}
or micro econometrics as in~\cite{HW2008} by adjunctions.
Forward looking agents, structural in the sense of rational expectations, can be formulated in categories by open games~\cite{GHWZ2018}.
The two flows in the adjunctions could then be simulations as in this paper from parameters to data and the backpropagation could be the other direction of adjunctions
where the data is used to update the parameters, poosibly also taking advantage of open games to also represent backpropagation as shown in~\cite{fong_backprop_2019}.
A structural model or theory is in economics an explicit decision model as the so called micro foundation of macroeconomics.

In \cite{EV2007} on Memory Evolutive Systems a categorical model is formulated for emergent effects, which can be used in economics in institutional design.
Elinore ~\cite{Ostrom2005Book} polycentrism, once categorified, can be included as open games for decentral, parallel endogenously optimal hierarchical control structures.
Macro and sectoral decisions in universals of money issuing central banks and government spending from debts and taxes can be integrated.
Interbanking market making in theory and models as programs in distributed computing environments are specifyable as universal Shapley values of liquidity pooling.
The categorical apperatus may also unify management science and economics by categorifying ERP systems like the object oriented perspective in~\cite{SCH2024} 
based on earlier work in the nineties on ARIS as a blue print for customizing SAPs R/3.
However, SAP's accounting was untyped and two dimensional for internal and external accounting only, whereas we have none of these constraints.

With the helper functions of the categorically typed functions and their mathematical types
we create, update, and relate economic objects and morphisms in the categories of accounts and flows.
To run an accounting system for MoMaT, we need to create the economic categorical state with 5 agents and accounts.
This extension can formally define agents as economy creators of the five agent economy $MoMaT^5$ and extending for a central bank $MoMaT^{cb}$ or government $MoMaT^{gov}$.
Hence, learning institutional designs and modelling structural breaks in a natural evolution is a generalised economic functionality of the compositionality of category theory.

The functionality of the operations are basic manipulations of the economic categorical state.
  Vertex Operations: Creating, retrieving, and updating economic accounts.
  Edge Operations: Creating and managing economic flows between accounts.
  Functor Operations: Establishing and applying transformations between categories.
  Natural Transformation Operations: Managing consistent changes across transformations.

\begin{definition}[Helper Functions]\label{def:helper_functions}
    The creation of the economic state is done by the helper functions.
    First we create three categories $\mathcal{C}_{accounts}$, flows $\mathcal{C}_{flows}$ and parameters $\mathcal{C}_{parameters}$.
    Then we create the accounts of the agents labor, resource, company, capitalist and bank.
    We add the parameter vertices to the category of parameters.
    We create the functor from nominal to real mapping $Lab^{Bank}$ to $Lab^{Lab}$, $Res^{Bank}$ to $Res^{Res}$, $Com^{Bank}$ to $Com^{Good}$.
    We create the functor from $Accounts$ to $Flows$.
    We initialize the memory vectors for the wage and repayments.
    
    The return of nothing () is because the purpose of update functions (names end in ! by style convention in Julia) is to change the first argument of the function
    as a side effect and not as usual to return by purpose some calculation without changing the arguments.
    The function \texttt{add\_vertex!} is used to add a vertex to the category, it returns the ID of the new vertex.
    The return of \texttt{add\_edge!} beside the side effect to change the category also returns the added edge.
    The \texttt{update\_vertex!} function returns nothing () beside the sideffect to update the amount of an account.
    The \texttt{add\_mapping!} function returns nothing () and is as well called for the side effect to add a mapping to the functor $\mathcal{F}$.
    The two functions on the natural transformations are defined but not used in the categorically typed program.
    They are implemented as the categorical program.
\begin{equation}\label{eq:helper_functions}
    \begin{array}{llll}
       \mathrm{Ob}(\mathcal{C}) & :: \mathrm{Ob}(\mathcal{C}_{par}),\mathrm{Ob}(\mathcal{C}_{flow}),\mathrm{Ob}(\mathcal{C}_{acc}) & \text{Parameters, Flows, Accounts} \\
       \mathrm{Hom}(\mathcal{C}) & :: \text{Morphisms of} & \mathcal{C}_{par},\mathcal{C}_{flow},\mathcal{C}_{acc}  \\ 
       \text{\bf Op(C)} \\
       \mathrm{get\_vertex} & :: \mathcal{C} \times \text{String} \to \mathrm{Ob}(\mathcal{C}) & \text{Retrieves vertex by name} \\
       \mathrm{add\_vertex!} & :: \mathcal{C}_{acc} \times \text{String} \times \text{String} \times \mathbb{R} \to \mathbb{N} & \text{Add account in } \mathcal{C}' \\ 
                                                                                                                                                 && \text{named::String} \\
                                                                                                                                                 && \text{typed::String} \\
                                                                                                                                                 && \text{with amount::} \mathbb{R} \\
                                                                                                                                                 && \text{return ID::} \mathbb{N} \text{ of new vertex} \\
      \mathrm{add\_edge!} & :: \mathcal{C} \times \mathbb{N} \times \mathbb{N} \times \mathbb{R} \to edge::Edge & \text{Adds edge with} \\
                                                                                                                                && \text{source::} \mathbb{N} \\
                                                                                                                                && \text{target::} \mathbb{N} \\
                                                                                                                                && \text{weight::} \mathbb{R} \\
                                                                                                                                && \text{return new edge::Edge} \\
      \mathrm{update\_vertex!} & :: \mathcal{C}_{acc} \times \text{String} \times \mathbb{R} \to () & \text{Updates an account::String}\\ 
                                                                                                                       && \text{amount::} \mathbb{R} \\
                                                                                                                       && \text{weight::} \mathbb{R} \\ 
                                                                                                                       && \text{return } nothing () \\
      \text{\bf Op(F)} \\
      \mathrm{add\_mapping!} & :: \mathcal{F} \times \mathbb{N} \times \mathbb{N} \to () & \text{Add map of morphisms to functor } \mathcal{F}' \\
                                                                                                                            && \text{source morphism::} \mathbb{N} \text{ in } \mathcal{C} \\
                                                                                                                            && \text{target morphism::} \mathbb{N} \text{ in } \mathcal{F}(\mathcal{C}) \\
                                                                                                                            && \text{return} nothing () \\
      \mathrm{apply\_functor} & :: \mathcal{F} \times \mathbb{N} \to \mathbb{N} & \text{Apply functor $\mathcal{F}$ to vertex::} \mathbb{N} \\
                                                                                   && \text{returns vertex ID::} \mathbb{N} \\
                                                                                   && \text{of } \texttt{vertex\_map} \text{ of } \mathcal{F}\\
      \text{\bf Op($\eta$)} \\                                                                                
      \mathrm{add\_component!} & :: \eta \times \mathbb{N} \times \text{Edge} \to \eta' & \text{Adds to } \eta \text{ edge::} \texttt{Edge} \text{with ID::} \mathbb{N} \\
      \mathrm{apply\_transformation} & :: \eta \times \mathbb{N} \to \eta' & \text{Applies } \eta \text{to vertex:: } \mathcal{N}\\
    \end{array}
\end{equation}    
\end{definition}

These functions satisfy the following mathematical properties:
\begin{itemize}
  \item Identity preservation: $\mathrm{get\_vertex}(\mathrm{add\_vertex!}(c, n, t, a)) = n$, account name
  \item Consistency preservation: All operations maintain the category structure
  \item Type safety: Operations respect the economic types of accounts and flows
\end{itemize}

Each function corresponds to a categorical operation:
\begin{itemize}
  \item Vertex operations correspond to object operations in the category
  \item Edge operations correspond to morphism operations in the category
  \item Update operations correspond to endomorphism operations in the category
\end{itemize}

\begin{figure}[t!]
    \centering
    \includegraphics[width=0.8\textwidth]{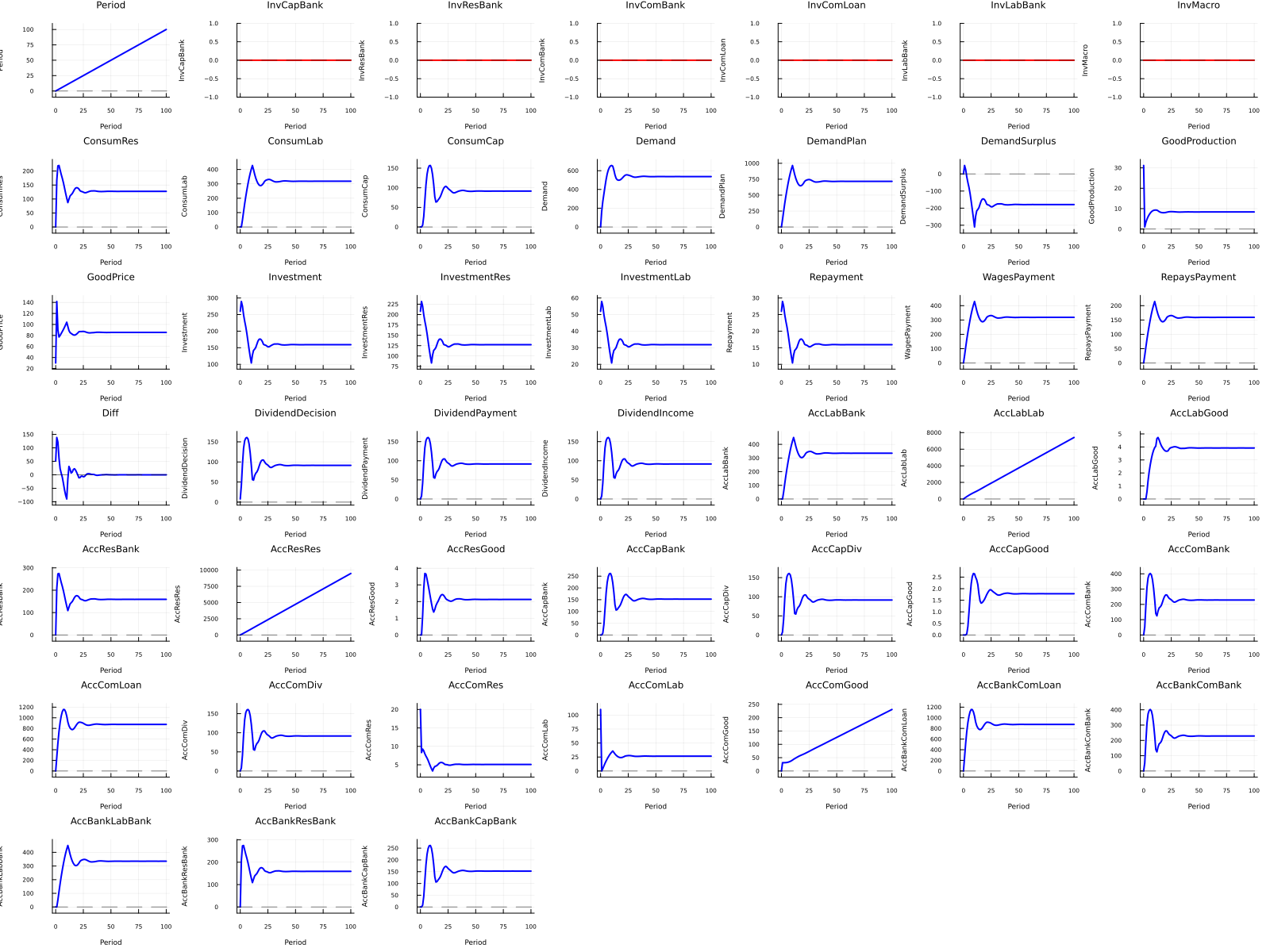}
    \caption{Time series plots of variables in the categorically typed recursive simulation}
    \label{q}
\end{figure}
   
\begin{definition}[Categorically Typed State]\label{def:categorical_state_implementation}
    The categorical state $S^{cat}$ is a categorical structure composed of three categories, two functors, two natural transformations, and historical vectors.
    
    \begin{equation}\label{eq:state:cat}
        S^{cat} = (\mathcal{C}_{acc}, \mathcal{C}_{flows}, \mathcal{C}_{pars}, \mathcal{F}_{price}, \mathcal{F}_{flow}, \eta_{price}, \eta_{flow}, \mathcal{H}_{wage}, \mathcal{H}_{repay})
    \end{equation}
    
    The categorical state is implemented in the categorically typed recursive state program.
    The economic simulation is programmed by annotating the calculated numbers with the categorical structures by the helper functions as categorical type (\texttt{struct}) generators.
    In the categorical program the natural transformations are implemented as the simulation of the evolution
    and the state category becomes the one of the endofunctor in section \ref{sec:categoricalstateevolution}.
    The categorical state $S^{cat}$ is then just one view on the category of the economy $C_{Economy}$ as the categorical state as a view on the categorical description of the evolution.
    The three categories $\mathcal{C}_{acc}$, $\mathcal{C}_{flows}$, $\mathcal{C}_{pars}$ organize accounts, flows, and parameters.
    The functors $\mathcal{F}_{price}: \mathcal{C}_{acc} \to \mathcal{C}_{acc}$ and $\mathcal{F}_{flow}: \mathcal{C}_{acc} \to \mathcal{C}_{flows}$ 
    connect nominal and real values as well as accounts with flows.
    The natural transformations $\eta_{price}: \mathcal{F}_{price} \Rightarrow \mathcal{F}_{price}$ and 
    $\eta_{flow}: \mathcal{F}_{flow} \Rightarrow \mathcal{F}_{flow}$ manage temporal changes.
    
    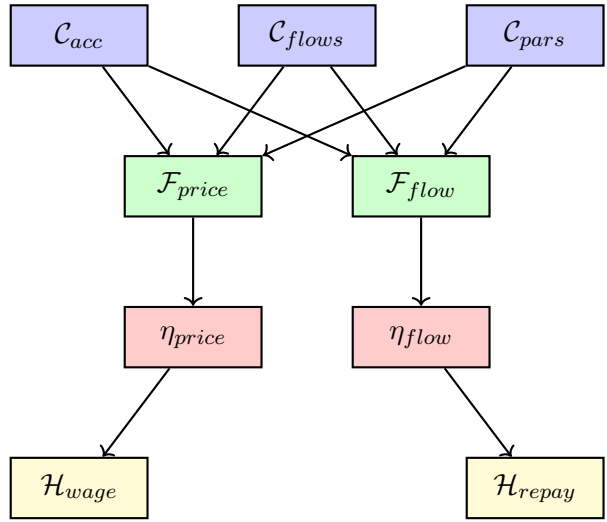
\begin{figure}[htbp]
        \centering
        \tikzset{
            stringdiagram/.style={
                baseline=(current bounding box.center),
                every node/.style={draw=none},
                every path/.style={thick}
            }
        }
        \begin{tikzpicture}[stringdiagram]
            \node[draw, rectangle, minimum width=1.8cm, minimum height=0.8cm, fill=blue!20] (accounts) at (0,3) {$\mathcal{C}_{acc}$};
            \node[draw, rectangle, minimum width=1.8cm, minimum height=0.8cm, fill=blue!20] (flows) at (3,3) {$\mathcal{C}_{flows}$};
            \node[draw, rectangle, minimum width=1.8cm, minimum height=0.8cm, fill=blue!20] (params) at (6,3) {$\mathcal{C}_{pars}$};
            \node[draw, rectangle, minimum width=1.8cm, minimum height=0.8cm, fill=green!20] (fprice) at (1.5,1) {$\mathcal{F}_{price}$};
            \node[draw, rectangle, minimum width=1.8cm, minimum height=0.8cm, fill=green!20] (fflow) at (4.5,1) {$\mathcal{F}_{flow}$}; 
            \node[draw, rectangle, minimum width=1.8cm, minimum height=0.8cm, fill=red!20] (etaprice) at (1.5,-1) {$\eta_{price}$};
            \node[draw, rectangle, minimum width=1.8cm, minimum height=0.8cm, fill=red!20] (etaflow) at (4.5,-1) {$\eta_{flow}$};
            \node[draw, rectangle, minimum width=1.8cm, minimum height=0.8cm, fill=yellow!20] (wage) at (0,-3) {$\mathcal{H}_{wage}$};
            \node[draw, rectangle, minimum width=1.8cm, minimum height=0.8cm, fill=yellow!20] (repay) at (6,-3) {$\mathcal{H}_{repay}$};
            \draw[thick,->] (accounts) -- (fprice);
            \draw[thick,->] (accounts) to[bend left=0] (fflow);
            \draw[thick,->] (flows) to[bend left=0] (fflow);
            \draw[thick,->] (flows) to[bend right=0] (fprice);
            \draw[thick,->] (params) to[bend left=-0] (fprice);
            \draw[thick,->] (params) to[bend right=00] (fflow);
            \draw[thick,->] (fprice) -- (etaprice);
            \draw[thick,->] (fflow) -- (etaflow);
            \draw[thick,->] (etaprice) -- (wage);
            \draw[thick,->] (etaflow) -- (repay);
        \end{tikzpicture}
        \caption{String Diagram: Categorical State Structure}
        \label{i}
    \end{figure}
    
    \begin{figure}[htbp]
    \centering
    \begin{tikzpicture}[node distance=2cm]
    \node[rectangle, fill=blue!20, minimum width=3cm, minimum height=1cm] (catstate) {Categorical State $S^{cat}$};
    \node[rectangle, fill=yellow!20, minimum width=2.5cm, minimum height=0.8cm, below left=of catstate] (accounts) {$\mathcal{C}_{acc}$};
    \node[rectangle, fill=yellow!20, minimum width=2.5cm, minimum height=0.8cm, below=of catstate] (flows) {$\mathcal{C}_{flows}$};
    \node[rectangle, fill=yellow!20, minimum width=2.5cm, minimum height=0.8cm, below right=of catstate] (params) {$\mathcal{C}_{pars}$};
    \node[rectangle, fill=green!20, minimum width=2.5cm, minimum height=0.8cm, below=of accounts] (fprice) {$\mathcal{F}_{price}$};
    \node[rectangle, fill=green!20, minimum width=2.5cm, minimum height=0.8cm, below=of flows] (fflow) {$\mathcal{F}_{flow}$};
    \node[rectangle, fill=red!20, minimum width=2.5cm, minimum height=0.8cm, below=of fprice] (etaprice) {$\eta_{price}$};
    \node[rectangle, fill=red!20, minimum width=2.5cm, minimum height=0.8cm, below=of fflow] (etaflow) {$\eta_{flow}$};
    
    \draw[->, thick] (catstate) -- (accounts);
    \draw[->, thick] (catstate) -- (flows);
    \draw[->, thick] (catstate) -- (params);
    \draw[->, thick] (accounts) -- (fprice);
    \draw[->, thick] (flows) -- (fflow);
    \draw[->, thick] (fprice) -- (etaprice);
    \draw[->, thick] (fflow) -- (etaflow);
    \end{tikzpicture}
    \caption{Categorical State Composition Structure}
    \label{j}
    \end{figure}
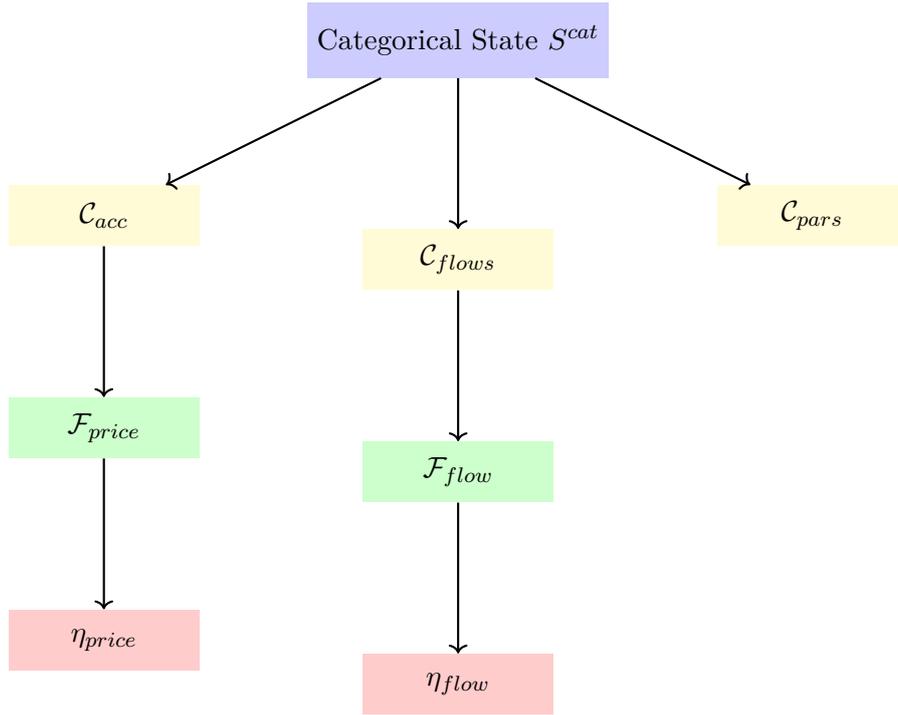
    
    \begin{figure}[htbp]
    \centering
    {\footnotesize
    \begin{tikzcd}[column sep=8em, row sep=7em, arrows={-stealth}]
    {\begin{matrix} \green{D} \\ \text{Categorical State} \\ S^{cat} \end{matrix}} \\
    & {\begin{matrix} \red{P} \\ \text{State Validation} \\ \text{(Pullback\_CatState)} \end{matrix}}
    & {\begin{matrix} B \\ \text{Flow Consistency} \\ \mathcal{C}_{flows} \end{matrix}} \\
    & {\begin{matrix} A \\ \text{Account Consistency} \\ \mathcal{C}_{acc} \end{matrix}}
    & {\begin{matrix} C \\ \text{Parameter Validation} \\ \mathcal{C}_{pars} \end{matrix}} \\
    \arrow["{\begin{matrix} \exists ! \; u \\ \text{State}_0: \text{Valid} \\ \text{State}_1: \text{Valid} \\ \text{State}_2: \text{Valid} \end{matrix}}", pos=0.3, color={rgb,255:red,92;green,214;blue,92}, dashed, from=1-1, to=2-2]
    \arrow["{\begin{matrix} d_A = p_A \circ u \\ \mathcal{F}_{price}: \mathcal{C}_{acc} \to \mathcal{C}_{acc} \\ \mathcal{F}_{flow}: \mathcal{C}_{acc} \to \mathcal{C}_{flows} \\ \text{Functor Validation} \end{matrix}}"{left=0.1cm}, pos=0.3, color={rgb,255:red,92;green,214;blue,92}, from=1-1, to=3-2]
    \arrow["{\begin{matrix} d_B = p_B \circ u \\ \eta_{price}: \mathcal{F}_{price} \Rightarrow \mathcal{F}_{price} \\ \eta_{flow}: \mathcal{F}_{flow} \Rightarrow \mathcal{F}_{flow} \\ \text{Natural Trans Validation} \end{matrix}}"{right=2.0cm}, pos=0.7, color={rgb,255:red,92;green,214;blue,92}, from=1-1, to=2-3]
    \arrow["{\begin{matrix} p_A \\ \text{Account} \\ \text{Check} \end{matrix}}"{left=0.1cm}, pos=0.3, color={rgb,255:red,214;green,92;blue,92}, from=2-2, to=3-2]
    \arrow["{\begin{matrix} p_B \\ \text{Flow} \\ \text{Check} \end{matrix}}"{below=0.2cm}, pos=0.4, color={rgb,255:red,214;green,92;blue,92}, from=2-2, to=2-3]
    \arrow["{\begin{matrix} f \\ \text{Account $\to$ Param} \\ \text{Type Consistency} \end{matrix}}"'{above=0.4cm}, from=3-2, to=3-3]
    \arrow["{\begin{matrix} g \\ \text{Flow $\to$ Param} \\ \text{Value Consistency} \end{matrix}}"{right=0.4cm}, pos=0.7, from=2-3, to=3-3]
    \end{tikzcd}
    }
    \vspace{-2.5cm}
    \caption{Categorical State Validation Pullback: The categorical state pullback ensures that all components $(\mathcal{C}_{acc}, \mathcal{C}_{flows}, \mathcal{C}_{pars}, \mathcal{F}_{price}, \mathcal{F}_{flow}, \eta_{price}, \eta_{flow})$ satisfy universal consistency properties through functor and natural transformation validation.}
    \label{k}
    \end{figure}
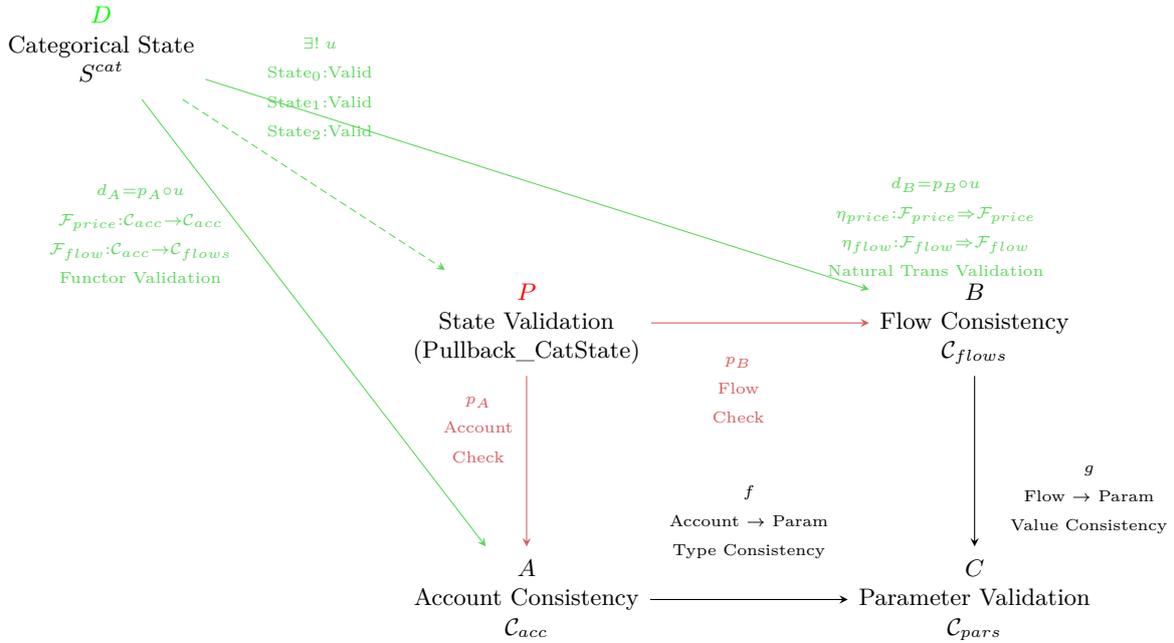
    
    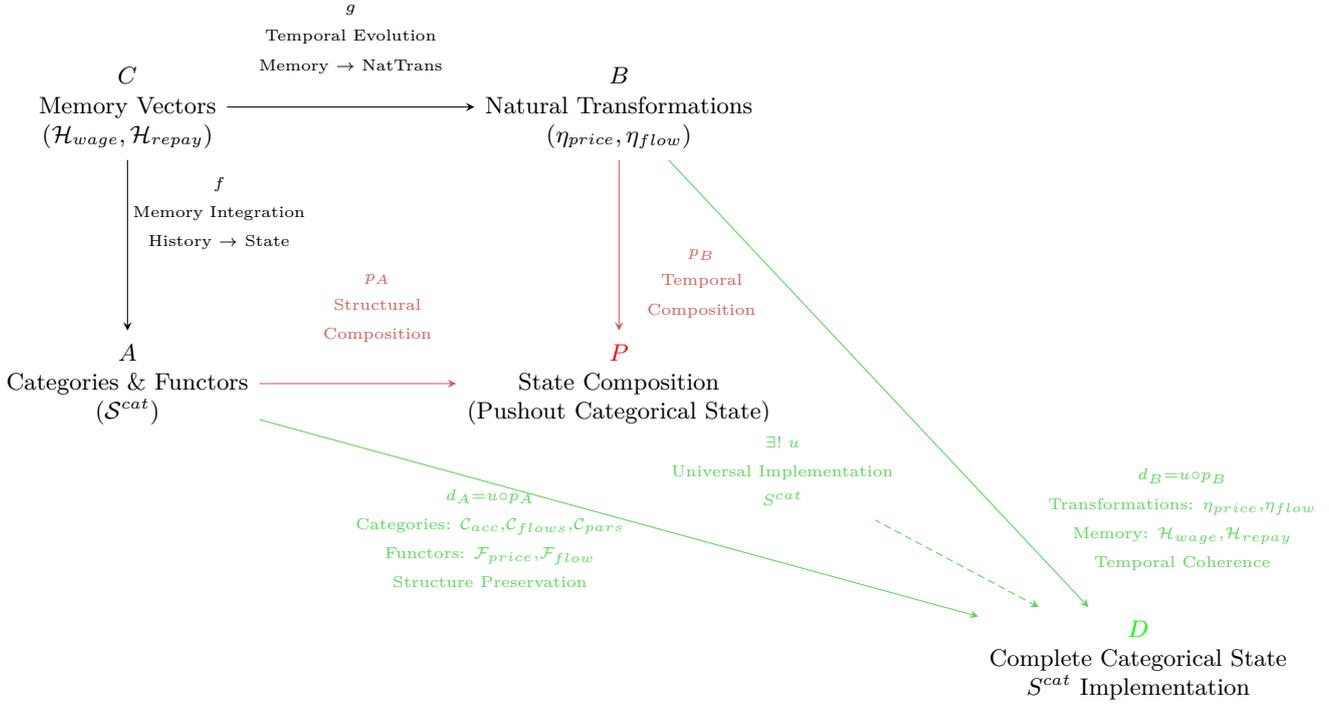
\begin{figure}[htbp]
    \centering
    {\footnotesize
    \begin{tikzcd}[column sep=8em, row sep=7em, arrows={-stealth}]
    {\begin{matrix} C \\ \text{Memory Vectors} \\ (\mathcal{H}_{wage}, \mathcal{H}_{repay}) \end{matrix}}
    & {\begin{matrix} B \\ \text{Natural Transformations} \\ (\eta_{price}, \eta_{flow}) \end{matrix}} \\
    {\begin{matrix} A \\ \text{Categories \& Functors} \\ (\mathcal{S}^{cat}) \end{matrix}} 
    & {\begin{matrix} \red{P} \\ \text{State Composition} \\ (\text{Pushout Categorical State}) \end{matrix}} & \\
    && {\begin{matrix} \green{D} \\ \text{Complete Categorical State} \\ S^{cat} \text{ Implementation} \end{matrix}} & \\
    \arrow["{\begin{matrix} f \\ \text{Memory Integration} \\ \text{History } \to \text{ State} \end{matrix}}"{right}, pos=0.3, from=1-1, to=2-1]
    \arrow["{\begin{matrix} g \\ \text{Temporal Evolution} \\ \text{Memory } \to \text{ NatTrans} \end{matrix}}"{above=0.3cm}, pos=0.5, from=1-1, to=1-2]
    \arrow["{\begin{matrix} p_A \\ \text{Structural} \\ \text{Composition} \end{matrix}}"{above=0.4cm}, pos=0.6, color={rgb,255:red,214;green,92;blue,92}, from=2-1, to=2-2]
    \arrow["{\begin{matrix} p_B \\ \text{Temporal} \\ \text{Composition} \end{matrix}}"{right=0.3cm}, pos=0.7, color={rgb,255:red,214;green,92;blue,92}, from=1-2, to=2-2]
    \arrow["{\begin{matrix} d_A = u \circ p_A \\ \text{Categories: } \mathcal{C}_{acc}, \mathcal{C}_{flows}, \mathcal{C}_{pars} \\ \text{Functors: } \mathcal{F}_{price}, \mathcal{F}_{flow} \\ \text{Structure Preservation} \end{matrix}}"{left=0.8cm}, pos=0.6, color={rgb,255:red,92;green,214;blue,92}, from=2-1, to=3-3]
    \arrow["{\begin{matrix} d_B = u \circ p_B \\ \text{Transformations: } \eta_{price}, \eta_{flow} \\ \text{Memory: } \mathcal{H}_{wage}, \mathcal{H}_{repay} \\ \text{Temporal Coherence} \end{matrix}}"{right=0.5cm}, pos=0.8, color={rgb,255:red,92;green,214;blue,92}, from=1-2, to=3-3]
    \arrow["{\begin{matrix} \exists ! \;u \\ \text{Universal Implementation} \\ S^{cat} \end{matrix}}"{description}, pos=0.2, color={rgb,255:red,92;green,214;blue,92}, dashed, from=2-2, to=3-3]
    \end{tikzcd}
    }
    \vspace{-2.5cm}
    \caption{Categorical State Composition Pushout: The categorical state pushout implements the universal composition of all categorical components into the complete economic simulation state $S^{cat}$, ensuring structural preservation and temporal coherence through the unique universal morphism $u$.}
    \label{l}
    \end{figure}
    
    These categorical constructions implement the entire economic simulation as categorical computations.
    Pullbacks validate transaction constraints through universal properties,
    while pushouts compute economic transitions through flow aggregation.
    The categorical machinery becomes from a typing system to the actual computational engine.
    Just like a compiler that derives an implementation for a specification in categorical types.

    In the categorical program we implement the natural transformations as actual code and not annotations of recursive calculation of the numbers 
    with wrapped around helper functions that generate the categories as Julia \texttt{struct}.
    The function \texttt{add\_mapping!} and \texttt{apply\_transformation} are implemented by the categorical program.
    In Figure~\ref{q} we see the time series plots of the categorically typed simulation to be the same as in Figure~(\ref{p}) of the recursive program.
    
\end{definition}

% --------------------------------------------------------------------------------------------------------------------------------------------------------------
%\clearpage
\subsubsection{Categorical Program}

The categorical simulation extends beyond just typing the recursive program as annotations by categorical types.
We use categorical structures as the actual computational machinery. 
While the categorically typed recursive program uses categorical types but still computes through iterations, 
the categorical simulation performs all computations through natural transformations between functors.

This opens the door to coalgebraic treatments or formally distinguish between theory and model through homomorphisms as in Tarski's logic and
we can use Lawvere's formulation as categories for theories, functors for models and natural transformations for homomorphisms in the econometrics of the Lucas Critique.
For an epistemological stance of the concepts {\it theory} and {\it model} in economics see \cite{Klein1998}.
A Lawvere inspired categorical approach might resolve issues of the so called {\it model communism} 
in rational expectations as locally and globally consistent theories and models of local or global agents.
Do the agents in the theory share the knowledge about the structures of the economy with the observing econometricians?
The observer versus observed duality of equipping agents in economic theories with estimation capabilities for estimating parameters from data might involve a gradual type of agents.
The more the agent observes of the full mathematical state of the system the more he is observer, economist or econometrician.
Econometric models are to be formulated compositionally showing how to use the informations of the subsystems in the information aggregation for the econometric informations 
about the whole macroeconomic system.
The agent's view is a functor and natural transformations tell the story how the observer and the observed interact.
However, a natural transformation might be not enough to represent the dynamically changing situation. 
The situation is more complex structured as the quantum effect of social systems where the observed system also changes the observer, he learns, and again we need endogenous econometricians.
The observers point of view changes are concept formation or colimit constructions, that manifest themselves in the memory of the agents, like now in the memory of the whole macroeconomy.
Theory becomes an infinitely composable structure of structural econometric theories and
of the real world implementation of finite models of data structures, computing machineries and information processors.
Theory is for defining models and improving them by learning.
However, it is too early to speculate how the categorical machinery and tool set can be used in old standing issues of the epistemology of economics.
In this paper, we want to categorify the seemingly simple recursive macroeconomic sectoral consistent model of MoMaT with proper types and implementation.
The rest is for future work and discussion and developments.

In the categorical set up each booking is represented as a natural transformation $\eta: \mathcal{F} \Rightarrow \mathcal{G}$ 
between functors mapping the account categories to updated states.
The components of these natural transformations encode the actual monetary flows and account updates. 
This allows the entire simulation to run by categorical composition of natural transformations, 
with pullbacks validating constraints and pushouts computing the economic transitions. 
The categorical machinery becomes from being a typing system to the specification of the actual implementation of the computations,
making the mathematical structure and the program for execution identical.
Category theory can serve not just as a framework for theory but as an actual implementation methodology for economic simulations and econometric estimations.
\begin{figure}[t!]
    \centering
    \includegraphics[width=0.8\textwidth]{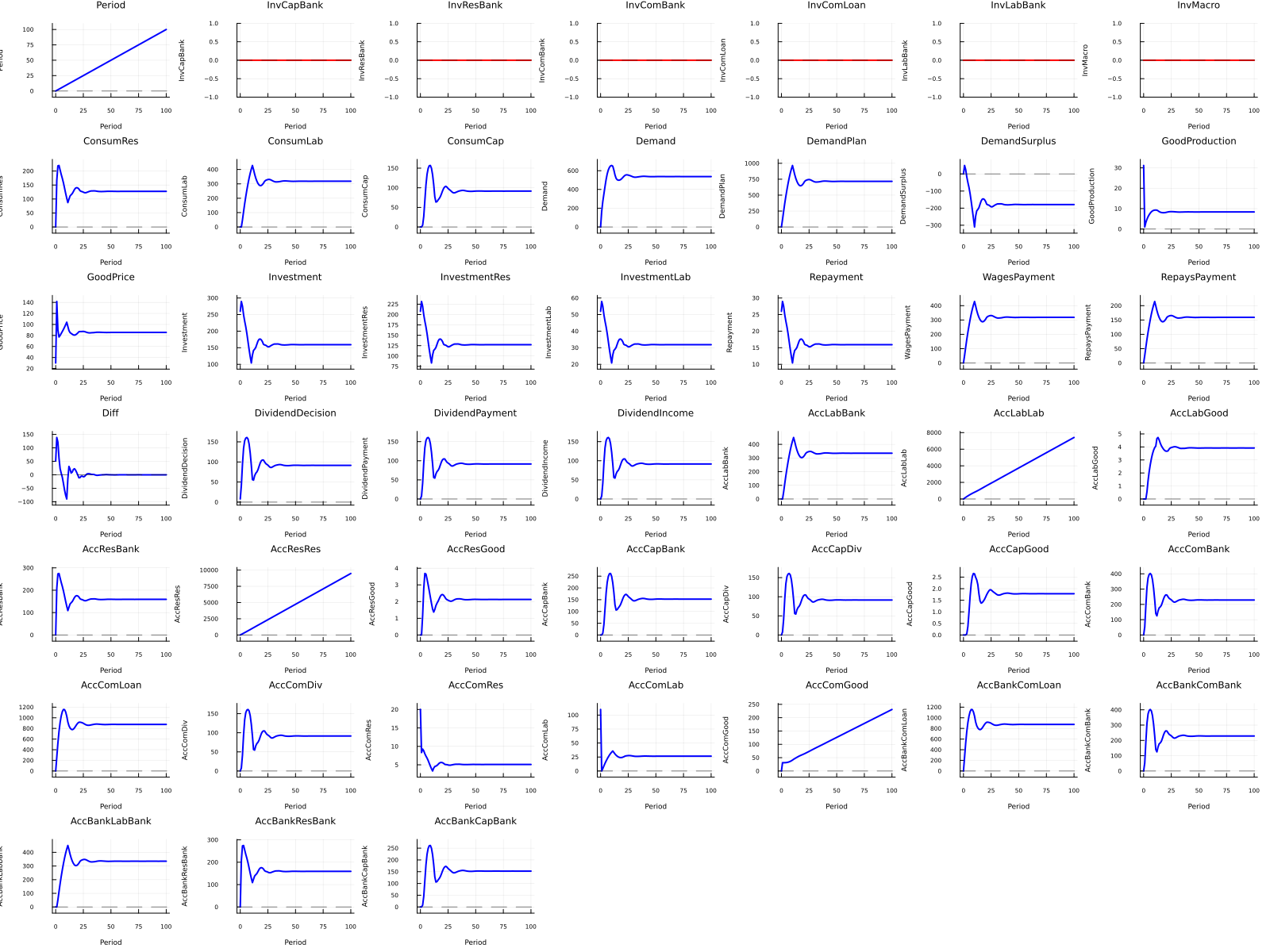}
    \caption{Time series plots of variables in the categorical simulation}
    \label{r}
\end{figure}
In Figure~\ref{r} we see the time series plots of the categorical simulation and that all figures, 
Figure~(\ref{p}), (\ref{q}) and (\ref{r}), all show the same results.

% -----------------------------------------------------------------------------------------------------------------------------------------------------------------------------------
\subsection{Consumption Decisions}

%\clearpage
\subsubsection{Consumption}

\begin{definition}[Consumption]
The consumption functor $\mathcal{F}_{cons}: \mathcal{C}_{acc} \rightarrow \mathcal{C}_{flow}$ transforms account balances to consumption flows:
\begin{align}
\mathcal{F}_{cons}(Res^{Bank}) &= Res^{Bank} \times \rho_r\\
\mathcal{F}_{cons}(Lab^{Bank}) &= Lab^{Bank} \times \rho_l\\
\mathcal{F}_{cons}(Cap^{Bank}) &= Cap^{Bank} \times \rho_c
\end{align}

\begin{align}
%               && \text{Type}                      &                                & \text{Morphism} \\ \hline
F_{ConsumRes} &:& \mathbb{R}_{\geq 0} \times [0,1] &\rightarrow \mathbb{R}_{\geq 0} & F_{ConsumRes}(x, \rho_r) &= x \times \rho_r\\
F_{ConsumLab} &:& \mathbb{R}_{\geq 0} \times [0,1] &\rightarrow \mathbb{R}_{\geq 0} & F_{ConsumLab}(x, \rho_l) &= x \times \rho_l\\
F_{ConsumCap} &:& \mathbb{R}_{\geq 0} \times [0,1] &\rightarrow \mathbb{R}_{\geq 0} & F_{ConsumCap}(x, \rho_c) &= x \times \rho_c
\end{align}
\hfill$\blacksquare$
\end{definition}

Evolution across \textbf{periods 0, 1, 2} with $\rho_r = 0.8$, $\rho_l = 0.95$, $\rho_c = 0.6$.

\begin{align}
\text{ConsumRes}_0 &= 0.0 \times 0.8 = 0.0 \nonumber\\
\text{ConsumRes}_1 &= 208.0 \times 0.8 = 166.4 \nonumber\\
\text{ConsumRes}_2 &= 273.19 \times 0.8 = 218.55 \nonumber
\end{align}

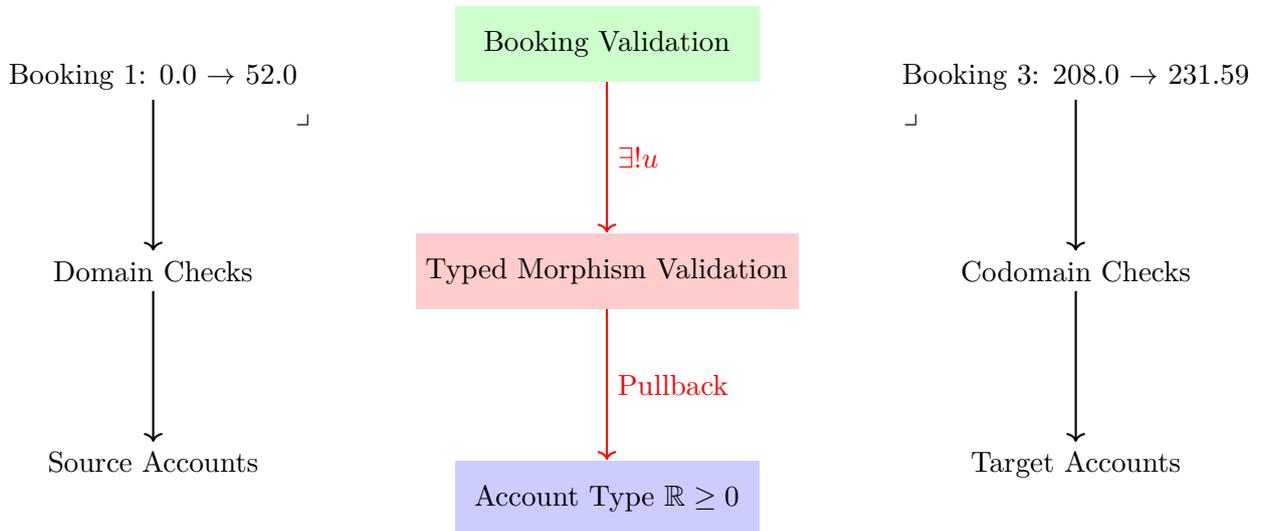
\begin{figure}[htbp]
\centering
\tikzset{
stringdiagram/.style={
baseline=(current bounding box.center),
every node/.style={draw=none},
every path/.style={thick}
}
}

\begin{tikzpicture}[stringdiagram]
\begin{scope}[shift={(0,0)}]
\node at (1.5,5.5) {$\mathcal{C}_{\text{BookingValidation}}$};
\node[draw, rectangle, minimum width=1.8cm, minimum height=0.8cm, fill=green!20] (validation) at (1.0,4.5) {Validation};
\node at (-0.5,3.5) {$\mathcal{C}_{\text{TypedMorphism}}$};
\node[draw, rectangle, minimum width=1.8cm, minimum height=0.8cm, fill=red!20] (morphism) at (1.0,2.5) {Morphism};
\node at (-0.5,1.5) {$\mathcal{C}_{\text{AccountType}}$};
\node[draw, rectangle, minimum width=1.8cm, minimum height=0.8cm, fill=blue!20] (acctype) at (1.0,0.5) {$\mathbb{R}_{\geq 0}$};
\node at (8.5,5.5) {$\mathcal{C}_{\text{Pullback}}$};
\node[draw, rectangle, minimum width=1.8cm, minimum height=0.8cm, fill=purple!20] (pullback) at (8.5,4.5) {$\exists ! u$};
\node at (7.0,3.5) {$\mathcal{C}_{\text{Conservation}}$};
\node[draw, rectangle, minimum width=1.8cm, minimum height=0.8cm, fill=orange!20] (conservation) at (8.5,2.5) {Conservation};
\node at (7.0,1.5) {$\mathcal{C}_{\text{Balance}}$};
\node[draw, rectangle, minimum width=1.8cm, minimum height=0.8cm, fill=yellow!20] (balance) at (8.5,0.5) {Balance};
\draw[->, thick] (validation) -- (morphism) node[midway,right=0.0cm] {$\pi_1$};
\draw[->, thick] (morphism) -- (acctype) node[midway,right=0.0cm] {$\pi_2$};
\draw[->, thick] (pullback) -- (conservation) node[midway,right=0.0cm] {$\phi_1$};
\draw[->, thick] (conservation) -- (balance) node[midway,right=0.0cm] {$\phi_2$};
\draw[->, thick, red] (validation) -- (pullback) node[midway,above] {$\eta_{\text{validation}}$};
\draw[->, thick, red] (morphism) -- (conservation) node[midway,above] {$\eta_{\text{morphism}}$};
\draw[->, thick, red] (acctype) -- (balance) node[midway,above] {$\eta_{\text{type}}$};
\node at (0.3,3.9) {$\lrcorner$};
\node at (7.8,3.9) {$\lrcorner$};
\node at (4.7,4.5) {$\otimes$};
\node at (4.7,2.5) {$\otimes$};
\node at (4.7,0.5) {$\otimes$};
\end{scope}
\end{tikzpicture}
\caption{Commutative Diagram: Booking Validation Pullback Flow}
\label{s}
\end{figure}

\begin{figure}[htbp]
\centering
\begin{tikzpicture}[node distance=2cm]
\node[rectangle, fill=green!20, minimum width=4cm, minimum height=1cm] (decision) {Consumption Decision};
\node[rectangle, fill=red!20, minimum width=4cm, minimum height=1cm, below=of decision] (function) {$\mathcal{F}_{cons}$: Balance $\times$ Ratio};
\node[left=of function] (mult1) {Multiplication};
\node[right=of function] (mult2) {Validation};
\node[above=of mult1] (res) {ConsumRes: 218.55};
\node[above=of mult2] (lab) {ConsumLab: 49.4};
\node[rectangle, fill=blue!20, minimum width=4cm, minimum height=1cm, below=of function] (accounts) {Accounts $\otimes$ Ratios};
\node[below=of mult1] (balance) {$\mathbb{R}\geq 0$ Balances};
\node[below=of mult2] (ratio) {[0,1] Ratios};
\draw[->, thick, red] (decision) -- (function) node[midway,right] {$\exists ! u$};
\draw[->, thick, red] (function) -- (accounts) node[midway,right] {Pullback};
\draw[->, thick] (res) -- (mult1);
\draw[->, thick] (lab) -- (mult2);
\draw[->, thick] (mult1) -- (balance);
\draw[->, thick] (mult2) -- (ratio);
\node at (-4,-1) {$\lrcorner$};
\node at (4,-1) {$\lrcorner$};
\node at (0,-4) {$\otimes$};
\end{tikzpicture}
\caption{Commutative Diagram: Consumption Function Pullback Flow}
\label{t}
\end{figure}

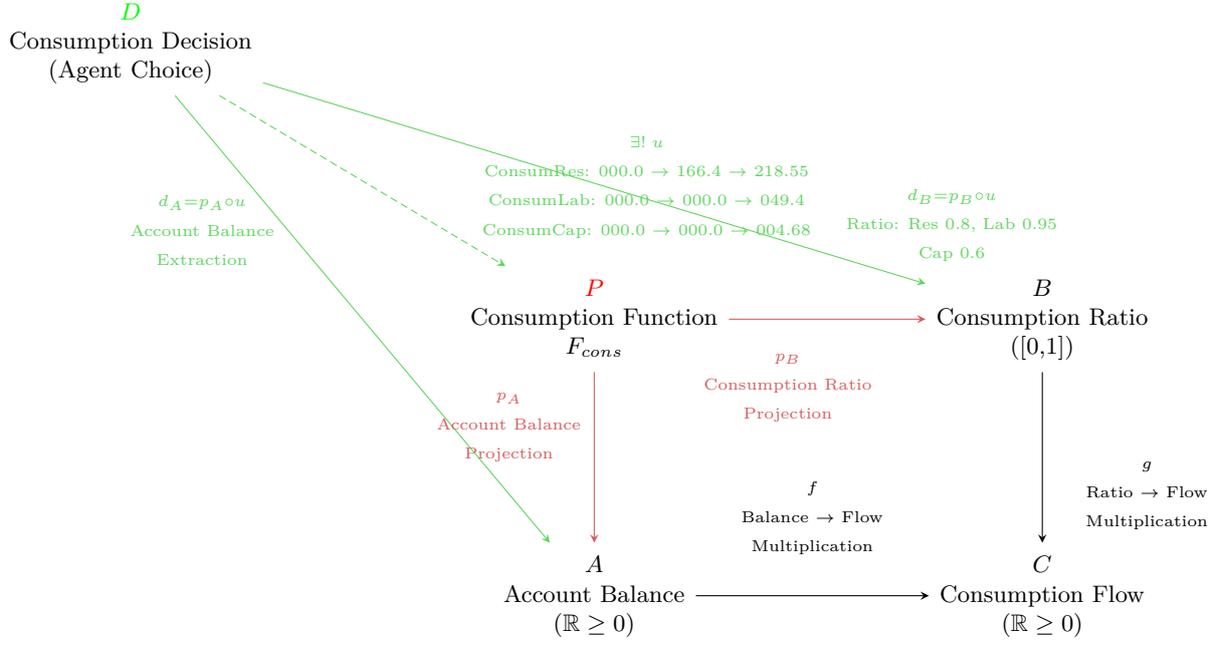
\begin{figure}[htbp]
    \centering
    {\footnotesize
    \begin{tikzcd}[column sep=8em, row sep=7em, arrows={-stealth}]
    {\begin{matrix} \green{D} \\ \text{Consumption Decision} \\ \text{(Agent Choice)} \end{matrix}} \\
    & {\begin{matrix} \red{P} \\ \text{Consumption Function} \\ F_{cons} \end{matrix}} & {\begin{matrix} B \\ \text{Consumption Ratio} \\ \text{([0,1])} \end{matrix}} \\
    & {\begin{matrix} A \\ \text{Account Balance} \\ \text{($\mathbb{R}\geq 0$)} \end{matrix}} & {\begin{matrix} C \\ \text{Consumption Flow} \\ \text{($\mathbb{R}\geq 0$)} \end{matrix}} \\
    \arrow["{\begin{matrix} \exists ! \; u \\ \text{ConsumRes: 000.0 $\to$ 166.4 $\to$ 218.55} \\ \text{ConsumLab: 000.0 $\to$ 000.0 $\to$ 049.4} \\ \text{ConsumCap: 000.0 $\to$ 000.0 $\to$ 004.68} \end{matrix}}", pos=0.9, color={rgb,255:red,92;green,214;blue,92}, dashed, from=1-1, to=2-2]
    \arrow["{\begin{matrix} d_A = p_A \circ u \\ \text{Account Balance} \\ \text{Extraction} \end{matrix}}"'{left=0.1cm}, pos=0.3, color={rgb,255:red,92;green,214;blue,92}, from=1-1, to=3-2]
    \arrow["{\begin{matrix} d_B = p_B \circ u \\ \text{Ratio: Res 0.8, Lab 0.95} \\ \text{Cap 0.6} \end{matrix}}"{right=1.5cm}, pos=0.7, color={rgb,255:red,92;green,214;blue,92}, from=1-1, to=2-3]
    \arrow["{\begin{matrix} p_A \\ \text{Account Balance} \\ \text{Projection} \end{matrix}}"'{left=0.1cm}, pos=0.3, color={rgb,255:red,214;green,92;blue,92}, from=2-2, to=3-2]
    \arrow["{\begin{matrix} p_B \\ \text{Consumption Ratio} \\ \text{Projection} \end{matrix}}"{below=0.2cm}, pos=0.3, color={rgb,255:red,214;green,92;blue,92}, from=2-2, to=2-3]
    \arrow["{\begin{matrix} f \\ \text{Balance $\to$ Flow} \\ \text{Multiplication} \end{matrix}}"'{above=0.4cm}, from=3-2, to=3-3]
    \arrow["{\begin{matrix} g \\ \text{Ratio $\to$ Flow} \\ \text{Multiplication} \end{matrix}}"{right=0.5cm}, pos=0.7, from=2-3, to=3-3]
    \end{tikzcd}
    }
    \vspace{-2.5cm}
    \caption{Consumption Function Pullback: The consumption decision pullback validates the multiplication of account balances with consumption ratios, ensuring proper economic behavior across agents.}
\label{u}
\end{figure}
        
\textbf{Economics: Consumer Decision-Making and Consumption Choice:}
Figure~(\ref{s}) shows the natural transformation between booking validation layers and their pullback universal property, ensuring type-safe accounting operations.
The consumption pullback captures the microeconomic foundation of consumer choice theory through categorical structures. 
The typed morphism $F_{\text{cons}}: \mathbb{R}_{\geq 0} \times [0,1] \rightarrow \mathbb{R}_{\geq 0}$ 
transforms account balances and consumption propensities into actual consumption decisions, implicitely respecting budget constraints and preference structures.

Economically, this represents how consumers allocate their values and wealth across goods (if we were to model several) based on their consumption ratios. 
The pullback ensures that consumption decisions are consistent with available resources: ConsumRes (218.55) 
emerges from combining resource account balances with $\rho_r$ (0.8), while ConsumLab (49.4) reflects labor's consumption with $\rho_l$ (0.95). 
The categorical structure guarantees that total consumption respects budget constraints.

The diagram reveals the tensor product structure $\text{Accounts} \otimes \text{Ratios}$ at the base layer, 
showing how consumption decisions emerge from the categorical product of available wealth and consumption preferences. 
The pullback corners ensure that individual consumption choices aggregate consistently into overall demand patterns. 
This categorical approach to consumption theory provides a rigorous foundation for understanding how microeconomic consumer decisions compose into macroeconomic demand structures, 
bridging individual choice theory with aggregate economic outcomes.

% -----------------------------------------------------------------------------------------------------------------------------------------------------------------------------------
%\clearpage
\subsubsection{Aggregate Demand}

\begin{definition}[Demand]
The demand pushout $\mathcal{D} = \text{ConsumRes} \sqcup \text{ConsumLab} \sqcup \text{ConsumCap}$ computes total demand:
$$\text{Demand} = \text{ConsumRes} + \text{ConsumLab} + \text{ConsumCap}$$
\begin{align}
\text{Pushout}_{\text{Demand}} &:& \mathbb{R}_{\geq 0} \times \mathbb{R}_{\geq 0} \times \mathbb{R}_{\geq 0} &\rightarrow \mathbb{R}_{\geq 0} & \text{Pushout}_{\text{Demand}}(c_r, c_l, c_c) &= c_r + c_l + c_c
\end{align}
\hfill$\blacksquare$
\end{definition}

Evolution across \textbf{periods 0, 1, 2}:
\begin{align}
Demand_0 &= 0.0 + 0.0 + 0.0 = 0.0 \nonumber\\
Demand_1 &= 166.4 + 0.0 + 0.0 = 166.4 \nonumber\\
Demand_2 &= 218.55 + 49.4 + 4.68 = 272.63 \nonumber
\end{align}

\begin{figure}[htbp]
\centering
\tikzset{
stringdiagram/.style={
baseline=(current bounding box.center),
every node/.style={draw=none},
every path/.style={thick}
}
}
\begin{tikzpicture}[stringdiagram]
\begin{scope}[shift={(0,0)}]
\node at (1.5,4.5) {$\mathcal{C}_{\text{Individual}}$};
\node[draw, rectangle, minimum width=2.0cm, minimum height=0.8cm, fill=yellow!20] (individual) at (1.0,3.5) {Consumption};
\node at (0.0,2.5) {$\mathcal{C}_{\text{Aggregation}}$};
\node[draw, rectangle, minimum width=2.0cm, minimum height=0.8cm, fill=red!20] (aggregation) at (1.0,1.5) {Demand Agg};
\node at (0.0,0.5) {$\mathcal{C}_{\text{Market}}$};
\node[draw, rectangle, minimum width=2.0cm, minimum height=0.8cm, fill=green!20] (market) at (1.0,-0.5) {Market Demand};
\node at (6.0,4.5) {$\mathcal{C}_{\text{Resource}}$};
\node[draw, rectangle, minimum width=2.0cm, minimum height=0.8cm, fill=yellow!20] (consumres) at (5.5,3.5) {218.55};
\node at (4.5,2.5) {$\mathcal{C}_{\text{Summing}}$};
\node[draw, rectangle, minimum width=2.0cm, minimum height=0.8cm, fill=red!20] (summing) at (5.5,1.5) {Sum Agg};
\node at (4.5,0.5) {$\mathcal{C}_{\text{Production}}$};
\node[draw, rectangle, minimum width=2.0cm, minimum height=0.8cm, fill=green!20] (production) at (5.5,-0.5) {Prod Signal};
\node at (11.5,4.5) {$\mathcal{C}_{\text{Labor}}$};
\node[draw, rectangle, minimum width=2.0cm, minimum height=0.8cm, fill=yellow!20] (consumlab) at (11.0,3.5) {49.4};
\node at (10.0,2.5) {$\mathcal{C}_{\text{Signal}}$};
\node[draw, rectangle, minimum width=2.0cm, minimum height=0.8cm, fill=red!20] (signal) at (11.0,1.5) {Market Signal};
\node at (10.0,0.5) {$\mathcal{C}_{\text{Pricing}}$};
\node[draw, rectangle, minimum width=2.0cm, minimum height=0.8cm, fill=green!20] (pricing) at (11.0,-0.5) {Price Signal};
\node at (16.5,4.5) {$\mathcal{C}_{\text{Total}}$};
\node[draw, rectangle, minimum width=2.0cm, minimum height=0.8cm, fill=yellow!20] (total) at (16.0,3.5) {272.63};
\node at (15.0,2.5) {$\mathcal{C}_{\text{Universal}}$};
\node[draw, rectangle, minimum width=2.0cm, minimum height=0.8cm, fill=red!20] (universal) at (16.0,1.5) {$\exists ! u$};
\node at (15.0,0.5) {$\mathcal{C}_{\text{Equilibrium}}$};
\node[draw, rectangle, minimum width=2.0cm, minimum height=0.8cm, fill=green!20] (equilibrium) at (16.0,-0.5) {Equilibrium};
\draw[->, thick] (individual) -- (aggregation) node[midway,right=0.2cm] {$\pi_1$};
\draw[->, thick] (aggregation) -- (market) node[midway,right=0.2cm] {$\pi_2$};
\draw[->, thick] (consumres) -- (summing) node[midway,right=0.2cm] {$\phi_1$};
\draw[->, thick] (summing) -- (production) node[midway,right=0.2cm] {$\phi_2$};
\draw[->, thick] (consumlab) -- (signal) node[midway,right=0.2cm] {$\psi_1$};
\draw[->, thick] (signal) -- (pricing) node[midway,right=0.2cm] {$\psi_2$};
\draw[->, thick] (total) -- (universal) node[midway,right=0.2cm] {$\zeta_1$};
\draw[->, thick] (universal) -- (equilibrium) node[midway,right=0.2cm] {$\zeta_2$};
\draw[->, thick, red] (individual) -- (consumres) node[midway,above] {$\eta_{\text{res}}$};
\draw[->, thick, red] (aggregation) -- (summing) node[midway,above] {$\eta_{\text{sum}}$};
\draw[->, thick, red] (market) -- (production) node[midway,above] {$\eta_{\text{prod}}$};
\draw[->, thick, red] (consumres) -- (consumlab) node[midway,above] {$\eta_{\text{lab}}$};
\draw[->, thick, red] (summing) -- (signal) node[midway,above] {$\eta_{\text{sig}}$};
\draw[->, thick, red] (production) -- (pricing) node[midway,above] {$\eta_{\text{price}}$};
\draw[->, thick, red] (consumres) -- (consumlab) node[midway,above] {$\eta_{\text{lab}}$};
\draw[->, thick, red] (consumlab) -- (total) node[midway,above] {$\eta_{\text{total}}$};
\draw[->, thick, red] (signal) -- (universal) node[midway,above] {$\eta_{\text{univ}}$};
\draw[->, thick, red] (pricing) -- (equilibrium) node[midway,above] {$\eta_{\text{eq}}$};
\node at (0.3,2.9) {$\ulcorner$};
\node at (6.3,2.9) {$\ulcorner$};
\node at (11.8,2.9) {$\ulcorner$};
\node at (16.8,2.9) {$\ulcorner$};
\node at (3.5,3.5) {$\otimes$};
\node at (3.5,1.5) {$\otimes$};
\node at (3.5,-0.5) {$\otimes$};
\node at (8.7,3.5) {$\otimes$};
\node at (8.7,1.5) {$\otimes$};
\node at (8.7,-0.5) {$\otimes$};        
\node at (14.2,3.5) {$\otimes$};
\node at (14.2,1.5) {$\otimes$};
\node at (14.2,-0.5) {$\otimes$};
\draw[->, thick, purple, dashed] (individual) to[bend left=20] (total) node at (8.6,5.5) {$\exists ! u$};
\end{scope}
\end{tikzpicture}
\caption{Commutative Diagram: Demand Aggregation Pushout Structure}
\label{v}
\end{figure}
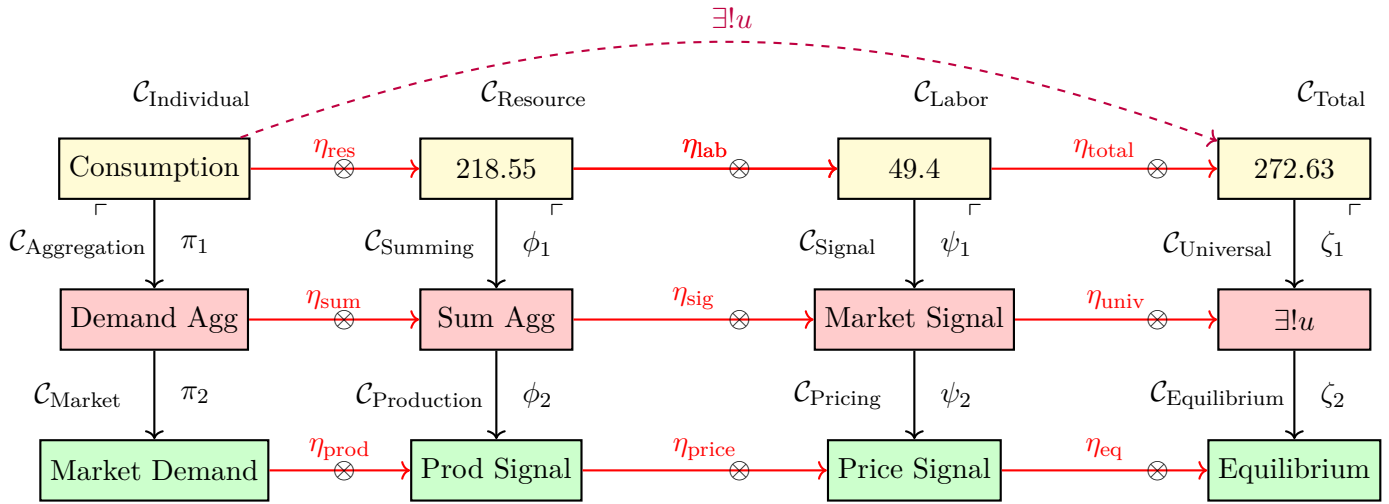

\begin{figure}[htbp]
\centering
\begin{tikzpicture}[node distance=2cm]
\node[rectangle, fill=yellow!20, minimum width=4cm, minimum height=1cm] (individual) {Individual Consumption};
\node[rectangle, fill=red!20, minimum width=4cm, minimum height=1cm, below=of individual] (aggregation) {Demand Aggregation};
\node[left=of aggregation] (summing) {Sum Aggregation};
\node[right=of aggregation] (signal) {Market Signal};
\node[above=of summing] (consumres) {ConsumRes: 218.55};
\node[above=of signal] (consumlab) {ConsumLab: 49.4};
\node[rectangle, fill=green!20, minimum width=4cm, minimum height=1cm, below=of aggregation] (market) {Market Demand: 272.63};
\node[below=of summing] (production) {Production Signal};
\node[below=of signal] (pricing) {Price Signal};
\draw[->, thick, red] (individual) -- (aggregation) node[midway,right] {Pushout};
\draw[->, thick, red] (aggregation) -- (market) node[midway,right] {$\exists ! u$};
\draw[->, thick] (consumres) -- (summing);
\draw[->, thick] (consumlab) -- (signal);
\draw[->, thick] (summing) -- (production);
\draw[->, thick] (signal) -- (pricing);
\node at (-4,-2) {$\ulcorner$};
\node at (4,-2) {$\ulcorner$};
\end{tikzpicture}
\caption{Commutative Diagram: Demand Aggregation Pushout Flow}
\label{w}
\end{figure}
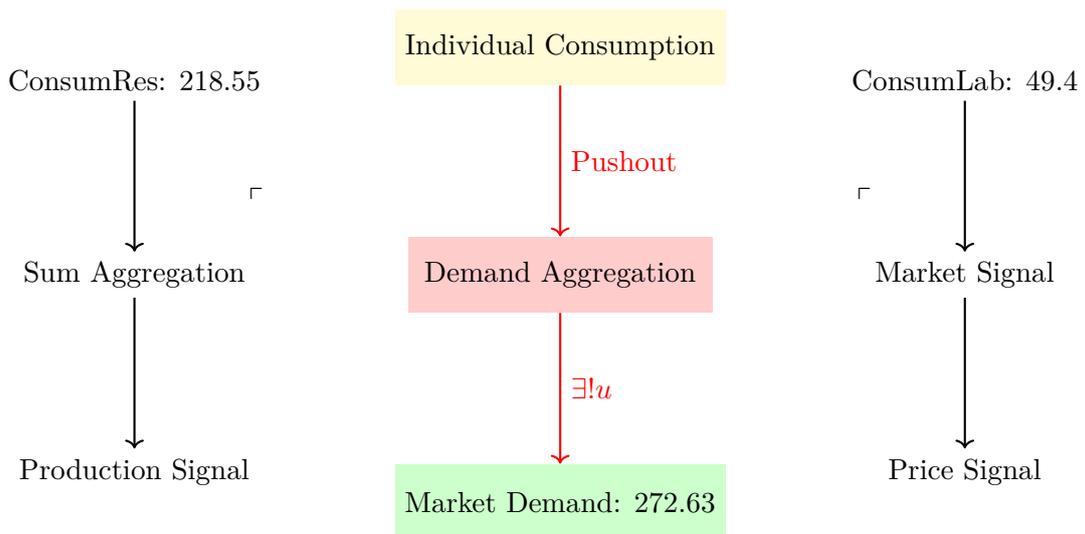

\begin{figure}[htbp]
    \centering
    {\footnotesize
    \begin{tikzcd}[column sep=8em, row sep=7em, arrows={-stealth}]
    {\begin{matrix} C \\ \text{Common Base} \\ \mathbb{R}\geq 0 \end{matrix}} & {\begin{matrix} B \\ \text{Labor Consumption} \\ \text{(ConsumLab)} \end{matrix}} \\
    {\begin{matrix} A \\ \text{Resource Consumption} \\ \text{(ConsumRes)} \end{matrix}} & {\begin{matrix} \red{P} \\ \text{Demand Aggregation} \\ \text{(Total Demand)} \end{matrix}} & \\
    && {\begin{matrix} \green{D} \\ \text{Market Demand} \\ \text{(Economic Signal)} \end{matrix}} & \\
    \arrow["{\begin{matrix} f \\ \text{ConsumRes Flow} \\ \text{0.0 $\to$ 166.4 $\to$ 218.55} \end{matrix}}"{right}, pos=0.2, from=1-1, to=2-1]
    \arrow["{\begin{matrix} g \\ \text{ConsumLab Flow} \\ \text{0.0 $\to$ 0.0 $\to$ 49.4} \end{matrix}}"{above=0.3cm}, pos=0.5, from=1-1, to=1-2]
    \arrow["{\begin{matrix} p_A \\ \text{Resource Consumption} \\ \text{Aggregation} \end{matrix}}"{above=0.4cm}, pos=0.6, color={rgb,255:red,214;green,92;blue,92}, from=2-1, to=2-2]
    \arrow["{\begin{matrix} p_B \\ \text{Labor Consumption} \\ \text{Aggregation} \end{matrix}}"{right=0.3cm}, pos=0.7, color={rgb,255:red,214;green,92;blue,92}, from=1-2, to=2-2]
    \arrow["{\begin{matrix} d_A \\ \text{Demand values} \\ \text{0.0, 166.4, 272.63} \end{matrix}}"{left=0.8cm}, pos=0.3, color={rgb,255:red,92;green,214;blue,92}, from=2-1, to=3-3]
    \arrow["{\begin{matrix} d_B \\ \text{Demand Signal} \\ \text{Aggregation} \end{matrix}}"{right=0.5cm}, pos=0.7, color={rgb,255:red,92;green,214;blue,92}, from=1-2, to=3-3]
    \arrow["{\begin{matrix} \exists ! \; u \\ \text{Sum Aggregation} \\ \text{Market Signal} \end{matrix}}"{description}, pos=0.3, color={rgb,255:red,92;green,214;blue,92}, dashed, from=2-2, to=3-3]
    \end{tikzcd}
    }
    \vspace{-2.5cm}
    \caption{Demand Aggregation Pushout: The demand pushout aggregates individual consumption flows (ConsumRes, ConsumLab, ConsumCap) into total market demand, providing the economic signal for production decisions.}
    \label{x}
\end{figure}
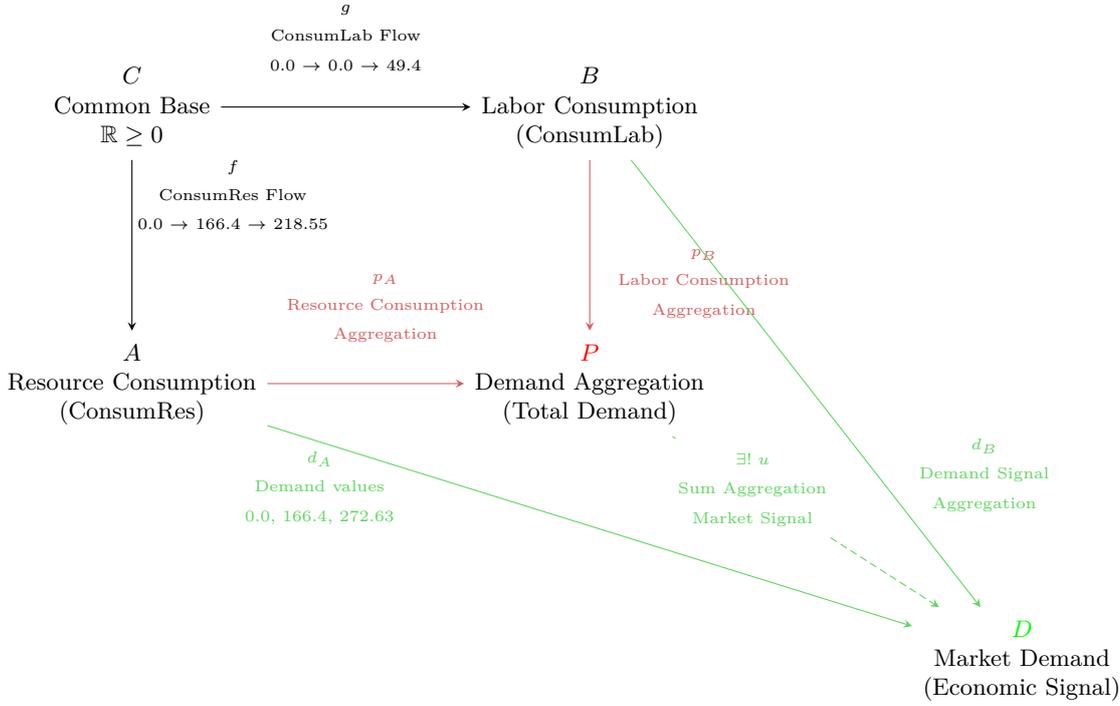
    
\textbf{Economics: Market Demand Formation and Aggregation Mechanisms}
Figure~(\ref{v}) shows the natural transformation sequence from individual consumption patterns (ConsumRes: 218.55, ConsumLab: 49.4) to market demand (272.63) through categorical pushout construction.
The demand aggregation pushout embodies the fundamental principle of market formation through categorical colimits. 
Individual consumption decisions by resource owners (ConsRes: 218.55) and labor (ConsumLab: 49.4) aggregate through the pushout universal property 
to create unified market demand (272.63), demonstrating how microeconomic choices compose into macroeconomic outcomes.

Economically, this captures the mechanism where individual consumption decisions, driven by personal utility maximisation, 
aggregate coherently into market signals that guide production decisions. 
The pushout structure ensures that the aggregation preserves economic information: the sum aggregation component maintains quantitative demand information, 
while the market signal component translates individual preferences into price relevant signals.

The three-layer structure of the diagram reveals the hierarchical nature of market formation: the yellow layer represents individual consumption choices, 
the red layer embodies the aggregation mechanism (demand aggregation), and the green layer shows the resulting market demand. 
The pushout corners mark where individual consumption morphisms compose into the universal market demand morphism. 
This categorical approach to demand aggregation provides mathematical foundations for understanding how decentralized individual decisions 
create coherent market outcomes without central coordination.
% -----------------------------------------------------------------------------------------------------------------------------------------------------------------------------------
\subsection{Production Decisions and Price Formation}
% -----------------------------------------------------------------------------------------------------------------------------------------------------------------------------------
%\clearpage
\subsubsection{Cobb-Douglas Production}

\begin{definition}[Production]
The production pushout aggregates labor and resource inputs according to a standard Cobb-Douglas function into the productive output of the economy:
$$\text{GoodProduction} = 1 + \alpha \times (Com^{Lab})^{\gamma} \times (Com^{Res})^{1-\gamma}$$

\begin{align}
\text{Pushout}_{\text{Production}}&:& \mathbb{R}_{\geq 0} \times \mathbb{R}_{\geq 0} \times \mathbb{R}_{> 0} \times (0,1) &\rightarrow \mathbb{R}_{> 0} &
\text{Pushout}_{\text{Production}}(L, R, \alpha, \gamma) &= 1 + \alpha \times L^{\gamma} \times R^{1-\gamma}
\end{align}
where $\alpha = 0.42$ and $\gamma = 0.75$.
\hfill$\blacksquare$
\end{definition}

Evolution across \textbf{periods 0, 1, 2}:
\begin{align}
GoodProduction_0 &= 1 + 0.42 \times 110.0^{0.75} \times 20.0^{0.25} = 31.17\nonumber \\
GoodProduction_1 &= 1 + 0.42 \times 0.0^{0.75} \times 8.32^{0.25} = 1.0\nonumber \\
GoodProduction_2 &= 1 + 0.42 \times 4.33^{0.75} \times 9.26^{0.25} = 3.20\nonumber
\end{align}

\begin{figure}[htbp]
\centering
\tikzset{
stringdiagram/.style={
baseline=(current bounding box.center),
every node/.style={draw=none},
every path/.style={thick}
}
}
\begin{tikzpicture}[stringdiagram]
\begin{scope}[shift={(0,0)}]
\node at (1.5,4.5) {$\mathcal{C}_{\text{Base}}$};
\node[draw, rectangle, minimum width=2.0cm, minimum height=0.8cm, fill=cyan!20] (base) at (1.0,3.5) {Constant=1};
\node at (0.0,2.5) {$\mathcal{C}_{\text{Function}}$};
\node[draw, rectangle, minimum width=2.0cm, minimum height=0.8cm, fill=red!20] (function) at (1.0,1.5) {Cobb-Douglas};
\node at (0.0,0.5) {$\mathcal{C}_{\text{Output}}$};
\node[draw, rectangle, minimum width=2.0cm, minimum height=0.8cm, fill=green!20] (output) at (1.0,-0.5) {Production};
\node at (6.5,4.5) {$\mathcal{C}_{\text{Labor}}$};
\node[draw, rectangle, minimum width=2.0cm, minimum height=0.8cm, fill=cyan!20] (labor) at (6.0,3.5) {110.0};
\node at (5.0,2.5) {$\mathcal{C}_{\text{LaborAgg}}$};
\node[draw, rectangle, minimum width=2.0cm, minimum height=0.8cm, fill=red!20] (laboragg) at (6.0,1.5) {$L^{0.75}$};
\node at (5.0,0.5) {$\mathcal{C}_{\text{LaborOut}}$};
\node[draw, rectangle, minimum width=2.0cm, minimum height=0.8cm, fill=green!20] (laborout) at (6.0,-0.5) {31.17};
\node at (11.0,4.5) {$\mathcal{C}_{\text{Resource}}$};
\node[draw, rectangle, minimum width=2.0cm, minimum height=0.8cm, fill=cyan!20] (resource) at (10.5,3.5) {20.0};
\node at (9.5,2.5) {$\mathcal{C}_{\text{ResourceAgg}}$};
\node[draw, rectangle, minimum width=2.0cm, minimum height=0.8cm, fill=red!20] (resourceagg) at (10.5,1.5) {$R^{0.25}$};
\node at (9.5,0.5) {$\mathcal{C}_{\text{ResourceOut}}$};
\node[draw, rectangle, minimum width=2.0cm, minimum height=0.8cm, fill=green!20] (resourceout) at (10.5,-0.5) {Scale: 0.42};
\node at (15.5,4.5) {$\mathcal{C}_{\text{Period1}}$};
\node[draw, rectangle, minimum width=2.0cm, minimum height=0.8cm, fill=cyan!20] (period1) at (15.0,3.5) {0.0 $\to$ 4.33};
\node at (14.0,2.5) {$\mathcal{C}_{\text{Universal}}$};
\node[draw, rectangle, minimum width=2.0cm, minimum height=0.8cm, fill=red!20] (universal) at (15.0,1.5) {$\exists ! u$};
\node at (14.0,0.5) {$\mathcal{C}_{\text{Evolution}}$};
\node[draw, rectangle, minimum width=2.0cm, minimum height=0.8cm, fill=green!20] (evolution) at (15.0,-0.5) {1.0 $\to$ 3.20};
\draw[->, thick] (base) -- (function) node[midway,right] {$\pi_1$};
\draw[->, thick] (function) -- (output) node[midway,right] {$\pi_2$};
\draw[->, thick] (labor) -- (laboragg) node[midway,right] {$\phi_1$};
\draw[->, thick] (laboragg) -- (laborout) node[midway,right] {$\phi_2$};
\draw[->, thick] (resource) -- (resourceagg) node[midway,right] {$\psi_1$};
\draw[->, thick] (resourceagg) -- (resourceout) node[midway,right] {$\psi_2$};
\draw[->, thick] (period1) -- (universal) node[midway,right] {$\zeta_1$};
\draw[->, thick] (universal) -- (evolution) node[midway,right] {$\zeta_2$};
\draw[->, thick, red] (base) -- (labor) node[midway,above] {$\eta_{\text{lab}}$};
\draw[->, thick, red] (function) -- (laboragg) node[midway,above] {$\eta_{\text{lagg}}$};
\draw[->, thick, red] (output) -- (laborout) node[midway,above] {$\eta_{\text{lout}}$};
\draw[->, thick, red] (labor) -- (resource) node[midway,above] {$\eta_{\text{res}}$};
\draw[->, thick, red] (laboragg) -- (resourceagg) node[midway,above] {$\eta_{\text{ragg}}$};
\draw[->, thick, red] (laborout) -- (resourceout) node[midway,above] {$\eta_{\text{rout}}$};
\draw[->, thick, red] (resource) -- (period1) node[midway,above] {$\eta_{\text{per1}}$};
\draw[->, thick, red] (resourceagg) -- (universal) node[midway,above] {$\eta_{\text{univ}}$};
\draw[->, thick, red] (resourceout) -- (evolution) node[midway,above] {$\eta_{\text{evol}}$};
\node at (0.3,2.8) {$\ulcorner$};
\node at (5.3,2.8) {$\ulcorner$};
\node at (9.8,2.8) {$\ulcorner$};
\node at (14.3,2.8) {$\ulcorner$};
\node at (3.5,3.5) {$\otimes$};
\node at (3.5,1.5) {$\otimes$};
\node at (3.5,-0.5) {$\otimes$};
\node at (8.2,3.5) {$\otimes$};
\node at (8.2,1.5) {$\otimes$};
\node at (8.2,-0.5) {$\otimes$};
\node at (12.7,3.5) {$\otimes$};
\node at (12.7,1.5) {$\otimes$};
\node at (12.7,-0.5) {$\otimes$};
\draw[->, thick, purple, dashed] (base) to[bend left=25] (period1) node at (8.5,6.0) {$1 + 0.42 \times L^{0.75} \times R^{0.25}$};
\end{scope}
\end{tikzpicture}
\caption{Commutative Diagram: Cobb-Douglas Production Pushout Structure}
\label{y}    
\end{figure}
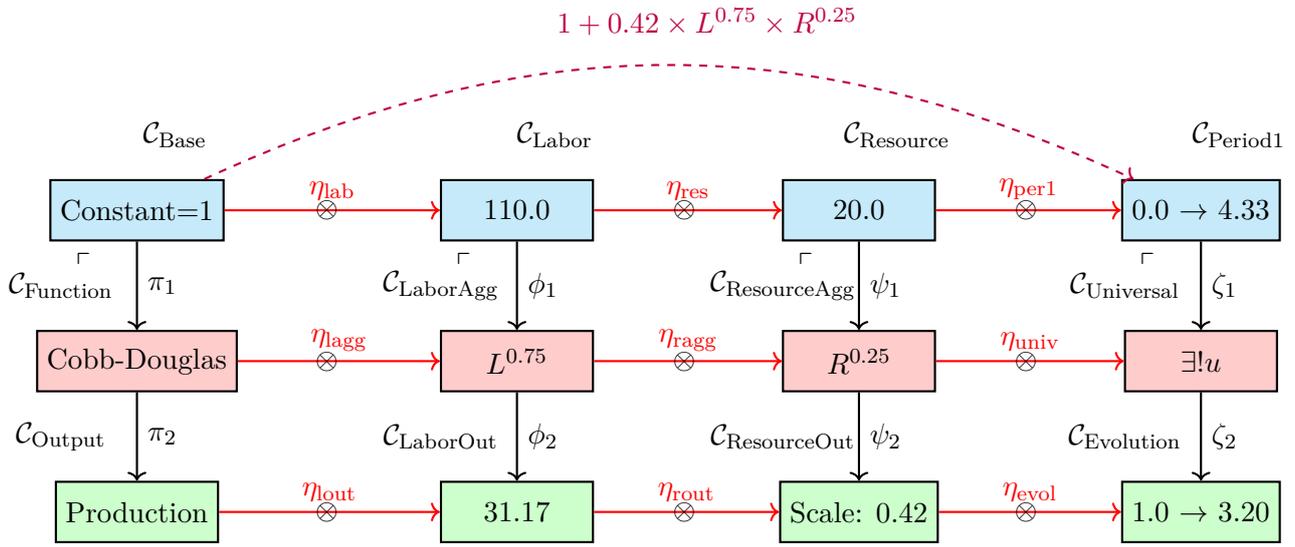

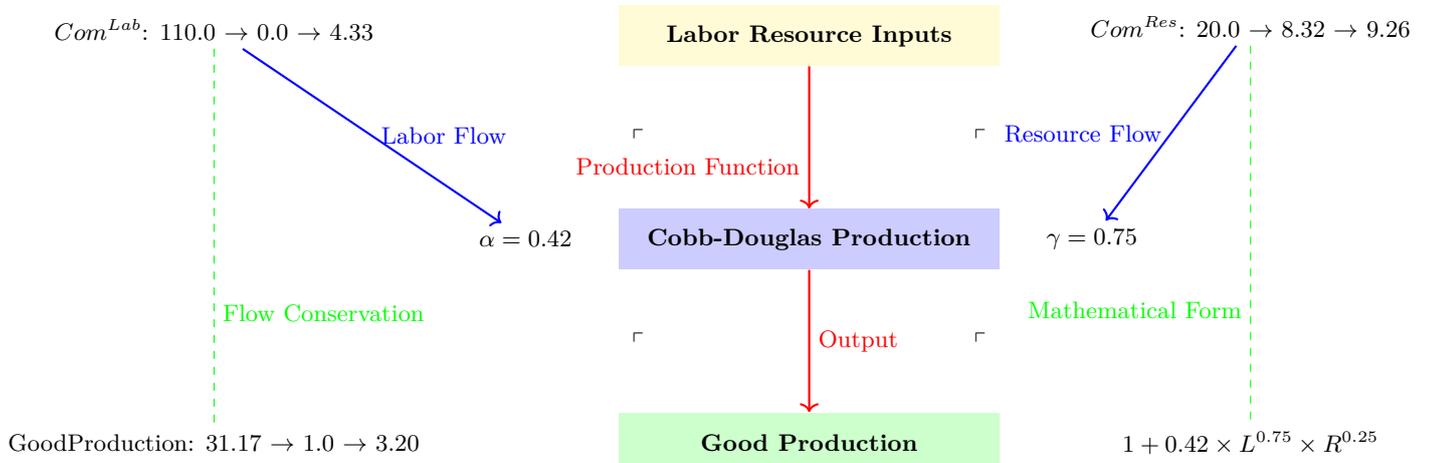
\begin{figure}[htbp]
\centering
{\footnotesize
\begin{tikzpicture}[scale=0.9]
\node[rectangle, fill=yellow!20, minimum width=5cm, minimum height=0.8cm] (inputs) at (0, 6) {\textbf{Labor Resource Inputs}};
\node[rectangle, fill=blue!20, minimum width=5cm, minimum height=0.8cm] (function) at (0, 3) {\textbf{Cobb-Douglas Production}};
\node[left=0.5cm of function] (scale) {$\alpha=0.42$};
\node[right=0.5cm of function] (elast) {$\gamma=0.75$};
\node[rectangle, fill=green!20, minimum width=5cm, minimum height=0.8cm] (production) at (0, 0) {\textbf{Good Production}};
\node[above=0.1cm, left=2.5cm of production] (output) {GoodProduction: 31.17 $\to$ 1.0 $\to$ 3.20};
\node[above=0.1cm, right=1.5cm of production] (formula) {$1 + 0.42 \times L^{0.75} \times R^{0.25}$};
\node[above=5.0cm of output] (labor) {$Com^{Lab}$: 110.0 $\to$ 0.0 $\to$ 4.33};
\node[above=5.0cm of formula] (resource) {$Com^{Res}$: 20.0 $\to$ 8.32 $\to$ 9.26};
\node at (-2.5, 4.5) {$\ulcorner$};
\node at (2.5, 4.5) {$\ulcorner$};
\node at (-2.5, 1.5) {$\ulcorner$};
\node at (2.5, 1.5) {$\ulcorner$};
\draw[->, thick, red] (inputs) -- (function) node[pos=0.7,left] {Production Function};
\draw[->, thick, red] (function) -- (production) node[midway,right] {Output};
\draw[->, thick, blue] (labor) -- (scale) node[midway,right] {Labor Flow};
\draw[->, thick, blue] (resource) -- (elast) node[midway,left] {Resource Flow};
\draw[dashed, green] (labor) -- (output) node[pos=0.7,right] {Flow Conservation};
\draw[dashed, green] (resource) -- (formula) node[pos=0.7,left] {Mathematical Form};
\end{tikzpicture}
}
\vspace{-1cm}
\caption{Commutative Diagram: Production Function Pushout Flow}
\label{z}
\end{figure}

\begin{figure}[htbp]
    \centering
    {\footnotesize
    \begin{tikzcd}[column sep=8em, row sep=7em, arrows={-stealth}]
    {\begin{matrix} C \\ \text{Production Base} \\ \text{(Constant = 1)} \end{matrix}} & {\begin{matrix} B \\ \text{Resource Input} \\ \text{(AccComRes)} \end{matrix}} \\
    {\begin{matrix} A \\ \text{Labor Input} \\ \text{(AccComLab)} \end{matrix}} & {\begin{matrix} \red{P} \\ \text{Production Function} \\ \text{(Cobb-Douglas)} \end{matrix}} & \\
    && {\begin{matrix} \green{D} \\ \text{Good Production} \\ \text{(Output)} \end{matrix}} & \\
    \arrow["{\begin{matrix} f \\ \text{AccComLab} \\ \text{110.0 $\to$ 0.0 $\to$ 4.33} \end{matrix}}"{right}, pos=0.2, from=1-1, to=2-1]
    \arrow["{\begin{matrix} g \\ \text{AccComRes} \\ \text{20.0 $\to$ 8.32 $\to$ 9.26} \end{matrix}}"{above=0.3cm}, pos=0.5, from=1-1, to=1-2]
    \arrow["{\begin{matrix} p_A \\ \text{Labor Input} \\ \text{Aggregation} \end{matrix}}"{above=0.3cm}, pos=0.6, color={rgb,255:red,214;green,92;blue,92}, from=2-1, to=2-2]
    \arrow["{\begin{matrix} p_B \\ \text{Resource Input} \\ \text{Aggregation} \end{matrix}}"{right=0.0cm}, pos=0.8, color={rgb,255:red,214;green,92;blue,92}, from=1-2, to=2-2]
    \arrow["{\begin{matrix} d_A \\ \text{GoodProduction} \\ \text{31.17, 1.0, 3.20} \end{matrix}}"{left=0.9cm}, pos=0.5, color={rgb,255:red,92;green,214;blue,92}, from=2-1, to=3-3]
    \arrow["{\begin{matrix} d_B \\ \text{Production Flow} \\ \text{Calculation} \end{matrix}}"{right=0.55cm}, pos=0.5, color={rgb,255:red,92;green,214;blue,92}, from=1-2, to=3-3]
    \arrow["{\begin{matrix} \exists ! \; u \\ \text{Cobb-Douglas} \\ \text{$1 + 0.42 \times L^{0.75} \times R^{0.25}$} \end{matrix}}"{description}, pos=0.15, color={rgb,255:red,92;green,214;blue,92}, dashed, from=2-2, to=3-3]
    \end{tikzcd}
    }
    \vspace{-2.5cm}
    \caption{Production Function Pushout: 
    The Cobb-Douglas production function pushout combines labor and resource inputs with scaling factor $\alpha=0.42$ and substitution elasticity $\gamma=0.75$, producing goods according to technology.}
    \label{A}
\end{figure}
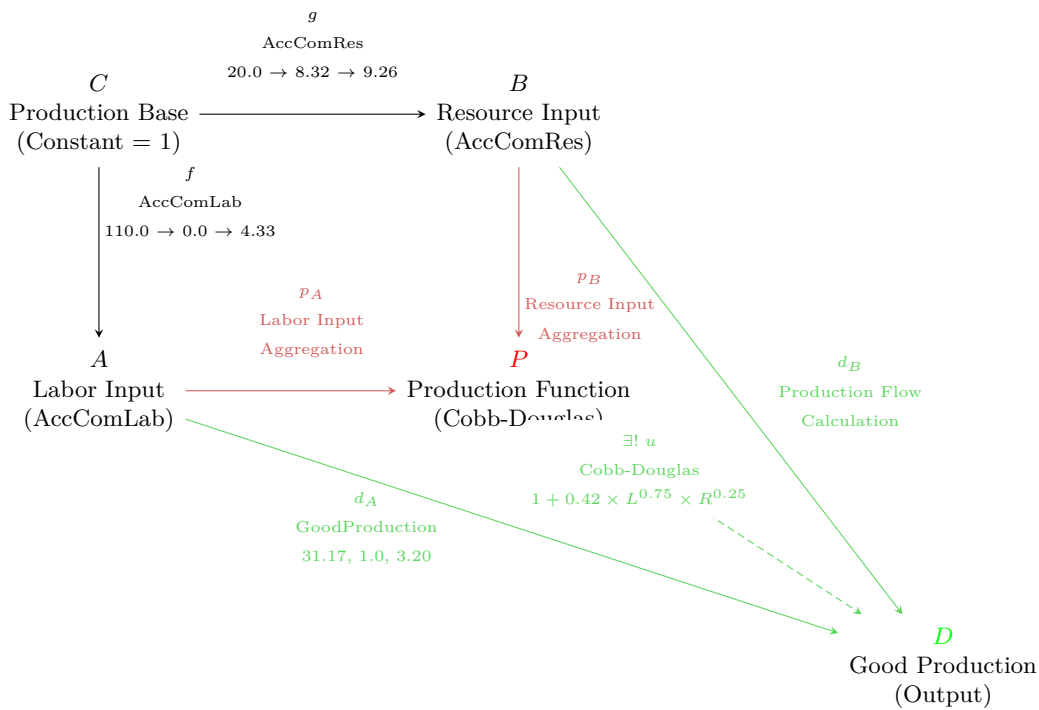

\textbf{Economics: Production Technology and Factor Combination:}
Figure~\ref{z} shows the natural transformation sequence from production base through labor and resource inputs (110.0 $\to$ 0.0 $\to$ 4.33, 20.0 $\to$ 8.32 $\to$ 9.26) 
to output evolution (31.17 $\to$ 1.0 $\to$ 3.20) as a categorical pushout construction.
The production function pushout encapsulates the fundamental microeconomic principle of productive transformation through categorical colimits. 
The Cobb-Douglas production function $1 + \alpha \times L^{\gamma} \times R^{1-\gamma}$ becomes a typed morphism 
$F_{\text{Production}}: \mathbb{R}_{\geq 0} \times \mathbb{R}_{\geq 0} \rightarrow \mathbb{R}_{> 0}$ 
that transforms labor and resource inputs into goods output while preserving technological relationships.

Economically, this captures how firms combine heterogeneous inputs ($Com^{Lab}$: 110.0 $\to$ 0.0 $\to$ 4.33 and $Com^{Res}$: 20.0 $\to$ 8.32 $\to$ 9.26) 
according to substitution possibilities characterized by the elasticity parameter (0.75) and scale factor (0.42). 
The pushout structure ensures that factor combination respects technological constraints while allowing for input substitution. 
The output varies ($GoodProduction$: 31.17 $\to$ 1.0 $\to$ 3.20) and reflects how production technology responds to input availability changes and consumption decisions 
consistently taking place within all other events.

The three-layer architecture of diagram~\ref{z} reveals the categorical nature of production technology: 
the yellow layer represents input factor endowments, the blue layer embodies the production technology (Cobb-Douglas transformation), 
and the green layer shows the resulting output levels. The pushout corners mark where individual factor inputs compose into the universal production output. 
This categorical approach to production theory provides rigorous foundations for understanding technological relationships in production, 
connecting microeconomic production theory with categorical mathematics and enabling formal analysis of technological progress and factor substitution patterns.
The consistent factorisations in the universal constructions are a key feature of the categorical approach to production and resource theory
and to economic evolution in parallel, sequential, hierarchical and universal structures with endofunctors.
Dynamics may include optimal adjustment of institutions as optimal hierarchical levels of control.
Institutional evolution in universals may involve learning constitutional logical rules for example by inductive logical programming.
Polycentric management of public AND private good production and consumption then forms a complete specification of economics.

% -----------------------------------------------------------------------------------------------------------------------------------------------------------------------------------
%\clearpage
\subsubsection{Cost-Based Planning and Market Equilibration}

The economic foundation of the price formation functor requires understanding two sequential categorical constructions: 
the company's cost-based revenue planning and the market's equilibration mechanism. These represent the fundamental tension between company cost structures 
and consumer demand patterns.
The company's demand planning functor transforms contractual obligations into revenue targets that ensure cost recovery and profit generation.

\begin{definition}[Cost-Based Planning]
The cost-based planning functor $F_{\text{CostPlan}}$ computes the company's required revenue based on contractual obligations:
$$\text{DemandPlan} = (\text{WagesPayment} + \text{RepaysPayment}) \times (1 + \mu)$$
The cost planning type signature is:
\begin{align}
F_{\text{CostPlan}} &:& \mathbb{R}_{\geq 0} \times \mathbb{R}_{\geq 0} \times \mathbb{R}_{\geq 0} \rightarrow \mathbb{R}_{\geq 0} && F_{\text{CostPlan}}(w, r, \mu) = (w + r) \times (1 + \mu)
\end{align}
where $w= \text{WagesPayment}, r = \text{RepaysPayment}, \mu = \text{MarkUp}$.
\hfill$\blacksquare$
\end{definition}

Evolution across \textbf{periods 0,1,2}:
\begin{align}
\text{DemandPlan}_0 &= (0.0 + 0.0) \times (1 + 0.5) = 0.0\nonumber \\
\text{DemandPlan}_1 &= (52.0 + 26.0) \times (1 + 0.5) = 117.0\nonumber \\
\text{DemandPlan}_2 &= (109.90 + 54.95) \times (1 + 0.5) = 247.27\nonumber
\end{align}

\begin{figure}[htbp]
    \centering
    \tikzset{
    stringdiagram/.style={
    baseline=(current bounding box.center),
    every node/.style={draw=none},
    every path/.style={thick}
    }
    }
    \vspace{-3.5cm}
    \begin{tikzpicture}[stringdiagram]
    \begin{scope}[shift={(0,0)}]
    \node at (1.5,4.5) {$\mathcal{C}_{\text{Memory}}$};
    \node[draw, rectangle, minimum width=2.0cm, minimum height=0.8cm, fill=orange!20] (memory) at (1.0,3.5) {Wage History};
    \node at (0.0,2.5) {$\mathcal{C}_{\text{Obligations}}$};
    \node[draw, rectangle, minimum width=2.0cm, minimum height=0.8cm, fill=red!20] (obligations) at (1.0,1.5) {Sum Wages};
    \node at (0.0,0.5) {$\mathcal{C}_{\text{Costs}}$};
    \node[draw, rectangle, minimum width=2.0cm, minimum height=0.8cm, fill=blue!20] (costs) at (1.0,-0.5) {Total Wage Cost};
    \node at (6.5,4.5) {$\mathcal{C}_{\text{DebtMem}}$};
    \node[draw, rectangle, minimum width=2.0cm, minimum height=0.8cm, fill=orange!20] (debtmem) at (6.0,3.5) {Repay History};
    \node at (5.0,2.5) {$\mathcal{C}_{\text{DebtObl}}$};
    \node[draw, rectangle, minimum width=2.0cm, minimum height=0.8cm, fill=red!20] (debtobl) at (6.0,1.5) {Sum Repayments};
    \node at (5.0,0.5) {$\mathcal{C}_{\text{DebtCosts}}$};
    \node[draw, rectangle, minimum width=2.0cm, minimum height=0.8cm, fill=blue!20] (debtcosts) at (6.0,-0.5) {Debt Service};
    \node at (11.0,4.5) {$\mathcal{C}_{\text{Planning}}$};
    \node[draw, rectangle, minimum width=2.0cm, minimum height=0.8cm, fill=orange!20] (planning) at (10.5,3.5) {Markup Policy};
    \node at (9.5,2.5) {$\mathcal{C}_{\text{Revenue}}$};
    \node[draw, rectangle, minimum width=2.0cm, minimum height=0.8cm, fill=red!20] (revenue) at (10.5,1.5) {Cost + Markup};
    \node at (9.5,0.5) {$\mathcal{C}_{\text{Target}}$};
    \node[draw, rectangle, minimum width=2.0cm, minimum height=0.8cm, fill=blue!20] (target) at (10.5,-0.5) {Revenue Target};
    \draw[->, thick] (memory) -- (obligations) node[midway,right=0.0cm] {$\Sigma$};
    \draw[->, thick] (obligations) -- (costs) node[midway,right=0.0cm] {$\iota_1$};
    \draw[->, thick] (debtmem) -- (debtobl) node[midway,right=0.0cm] {$\Sigma$};
    \draw[->, thick] (debtobl) -- (debtcosts) node[midway,right=0.0cm] {$\iota_2$};
    \draw[->, thick] (planning) -- (revenue) node[midway,right=0.0cm] {$\times$};
    \draw[->, thick] (revenue) -- (target) node[midway,right=0.0cm] {$\pi$};
    \draw[->, thick, red] (memory) -- (debtmem) node[midway,above] {$\eta_{\text{contract}}$};
    \draw[->, thick, red] (obligations) -- (debtobl) node[midway,above] {$\eta_{\text{sum}}$};
    \draw[->, thick, red] (costs) -- (debtcosts) node[midway,above] {$\eta_{\text{cost}}$};
    \draw[->, thick, red] (debtmem) -- (planning) node[midway,above] {$\eta_{\text{plan}}$};
    \draw[->, thick, red] (debtobl) -- (revenue) node[midway,above] {$\eta_{\text{rev}}$};
    \draw[->, thick, red] (debtcosts) -- (target) node[midway,above] {$\eta_{\text{target}}$};
    \node at (0.3,2.9) {$\lrcorner$};
    \node at (5.3,2.9) {$\lrcorner$};
    \node at (9.8,2.9) {$\lrcorner$};
    \node at (3.5,3.5) {$\oplus$};
    \node at (3.5,1.5) {$\oplus$};
    \node at (3.5,-0.5) {$\oplus$};
    \node at (8.2,3.5) {$\otimes$};
    \node at (8.2,1.5) {$\otimes$};
    \node at (8.2,-0.5) {$\otimes$};
    \draw[->, thick, purple, dashed] (memory) to[bend left=145, looseness=1.75] (target) node at (8.5,5.5) {$(w + r) \times (1 + \mu)$};
    \end{scope}
    \end{tikzpicture}
    \caption{Commutative Diagram: Cost-Based Planning Structure}
\label{B}    
\end{figure}
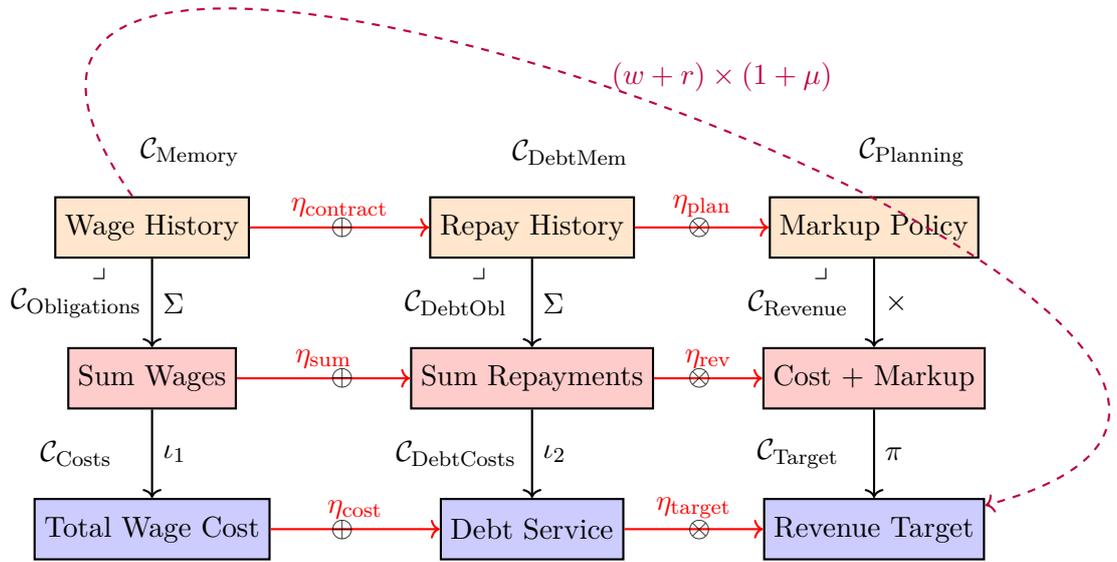
    
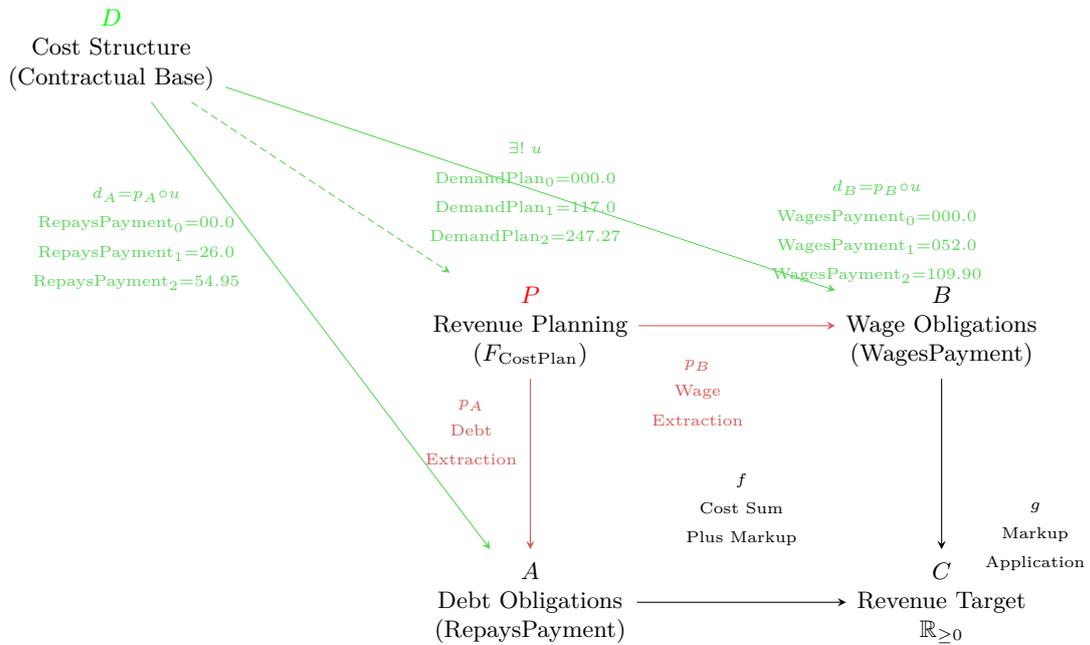
\begin{figure}[htbp]
\centering
{\footnotesize
\begin{tikzcd}[column sep=8em, row sep=7em, arrows={-stealth}]
{\begin{matrix} \green{D} \\ \text{Cost Structure} \\ \text{(Contractual Base)} \end{matrix}} \\
& {\begin{matrix} \red{P} \\ \text{Revenue Planning} \\ (F_{\text{CostPlan}}) \end{matrix}}
& {\begin{matrix} B \\ \text{Wage Obligations} \\ \text{(WagesPayment)} \end{matrix}} \\
& {\begin{matrix} A \\ \text{Debt Obligations} \\ \text{(RepaysPayment)} \end{matrix}}
& {\begin{matrix} C \\ \text{Revenue Target} \\ \mathbb{R}_{\geq 0} \end{matrix}} \\
\arrow["{\begin{matrix} \exists ! \; u \\ \text{DemandPlan}_0 = 000.0 \\ \text{DemandPlan}_1 = 117.0 \\ \text{DemandPlan}_2 = 247.27 \end{matrix}}", pos=0.9, color={rgb,255:red,92;green,214;blue,92}, dashed, from=1-1, to=2-2]
\arrow["{\begin{matrix} d_A = p_A \circ u \\ \text{RepaysPayment}_0 = 00.0 \\ \text{RepaysPayment}_1 = 26.0 \\ \text{RepaysPayment}_2 = 54.95 \end{matrix}}"{left=0.1cm}, pos=0.3, color={rgb,255:red,92;green,214;blue,92}, from=1-1, to=3-2]
\arrow["{\begin{matrix} d_B = p_B \circ u \\ \text{WagesPayment}_0 = 000.0 \\ \text{WagesPayment}_1 = 052.0 \\ \text{WagesPayment}_2 = 109.90 \end{matrix}}"{right=1.5cm}, pos=0.7, color={rgb,255:red,92;green,214;blue,92}, from=1-1, to=2-3]
\arrow["{\begin{matrix} p_A \\ \text{Debt} \\ \text{Extraction} \end{matrix}}"{left=0.1cm}, pos=0.3, color={rgb,255:red,214;green,92;blue,92}, from=2-2, to=3-2]
\arrow["{\begin{matrix} p_B \\ \text{Wage} \\ \text{Extraction} \end{matrix}}"{below=0.2cm}, pos=0.3, color={rgb,255:red,214;green,92;blue,92}, from=2-2, to=2-3]
\arrow["{\begin{matrix} f \\ \text{Cost Sum} \\ \text{Plus Markup} \end{matrix}}"'{above=0.6cm}, from=3-2, to=3-3]
\arrow["{\begin{matrix} g \\ \text{Markup} \\ \text{Application} \end{matrix}}"{right=0.5cm}, pos=0.9, from=2-3, to=3-3]
\end{tikzcd}
}
\vspace{-2.5cm}
\caption{Cost-Based Planning Pullback: The cost planning pullback ensures revenue targets cover all contractual obligations (wages, debt service) plus desired profit markup.}
\label{C}
\end{figure}

Figure~(\ref{B}) shows the transformation from historical contractual obligations (wage and debt service memories) through cost aggregation (0.0 $\to$ 78.0 $\to$ 164.85) 
to revenue targets (0.0 $\to$ 117.0 $\to$ 247.27) via categorical construction.    
The market disequilibrium functor identifies the tension between consumer demand and company revenue requirements.

\begin{definition}[Market Disequilibrium]
The market disequilibrium functor $F_{\text{Disequilibrium}}$ computes the gap between actual and planned demand:
$$\text{DemandSurplus} = \text{Demand} - \text{DemandPlan}$$

The disequilibrium type signature is:
\begin{align}
F_{\text{Disequilibrium}}&:& \mathbb{R}_{\geq 0} \times \mathbb{R}_{\geq 0} &\rightarrow \mathbb{R} &&
F_{\text{Disequilibrium}}(d, d^*) &= d - d^*\\
\end{align}
where $d = \text{Demand}, d^* = \text{DemandPlan}$
\hfill$\blacksquare$
\end{definition}

Evolution across periods:
\begin{align}
\text{DemandSurplus}_0 &= 0.0 - 0.0 = 0.0\nonumber\\
\text{DemandSurplus}_1 &= 166.4 - 117.0 = 49.4\nonumber\\
\text{DemandSurplus}_2 &= 272.63 - 247.27 = 25.36\nonumber
\end{align}

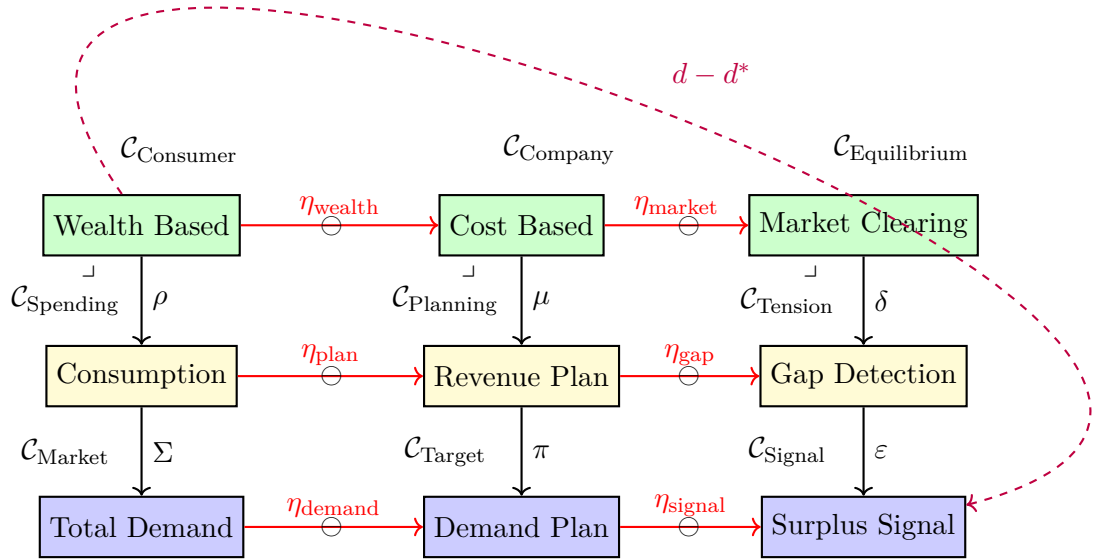
\begin{figure}[htbp]
\centering
\tikzset{
stringdiagram/.style={
baseline=(current bounding box.center),
every node/.style={draw=none},
every path/.style={thick}
}
}
\vspace{-3.5cm}
\begin{tikzpicture}[stringdiagram]
\begin{scope}[shift={(0,0)}]
\node at (1.5,4.5) {$\mathcal{C}_{\text{Consumer}}$};
\node[draw, rectangle, minimum width=2.0cm, minimum height=0.8cm, fill=green!20] (consumer) at (1.0,3.5) {Wealth Based};
\node at (0.0,2.5) {$\mathcal{C}_{\text{Spending}}$};
\node[draw, rectangle, minimum width=2.0cm, minimum height=0.8cm, fill=yellow!20] (spending) at (1.0,1.5) {Consumption};
\node at (0.0,0.5) {$\mathcal{C}_{\text{Market}}$};
\node[draw, rectangle, minimum width=2.0cm, minimum height=0.8cm, fill=blue!20] (market) at (1.0,-0.5) {Total Demand};
\node at (6.5,4.5) {$\mathcal{C}_{\text{Company}}$};
\node[draw, rectangle, minimum width=2.0cm, minimum height=0.8cm, fill=green!20] (company) at (6.0,3.5) {Cost Based};
\node at (5.0,2.5) {$\mathcal{C}_{\text{Planning}}$};
\node[draw, rectangle, minimum width=2.0cm, minimum height=0.8cm, fill=yellow!20] (companyplan) at (6.0,1.5) {Revenue Plan};
\node at (5.0,0.5) {$\mathcal{C}_{\text{Target}}$};
\node[draw, rectangle, minimum width=2.0cm, minimum height=0.8cm, fill=blue!20] (comptarget) at (6.0,-0.5) {Demand Plan};
\node at (11.0,4.5) {$\mathcal{C}_{\text{Equilibrium}}$};
\node[draw, rectangle, minimum width=2.0cm, minimum height=0.8cm, fill=green!20] (equilibrium) at (10.5,3.5) {Market Clearing};
\node at (9.5,2.5) {$\mathcal{C}_{\text{Tension}}$};
\node[draw, rectangle, minimum width=2.0cm, minimum height=0.8cm, fill=yellow!20] (tension) at (10.5,1.5) {Gap Detection};
\node at (9.5,0.5) {$\mathcal{C}_{\text{Signal}}$};
\node[draw, rectangle, minimum width=2.0cm, minimum height=0.8cm, fill=blue!20] (signal) at (10.5,-0.5) {Surplus Signal};
\draw[->, thick] (consumer) -- (spending) node[midway,right=0.0cm] {$\rho$};
\draw[->, thick] (spending) -- (market) node[midway,right=0.0cm] {$\Sigma$};
\draw[->, thick] (company) -- (companyplan) node[midway,right=0.0cm] {$\mu$};
\draw[->, thick] (companyplan) -- (comptarget) node[midway,right=0.0cm] {$\pi$};
\draw[->, thick] (equilibrium) -- (tension) node[midway,right=0.0cm] {$\delta$};
\draw[->, thick] (tension) -- (signal) node[midway,right=0.0cm] {$\varepsilon$};
\draw[->, thick, red] (consumer) -- (company) node[midway,above] {$\eta_{\text{wealth}}$};
\draw[->, thick, red] (spending) -- (companyplan) node[midway,above] {$\eta_{\text{plan}}$};
\draw[->, thick, red] (market) -- (comptarget) node[midway,above] {$\eta_{\text{demand}}$};
\draw[->, thick, red] (company) -- (equilibrium) node[midway,above] {$\eta_{\text{market}}$};
\draw[->, thick, red] (companyplan) -- (tension) node[midway,above] {$\eta_{\text{gap}}$};
\draw[->, thick, red] (comptarget) -- (signal) node[midway,above] {$\eta_{\text{signal}}$};
\node at (0.3,2.9) {$\lrcorner$};
\node at (5.3,2.9) {$\lrcorner$};
\node at (9.8,2.9) {$\lrcorner$};
\node at (3.5,3.5) {$\ominus$};
\node at (3.5,1.5) {$\ominus$};
\node at (3.5,-0.5) {$\ominus$};
\node at (8.2,3.5) {$\ominus$};
\node at (8.2,1.5) {$\ominus$};
\node at (8.2,-0.5) {$\ominus$};
\draw[->, thick, purple, dashed] (consumer) to[bend left=145, looseness=1.75] (signal) node at (8.5,5.5) {$d - d^*$};
\end{scope}
\end{tikzpicture}
\caption{Commutative Diagram: Market Equilibration Structure}
\label{D}
\end{figure}

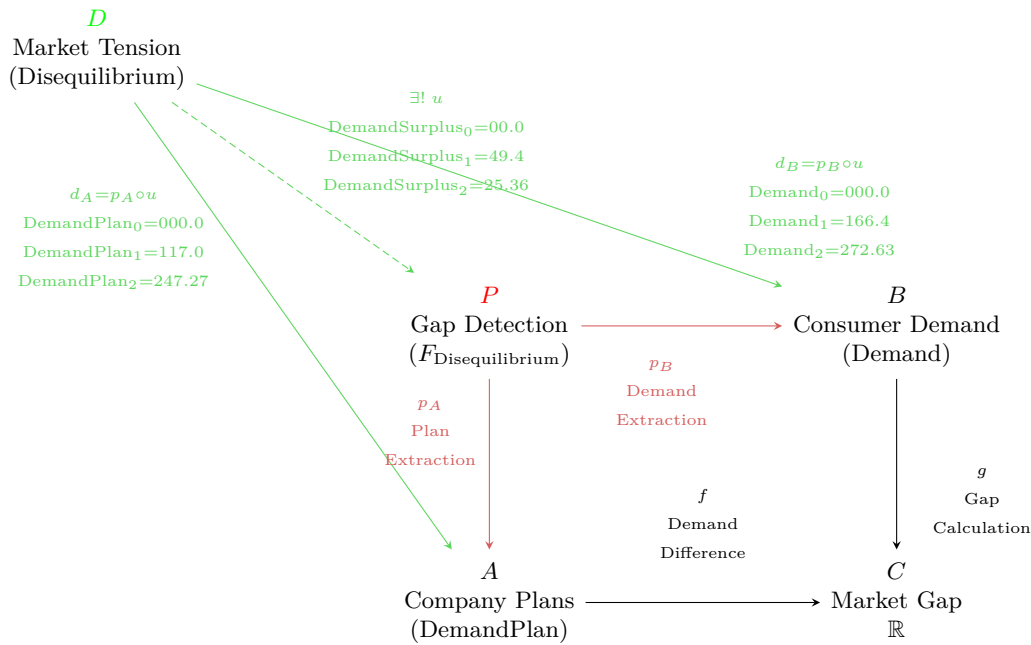
\begin{figure}[htbp]
    \centering
    {\footnotesize
    \begin{tikzcd}[column sep=8em, row sep=7em, arrows={-stealth}]
    {\begin{matrix} \green{D} \\ \text{Market Tension} \\ \text{(Disequilibrium)} \end{matrix}} \\
    & {\begin{matrix} \red{P} \\ \text{Gap Detection} \\ (F_{\text{Disequilibrium}}) \end{matrix}}
    & {\begin{matrix} B \\ \text{Consumer Demand} \\ \text{(Demand)} \end{matrix}} \\
    & {\begin{matrix} A \\ \text{Company Plans} \\ \text{(DemandPlan)} \end{matrix}}
    & {\begin{matrix} C \\ \text{Market Gap} \\ \mathbb{R} \end{matrix}} \\
    \arrow["{\begin{matrix} \exists ! \; u \\ \text{DemandSurplus}_0 = 00.0 \\ \text{DemandSurplus}_1 = 49.4 \\ \text{DemandSurplus}_2 = 25.36 \end{matrix}}", pos=0.6, color={rgb,255:red,92;green,214;blue,92}, dashed, from=1-1, to=2-2]
    \arrow["{\begin{matrix} d_A = p_A \circ u \\ \text{DemandPlan}_0 = 000.0 \\ \text{DemandPlan}_1 = 117.0 \\ \text{DemandPlan}_2 = 247.27 \end{matrix}}"{left=0.2cm}, pos=0.3, color={rgb,255:red,92;green,214;blue,92}, from=1-1, to=3-2]
    \arrow["{\begin{matrix} d_B = p_B \circ u \\ \text{Demand}_0 = 000.0 \\ \text{Demand}_1 = 166.4 \\ \text{Demand}_2 = 272.63 \end{matrix}}"{right=2.5cm}, pos=0.6, color={rgb,255:red,92;green,214;blue,92}, from=1-1, to=2-3]
    \arrow["{\begin{matrix} p_A \\ \text{Plan} \\ \text{Extraction} \end{matrix}}"{left=0.1cm}, pos=0.3, color={rgb,255:red,214;green,92;blue,92}, from=2-2, to=3-2]
    \arrow["{\begin{matrix} p_B \\ \text{Demand} \\ \text{Extraction} \end{matrix}}"{below=0.2cm}, pos=0.4, color={rgb,255:red,214;green,92;blue,92}, from=2-2, to=2-3]
    \arrow["{\begin{matrix} f \\ \text{Demand} \\ \text{Difference} \end{matrix}}"'{above=0.4cm}, from=3-2, to=3-3]
    \arrow["{\begin{matrix} g \\ \text{Gap} \\ \text{Calculation} \end{matrix}}"{right=0.4cm}, pos=0.7, from=2-3, to=3-3]
    \end{tikzcd}
    }
    \vspace{-2.5cm}
    \caption{Market Disequilibrium Pullback: The disequilibrium pullback measures market tension between consumer spending capacity and company revenue requirements.}
    \label{E}
\end{figure}

\textbf{Economics: Cost Structure and Market Coordination}
Figure~\ref{D} shows the tension resolution between consumer demand (0.0 $\to$ 166.4 $\to$ 272.63) 
and company revenue plans (0.0 $\to$ 117.0 $\to$ 247.27) producing market signals (0.0 $\to$ 49.4 $\to$ 25.36).

The cost-based planning and market equilibration functors embody the fundamental economic tension between company survival constraints and market demand patterns. 
The cost planning functor $(w + r) \times (1 + \mu)$ represents the minimum revenue requirement for company viability, 
transforming contractual obligations (wages: 0.0 $\to$ 52.0 $\to$ 109.90, debt service: 0.0 $\to$ 26.0 $\to$ 54.95) into survival-critical revenue targets (0.0 $\to$ 117.0 $\to$ 247.27).

The market disequilibrium functor $d - d^*$ detects coordination failures between consumer spending capacity (0.0 $\to$ 166.4 $\to$ 272.63) 
and company revenue requirements. The resulting demand surplus (0.0 $\to$ 49.4 $\to$ 25.36) becomes the crucial signal for 
investment decisions and price adjustments, representing the economy's adaptive mechanism for resolving mismatches of demand and supply.

This categorical decomposition reveals the sequential nature of market coordination: companies first determine survival critical revenue targets 
based on contractual obligations, then markets generate equilibration signals through demand plan gaps. 
The three-layer architecture of diagram~\ref{D} shows how historical contractual commitments (orange layer) flow through 
current planning decisions (yellow layer) to generate market coordination signals (blue layer). 
This provides mathematical foundations for understanding dynamics like business cycles as categorical limit constructions, 
connecting microeconomic cost structures with macroeconomic coordination mechanisms.

% -----------------------------------------------------------------------------------------------------------------------------------------------------------------------------------
%\clearpage
\subsubsection{Price Formation}

\begin{definition}[Price]
The price functor $F_{price}$ 
transforms demand and supply into prices.
The price coordinates the turnover plan of the company with the consumption plan of the consumers mediated through production:
$$\text{GoodPrice} = \frac{\text{DemandPlan}}{\text{GoodProduction}} + \mu$$

\begin{align}
F_{price} &:& \mathbb{R}_{\geq 0} \times \mathbb{R}_{> 0} \times \mathbb{R}_{\geq 0} &\rightarrow \mathbb{R}_{> 0} 
&& F_{\text{PriceFormation}}(d, q, m) &= \frac{d}{q} + \mu
\end{align}
where $d = \text{DemandPlan}, q = \text{GoodProduction}, \mu = \text{PriceMarkup}$.
\hfill$\blacksquare$
\end{definition}

Evolution across \textit{periods 0, 1, 2}:
\begin{align}
\text{GoodPrice}_0 &= \frac{0.0}{31.17} + 30.0 = 30.0\nonumber\\
\text{GoodPrice}_1 &= \frac{117.0}{1.0} + 24.7 = 141.7\nonumber\\
\text{GoodPrice}_2 &= \frac{247.27}{3.20} + 12.65 = 89.94\nonumber
\end{align}

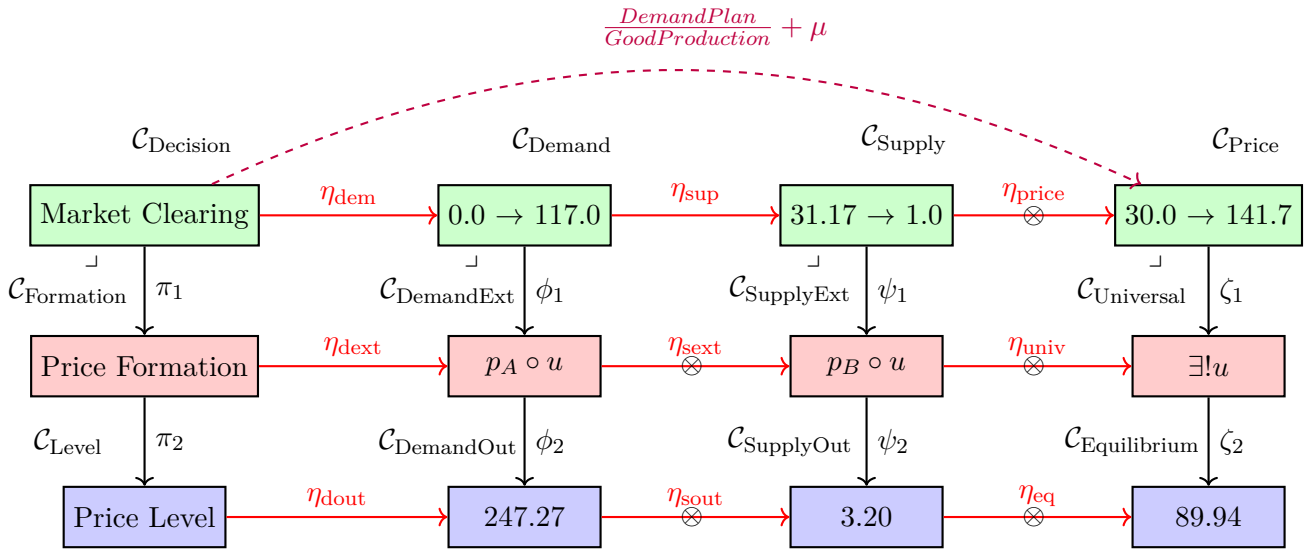
\begin{figure}[htbp]
\centering
\tikzset{
stringdiagram/.style={
baseline=(current bounding box.center),
every node/.style={draw=none},
every path/.style={thick}
}
}
\begin{tikzpicture}[stringdiagram]
\begin{scope}[shift={(0,0)}]
\node at (1.5,4.5) {$\mathcal{C}_{\text{Decision}}$};
\node[draw, rectangle, minimum width=2.0cm, minimum height=0.8cm, fill=green!20] (decision) at (1.0,3.5) {Market Clearing};
\node at (0.0,2.5) {$\mathcal{C}_{\text{Formation}}$};
\node[draw, rectangle, minimum width=2.0cm, minimum height=0.8cm, fill=red!20] (formation) at (1.0,1.5) {Price Formation};
\node at (0.0,0.5) {$\mathcal{C}_{\text{Level}}$};
\node[draw, rectangle, minimum width=2.0cm, minimum height=0.8cm, fill=blue!20] (level) at (1.0,-0.5) {Price Level};
\node at (6.5,4.5) {$\mathcal{C}_{\text{Demand}}$};
\node[draw, rectangle, minimum width=2.0cm, minimum height=0.8cm, fill=green!20] (demand) at (6.0,3.5) {0.0 $\to$ 117.0};
\node at (5.0,2.5) {$\mathcal{C}_{\text{DemandExt}}$};
\node[draw, rectangle, minimum width=2.0cm, minimum height=0.8cm, fill=red!20] (demandext) at (6.0,1.5) {$p_A \circ u$};
\node at (5.0,0.5) {$\mathcal{C}_{\text{DemandOut}}$};
\node[draw, rectangle, minimum width=2.0cm, minimum height=0.8cm, fill=blue!20] (demandout) at (6.0,-0.5) {247.27};
\node at (11.0,4.5) {$\mathcal{C}_{\text{Supply}}$};
\node[draw, rectangle, minimum width=2.0cm, minimum height=0.8cm, fill=green!20] (supply) at (10.5,3.5) {31.17 $\to$ 1.0};
\node at (9.5,2.5) {$\mathcal{C}_{\text{SupplyExt}}$};
\node[draw, rectangle, minimum width=2.0cm, minimum height=0.8cm, fill=red!20] (supplyext) at (10.5,1.5) {$p_B \circ u$};
\node at (9.5,0.5) {$\mathcal{C}_{\text{SupplyOut}}$};
\node[draw, rectangle, minimum width=2.0cm, minimum height=0.8cm, fill=blue!20] (supplyout) at (10.5,-0.5) {3.20};
\node at (15.5,4.5) {$\mathcal{C}_{\text{Price}}$};
\node[draw, rectangle, minimum width=2.0cm, minimum height=0.8cm, fill=green!20] (price) at (15.0,3.5) {30.0 $\to$ 141.7};
\node at (14.0,2.5) {$\mathcal{C}_{\text{Universal}}$};
\node[draw, rectangle, minimum width=2.0cm, minimum height=0.8cm, fill=red!20] (universal) at (15.0,1.5) {$\exists ! u$};
\node at (14.0,0.5) {$\mathcal{C}_{\text{Equilibrium}}$};
\node[draw, rectangle, minimum width=2.0cm, minimum height=0.8cm, fill=blue!20] (equilibrium) at (15.0,-0.5) {89.94};
\draw[->, thick] (decision) -- (formation) node[midway,right=0.0cm] {$\pi_1$};
\draw[->, thick] (formation) -- (level) node[midway,right=0.0cm] {$\pi_2$};
\draw[->, thick] (demand) -- (demandext) node[midway,right=0.0cm] {$\phi_1$};
\draw[->, thick] (demandext) -- (demandout) node[midway,right=0.0cm] {$\phi_2$};
\draw[->, thick] (supply) -- (supplyext) node[midway,right=0.0cm] {$\psi_1$};
\draw[->, thick] (supplyext) -- (supplyout) node[midway,right=0.0cm] {$\psi_2$};
\draw[->, thick] (price) -- (universal) node[midway,right=0.0cm] {$\zeta_1$};
\draw[->, thick] (universal) -- (equilibrium) node[midway,right=0.0cm] {$\zeta_2$};
\draw[->, thick, red] (decision) -- (demand) node[midway,above] {$\eta_{\text{dem}}$};
\draw[->, thick, red] (formation) -- (demandext) node[midway,above] {$\eta_{\text{dext}}$};
\draw[->, thick, red] (level) -- (demandout) node[midway,above] {$\eta_{\text{dout}}$};
\draw[->, thick, red] (demand) -- (supply) node[midway,above] {$\eta_{\text{sup}}$};
\draw[->, thick, red] (demandext) -- (supplyext) node[midway,above] {$\eta_{\text{sext}}$};
\draw[->, thick, red] (demandout) -- (supplyout) node[midway,above] {$\eta_{\text{sout}}$};
\draw[->, thick, red] (supply) -- (price) node[midway,above] {$\eta_{\text{price}}$};
\draw[->, thick, red] (supplyext) -- (universal) node[midway,above] {$\eta_{\text{univ}}$};
\draw[->, thick, red] (supplyout) -- (equilibrium) node[midway,above] {$\eta_{\text{eq}}$};
\node at (0.3,2.9) {$\lrcorner$};
\node at (5.3,2.9) {$\lrcorner$};
\node at (9.8,2.9) {$\lrcorner$};
\node at (14.3,2.9) {$\lrcorner$};
\node at (8.2,1.5) {$\otimes$};
\node at (8.2,-0.5) {$\otimes$};
\node at (12.7,3.5) {$\otimes$};
\node at (12.7,1.5) {$\otimes$};
\node at (12.7,-0.5) {$\otimes$};
\draw[->, thick, purple, dashed] (decision) to[bend left=25] (price) node at (8.5,6.0) {$\frac{DemandPlan}{GoodProduction} + \mu$};
\end{scope}
\end{tikzpicture}
\caption{Commutative Diagram: Price Formation Pullback Structure}
\label{F}
\end{figure}

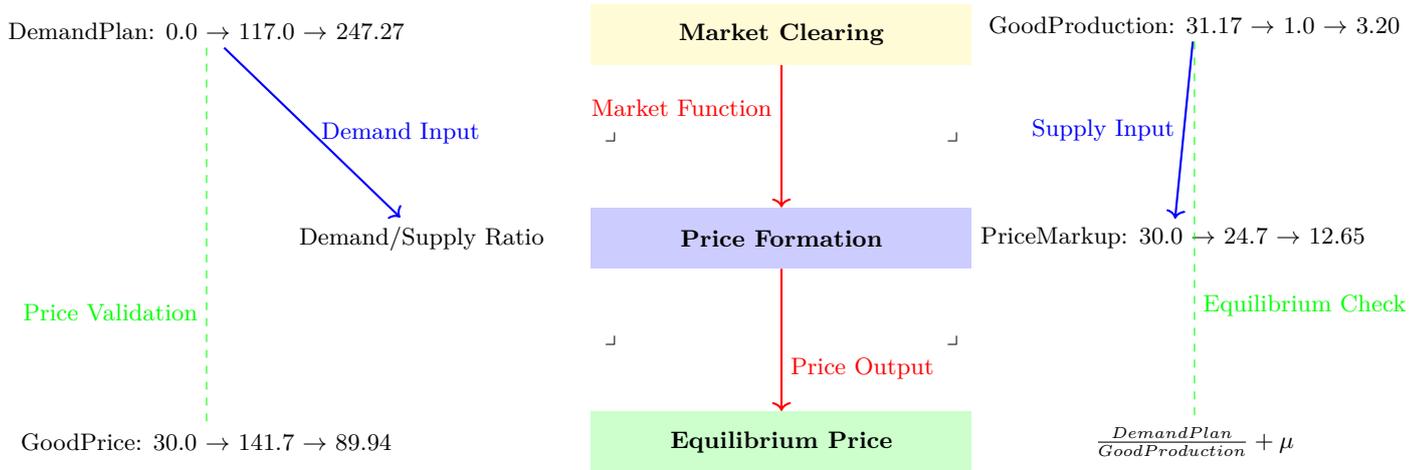
\begin{figure}[htbp]
\centering
{\footnotesize
\begin{tikzpicture}[scale=0.9]
\node[rectangle, fill=yellow!20, minimum width=5cm, minimum height=0.8cm] (clearing) at (0, 6) {\textbf{Market Clearing}};
\node[rectangle, fill=blue!20, minimum width=5cm, minimum height=0.8cm] (formation) at (0, 3) {\textbf{Price Formation}};
\node[above=0.5cm, left=0.5cm of formation] (ratio) {Demand/Supply Ratio};
\node[above=0.5cm, right=0.0cm of formation] (markup) {PriceMarkup: 30.0 $\to$ 24.7 $\to$ 12.65};
\node[rectangle, fill=green!20, minimum width=5cm, minimum height=0.8cm] (price) at (0, 0) {\textbf{Equilibrium Price}};
\node[above=0.1cm of price, left=2.5cm of price] (goodprice) {GoodPrice: 30.0 $\to$ 141.7 $\to$ 89.94};
\node[above=0.1cm of price, right=1.5cm of price] (equation) {$\frac{DemandPlan}{GoodProduction} + \mu$};
\node[above=5.0cm of goodprice] (demand) {DemandPlan: 0.0 $\to$ 117.0 $\to$ 247.27};
\node[above=5.0cm of equation] (supply) {GoodProduction: 31.17 $\to$ 1.0 $\to$ 3.20};
\node at (-2.5, 4.5) {$\lrcorner$};
\node at (2.5, 4.5) {$\lrcorner$};
\node at (-2.5, 1.5) {$\lrcorner$};
\node at (2.5, 1.5) {$\lrcorner$};
\draw[->, thick, red] (clearing) -- (formation) node[pos=0.3,left] {Market Function};
\draw[->, thick, red] (formation) -- (price) node[pos=0.7,right] {Price Output};
\draw[->, thick, blue] (demand) -- (ratio) node[midway,right] {Demand Input};
\draw[->, thick, blue] (supply) -- (markup) node[midway,left] {Supply Input};
\draw[dashed, green] (demand) -- (goodprice) node[pos=0.7,left] {Price Validation};
\draw[dashed, green] (supply) -- (equation) node[pos=0.7,right] {Equilibrium Check};
\node at (0,-1.0) {};
\end{tikzpicture}
}
\vspace{-1cm}
\caption{Commutative Diagram: Price Formation Pullback Flow}
\label{G}
\end{figure}

\begin{figure}[htbp]
\centering
{\footnotesize
\begin{tikzcd}[column sep=8em, row sep=7em, arrows={-stealth}]
{\begin{matrix} \green{D} \\ \text{Price Decision} \\ \text{(Market Clearing)} \end{matrix}} \\
& {\begin{matrix} \red{P} \\ \text{Price Formation} \\ (F_{price}) \end{matrix}}
& {\begin{matrix} B \\ \text{Good Production} \\ \text{(Supply)} \end{matrix}} \\
& {\begin{matrix} A \\ \text{Demand Plan} \\ \text{(Demand)} \end{matrix}}
& {\begin{matrix} C \\ \text{Price Level} \\ \mathbb{R}_{> 0} \end{matrix}} \\
\arrow["{\begin{matrix} \exists ! \; u \\ \text{GoodPrice}_0 = 030.0 \\ \text{GoodPrice}_1 = 141.7 \\ \text{GoodPrice}_2 = 089.94 \end{matrix}}", pos=0.9, color={rgb,255:red,92;green,214;blue,92}, dashed, from=1-1, to=2-2]
\arrow["{\begin{matrix} d_A = p_A \circ u \\ \text{DemandPlan}_0 = 000.0 \\ \text{DemandPlan}_1 = 117.0 \\ \text{DemandPlan}_2 = 247.27 \end{matrix}}"{left=0.1cm}, pos=0.3, color={rgb,255:red,92;green,214;blue,92}, from=1-1, to=3-2]
\arrow["{\begin{matrix} d_B = p_B \circ u \\ \text{GoodProduction}_0 = 31.17 \\ \text{GoodProduction}_1 = 1.0 \\ \text{GoodProduction}_2 = 3.20 \end{matrix}}"{right=1.5cm}, pos=0.7, color={rgb,255:red,92;green,214;blue,92}, from=1-1, to=2-3]
\arrow["{\begin{matrix} p_A \\ \text{Demand} \\ \text{Extraction} \end{matrix}}"{left=0.1cm}, pos=0.1, color={rgb,255:red,214;green,92;blue,92}, from=2-2, to=3-2]
\arrow["{\begin{matrix} p_B \\ \text{Supply} \\ \text{Extraction} \end{matrix}}"{below=0.4cm}, pos=0.6, color={rgb,255:red,214;green,92;blue,92}, from=2-2, to=2-3]
\arrow["{\begin{matrix} f \\ \text{Demand/Supply} \\ \text{Ratio + Markup} \end{matrix}}"'{above=0.6cm}, from=3-2, to=3-3]
\arrow["{\begin{matrix} g \\ \text{Supply/Demand} \\ \text{Ratio + Markup} \end{matrix}}"{right=0.5cm}, pos=0.9, from=2-3, to=3-3]
\end{tikzcd}
}
\vspace{-2.5cm}
\caption{Price Formation Pullback: The price formation pullback validates the equilibrium price calculation as the ratio of demand to supply plus markup, ensuring market clearing across periods.}
\label{H}
\end{figure}
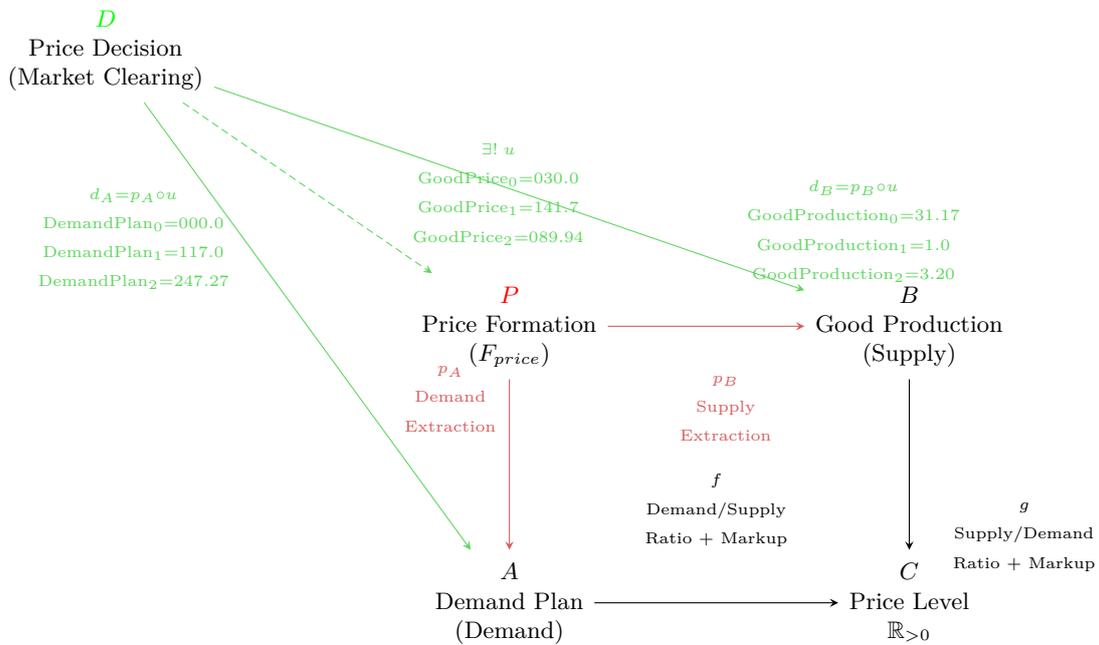

\textbf{Economics: Market Clearing and Price Discovery Mechanisms:}
Figure~\ref{F} shows the natural transformation sequence from market clearing decision through demand and supply coordination (0.0 $\to$ 117.0 $\to$ 247.27, 31.17 $\to$ 1.0 $\to$ 3.20) 
to price equilibrium (30.0$ \to$ 141.7 $\to$ 89.94) via categorical pullback construction.

The price formation pullback embodies the fundamental principle of market equilibrium through categorical limits. 
The price formation function $DemandPlan / GoodProduction + \mu$ becomes a 
typed morphism $F_{price}: \mathbb{R}_{\geq 0} \times \mathbb{R}_{> 0} \times \mathbb{R}_{\geq 0} \rightarrow \mathbb{R}_{> 0}$ 
that transforms demand-supply relationships into equilibrium prices while ensuring market clearing conditions.

Economically, this captures the essential market mechanism where prices emerge from the interaction of demand plan (DemandPlan: 0.0 $\to$ 117.0 $\to$ 247.27) 
and supply (GoodProduction: 31.17 $\to$ 1.0 $\to$ 3.20) forces, mediated by markup policies (PriceMarkup: 30.0 $\to$ 24.7 $\to$ 12.65). 
The pullback constraint ensures that price formation respects both demand constraints and supply capacities, generating equilibrium prices (GoodPrice: 30.0 $\to$ 141.7 $\to$ 89.94) 
that clear markets, resulting in the price evolution.

The three-layer architecture of diagram~\ref{G} reveals the categorical nature of price discovery: 
the yellow layer represents market clearing conditions, the blue layer embodies the price formation mechanism, and the green layer shows the resulting equilibrium prices. 
The pullback corners ensure that price formation respects both demand-side constraints and supply-side constraints simultaneously. 
This categorical approach to price theory provides mathematical foundations for understanding market equilibrium as a categorical limit, 
connecting to general equilibrium theory by categorical mathematics for a formal analysis of price stability, market efficiency and
the compositional, universal and consistency properties of the {\it general} of general equilibrium theory.
%---------------------------------------------------------------------------------------------------------------------------------------------------------------------------------
\subsection{Investment Decisions}

%\clearpage
\subsubsection{Investment Decision Sigmoid}

\begin{definition}[Investment Learning]
The investment morphism is specified to implement a sigmoid activation function as the sectoral and overall systems learning and adaptation technology.
The sigmoid function transforms $DemandSurplus$ into investment:
$$\text{Investment} = \sigma_A + \frac{\sigma_B}{1 + e^{-\frac{DemandSurplus}{\sigma_C}}}$$
\begin{align}
F_{\text{Investment}}: \mathbb{R} \times \mathbb{R}_{> 0} \times \mathbb{R}_{> 0} \times \mathbb{R}_{> 0} &\rightarrow \mathbb{R}_{> 0}\\
F_{\text{Investment}}(\text{DemandSurplus}, \sigma_A, \sigma_B, \sigma_C) &= \sigma_A + \frac{\sigma_B}{1 + e^{-\frac{DemandSurplus}{\sigma_C}}}
\end{align}
with $\sigma_A = 20.0$, $\sigma_B = 480.0$, $\sigma_C = 200.0$:
\hfill$\blacksquare$
\end{definition}

Evolution across \textit{periods 0, 1, 2}:
\begin{align}
\text{Investment}_0 &= 20.0 + \frac{480.0}{1 + \exp(-0.0 / 200.0)} = 260.0\nonumber\\
\text{Investment}_1 &= 20.0 + \frac{480.0}{1 + \exp(-49.4 / 200.0)} = 289.49\nonumber\\
\text{Investment}_2 &= 20.0 + \frac{480.0}{1 + \exp(-25.36 / 200.0)} = 275.20\nonumber
\end{align}

\begin{figure}[htbp]
\centering
\tikzset{
stringdiagram/.style={
baseline=(current bounding box.center),
every node/.style={draw=none},
every path/.style={thick}
}
}
\begin{tikzpicture}[stringdiagram]
\begin{scope}[shift={(0,0)}]
\node at (1.5,4.5) {$\mathcal{C}_{\text{Decision}}$};
\node[draw, rectangle, minimum width=2.0cm, minimum height=0.8cm, fill=yellow!20] (decision) at (1.0,3.5) {Investment};
\node at (0.0,2.5) {$\mathcal{C}_{\text{Sigmoid}}$};
\node[draw, rectangle, minimum width=2.0cm, minimum height=0.8cm, fill=blue!20] (sigmoid) at (1.0,1.5) {Sigmoid};
\node at (0.0,0.5) {$\mathcal{C}_{\text{Level}}$};
\node[draw, rectangle, minimum width=2.0cm, minimum height=0.8cm, fill=green!20] (level) at (1.0,-0.5) {Investment};
\node at (6.5,4.5) {$\mathcal{C}_{\text{Surplus}}$};
\node[draw, rectangle, minimum width=2.0cm, minimum height=0.8cm, fill=yellow!20] (surplus) at (6.0,3.5) {0.0 $\to$ 49.4};
\node at (5.0,2.5) {$\mathcal{C}_{\text{Activation}}$};
\node[draw, rectangle, minimum width=2.0cm, minimum height=0.8cm, fill=blue!20] (activation) at (6.0,1.5) {Activation};
\node at (5.0,0.5) {$\mathcal{C}_{\text{Response}}$};
\node[draw, rectangle, minimum width=2.0cm, minimum height=0.8cm, fill=green!20] (response) at (6.0,-0.5) {260.0};
\node at (11.0,4.5) {$\mathcal{C}_{\text{Parameters}}$};
\node[draw, rectangle, minimum width=2.0cm, minimum height=0.8cm, fill=yellow!20] (params) at (10.5,3.5) {sigA=20.0};
\node at (9.5,2.5) {$\mathcal{C}_{\text{Transform}}$};
\node[draw, rectangle, minimum width=2.0cm, minimum height=0.8cm, fill=blue!20] (transform) at (10.5,1.5) {Nonlinear};
\node at (9.5,0.5) {$\mathcal{C}_{\text{Output}}$};
\node[draw, rectangle, minimum width=2.0cm, minimum height=0.8cm, fill=green!20] (output) at (10.5,-0.5) {289.49};
\node at (15.5,4.5) {$\mathcal{C}_{\text{Evolution}}$};
\node[draw, rectangle, minimum width=2.0cm, minimum height=0.8cm, fill=yellow!20] (evolution) at (15.0,3.5) {25.36};
\node at (14.0,2.5) {$\mathcal{C}_{\text{Universal}}$};
\node[draw, rectangle, minimum width=2.0cm, minimum height=0.8cm, fill=blue!20] (universal) at (15.0,1.5) {$\exists ! u$};
\node at (14.0,0.5) {$\mathcal{C}_{\text{Final}}$};
\node[draw, rectangle, minimum width=2.0cm, minimum height=0.8cm, fill=green!20] (final) at (15.0,-0.5) {275.20};
\draw[->, thick] (decision) -- (sigmoid) node[midway,right=0.0cm] {$\pi_1$};
\draw[->, thick] (sigmoid) -- (level) node[midway,right=0.0cm] {$\pi_2$};
\draw[->, thick] (surplus) -- (activation) node[midway,right=0.0cm] {$\phi_1$};
\draw[->, thick] (activation) -- (response) node[midway,right=0.0cm] {$\phi_2$};
\draw[->, thick] (params) -- (transform) node[midway,right=0.0cm] {$\psi_1$};
\draw[->, thick] (transform) -- (output) node[midway,right=0.0cm] {$\psi_2$};
\draw[->, thick] (evolution) -- (universal) node[midway,right=0.0cm] {$\zeta_1$};
\draw[->, thick] (universal) -- (final) node[midway,right=0.0cm] {$\zeta_2$};
\draw[->, thick, red] (decision) -- (surplus) node[midway,above] {$\eta_{\text{sur}}$};
\draw[->, thick, red] (sigmoid) -- (activation) node[midway,above] {$\eta_{\text{act}}$};
\draw[->, thick, red] (level) -- (response) node[midway,above] {$\eta_{\text{res}}$};
\draw[->, thick, red] (surplus) -- (params) node[midway,above] {$\eta_{\text{par}}$};
\draw[->, thick, red] (activation) -- (transform) node[midway,above] {$\eta_{\text{trans}}$};
\draw[->, thick, red] (response) -- (output) node[midway,above] {$\eta_{\text{out}}$};
\draw[->, thick, red] (params) -- (evolution) node[midway,above] {$\eta_{\text{evol}}$};
\draw[->, thick, red] (transform) -- (universal) node[midway,above] {$\eta_{\text{univ}}$};
\draw[->, thick, red] (output) -- (final) node[midway,above] {$\eta_{\text{fin}}$};
\node at (0.3,2.9) {$\lrcorner$};
\node at (5.3,2.9) {$\lrcorner$};
\node at (9.8,2.9) {$\lrcorner$};
\node at (14.3,2.9) {$\lrcorner$};
\node at (3.5,3.5) {$\otimes$};
\node at (3.5,1.5) {$\otimes$};
\node at (3.5,-0.5) {$\otimes$};
\node at (8.2,3.5) {$\otimes$};
\node at (8.2,1.5) {$\otimes$};
\node at (8.2,-0.5) {$\otimes$};
\node at (12.7,3.5) {$\otimes$};
\node at (12.7,1.5) {$\otimes$};
\node at (12.7,-0.5) {$\otimes$};
\draw[->, thick, purple, dashed] (decision) to[bend left=25] (evolution) node at (10.0, 6.0) {$\scriptsize \sigma_A + \sigma_B / (1 + exp(-\text{DemandSurplus}/\sigma_C ))$};
\end{scope}
\end{tikzpicture}
\caption{Commutative Diagram: Investment Sigmoid Pullback Structure}
\label{I}
\end{figure}
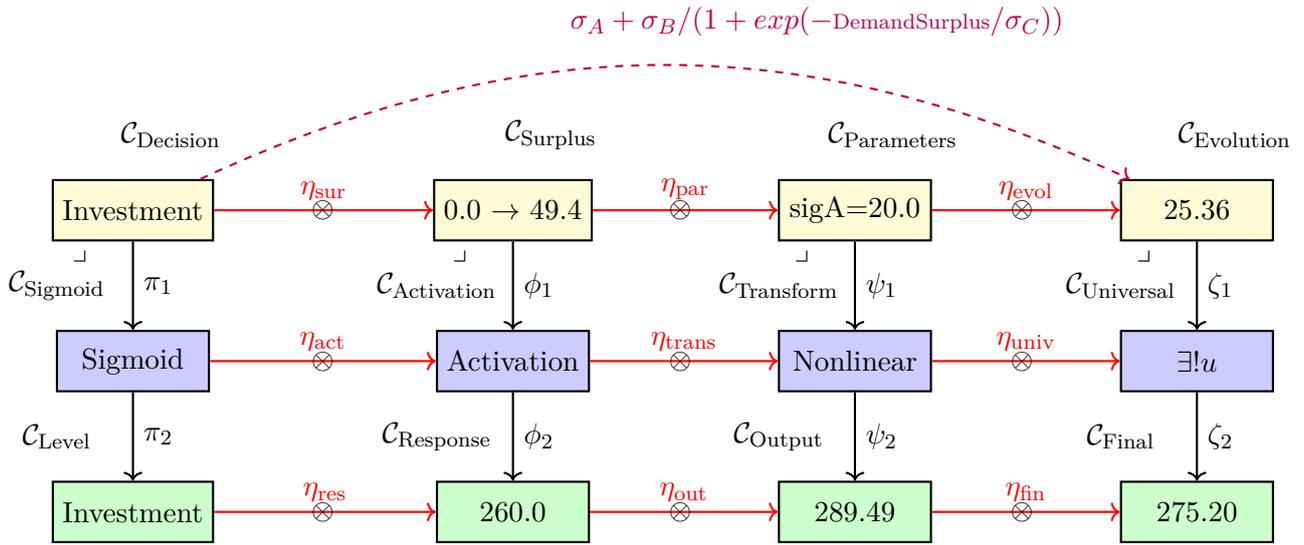

\begin{figure}[htbp]
\centering
{\footnotesize
\begin{tikzpicture}[scale=0.9]
\node[rectangle, fill=yellow!20, minimum width=5cm, minimum height=0.8cm] (decision) at (0, 6) {\textbf{Investment Decision}};
\node[rectangle, fill=blue!20, minimum width=5cm, minimum height=0.8cm] (sigmoid) at (0, 3) {\textbf{Sigmoid Function}};
\node[left=0.5cm of sigmoid] (activation) {Sigmoid Activation};
\node[right=0.5cm of sigmoid] (transform) {Nonlinear Transform};
\node[rectangle, fill=green!20, minimum width=5cm, minimum height=0.8cm] (investment) at (0, 0) {\textbf{Investment Level}};
\node[above=0.1cm, left=1.5cm of investment] (output) {Investment: 260.0 $\to$ 289.49 $\to$ 275.20};
\node[above=0.1cm, right=1.5cm of investment] (formula) {$\sigma_A + \frac{\sigma_B}{1 + e^{-\frac{DemandSruplus}{\sigma_C}}}$};
\node[above=5.0cm of output] (surplus) {DemandSurplus: 0.0 $\to$ 49.4 $\to$ 25.36};
\node[above=5.0cm of formula] (params) {sigA=20.0, sigB=480.0, sigC=200.0};
\node at (-2.5, 4.5) {$\lrcorner$};
\node at (2.5, 4.5) {$\lrcorner$};
\node at (-2.5, 1.5) {$\lrcorner$};
\node at (2.5, 1.5) {$\lrcorner$};
\draw[->, thick, red] (decision) -- (sigmoid) node[pos=0.7,left] {Sigmoid Function};
\draw[->, thick, red] (sigmoid) -- (investment) node[pos=0.3,right] {Investment Output};
\draw[->, thick, blue] (surplus) -- (activation) node[midway,right] {Surplus Input};
\draw[->, thick, blue] (params) -- (transform) node[midway,left] {Parameter Input};
\draw[dashed, green] (surplus) -- (output) node[pos=0.7,left] {Investment Response};
\draw[dashed, green] (params) -- (formula) node[pos=0.7,right] {Mathematical Form};
\node at (0.0,-1.0) {};
\end{tikzpicture}
}
\vspace{-1cm}
\caption{Commutative Diagram: Investment Sigmoid Pullback Flow}
\label{J}
\end{figure}
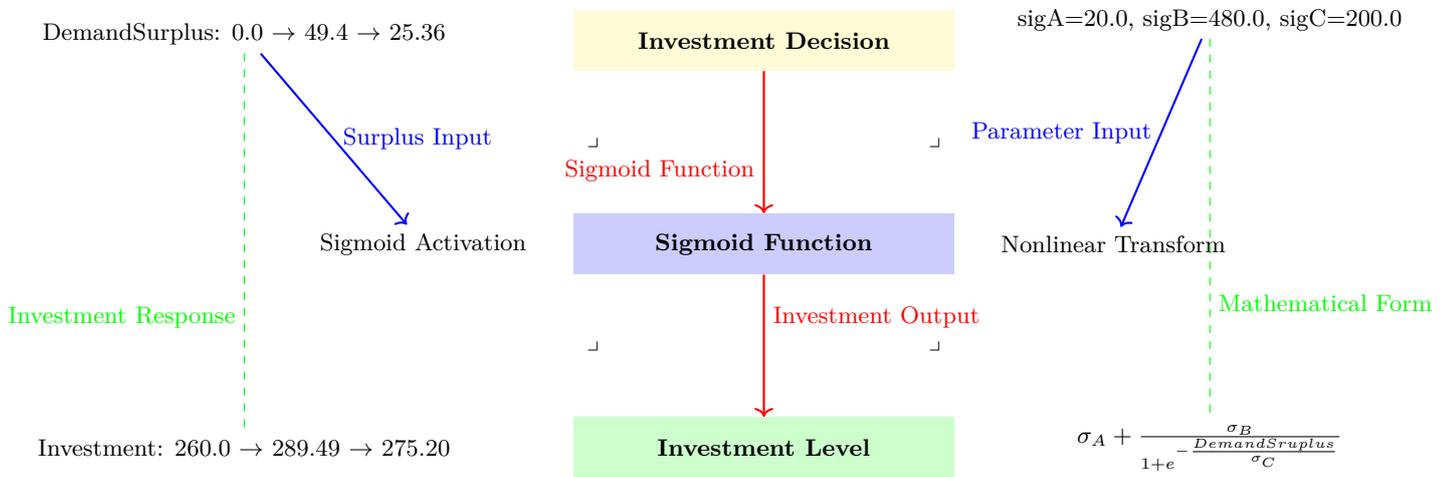

\begin{figure}[htbp]
\centering
{\footnotesize
\begin{tikzcd}[column sep=8em, row sep=7em, arrows={-stealth}]
{\begin{matrix} \green{D} \\ \text{Investment Decision} \\ \text{(Sigmoid Response)} \end{matrix}} \\
& {\begin{matrix} \red{P} \\ \text{Investment Function} \\ \text{(Sigmoid)} \end{matrix}}
& {\begin{matrix} B \\ \text{Sigmoid Parameters} \\ \text{(a,b,c)} \end{matrix}} \\
& {\begin{matrix} A \\ \text{Demand Surplus} \\ \mathbb{R} \end{matrix}}
& {\begin{matrix} C \\ \text{Investment Level} \\ \mathbb{R}_{> 0} \end{matrix}} \\
\arrow["{\begin{matrix} \exists ! \; u \\ \text{Investment}_0 = 260.0 \\ \text{Investment}_1 = 289.49 \\ \text{Investment}_2 = 275.20 \end{matrix}}", pos=0.9, color={rgb,255:red,92;green,214;blue,92}, dashed, from=1-1, to=2-2]
\arrow["{\begin{matrix} d_A = p_A \circ u \\ \text{DemandSurplus}_0 = 00.0 \\ \text{DemandSurplus}_1 = 49.4 \\ \text{DemandSurplus}_2 = 25.36 \end{matrix}}"{left=0.1cm}, pos=0.3, color={rgb,255:red,92;green,214;blue,92}, from=1-1, to=3-2]
\arrow["{\begin{matrix} d_B = p_B \circ u \\ \text{sigA} = 20.0 \\ \text{sigB} = 480.0 \\ \text{sigC} = 200.0 \end{matrix}}"{right=1.5cm}, pos=0.7, color={rgb,255:red,92;green,214;blue,92}, from=1-1, to=2-3]
\arrow["{\begin{matrix} p_A \\ \text{Surplus} \\ \text{Input} \end{matrix}}"{left=0.1cm}, pos=0.3, color={rgb,255:red,214;green,92;blue,92}, from=2-2, to=3-2]
\arrow["{\begin{matrix} p_B \\ \text{Parameter} \\ \text{Application} \end{matrix}}"{below=0.2cm}, pos=0.4, color={rgb,255:red,214;green,92;blue,92}, from=2-2, to=2-3]
\arrow["{\begin{matrix} f \\ \sigma_A + \sigma_B/(1+exp(-\text{DemandSurplus}/\sigma_C)) \\ \text{Sigmoid Transform} \end{matrix}}"'{above=0.4cm}, from=3-2, to=3-3]
\arrow["{\begin{matrix} g \\ \text{Parameter} \\ \text{Sigmoid} \end{matrix}}"{right=0.3cm}, pos=0.7, from=2-3, to=3-3]
\end{tikzcd}
}
\vspace{-2.5cm}
\caption{Investment Sigmoid Pullback: The investment decision pullback applies the sigmoid function to demand surplus, with parameters ($\sigma_A=20.0, \sigma_B=480.0, \sigma_C=200.0$) governing the response curve.}
\label{K}
\end{figure}
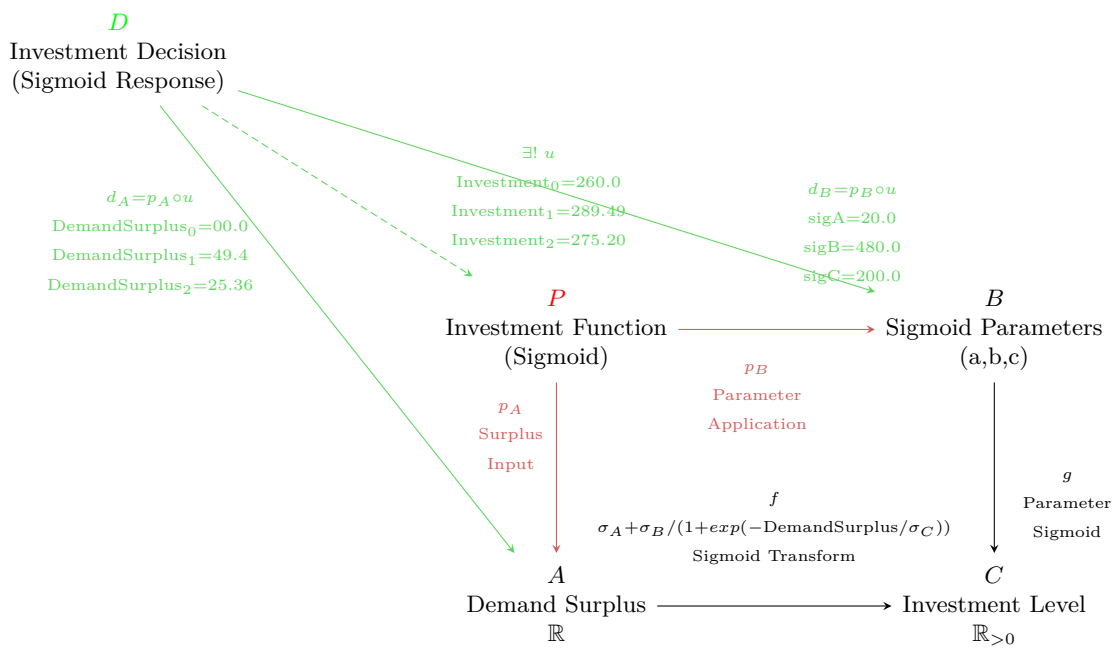

\textbf{Economics: Investment Decision-Making:}
Figure~\ref{K} shows the natural transformation sequence from investment decision through demand surplus (0.0 $\to$ 49.4 $\to$ 25.36) 
and parameters ($\sigma_A=20.0, \sigma_B=480.0, \sigma_C=200.0$) to investment levels (260.0 $\to$ 289.49 $\to$ 275.20) via categorical pullback construction.

The investment sigmoid pullback captures the corporate investment decisions through categorical structures. 
The sigmoid transformation $\sigma_A + \sigma_B / (1 + exp(-\text{DemandSurplus}/\sigma_C))$ 
becomes a typed morphism $F_{\text{Investment}}: \mathbb{R} \times \mathbb{R}_{> 0} \times \mathbb{R}_{> 0} \times \mathbb{R}_{> 0} \rightarrow \mathbb{R}_{> 0}$ 
that transforms demand surplus signals into investment levels while incorporating behavioral response patterns.

Economically, this represents how firms respond to market signals (DemandSurplus: 0.0 $\to$ 49.4 $\to$ 25.36) 
through some rationality and gradual adjustment and learning mechanisms captured by the sigmoid parameters ($\sigma_A=20.0, \sigma_B=480.0, \sigma_C=200.0$). 
The pullback ensures that investment decisions emerge from the interaction of market conditions and firm behavioral parameters, 
generating investment patterns (Investment: 260.0 $\to$ 289.49 $\to$ 275.20) that avoid extreme responses and maintaining sensitivity to economic conditions.

The three-layer architecture of diagram~\ref{J} reveals the categorical nature of behavioral investment theory: the yellow layer represents investment decision triggers, 
the blue layer embodies the behavioral response mechanism (sigmoid activation), and the green layer shows the resulting investment levels. 
The pullback corners ensure that investment responses respect both market signal constraints and behavioral parameter constraints. 
This categorical approach to investment theory bridges microeconomic investment theory with behavioral finance, 
providing mathematical foundations for understanding how market signals translate into capital formation through firm behavior patterns 
that may incorporate bounded rationality or adjustment costs.

%---------------------------------------------------------------------------------------------------------------------------------------------------------------------------------
%\clearpage
\subsubsection{Investment Allocation}

\begin{definition}[Investment Allocation]
The investment allocation functors distribute investment between resources and labor.
\begin{align}
\text{InvestmentRes} &= \text{Investment} \times \lambda\\
\text{InvestmentLab} &= \text{Investment} \times (1 - \lambda)
\end{align}
\begin{align}
F_{InvestRes}&:& \mathbb{R}_{> 0} \times [0,1] &\rightarrow \mathbb{R}_{\geq 0} && F_{InvestRes}(i, s) &= i \times s\\
F_{InvestLab}&:& \mathbb{R}_{> 0} \times [0,1] &\rightarrow \mathbb{R}_{\geq 0} && F_{InvestLab}(i, s) &= i \times (1 - s)
\end{align}
with constraint $s_{\text{res}} + s_{\text{lab}} = 1$ and $\lambda = 0.8$ and $1 - \lambda = 0.2$:
\hfill$\blacksquare$
\end{definition}

Evolution across \textbf{periods 0, 1, 2}:
\begin{align}
\text{InvestmentRes}_0 &= 260.0 \times 0.8 = 208.0\nonumber\\
\text{InvestmentRes}_1 &= 289.49 \times 0.8 = 231.59\nonumber\\
\text{InvestmentRes}_2 &= 275.20 \times 0.8 = 220.16\nonumber
\end{align}

\begin{figure}[htbp]
\centering
\tikzset{
stringdiagram/.style={
baseline=(current bounding box.center),
every node/.style={draw=none},
every path/.style={thick}
}
}
\begin{tikzpicture}[stringdiagram]
\begin{scope}[shift={(0,0)}]
\node at (1.5,4.5) {$\mathcal{C}_{\text{Base}}$};
\node[draw, rectangle, minimum width=2.0cm, minimum height=0.8cm, fill=yellow!20] (base) at (1.0,3.5) {Investment};
\node at (0.0,2.5) {$\mathcal{C}_{\text{Allocation}}$};
\node[draw, rectangle, minimum width=2.0cm, minimum height=0.8cm, fill=blue!20] (allocation) at (1.0,1.5) {Allocation};
\node at (0.0,0.5) {$\mathcal{C}_{\text{Allocated}}$};
\node[draw, rectangle, minimum width=2.0cm, minimum height=0.8cm, fill=green!20] (allocated) at (1.0,-0.5) {Allocated};
\node at (6.5,4.5) {$\mathcal{C}_{\text{Total}}$};
\node[draw, rectangle, minimum width=2.0cm, minimum height=0.8cm, fill=yellow!20] (total) at (6.0,3.5) {260.0};
\node at (5.0,2.5) {$\mathcal{C}_{\text{ResShare}}$};
\node[draw, rectangle, minimum width=2.0cm, minimum height=0.8cm, fill=blue!20] (resshare) at (6.0,1.5) {0.8};
\node at (5.0,0.5) {$\mathcal{C}_{\text{InvestRes}}$};
\node[draw, rectangle, minimum width=2.0cm, minimum height=0.8cm, fill=green!20] (investres) at (6.0,-0.5) {208.0};
\node at (11.0,4.5) {$\mathcal{C}_{\text{Constraint}}$};
\node[draw, rectangle, minimum width=2.0cm, minimum height=0.8cm, fill=yellow!20] (constraint) at (10.5,3.5) {0.8+0.2=1};
\node at (9.5,2.5) {$\mathcal{C}_{\text{LabShare}}$};
\node[draw, rectangle, minimum width=2.0cm, minimum height=0.8cm, fill=blue!20] (labshare) at (10.5,1.5) {0.2};
\node at (9.5,0.5) {$\mathcal{C}_{\text{InvestLab}}$};
\node[draw, rectangle, minimum width=2.0cm, minimum height=0.8cm, fill=green!20] (investlab) at (10.5,-0.5) {52.0};
\node at (15.5,4.5) {$\mathcal{C}_{\text{Evolution}}$};
\node[draw, rectangle, minimum width=2.0cm, minimum height=0.8cm, fill=yellow!20] (evolution) at (15.0,3.5) {289.49$\to$275.20};
\node at (14.0,2.5) {$\mathcal{C}_{\text{Universal}}$};
\node[draw, rectangle, minimum width=2.0cm, minimum height=0.8cm, fill=blue!20] (universal) at (15.0,1.5) {$\exists ! u$};
\node at (14.0,0.5) {$\mathcal{C}_{\text{Final}}$};
\node[draw, rectangle, minimum width=2.0cm, minimum height=0.8cm, fill=green!20] (final) at (15.0,-0.5) {231.59$\to$220.16};
\draw[->, thick] (base) -- (allocation) node[midway,right=0.0cm] {$\pi_1$};
\draw[->, thick] (allocation) -- (allocated) node[midway,right=0.0cm] {$\pi_2$};
\draw[->, thick] (total) -- (resshare) node[midway,right=0.0cm] {$\phi_1$};
\draw[->, thick] (resshare) -- (investres) node[midway,right=0.0cm] {$\phi_2$};
\draw[->, thick] (constraint) -- (labshare) node[midway,right=0.0cm] {$\psi_1$};
\draw[->, thick] (labshare) -- (investlab) node[midway,right=0.0cm] {$\psi_2$};
\draw[->, thick] (evolution) -- (universal) node[midway,right=0.0cm] {$\zeta_1$};
\draw[->, thick] (universal) -- (final) node[midway,right=0.0cm] {$\zeta_2$};
\draw[->, thick, red] (base) -- (total) node[midway,above] {$\eta_{\text{tot}}$};
\draw[->, thick, red] (allocation) -- (resshare) node[midway,above] {$\eta_{\text{res}}$};
\draw[->, thick, red] (allocated) -- (investres) node[midway,above] {$\eta_{\text{ires}}$};
\draw[->, thick, red] (total) -- (constraint) node[midway,above] {$\eta_{\text{con}}$};
\draw[->, thick, red] (resshare) -- (labshare) node[midway,above] {$\eta_{\text{lab}}$};
\draw[->, thick, red] (investres) -- (investlab) node[midway,above] {$\eta_{\text{ilab}}$};
\draw[->, thick, red] (constraint) -- (evolution) node[midway,above] {$\eta_{\text{evol}}$};
\draw[->, thick, red] (labshare) -- (universal) node[midway,above] {$\eta_{\text{univ}}$};
\draw[->, thick, red] (investlab) -- (final) node[midway,above] {$\eta_{\text{fin}}$};
\node at (0.3,2.9) {$\ulcorner$};
\node at (5.3,2.9) {$\ulcorner$};
\node at (9.8,2.9) {$\ulcorner$};
\node at (14.3,2.9) {$\ulcorner$};
\node at (3.5,3.5) {$\otimes$};
\node at (3.5,1.5) {$\otimes$};
\node at (3.5,-0.5) {$\otimes$};
\node at (8.2,1.5) {$\otimes$};
\node at (8.2,-0.5) {$\otimes$};
\node at (12.7,3.5) {$\otimes$};
\node at (12.7,1.5) {$\otimes$};
\node at (12.7,-0.5) {$\otimes$};
\draw[->, thick, purple, dashed] (base) to[bend left=25] (evolution) node at (8.0, 6.0) {$0.8 \times I + 0.2 \times I$};
\end{scope}
\end{tikzpicture}
\caption{Commutative Diagram: Investment Allocation Pushout Structure}
\label{L}
\end{figure}
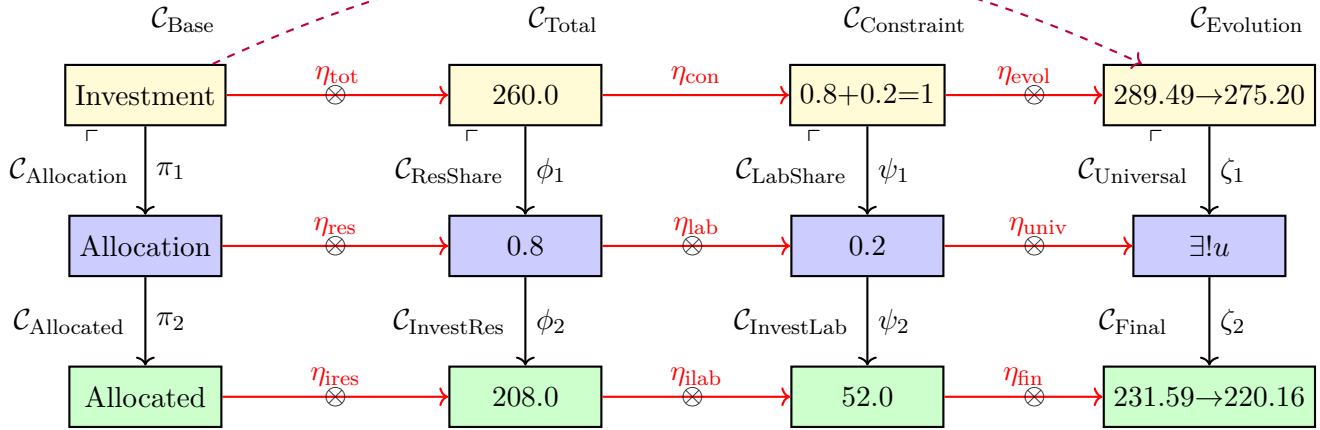

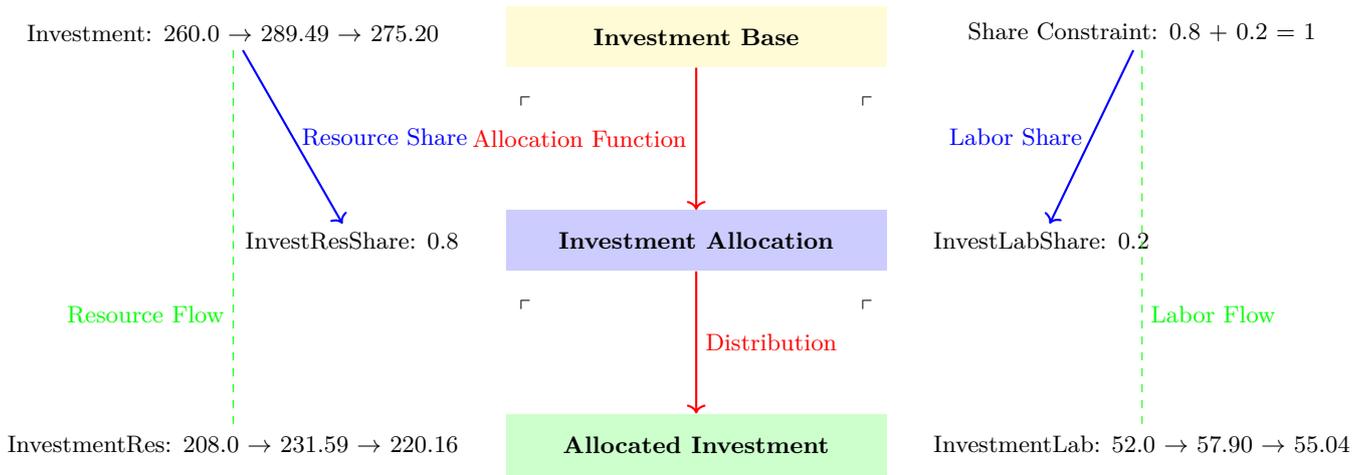
\begin{figure}[htbp]
\centering
{\footnotesize
\begin{tikzpicture}[scale=0.9]
\node[rectangle, fill=yellow!20, minimum width=5cm, minimum height=0.8cm] (base) at (0, 6) {\textbf{Investment Base}};
\node[rectangle, fill=blue!20, minimum width=5cm, minimum height=0.8cm] (allocation) at (0, 3) {\textbf{Investment Allocation}};
\node[left=0.5cm of allocation] (resshare) {InvestResShare: 0.8};
\node[right=0.5cm of allocation] (labshare) {InvestLabShare: 0.2};
\node[rectangle, fill=green!20, minimum width=5cm, minimum height=0.8cm] (allocated) at (0, 0) {\textbf{Allocated Investment}};
\node[left=0.5cm of allocated] (investres) {InvestmentRes: 208.0 $\to$ 231.59 $\to$ 220.16};
\node[right=0.5cm of allocated] (investlab) {InvestmentLab: 52.0 $\to$ 57.90 $\to$ 55.04};
\node[above=5.0cm of investres] (total) {Investment: 260.0 $\to$ 289.49 $\to$ 275.20};
\node[above=5.0cm of investlab] (constraint) {Share Constraint: 0.8 + 0.2 = 1};
\node at (-2.5, 5) {$\ulcorner$};
\node at (2.5, 5) {$\ulcorner$};
\node at (-2.5, 2) {$\ulcorner$};
\node at (2.5, 2) {$\ulcorner$};
\draw[->, thick, red] (base) -- (allocation) node[midway,left] {Allocation Function};
\draw[->, thick, red] (allocation) -- (allocated) node[midway,right] {Distribution};
\draw[->, thick, blue] (total) -- (resshare) node[midway,right] {Resource Share};
\draw[->, thick, blue] (constraint) -- (labshare) node[midway,left] {Labor Share};
\draw[dashed, green] (total) -- (investres) node[pos=0.7,left] {Resource Flow};
\draw[dashed, green] (constraint) -- (investlab) node[pos=0.7,right] {Labor Flow};
\node at (0.0,-1.0) {};
\end{tikzpicture}
}
\vspace{-1cm}
\caption{Commutative Diagram: Investment Allocation Pushout Flow}
\label{M}
\end{figure}

\begin{figure}[htbp]
\centering
{\footnotesize
\begin{tikzcd}[column sep=8em, row sep=7em, arrows={-stealth}]
    {\begin{matrix} C \\ \text{Investment Base} \\ \text{(Total Investment)} \end{matrix}}
    & {\begin{matrix} B \\ \text{Labor Share} \\ (1-\lambda) \end{matrix}} \\
    {\begin{matrix} A \\ \text{Resource Share} \\ \lambda \end{matrix}} 
    & {\begin{matrix} \red{P} \\ \text{Investment Allocation} \\ \text{(Distribution)} \end{matrix}} & \\
    && {\begin{matrix} \green{D} \\ \text{Allocated Investment} \\ \text{(Res + Lab)} \end{matrix}} & \\
    \arrow["{\begin{matrix} f \\ \text{InvestResShare} \\ \text{0.8} \end{matrix}}"{right}, pos=0.3, from=1-1, to=2-1]
    \arrow["{\begin{matrix} g \\ \text{InvestLabShare} \\ \text{0.2} \end{matrix}}"{above=0.3cm}, pos=0.5, from=1-1, to=1-2]
    \arrow["{\begin{matrix} p_A \\ \text{Resource} \\ \text{Allocation} \end{matrix}}"{above=0.4cm}, pos=0.6, color={rgb,255:red,214;green,92;blue,92}, from=2-1, to=2-2]
    \arrow["{\begin{matrix} p_B \\ \text{Labor} \\ \text{Allocation} \end{matrix}}"{right=0.3cm}, pos=0.7, color={rgb,255:red,214;green,92;blue,92}, from=1-2, to=2-2]
    \arrow["{\begin{matrix} d_A = u \circ p_A \\ \text{InvestmentRes}_0 = 208.0 \\ \text{InvestmentRes}_1 = 231.6 \\ \text{InvestmentRes}_2 = 220.2 \end{matrix}}"{left=0.8cm}, pos=0.6, color={rgb,255:red,92;green,214;blue,92}, from=2-1, to=3-3]
    \arrow["{\begin{matrix} d_B = u \circ p_B \\ \text{InvestmentLab}_0 = 52.0 \\ \text{InvestmentLab}_1 = 57.9 \\ \text{InvestmentLab}_2 = 55.0 \end{matrix}}"{right=0.5cm}, pos=0.8, color={rgb,255:red,92;green,214;blue,92}, from=1-2, to=3-3]
    \arrow["{\begin{matrix} \exists ! \;u \\ \text{Constraint:} \\ \text{0.8 + 0.2 = 1} \end{matrix}}"{description}, pos=0.2, color={rgb,255:red,92;green,214;blue,92}, dashed, from=2-2, to=3-3]
\end{tikzcd}
}
\vspace{-2.5cm}
\caption{Investment Allocation Pushout: The investment allocation pushout distributes total investment between resource and labor shares (0.8 and 0.2), ensuring the constraint that shares sum to 1.}
\label{N}
\end{figure}
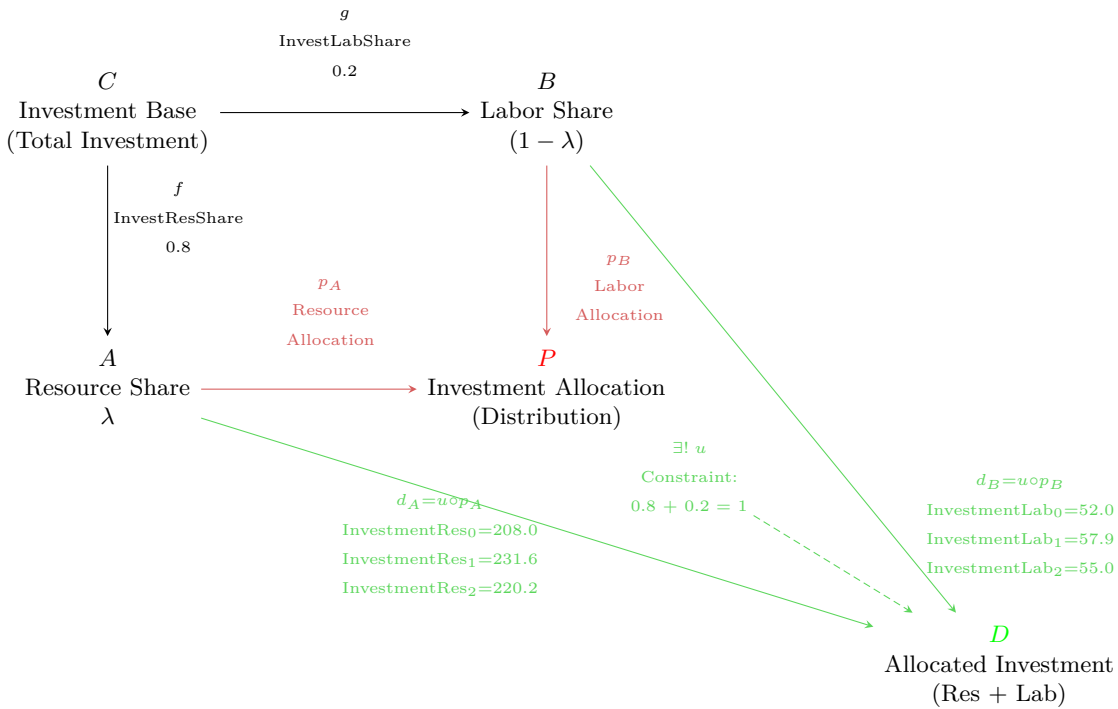

\textbf{Economics: Capital Allocation and Portfolio Theory:}
Diagram~\ref{L} shows the natural transformation sequence from investment base through total investment (260.0 $\to$ 289.49 $\to$ 275.20) and share constraint (0.8+0.2=1) 
to resource and labor allocation (208.0 $\to$ 231.59 $\to$ 220.16, 52.0 $\to$ 57.90 $\to$ 55.04) via the categorical pushout construction.

The investment allocation pushout embodies the fundamental principle of capital budgeting through categorical colimits. 
The allocation functions $F_{\text{InvestRes}}$ and $F_{\text{InvestLab}}$ become typed morphisms that distribute total investment 
capital (Investment: 260.0 $\to$ 289.49 $\to$ 275.20) across factor-specific investments while respecting portfolio constraint that allocation shares sum to unity (0.8 + 0.2 = 1).

Economically, this captures how firms allocate scarce investment capital across heterogeneous productive factors based on factor-specific 
investment shares ($\lambda=0.8$, $1-\lambda=0.2$). The pushout structure ensures that allocation decisions respect both capital 
constraints and factor complementarity relationships, generating balanced investment flows (InvestmentRes: 208.0 $\to$ 231.59 $\to$ 220.16 and 
InvestmentLab: 52.0 $\to$ 57.90 $\to$ 55.04) that maintain productive factor balance while adapting to changing economic conditions.

The three-layer architecture of diagram~\ref{M} reveals the categorical nature of capital allocation theory: the yellow layer 
represents total investment capacity, the blue layer embodies the allocation mechanism, and the green layer shows the resulting 
factor-specific investment flows. The pushout corners mark where total investment morphisms compose into the universal factor allocation morphism. 
This categorical approach to capital budgeting connects microeconomic investment theory with portfolio optimisation theory, providing 
mathematical foundations for understanding how firms balance competing investment opportunities while maintaining operational efficiency 
and responding to changing factor price relationships.

%---------------------------------------------------------------------------------------------------------------------------------------------------------------------------------
%\clearpage
\subsubsection{Intertemporal Contract Memory Updates}

Investment decisions create future contractual obligations that must be tracked through categorical state updates. 
This memory mechanism represents the intertemporal dimension of economic contracts, where current investment decisions generate future payment obligations.

The memory update functors manage the intertemporal contract obligations created by investment decisions.

\begin{definition}[Memory Update]
The memory update functors $F_{\text{MemoryUpdate}}$ maintain FILO (First In, Last Out) stacks of contractual obligations:
\begin{align}
\mathcal{H}_{wage}^{t+1} &= \texttt{p2H}(\mathcal{H}_{wage}^t, \text{InvestmentLab}^t)\\
\mathcal{H}_{repay}^{t+1} &= \texttt{p2H}(\mathcal{H}_{repay}^t, \text{Repayment}^t)
\end{align}

The FILO update operation is defined as:
\begin{align}
F_{\text{FILO}} &:& \mathbb{R}^{\tau} \times \mathbb{R} &\rightarrow \mathbb{R}^{\tau}
&& F_{\text{FILO}}([h_1, h_2, \ldots, h_{\tau}], x) &= [x, h_1, h_2, \ldots, h_{\tau-1}]
\end{align}
where $\tau = 10$ is the memory length.

The memory update is implemented as imperative state modification:
\texttt{p2H(hist, newelem) = [newelem; hist[1:end-1]]}\\
\texttt{stateNew.wageHist = p2H(state.wageHist, InvestmentLab)}\\
\texttt{stateNew.repayHist = p2H(state.repayHist, Repayment)}\\
\hfill$\blacksquare$
\end{definition}

The categorical formulation integrates memory updates as natural transformations 
that preserve temporal ordering while maintaining categorical structure through state morphisms.
Evolution of wage memory across \textbf{periods 0, 1, 2}:
\begin{align}
\mathcal{H}_{wage,0} &= [0.0, 0.0, 0.0, 0.0, 0.0, 0.0, 0.0, 0.0, 0.0, 0.0]\nonumber\\
\mathcal{H}_{wage,1} &= [52.0, 0.0, 0.0, 0.0, 0.0, 0.0, 0.0, 0.0, 0.0, 0.0]\nonumber\\
\mathcal{H}_{wage,2} &= [57.90, 52.0, 0.0, 0.0, 0.0, 0.0, 0.0, 0.0, 0.0, 0.0]\nonumber
\end{align}

Evolution of repayment memory across \textbf{periods 0, 1, 2}:
\begin{align}
\mathcal{H}_{repay,0} &= [0.0, 0.0, 0.0, 0.0, 0.0, 0.0, 0.0, 0.0, 0.0, 0.0]\nonumber\\
\mathcal{H}_{repay,1} &= [26.0, 0.0, 0.0, 0.0, 0.0, 0.0, 0.0, 0.0, 0.0, 0.0]\nonumber\\
\mathcal{H}_{repay,2} &= [28.95, 26.0, 0.0, 0.0, 0.0, 0.0, 0.0, 0.0, 0.0, 0.0]\nonumber
\end{align}

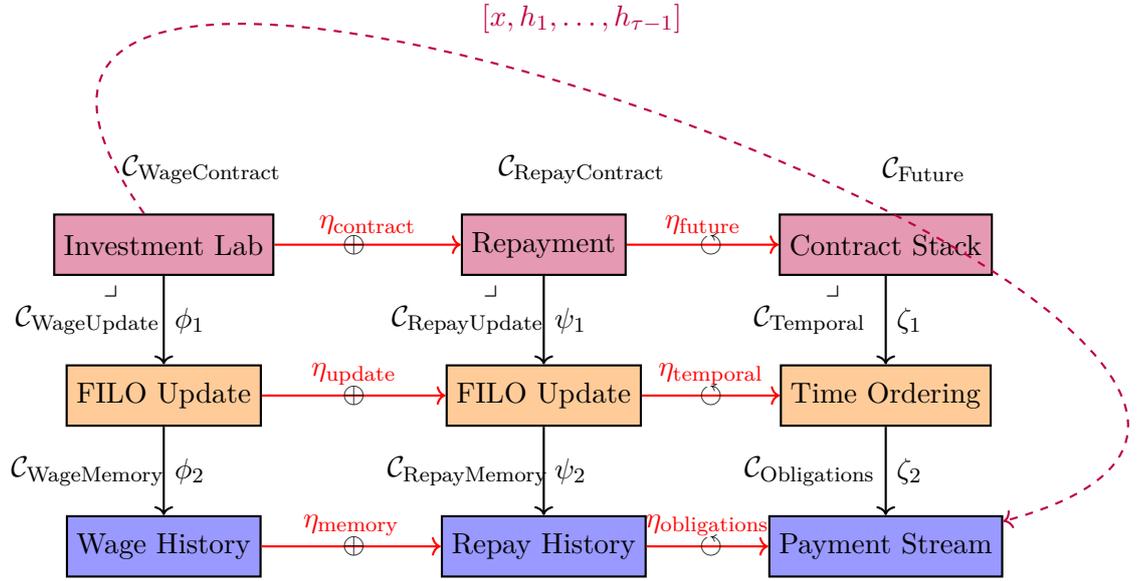
\begin{figure}[htbp]
\centering
\tikzset{
stringdiagram/.style={
baseline=(current bounding box.center),
every node/.style={draw=none},
every path/.style={thick}
}
}
\begin{tikzpicture}[stringdiagram]
\begin{scope}[shift={(0,0)}]
\node at (1.5,4.5) {$\mathcal{C}_{\text{WageContract}}$};
\node[draw, rectangle, minimum width=2.0cm, minimum height=0.8cm, fill=purple!40] (wagecontract) at (1.0,3.5) {Investment Lab};
\node at (0.0,2.5) {$\mathcal{C}_{\text{WageUpdate}}$};
\node[draw, rectangle, minimum width=2.0cm, minimum height=0.8cm, fill=orange!40] (wageupdate) at (1.0,1.5) {FILO Update};
\node at (0.0,0.5) {$\mathcal{C}_{\text{WageMemory}}$};
\node[draw, rectangle, minimum width=2.0cm, minimum height=0.8cm, fill=blue!40] (wagememory) at (1.0,-0.5) {Wage History};
\node at (6.5,4.5) {$\mathcal{C}_{\text{RepayContract}}$};
\node[draw, rectangle, minimum width=2.0cm, minimum height=0.8cm, fill=purple!40] (repaycontract) at (6.0,3.5) {Repayment};
\node at (5.0,2.5) {$\mathcal{C}_{\text{RepayUpdate}}$};
\node[draw, rectangle, minimum width=2.0cm, minimum height=0.8cm, fill=orange!40] (repayupdate) at (6.0,1.5) {FILO Update};
\node at (5.0,0.5) {$\mathcal{C}_{\text{RepayMemory}}$};
\node[draw, rectangle, minimum width=2.0cm, minimum height=0.8cm, fill=blue!40] (repaymemory) at (6.0,-0.5) {Repay History};
\node at (11.0,4.5) {$\mathcal{C}_{\text{Future}}$};
\node[draw, rectangle, minimum width=2.0cm, minimum height=0.8cm, fill=purple!40] (future) at (10.5,3.5) {Contract Stack};
\node at (9.5,2.5) {$\mathcal{C}_{\text{Temporal}}$};
\node[draw, rectangle, minimum width=2.0cm, minimum height=0.8cm, fill=orange!40] (temporal) at (10.5,1.5) {Time Ordering};
\node at (9.5,0.5) {$\mathcal{C}_{\text{Obligations}}$};
\node[draw, rectangle, minimum width=2.0cm, minimum height=0.8cm, fill=blue!40] (obligations) at (10.5,-0.5) {Payment Stream};
\draw[->, thick] (wagecontract) -- (wageupdate) node[midway,right=0.0cm] {$\phi_1$};
\draw[->, thick] (wageupdate) -- (wagememory) node[midway,right=0.0cm] {$\phi_2$};
\draw[->, thick] (repaycontract) -- (repayupdate) node[midway,right=0.0cm] {$\psi_1$};
\draw[->, thick] (repayupdate) -- (repaymemory) node[midway,right=0.0cm] {$\psi_2$};
\draw[->, thick] (future) -- (temporal) node[midway,right=0.0cm] {$\zeta_1$};
\draw[->, thick] (temporal) -- (obligations) node[midway,right=0.0cm] {$\zeta_2$};
\draw[->, thick, red] (wagecontract) -- (repaycontract) node[midway,above] {$\eta_{\text{contract}}$};
\draw[->, thick, red] (wageupdate) -- (repayupdate) node[midway,above] {$\eta_{\text{update}}$};
\draw[->, thick, red] (wagememory) -- (repaymemory) node[midway,above] {$\eta_{\text{memory}}$};
\draw[->, thick, red] (repaycontract) -- (future) node[midway,above] {$\eta_{\text{future}}$};
\draw[->, thick, red] (repayupdate) -- (temporal) node[midway,above] {$\eta_{\text{temporal}}$};
\draw[->, thick, red] (repaymemory) -- (obligations) node[midway,above] {$\eta_{\text{obligations}}$};
\node at (0.3,2.9) {$\lrcorner$};
\node at (5.3,2.9) {$\lrcorner$};
\node at (9.8,2.9) {$\lrcorner$};
\node at (3.5,3.5) {$\oplus$};
\node at (3.5,1.5) {$\oplus$};
\node at (3.5,-0.5) {$\oplus$};
\node at (8.2,3.5) {$\circlearrowleft$};
\node at (8.2,1.5) {$\circlearrowleft$};
\node at (8.2,-0.5) {$\circlearrowleft$};
\draw[->, thick, purple, dashed] (wagecontract) to[bend left=145, looseness=1.75] (obligations) node at (6.5,6.5) {$[x, h_1, \ldots, h_{\tau-1}]$};
\end{scope}
\end{tikzpicture}
\caption{Commutative Diagram: Intertemporal Contract Memory Structure}
\label{O}
\end{figure}

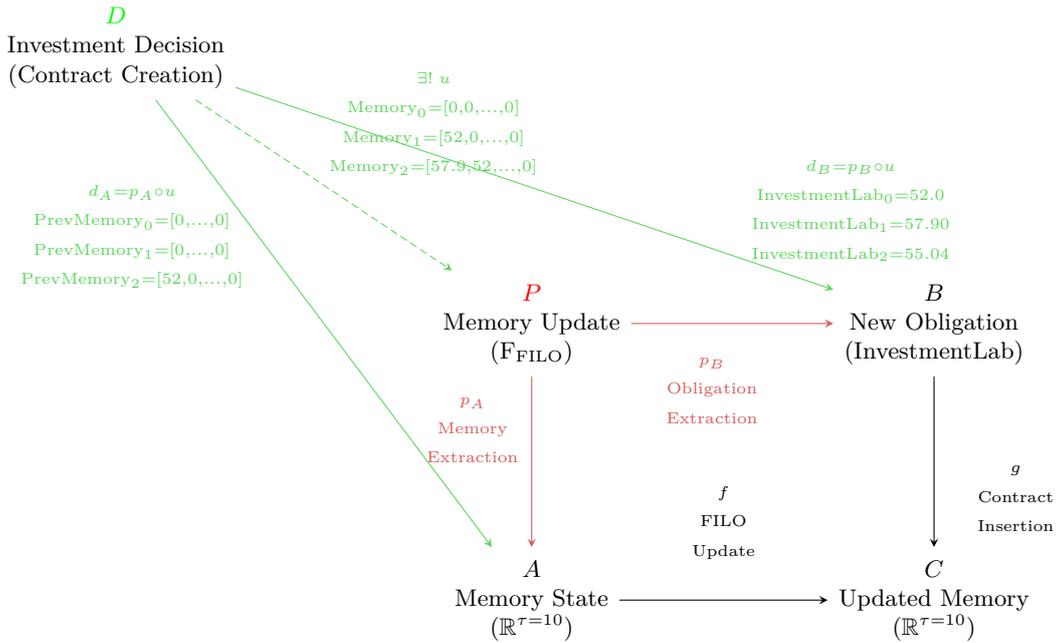
\begin{figure}[htbp]
\centering
{\footnotesize
\begin{tikzcd}[column sep=8em, row sep=7em, arrows={-stealth}]
{\begin{matrix} \green{D} \\ \text{Investment Decision} \\ \text{(Contract Creation)} \end{matrix}} \\
& {\begin{matrix} \red{P} \\ \text{Memory Update} \\ \text{(F$_{\text{FILO}}$)} \end{matrix}}
& {\begin{matrix} B \\ \text{New Obligation} \\ \text{(InvestmentLab)} \end{matrix}} \\
& {\begin{matrix} A \\ \text{Memory State} \\ \text{($\mathbb{R}^{\tau=10}$)} \end{matrix}}
& {\begin{matrix} C \\ \text{Updated Memory} \\ \text{($\mathbb{R}^{\tau=10}$)} \end{matrix}} \\
\arrow["{\begin{matrix} \exists ! \; u \\ \text{Memory}_0 = [0,0,...,0] \\ \text{Memory}_1 = [52,0,...,0] \\ \text{Memory}_2 = [57.9,52,...,0] \end{matrix}}", pos=0.5, color={rgb,255:red,92;green,214;blue,92}, dashed, from=1-1, to=2-2]
\arrow["{\begin{matrix} d_A = p_A \circ u \\ \text{PrevMemory}_0 = [0,...,0] \\ \text{PrevMemory}_1 = [0,...,0] \\ \text{PrevMemory}_2 = [52,0,...,0] \end{matrix}}"{left=0.1cm}, pos=0.3, color={rgb,255:red,92;green,214;blue,92}, from=1-1, to=3-2]
\arrow["{\begin{matrix} d_B = p_B \circ u \\ \text{InvestmentLab}_0 = 52.0 \\ \text{InvestmentLab}_1 = 57.90 \\ \text{InvestmentLab}_2 = 55.04 \end{matrix}}"{right=2.0cm}, pos=0.6, color={rgb,255:red,92;green,214;blue,92}, from=1-1, to=2-3]
\arrow["{\begin{matrix} p_A \\ \text{Memory} \\ \text{Extraction} \end{matrix}}"{left=0.1cm}, pos=0.3, color={rgb,255:red,214;green,92;blue,92}, from=2-2, to=3-2]
\arrow["{\begin{matrix} p_B \\ \text{Obligation} \\ \text{Extraction} \end{matrix}}"{below=0.2cm}, pos=0.4, color={rgb,255:red,214;green,92;blue,92}, from=2-2, to=2-3]
\arrow["{\begin{matrix} f \\ \text{FILO} \\ \text{Update} \end{matrix}}"'{above=0.4cm}, from=3-2, to=3-3]
\arrow["{\begin{matrix} g \\ \text{Contract} \\ \text{Insertion} \end{matrix}}"{right=0.5cm}, pos=0.7, from=2-3, to=3-3]
\end{tikzcd}
}
\vspace{-2.5cm}
\caption{Memory Update Pullback: The memory update pullback ensures intertemporal contract obligations are correctly inserted into FILO stacks, preserving temporal ordering of contractual commitments.}
\label{P}
\end{figure}

\textbf{Economics: Temporal Consistency and Contract Theory:}
The memory update mechanism creates a crucial {\it intertemporal closure and feedback loop} in the economic system.
Today's investment decisions become tomorrow's cost structure. 
This closure arises from the closure of debt relations in loans and repayments and the closure of production and consumption relations.
Figure~\ref{O} shows how investment decisions create future contractual obligations through FILO memory updates, 
transforming current investment flows (52.0 $\to$ 57.90 $\to$ 55.04, 26.0 $\to$ 28.95 $\to$ 27.52) 
into temporally ordered payment streams via categorical memory functors.

\begin{align}
\text{Investment}^t &\rightarrow \text{FutureObligation}^{t+1,\ldots,t+\tau}\label{eq:investmenttofutureobligation}\\
\text{FutureObligation}^{t+1} &\rightarrow \text{CostStructure}^{t+1}\label{eq:futureobligationtocoststructure}\\
\text{CostStructure}^{t+1} &\rightarrow \text{DemandPlan}^{t+1}\label{eq:coststructuretodemandplan}
\end{align}

The two intertemporal memory updates on the labor and investment contracts 
embody an important economic principle that current investment decisions create future contractual obligations. 
The FILO structure $F_{\text{FILO}}: \mathbb{R}^{\tau} \times \mathbb{R} \rightarrow \mathbb{R}^{\tau}$ represents how investment commitments 
(InvestmentLab: 52.0 $\to$ 57.90 $\to$ 55.04, Repayment: 26.0 $\to$ 28.95 $\to$ 27.52) generate temporally ordered payment obligations 
that constrain future economic decisions.

The types of the memory reveal how economic systems maintain {\it temporal consistency} through contract enforcement mechanisms. 
The memory updates ensure that investment decisions today (generating investment flows) automatically become cost constraints tomorrow 
(through WagesPayment and RepaysPayment aggregation), creating an essential feedback mechanism that connects investment cycles 
with cost structures across time periods.
The diagrams show how categorical memory functors preserve temporal ordering while maintaining economic consistency: 
investment decisions (purple layer) flow through FILO updates (orange layer) to create structured payment obligations (blue layer). 
This provides mathematical foundations for understanding how modern economies balance current investment opportunities with future 
payment obligations, connecting investment theory with contract theory through categorical time structures.
It also shows how to construct contracts as executed in verified and consistently aggregated macroeconomic accounting systems.

The modal logical aspects of contracts over time can be modelled in coalgebras for equation~(\ref{eq:investmenttofutureobligation})
for finit and infinite $\tau$ for institutions, companies, corporates or sectors.
In computer science the categorical coalgebraic approach is used for programming language design and their semantics (DSL).
A fundamental paper by~\cite{rutten_universal_2000} shows the categorical constructions used for dynamical systems from a computer scientific point of view.
In~\cite{AW2015} the coalgebraic approach is used for infinite game theory.
Coalgebras are also useful for the stochastic extension and game theoretical branching logics,
like trees and branching time in Kripke structures or modalities of deontic, temporal or other forms of logic.
An economic logic can be added on top of the memory of the system or subsystems.
On the practical side coalgebras are infinite data structures that allow to express infinite streams as in infinite horizon econometric theories like~(\cite{WK2010}).
The extraction of calculations from these obviously never completely computable structures,
is to evaluate them in a lazy way where calculations are only promissed to be done, and only calculated if needed.
Calculations is alwasys an extraction of some first finite numbers from the infinite streams located in final coalgebras.
The expressivity of the DSL increases very much with coalgebraic operations for example provided as stream calculus.
The calculus allows to express high level economic concepts in a compositional way
and let the compiler compose the meaning as the calculation of the whole program from the behavioural meaning of the calculated parts. 
This reduces the high level concepts into the low level implementation, which we do not want to be concerned with, 
with the consistency of the calculations being enforced by the compiler.
The lazily evaluated finite parts of the infinite stream are then the simulations of a finite sample from an infinite stream to which the parameters of the simulations map in the coalgebras.
Convergence in numerical methods and estimators or Monte Carlo simulations are mathematically typed as coalgebras.
So, it is useful to accompany the initial algebras of usual calculations and recursions by the coalgebraic final coalgebras.

The legal abstraction principle used in~\cite{MW2025} to disentangle the nature and type of money will be typed as contra variant and covariant functors
as the {\it tit for tat} in legal value exchanges.
The economic and legal content of contracts as programs in high level DSL languages 
are constructions on final structures complementing initial algebras.
The products of real and nominal categories should be analysed from the points of view and functors of
denotic (denatic modalties formalise legal musts and can haves) and temporal (like eventually, surely, maybe, once ...) modalities of legal contracts 
in equations (\ref{eq:investmenttofutureobligation}), (\ref{eq:futureobligationtocoststructure}) and (\ref{eq:coststructuretodemandplan}).

The memory of MoMaT can be extended in many more ways towards the ultimate goal of a consciousness of the economy as a whole.
It can contain a categorical state as a current theory an agent maintains about the economy as a whole or subparts.
The informational state of agents and aggregated levels entities is a functor that is kept consistent in natural transformations with the overall system information state.
The category of the economy $\mathcal{C}_{Economy}$ can be modularised in any subbranching depths and aspects of reflexive agents and 
levels, see~\cite{Dagstuhl2015} for a categorical semantics of reflexive economics.
In this workshop, we have laid out the foundations of this paper's categorical types of macroeconomic systems, wherein econometric agents can be modelled.

A memory that emerges into consciousness of biological and cognitive systems has been modelled categorically in terms of colimits as binding and emergence of memory evolutive systems (MES) by~\cite{EV2007}.
In this paper we present a memory of contracts, macroeconomics as a colimit (albeit not as colimits arranged which create emergent properties in the aggregated entity as in MES),
and an endofunctor implementing a categorical evolution.
The coregulators of MES can be modelled as open games of~\cite{GHWZ2018} for a polycentric management which needs a compositional econometric approach as (sub)system learners.
Emergence in MES is about reusing objects and concepts with emergent properties in various contexts as their defining property.
In economics of course this can be interpreted as reusage of creatable, copyable and deletable resources in initial algebras and final coalgebras.
The multiplicity property of complex objects with emergent properties in MES models the reusage of concepts as colimits.
The biological concepts of MES of binding, emergence and concept formation are useful for economic policy implementations and institutional design.
The multiplicity property is about reusage of concepts in different contexts which is what makes them valuable.
There is an economic value of pulling concepts, as in biological conscious systems, in front of the paranthesis.
This is very naturally an economic principle or even the economic principle, that a composition, the whole, is more valuable, i.e. is more than the parts.
Coregulators of MES formalise something very simular to polycentrism of optimal hierarchical control levels as a top level systemic economic policy question on decentralisation.
The memory can encompass as concept formation, as in MES, the causal detections of structural econometric theories as in~\cite{WK2010} which can be categorified as adjunctions.

%---------------------------------------------------------------------------------------------------------------------------------------------------------------------------------
%\clearpage
\subsubsection{Dividend Decision}

\begin{definition}[Dividend]
The dividend morphism transforms company profits into dividend payments:
\begin{align}
\text{Diff} &= \text{GoodProduction} \times \text{GoodPrice} - \text{WagesPayment} - \text{RepaysPayment}\\
\text{DividendDecision} &= \text{max}(0, \text{Diff} \times \text{DivRateDiff}) + \text{AccComBank} \times \text{DivRateBank}
\end{align}
\begin{align}
F_{\text{Diff}}&:& \mathbb{R}_{> 0} \times \mathbb{R}_{> 0} \times \mathbb{R}_{\geq 0} \times \mathbb{R}_{\geq 0} &\rightarrow \mathbb{R}
&& F_{\text{Diff}}(q, p, w, r)         &=& q \times p - w - r\\
F_{\text{Dividend}}&:& \mathbb{R} \times \mathbb{R}_{\geq 0} \times [0,1] \times [0,1] &\rightarrow \mathbb{R}_{\geq 0}
&& F_{\text{Dividend}}(d, b, r_d, r_b) &=& \max(0, d \times r_d) + b \times r_b
\end{align}
\hfill$\blacksquare$
\end{definition}

With $\delta_c = 0.15$ and $\delta_b = 0.4$:
\begin{align}
\text{Diff}_0 &= 52.0\\
\text{Diff}_1 &= 138.50\\
\text{Diff}_2 &= 121.25
\end{align}

The dividend payments across \textbf{periods 0, 1, 2}:
\begin{align}
\text{DividendDecision}_0 &= \text{max}(0, 52.0 \times 0.15) + 0.0 \times 0.4 = 7.8 \nonumber\\
\text{DividendDecision}_1 &= \text{max}(0, 138.5 \times 0.15) + 52.0 \times 0.4 = 41.6 \nonumber\\
\text{DividendDecision}_2 &= \text{max}(0, 121.3 \times 0.15) + 190.5 \times 0.4 = 94.4 \nonumber
\end{align}

\begin{figure}[htbp]
\centering
\vspace{-1cm}
%{\footnotesize
\begin{tikzcd}[column sep=3.5cm, row sep=1.8cm]
& \textcolor{blue}{\text{Profit Differential}} \arrow[d, "\eta_{\text{div}}"', red] & \textcolor{blue}{\text{Bank Balance}} \arrow[d, "\eta_{\text{div}}"', red] \\
\textcolor{green}{\text{Dividend Base}} \arrow[r, "F_{\text{profit}}", blue] \arrow[d, "\oplus"', blue] & \textcolor{green}{\text{52.0 $\to$ 138.50 $\to$ 121.25}} \arrow[r, "F_{\text{bank}}", blue] \arrow[d, "\oplus"', blue] & \textcolor{green}{\text{0.0 $\to$ 52.0 $\to$ 190.50}} \arrow[d, "\oplus"', blue] \\
\textcolor{orange}{\text{Rate Application}} \arrow[r, "r_d: 0.15"', orange] \arrow[d, "\lrcorner"', purple] & \textcolor{orange}{\text{7.8 $\to$ 20.78 $\to$ 18.19}} \arrow[r, "r_b: 0.4"', orange] \arrow[d, "\lrcorner"', purple] & \textcolor{orange}{\text{0.0 $\to$ 20.8 $\to$ 76.20}} \arrow[d, "\lrcorner"', purple] \\
\textcolor{red}{\text{Dividend Output}} \arrow[r, "formula"', dashed, purple] & \textcolor{red}{\text{7.8 $\to$ 41.57 $\to$ 94.39}} \arrow[r, "="', dashed, purple] & \textcolor{red}{\text{Dividend Decision}}
\end{tikzcd}
%}
\vspace{0.5cm}
\caption{Commutative Diagram: Dividend Decision Natural Transformation with pullback structure}
\label{Q}
\end{figure}
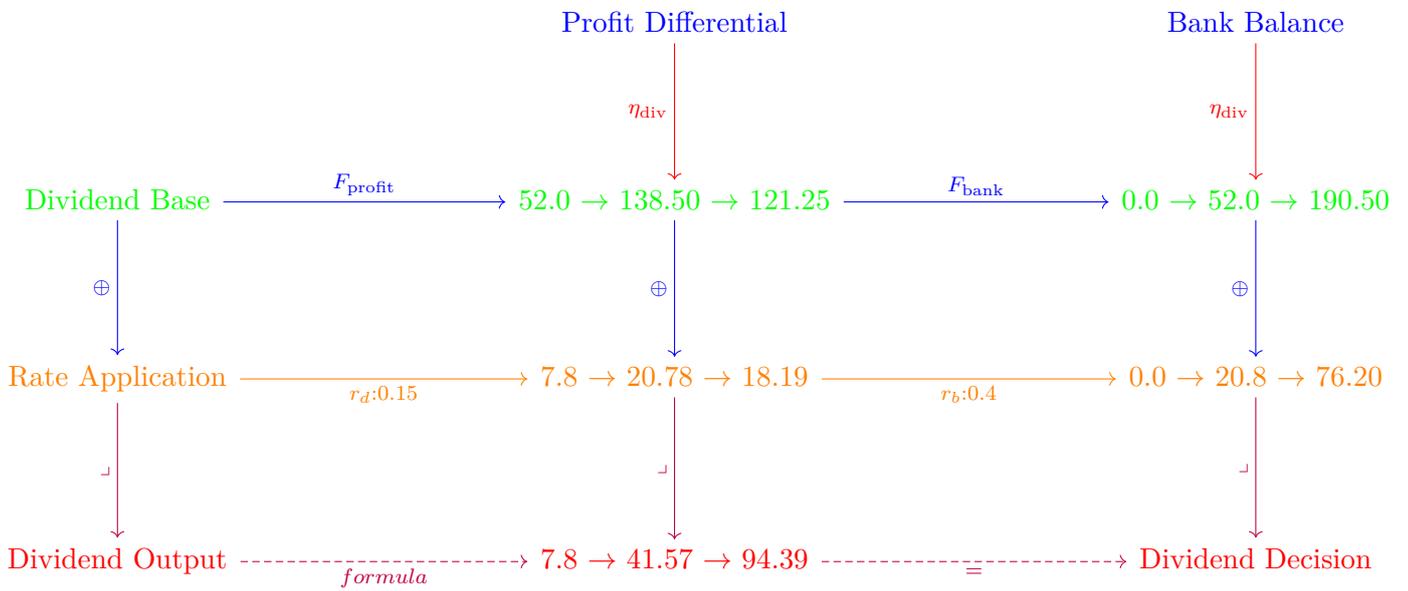
    
\begin{figure}[htbp]
\centering
{\footnotesize
\begin{tikzpicture}[scale=0.5]
\node[rectangle, fill=yellow!20, minimum width=5cm, minimum height=0.8cm] (policy) at (0, 11) {\textbf{Dividend Policy}};
\node[rectangle, fill=blue!20, minimum width=5cm, minimum height=0.8cm] (calculation) at (0, 4.5) {\textbf{Dividend Calculation}};
\node[left=1.0cm of calculation] (profitrate) {DivRateDiff: 0.15};
\node[right=0.5cm of calculation] (bankrate) {DivRateBank: 0.4};
\node[rectangle, fill=green!20, minimum width=5cm, minimum height=0.8cm] (payment) at (0, 0) {\textbf{Dividend Payment}};
\node[above=0.1cm, left=2.0cm of payment] (decision) {DividendDecision: 7.8 $\to$ 41.57 $\to$ 94.39};
\node[above=0.1cm, right=1.5cm of payment] (formula) {$\max(0, d \times r_d) + b \times r_b$};
\node[above=5.0cm of decision] (diff) {Diff: 52.0 $\to$ 138.50 $\to$ 121.25};
\node[above=5.0cm of formula] (bank) {AccComBank: 0.0 $\to$ 52.0 $\to$ 190.5};
\node at (-2.5, 7.5) {$\lrcorner$};
\node at (2.5, 7.5) {$\lrcorner$};
\node at (-2.5, 3.0) {$\lrcorner$};
\node at (2.5, 3.0) {$\lrcorner$};
\draw[->, thick, red] (policy) -- (calculation) node[pos=0.7,left] {Dividend Function};
\draw[->, thick, red] (calculation) -- (payment) node[pos=0.5,right] {Payment Output};
\draw[->, thick, blue] (diff) -- (profitrate) node[midway,right] {Profit Component};
\draw[->, thick, blue] (bank) -- (bankrate) node[midway,left] {Bank Component};
\draw[dashed, green] (diff) -- (decision) node[pos=0.7,left] {Dividend Flow};
\draw[dashed, green] (bank) -- (formula) node[pos=0.7,right] {Formula};
\node at (0.0,-1.0) {};
\end{tikzpicture}
}
\vspace{-1cm}
\caption{Commutative Diagram: Dividend Decision Pullback Flow}
\label{R}
\end{figure}
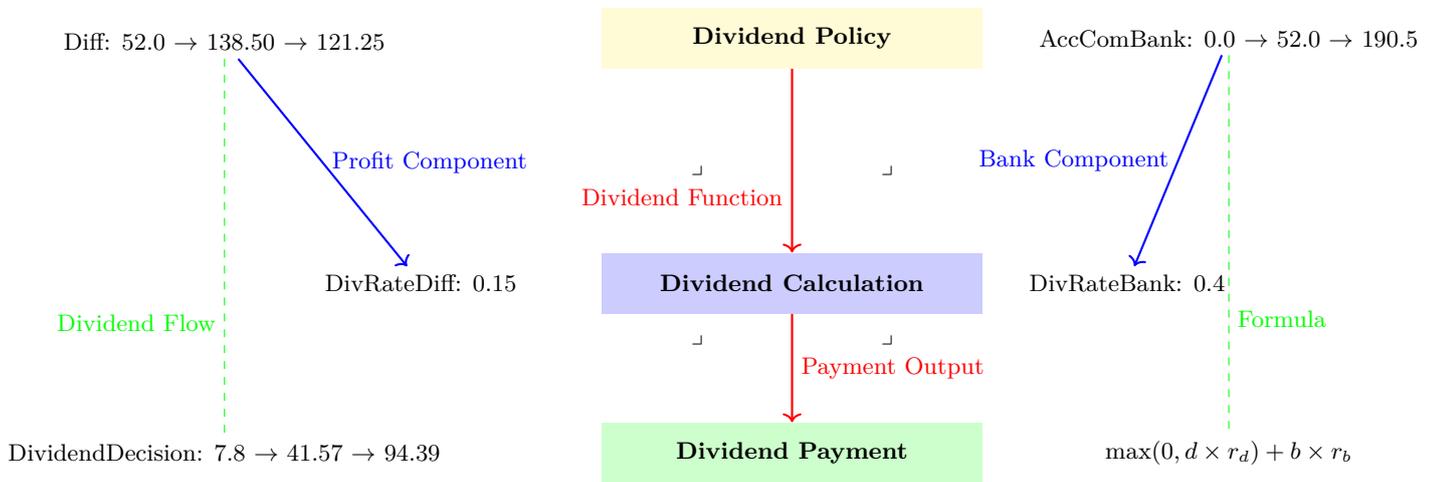

\begin{figure}[htbp]
\centering
{\footnotesize
    \begin{tikzcd}[column sep=8em, row sep=7em, arrows={-stealth}]
    {\begin{matrix} \green{D} \\ \text{Dividend Policy} \\ \text{(Corporate Finance)} \end{matrix}} \\
    & {\begin{matrix} \red{P} \\ \text{Dividend Calculation} \\ F_{\text{Dividend}} \end{matrix}}
    & {\begin{matrix} B \\ \text{Bank Balance} \\ \text{(AccComBank)} \end{matrix}} \\
    & {\begin{matrix} A \\ \text{Profit Differential} \\ \text{(Diff)} \end{matrix}}
    & {\begin{matrix} C \\ \text{Dividend Payment} \\ \mathbb{R}_{\geq 0} \end{matrix}} \\
    \arrow["{\begin{matrix} \exists ! \; u \\ \text{DividendDecision}_0 = 07.8 \\ \text{DividendDecision}_1 = 41.6 \\ \text{DividendDecision}_2 = 94.4 \end{matrix}}", pos=0.4, color={rgb,255:red,92;green,214;blue,92}, dashed, from=1-1, to=2-2]
    \arrow["{\begin{matrix} d_A = p_A \circ u \\ \text{Diff}_0 = 052.0 \\ \text{Diff}_1 = 138.5 \\ \text{Diff}_2 = 121.3 \end{matrix}}"{left=0.3cm}, pos=0.3, color={rgb,255:red,92;green,214;blue,92}, from=1-1, to=3-2]
    \arrow["{\begin{matrix} d_B = p_B \circ u \\ \text{AccComBank}_0 = 000.0 \\ \text{AccComBank}_1 = 052.0 \\ \text{AccComBank}_2 = 190.5 \end{matrix}}"{right=2.5cm}, pos=0.5, color={rgb,255:red,92;green,214;blue,92}, from=1-1, to=2-3]
    \arrow["{\begin{matrix} p_A \\ \text{Profit} \\ \text{Component} \end{matrix}}"{left=0.1cm}, pos=0.3, color={rgb,255:red,214;green,92;blue,92}, from=2-2, to=3-2]
    \arrow["{\begin{matrix} p_B \\ \text{Bank} \\ \text{Component} \end{matrix}}"{below=0.2cm}, pos=0.4, color={rgb,255:red,214;green,92;blue,92}, from=2-2, to=2-3]
    \arrow["{\begin{matrix} f \\ \max(0, \text{Diff} \times 0.15) \\ + \text{Bank} \times 0.4 \end{matrix}}"{above=0.4cm}, from=3-2, to=3-3]
    \arrow["{\begin{matrix} g \\ \text{Bank Balance} \\ \text{Dividend Rate} \end{matrix}}"{right=0.5cm}, pos=0.7, from=2-3, to=3-3]
    \end{tikzcd}
}
\vspace{-2.5cm}
\caption{Dividend Decision Pullback: The dividend decision pullback combines profit differential and bank balance with respective dividend rates, ensuring non-negative dividend payments.}
\label{S}
\end{figure}
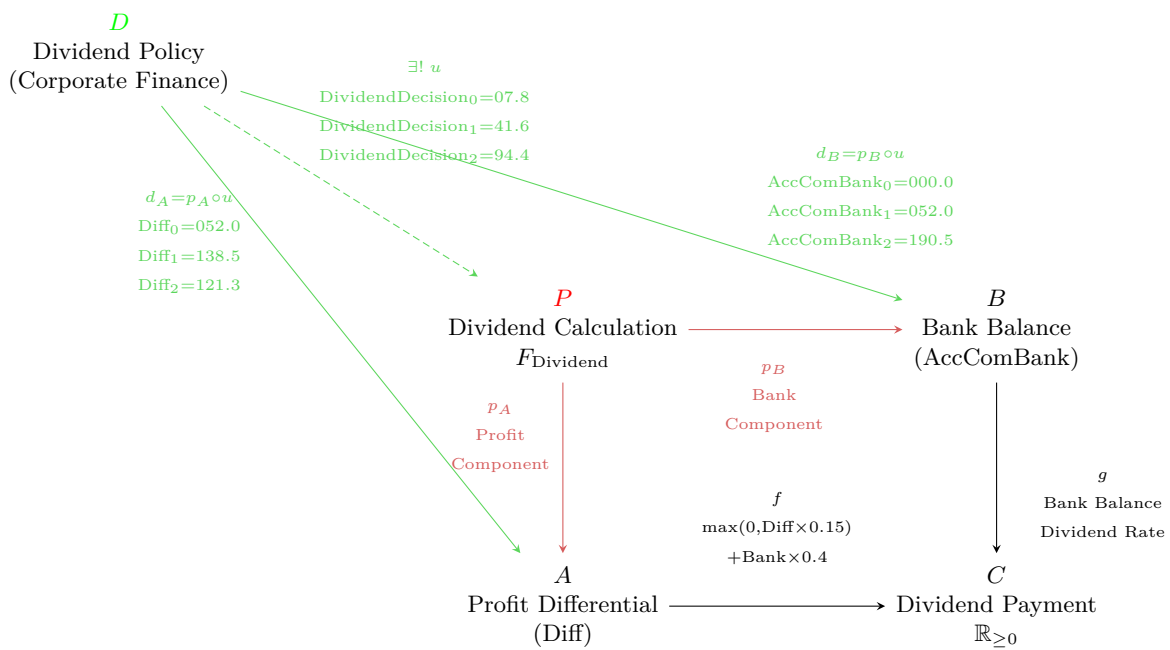

\textbf{Economics: Corporate Finance and Dividend Policy:}
The dividend decision pullback embodies the fundamental principle of corporate finance through categorical limits. 
The dividend function $\max(0, d \times r_d) + b \times r_b$ 
becomes a typed morphism $F_{\text{Dividend}}: \mathbb{R} \times \mathbb{R}_{\geq 0} \times [0,1] \times [0,1] \rightarrow \mathbb{R}_{\geq 0}$ 
that transforms profit differentials and bank balances into dividend payments while respecting non-negativity constraints and corporate finance policy.

Economically, this captures how firms balance profit distribution to shareholders with operational cash management through systematic dividend 
policies (DivRateDiff: 0.15, DivRateBank: 0.4). 
The pullback ensures that dividend decisions emerge from the interaction of profit performance (Diff: 52.0 $\to$ 138.50 $\to$ 121.25) 
and financial reserves (AccComBank: 0.0 $\to$ 52.0 $\to$ 190.50), generating dividend patterns (DividendDecision: 7.8 $\to$ 41.57 $\to$ 94.39) 
that reflect both profitability and financial prudence.

The three-layer architecture of diagram~\ref{R} reveals the categorical nature of corporate finance theory: 
the yellow layer represents dividend policy foundations, the blue layer embodies the dividend calculation mechanism, 
and the green layer shows the resulting dividend payments. 
The pullback corners ensure that dividend decisions respect both profit constraints and liquidity constraints simultaneously. 
This categorical approach to corporate finance provides mathematical foundations for understanding 
how firms balance shareholder returns with operational requirements, connecting corporate finance theory 
with categorical mathematics and enabling formal analysis of optimal dividend policies under uncertainty.

%---------------------------------------------------------------------------------------------------------------------------------------------------------------------------------
\subsection{Universal Constructions}\label{sec:universalconstructions}

Universal constructions in mathematics and category theory are what in economics often the argmax of a maximisation is used for.
For an intuition of the vast generalisation universals are over maximisation, we can think about them to also encompass what economists call institutions:
structural contextual rules of finding good rules like constitutions for politics or institutional dynamics and diversity, like in polycentrism~\cite{Ostrom2005Book},
or the selection functions of open games in~\cite{GHWZ2018} as higher order functions where the max operator is one of any possible.
Institutions are what in economics is a top level structure and is is rather awkward to understand and to formulate them as the result of an simple maximisation of some function over reals only.
Of course in economics, we discuss {\it optimal institutions} as an outcome of optimisation but in what sense optimal, what types are involved is to be worked out as part of the theory maintained.
A maximisation of a function over reals or vector spaces can take advantage of universal constructions as a first step towards adjunctions.
By that we get for economics a general approach to formulate outcomes having some best, maximal, most congruent, unique or whatever universality we need to use to describe reality in theory.
Universals as pullbacks and pushouts can be derived from an even more general principle of adjoint functors which are said to be a most important contribution of category theory to mathematics.
We do not use adjointness of functors explicitely but only as pushouts and pullbacks but adjoint functors as expressing the optimality itself should be kept in mind when
looking at the universal constructions from an economic perspective.
For example, for modelling an agent in the theory being an econometrician using a local or globally valid theory of the economic data.
The very fact that the universal constructions are themselves constructions of more general categorical patterns hints into the possibility to formulate the 
economists being in the theory an agent with the same abilities as the agent in the theory, namely building theory including constructing the optimisation itself.
Endogenous and more importantly, compositional econometric models are the key to the compositional and endogenous nature of hierarchical control.
The compositional properties of econometric theories have to show the available inverse informations of the data in and about the hierachical levels of the system.

%\clearpage
\subsubsection{Pullbacks for Constraint Validation}

Pullbacks implement constraint validation before economic calculations.
In case of the investment decision, we categorify the validation property of the banking sector being responsible for granting loans to companies 
and making sure to get the repayments from the hopefully successful production by the producer and purchase and payment by their customers.

\begin{definition}[Investment]
The investment pullback as a universal construction validates loan creation by the bank for the company. The production process of banking is a universal optimisation
in the following diagram.

\begin{tikzcd}
\\
\text{InvestmentDecision} \arrow[r, "p_2"] \arrow[d, "p_1"] & \text{BankLiquidity} \arrow[d, "g"] \\
\text{CompanyCapacity} \arrow[r, "f"] & \text{ValidInvestment} \\
\end{tikzcd}

The validation of the pullback is part of the design of investment contracts of the banking and financial sector.
This may include econometric or AI methods to validate the properties of the companies and banks involved as a more complex algorithms then the sigmoid learning we have for now. 
Similarly, sensitivity analysis like in~\cite{HMSW2019} is part of validation of regulatory needs.
The universal construction allows to formulate validation rules for the company like looking at its bank account.
Also bank's liquidity contracts can be a complex structure of validations of international business and financial plans in the supply decision of liquidity at the bank.
With central banks and governments the validation rules become even more complex.
Any algorithm that improves the validation process can be build in as a maximisation, universal construction or adjunction as needed.

\begin{align}
\text{Pullback}_{\text{Investment}}&: \mathbb{R}_{> 0} \times \mathbb{R}_{\geq 0} \rightarrow \{0, 1\} 
& \text{Pullback}_{\text{Investment}}(i, c) 
& = 
\begin{cases}
& 1 \quad \text{ if } i \leq c + \text{CreditLimit} \\
& 0 \quad \text{otherwise}
\end{cases}
\end{align}
where 
\begin{align}
p_1: \text{InvestmentDecision} &\rightarrow \text{CompanyCapacity} \nonumber \\
p_2: \text{InvestmentDecision} &\rightarrow \text{BankLiquidity} \nonumber \\
f: \text{CompanyCapacity} &\rightarrow \text{ValidInvestment} \nonumber \\
g: \text{BankLiquidity} &\rightarrow \text{ValidInvestment} \nonumber
\end{align}
\hfill$\blacksquare$
\end{definition}

Investment validation evolution in \textbf{periods 0, 1, 2}.
\begin{align}
\text{Investment}_0 &= 260.0 \text{ (validates against initial state)}\nonumber\\
\text{Investment}_1 &= 289.49 \text{ (validates against period 1 liuidity)}\nonumber\\
\text{Investment}_2 &= 275.20 \text{ (validates against period 2 liquidity)}\nonumber
\end{align}

\begin{figure}[htbp]
\centering
\vspace{-1cm}
\begin{tikzcd}[column sep=3.0cm, row sep=1.8cm]\\
    & \textcolor{blue}{\text{Company Capacity}} 
        \arrow[d, "\eta_{\text{inv}}"', red] 
    & \textcolor{blue}{\text{Bank Liquidity}} 
        \arrow[d, "\eta_{\text{inv}}"', red] \\
    \textcolor{green}{\text{Investment Base}} 
        \arrow[r, "F_{\text{capacity}}", blue] 
        \arrow[d, "\oplus"', blue] 
    & \textcolor{green}{\text{260.0 $\to$ 289.49 $\to$ 275.20}} 
        \arrow[r, "F_{\text{liquidity}}", blue] 
        \arrow[d, "\oplus"', blue] 
    & \textcolor{green}{\text{CreditLimit Available}} 
        \arrow[d, "\oplus"', blue] \\
    \textcolor{orange}{\text{Constraint Check}} 
        \arrow[r, "capacity"', orange] 
        \arrow[d, "\lrcorner"', purple] 
    & \textcolor{orange}{\text{Capacity Validation}} 
        \arrow[r, "credit"', orange] 
        \arrow[d, "\lrcorner"', purple] 
    & \textcolor{orange}{\text{Liquidity Validation}} 
        \arrow[d, "\lrcorner"', purple] \\
    \textcolor{red}{\text{Valid Investment}} 
        \arrow[r, "constraint"', dashed, purple] 
    & \textcolor{red}{\text{Investment Decision}} 
        \arrow[r, "="', dashed, purple] 
    & \textcolor{red}{\text{Validation Result}}
\end{tikzcd}
\vspace{0.5cm}
\caption{Commutative Diagram: Investment Validation Natural Transformation with pullback structure}
\label{T}
\end{figure}
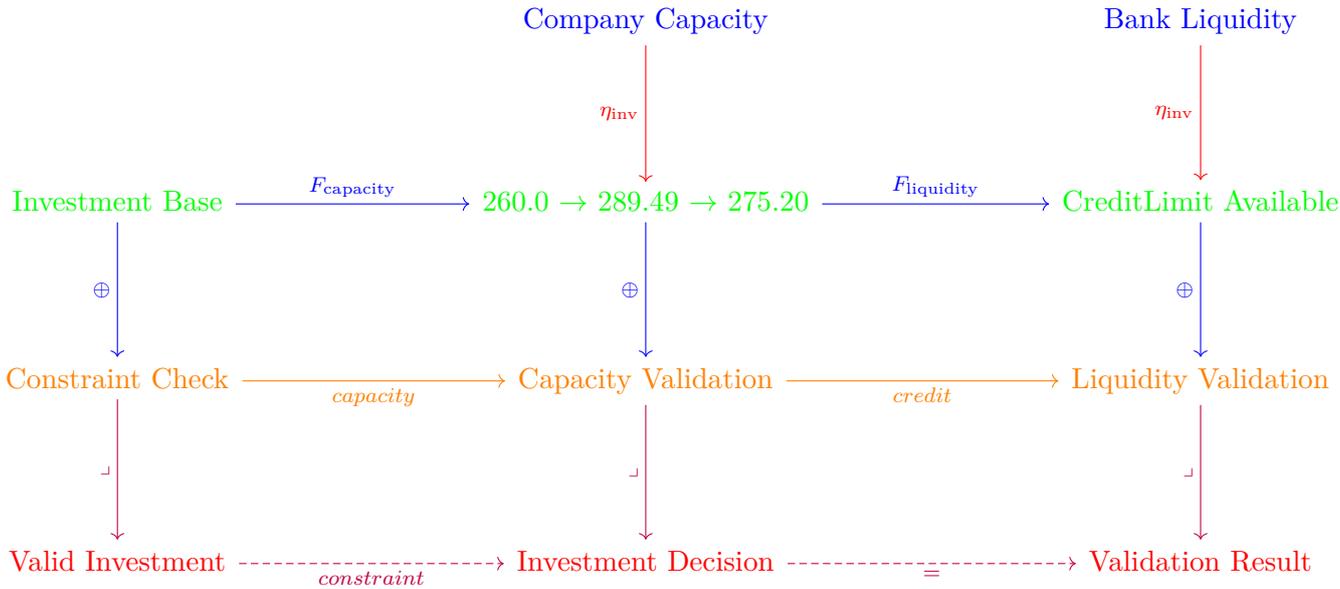

\begin{figure}[htbp]
\centering
{\footnotesize
\begin{tikzpicture}[scale=0.9]
\node[rectangle, fill=yellow!20, minimum width=5cm, minimum height=0.8cm] (validation) at (0, 6) {\textbf{Investment Validation}};
\node[rectangle, fill=blue!20, minimum width=5cm, minimum height=0.8cm] (constraint) at (0, 3) {\textbf{Constraint Validation}};
\node[left=0.5cm of constraint] (capacitycheck) {Capacity Check};
\node[right=0.5cm of constraint] (liquiditycheck) {Liquidity Check};
\node[rectangle, fill=green!20, minimum width=5cm, minimum height=0.8cm] (valid) at (0, 0) {\textbf{Valid Investment}};
\node[above=0.1cm,left=1.5cm of valid] (investment) {Investment: 260.0 $\to$ 289.49 $\to$ 275.20};
\node[above=0.1cm,right=1.5cm of valid] (validation_rule) {$i \leq c + \text{CreditLimit}$};
\node[above=5.0cm of investment] (capacity) {CompanyCapacity: Investment Need};
\node[above=5.0cm of validation_rule] (liquidity) {BankLiquidity: Available Credit};
\node at (-2.5, 4.5) {$\lrcorner$};
\node at (2.5, 4.5) {$\lrcorner$};
\node at (-2.5, 1.5) {$\lrcorner$};
\node at (2.5, 1.5) {$\lrcorner$};
\draw[->, thick, red] (validation) -- (constraint) node[pos=0.2,right] {Validation Function};
\draw[->, thick, red] (constraint) -- (valid) node[pos=0.7,right] {Validation Result};
\draw[->, thick, blue] (capacity) -- (capacitycheck) node[midway,right] {Capacity Input};
\draw[->, thick, blue] (liquidity) -- (liquiditycheck) node[midway,left] {Liquidity Input};
\draw[dashed, green] (capacity) -- (investment) node[pos=0.7,left] {Investment Flow};
\draw[dashed, green] (liquidity) -- (validation_rule) node[pos=0.7 ,right] {Credit Constraint};
\node at (0,-1.0) {};
\end{tikzpicture}
}
\vspace{-1cm}
\caption{Commutative Diagram: Investment Validation Pullback Flow}
\label{U}
\end{figure}
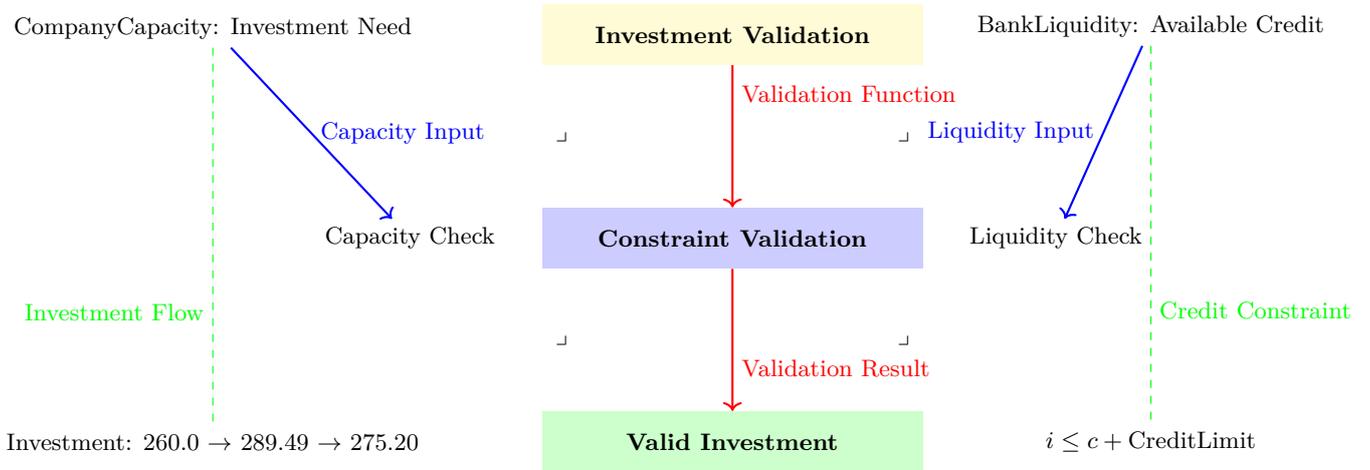

The calculation taking place in Figure~\ref{T} in the commutative diagram of the natural transformation are the investment decisions and validation results.
In the hierarchical commutative diagram~\ref{U} we can see the investment validation pullback corners, as another point of view on the compositions of the processes involved.

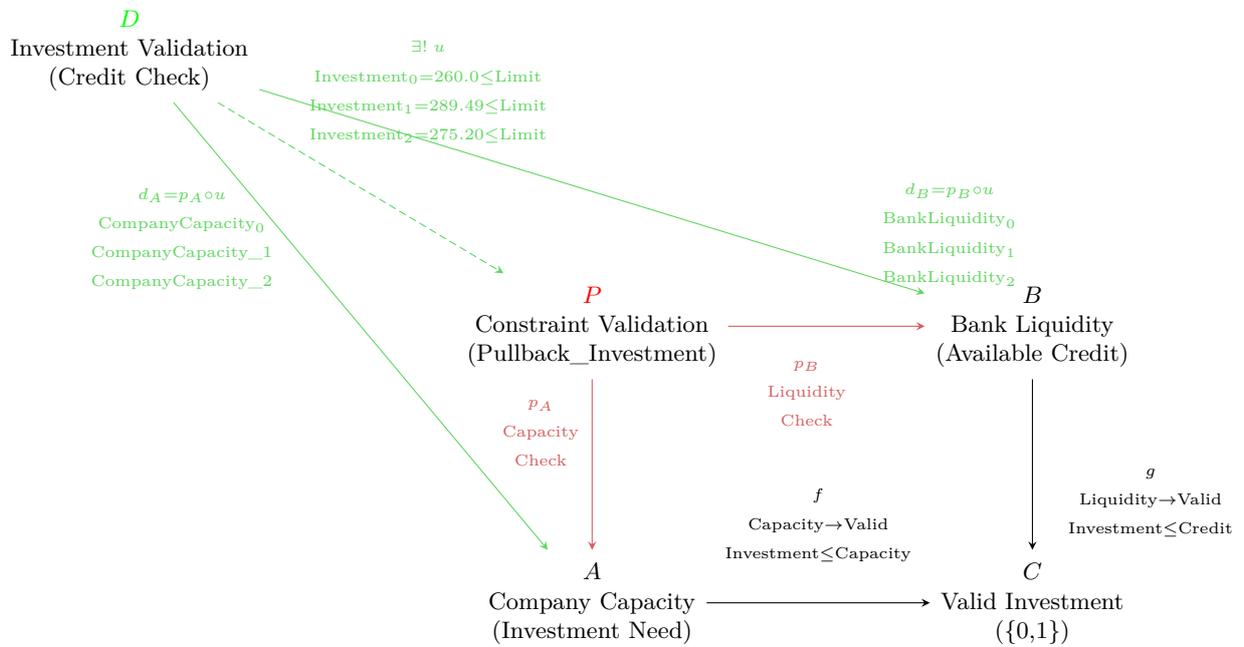
\begin{figure}[htbp]
    \centering
    {\footnotesize
    \begin{tikzcd}[column sep=8em, row sep=7em, arrows={-stealth}]
    {\begin{matrix} \green{D} \\ \text{Investment Validation} \\ \text{(Credit Check)} \end{matrix}} & & \\
    & {\begin{matrix} \red{P} \\ \text{Constraint Validation} \\ \text{(Pullback\_Investment)} \end{matrix}} & {\begin{matrix} B \\ \text{Bank Liquidity} \\ \text{(Available Credit)} \end{matrix}} \\
    & {\begin{matrix} A \\ \text{Company Capacity} \\ \text{(Investment Need)} \end{matrix}} & {\begin{matrix} C \\ \text{Valid Investment} \\ \text{(\{0,1\})} \end{matrix}} \\
    \arrow["{\begin{matrix} \exists ! \; u \\ \text{Investment}_0 = 260.0 \leq \text{Limit} \\ \text{Investment}_1 = 289.49 \leq \text{Limit} \\ \text{Investment}_2 = 275.20 \leq \text{Limit} \end{matrix}}", pos=0.3, color={rgb,255:red,92;green,214;blue,92}, dashed, from=1-1, to=2-2]
    \arrow["{\begin{matrix} d_A = p_A \circ u \\ \text{CompanyCapacity}_0 \\ \text{CompanyCapacity}\_1 \\ \text{CompanyCapacity}\_2 \end{matrix}}"{left=0.1cm}, pos=0.3, color={rgb,255:red,92;green,214;blue,92}, from=1-1, to=3-2]
    \arrow["{\begin{matrix} d_B = p_B \circ u \\ \text{BankLiquidity}_0 \\ \text{BankLiquidity}_1 \\ \text{BankLiquidity}_2 \end{matrix}}"{right=2.0cm}, pos=0.7, color={rgb,255:red,92;green,214;blue,92}, from=1-1, to=2-3]
    \arrow["{\begin{matrix} p_A \\ \text{Capacity} \\ \text{Check} \end{matrix}}"{left=0.1cm}, pos=0.3, color={rgb,255:red,214;green,92;blue,92}, from=2-2, to=3-2]
    \arrow["{\begin{matrix} p_B \\ \text{Liquidity} \\ \text{Check} \end{matrix}}"{below=0.2cm}, pos=0.4, color={rgb,255:red,214;green,92;blue,92}, from=2-2, to=2-3]
    \arrow["{\begin{matrix} f \\ \text{Capacity} \to \text{Valid} \\ \text{Investment} \leq \text{Capacity} \end{matrix}}"{above=0.4cm}, from=3-2, to=3-3]
    \arrow["{\begin{matrix} g \\ \text{Liquidity} \to \text{Valid} \\ \text{Investment} \leq \text{Credit} \end{matrix}}"{right=0.4cm}, pos=0.7, from=2-3, to=3-3]
    \end{tikzcd}
    }
    \vspace{-2.5cm}
    \caption{Investment Validation Pullback: The investment validation pullback ensures that investment decisions satisfy both company capacity constraints and bank liquidity limits across all periods.}
    \label{V}
\end{figure}
    
\textbf{Economics: Credit Constraints and Financial Intermediation}
The investment validation pullback embodies the fundamental principle of credit constraint enforcement through categorical limits. 
A validation function implementing $i \leq c + \text{CreditLimit}$ 
becomes a typed morphism $\text{Pullback}_{\text{Investment}}: \mathbb{R}_{> 0} \times \mathbb{R}_{\geq 0} \rightarrow \{0, 1\}$ 
that ensures investment decisions respect both company capacity constraints and bank liquidity limits before execution.

Economically, this captures the essential role of financial intermediation in validating economic feasibility of 
investment projects (Investment: 260.0 $\to$ 289.49 $\to$ 275.20) against available credit capacity and company financial capacity. 
The pullback ensures that investment validation respects both microeconomic level firms' constraints and macroeconomic banking system's constraints, 
preventing excessive leverage while enabling productive capital formation within resource aware realistic limits.

The three-layer architecture of diagram~\ref{U} reveals the categorical nature of financial constraint validation.
The yellow layer represents investment validation requirements, the blue layer embodies the constraint validation mechanism.
The green layer shows the resulting validation decisions. 
The pullback corners ensure that validation respects both company capacity constraints and bank liquidity constraints simultaneously. 
This categorical approach to credit constraint theory provides mathematical foundations 
for understanding financial intermediation as a categorical limit process, connecting banking theory with investment theory.
It enables formal analysis of credit cycles and financial stability mechanism for the issues of Marx, Keynes or as in MoMaT,
the desynchronized payments of the producer to suppliers now and only later back from clients, synchronised by a loan now and repayments later.
The universal constructions shows this to be the macroeconomic accounting solution for the banking and monetary sector.
The real accounts can be used to record the production and distribution of the common product, the GDP.
Hence, the hierarchical risk-sharing solution in this paper for the labor sharing decentralized and parallel economies can be extended to also implement 
the GDP sharing in the product, initial and final objects of categories of real and nominal accounts.
The macroeconomic accounting is the institutional answer to trace hierarchically risk from companies, to banks and ultimately to central banks and governments.
The micro- and macroeconomic levels in the five agent MoMaT theory can be extended to include the central banks at the macroeconomic level and the banks at the then mesoeconomic level.
Such a system includes gold based warehouse receipts as bank notes from former centuries, to paper money since 1970ies up to digital and crypto and quantum validated IDs 
of money being a means of loan repayments or dedebting in a validation process.
This includes fractional and other financial contracts of macroeconomic accounting agents like central banks, interbanking markets or governments.

%---------------------------------------------------------------------------------------------------------------------------------------------------------------------------------
%\clearpage
\subsubsection{Pushouts for Economic Computation}

\begin{definition}[Accounts Update]
Pushouts perform the actual economic calculations and aggregations.
The account update pushout implements the 8 bookings.

\begin{tikzcd}
\\
\text{Transaction} \arrow[r, "g"] \arrow[d, "f"] & \text{Target} \arrow[d, "q_2"]\\
\text{Source} \arrow[r, "q_1"] & \text{UpdatedAccounts}\\
\end{tikzcd}

\begin{align}
\text{Pushout}_{\text{Account}}&:& \mathbb{R}_{\geq 0} \times \mathbb{R} &\rightarrow \mathbb{R}_{\geq 0}
&&\text{Pushout}_{\text{Account}}(a, \Delta) = \max(0, a + \Delta)\\
\end{align}
where 
\begin{align}
f: \text{Transaction} &\rightarrow \text{Source} \nonumber\\
g: \text{Transaction} &\rightarrow \text{Target} \nonumber\\
q_1: \text{Source} &\rightarrow \text{UpdatedAccounts} \nonumber\\
q_2: \text{Target} &\rightarrow \text{UpdatedAccounts} \nonumber\\
\text{with constraint} \quad \Sigma_i \Delta_i &= 0 \text{ (conservation)} \nonumber\\
\end{align}
\hfill$\blacksquare$
\end{definition}

Account evolution through pushout computations:
\begin{align}
\text{Period 0$\to$1:} \quad &\text{AccResBank}: 0.0 \rightarrow 208.0\nonumber\\
&\text{AccComBank}: 0.0 \rightarrow 52.0\nonumber\\
&\text{AccComLoan}: 0.0 \rightarrow 260.0\nonumber
\end{align}

%\clearpage
\subsubsection{Categorical Types of the Eight Macro Bookings}\label{sec:categoricaltypes}
The categorical typing of the 8 macroeconomic bookings reveals how category theory disentangles the complex bidirectional flows between real and nominal accounts 
in the eight macroeconomic bookings of equations~(\ref{eq:bookingsall}).
Each booking is a morphism in the product category $\mathcal{C}_{\text{Real}} \times \mathcal{C}_{\text{Nominal}}$ where real flows and nominal flows move in opposite directions through categorical composition:

\begin{table}[htbp]
\centering
\small
\begin{tabular}{|c|l|c|l|l|}
\hline
\textbf{Booking} & \textbf{Economic Exchange} & \textbf{Agents} & \textbf{Real Flow Category} & \textbf{Nominal Flow Category} \\ \hline
1 & Lab sells Lab to Com & 3 & $Lab^{Lab}_{\mathcal{R}} \rightarrow Com^{Lab}_{\mathcal{R}}$ & $Com^{Bank}_{\mathcal{N}} \rightarrow Lab^{Bank}_{\mathcal{N}}$ \\ \hline
2 & Lab buys Good from Com & 3 & $Com^{Good}_{\mathcal{R}} \rightarrow Lab^{Good}_{\mathcal{R}}$ & $Lab^{Bank}_{\mathcal{N}} \rightarrow Com^{Bank}_{\mathcal{N}}$ \\ \hline
3 & Res sells Res to Com & 3 & $Res^{Res}_{\mathcal{R}} \rightarrow Com^{Res}_{\mathcal{R}}$ & $Com^{Bank}_{\mathcal{N}} \rightarrow Res^{Bank}_{\mathcal{N}}$ \\ \hline
4 & Res buys Good from Com & 3 & $Com^{Good}_{\mathcal{R}} \rightarrow Res^{Good}_{\mathcal{R}}$ & $Res^{Bank}_{\mathcal{N}} \rightarrow Com^{Bank}_{\mathcal{N}}$ \\ \hline
5 & Com gets Loan from Bank & 2 & $\emptyset$ & $Bank^{Loan}_{\mathcal{N}} \leftrightarrow Com^{Loan}_{\mathcal{N}}$ \\ \hline
6 & Com pays Div to Cap & 3 & $\emptyset$ & $Com^{Bank}_{\mathcal{N}} \rightarrow Cap^{Bank}_{\mathcal{N}}$ \\ \hline
7 & Com repays Loan to Bank & 2 & $\emptyset$ & $Com^{Bank}_{\mathcal{N}} \rightarrow Bank^{Loan}_{\mathcal{N}}$ \\ \hline
8 & Cap buys Good from Com & 3 & $Com^{Good}_{\mathcal{R}} \rightarrow Cap^{Good}_{\mathcal{R}}$ & $Cap^{Bank}_{\mathcal{N}} \rightarrow Com^{Bank}_{\mathcal{N}}$ \\ \hline
\end{tabular}
\caption{Categorical Types of the 8 Macroeconomic Bookings: Real vs Nominal Flow Categories}
\end{table}

\textbf{Categorical Commentary on Real/Nominal Disentanglement:}
The categorical approach resolves the typing complexity by recognizing that each booking operates simultaneously in two categories:
\begin{enumerate}
\item \textbf{Real Flow Category} $\mathcal{C}_{\text{Real}}$: Objects are real accounts (Lab, Res, Good), morphisms are physical transfers measured in units [h], [kg], [G]
\item \textbf{Nominal Flow Category} $\mathcal{C}_{\text{Nominal}}$: Objects are monetary accounts (Bank, Loan, Div), morphisms are monetary transfers measured in [EU]
\end{enumerate}

The key categorical insight is that bookings 1-4,6,8 are contravariant functors $F: \mathcal{C}_{\text{Real}}^{op} \rightarrow \mathcal{C}_{\text{Nominal}}$ 
where real flows and nominal flows move in opposite directions, preserving the double-entry principle categorically. 
Bookings 5,7 are pure nominal morphisms representing financial contracts without real commodity exchange.
The three-agent bookings (1,2,3,4,6,8) involve Bank as a mediating functor that ensures the categorical composition $Lab \leftrightarrow Com$ 
factors through $Lab \rightarrow Bank \rightarrow Com$, maintaining the categorical pullback property of banking intermediation. 
The two-agent bookings 5,7 represent morphisms in the loan contract category.
This categorical typing eliminates the confusions of the recursive formulation of the eight bookings about
the types of the bookings and the types of the variables in the bookings by properly recognizing that each booking is not a simple morphism 
but a natural transformation between functors that preserves both real and nominal categorical structures while ensuring conservation laws through categorical limits.

Following the categorical typing from equation~\ref{eq:bookingsall}, each booking is a typed morphism:

\begin{align}
\text{Booking}_1&:& Lab^{Lab}_{\mathcal{R}} \times Com^{Bank}_{\mathcal{N}} &\rightarrow Lab^{Bank}_{\mathcal{N}} \times Com^{Lab}_{\mathcal{R}} \nonumber\\
\text{Booking}_2&:& Lab^{Bank}_{\mathcal{N}} \times Com^{Good}_{\mathcal{R}} &\rightarrow Lab^{Good}_{\mathcal{R}} \times Com^{Bank}_{\mathcal{N}} \nonumber\\
\text{Booking}_3&:& Res^{Res}_{\mathcal{R}} \times Com^{Bank}_{\mathcal{N}} &\rightarrow Res^{Bank}_{\mathcal{N}} \times Com^{Res}_{\mathcal{R}} \nonumber\\
\text{Booking}_4&:& Res^{Bank}_{\mathcal{N}} \times Com^{Good}_{\mathcal{R}} &\rightarrow Res^{Good}_{\mathcal{R}} \times Com^{Bank}_{\mathcal{N}} \nonumber\\
\text{Booking}_5&:& Bank^{Loan}_{\mathcal{N}} &\rightarrow Com^{Loan}_{\mathcal{N}} \times Bank^{Loan}_{\mathcal{N}} \nonumber\\
\text{Booking}_6&:& Com^{Bank}_{\mathcal{N}} &\rightarrow Cap^{Div}_{\mathcal{N}} \times Bank^{Cap}_{\mathcal{N}} \nonumber\\
\text{Booking}_7&:& Com^{Bank}_{\mathcal{N}} \times Bank^{Loan}_{\mathcal{N}} &\rightarrow Com^{Loan}_{\mathcal{N}} \times Bank^{Loan}_{\mathcal{N}} \nonumber\\
\text{Booking}_8&:& Cap^{Bank}_{\mathcal{N}} \times Com^{Good}_{\mathcal{R}} &\rightarrow Cap^{Good}_{\mathcal{R}} \times Com^{Bank}_{\mathcal{N}} \nonumber
\end{align}
where $\mathcal{R}$ denotes real flow categories and $\mathcal{N}$ denotes nominal flow categories, 
with each morphism preserving the contravariant relationship between real and nominal flows as categorical functors.

\textbf{Economic Interpretation of Categorical Booking Types:}
The categorical morphism types reveal the fundamental economic structure of market exchanges and financial intermediation. 
Bookings 1 and 3 represent factor market transactions where labor $Lab^{Lab}_{\mathcal{R}}$ 
and resources $Res^{Res}_{\mathcal{R}}$ flow to the company while monetary compensation flows in the opposite direction, 
implementing the economic principle of quid pro quo through contravariant functors. 
These flows are to be typed in a legal sense and modalities as discussed in~\cite{MW2025} as the abstraction principle to 
disentangle into two opposingly flowing values of exchange in a quid pro quo game.
Bookings 2, 4, and 8 represent product market transactions where goods $Com^{Good}_{\mathcal{R}}$ 
flow from company to consumers $Lab, Res, Cap$ while payments flow back, completing the circular flow of the economy. 
The pure nominal bookings 5 and 7 capture the temporal dimension of financial intermediation, 
where Booking 5 creates future obligations $Com^{Loan}_{\mathcal{N}}$ in exchange for present liquidity, 
while Booking 7 extinguishes those obligations through repayment—these represent the banking sector's core function of intertemporal coordination. 
Booking 6 implements profit distribution, transferring corporate earnings to capitalists through dividend payments, 
representing the return to risk-taking and capital provision, which we have not modelled as a tit for tat flow of values. 
The categorical typing thus captures the complete economic cycle: production factor acquisition (1,3), 
consumption fulfillment (2,4,8), intertemporal financial intermediation (5,7), and profit distribution (6), 
with each morphism preserving the fundamental economic law that real and monetary flows 
must move in opposite directions to maintain market equilibrium, quid pro quo games and double-entry accounting consistency
as verified properties in the pullback.
Combining the possibility to protocoll the processes in categorical terms opens a way to rewind the debt processes during liquidation of banks.
The restitution in German law is a contra variant functor for rewinding debt contracts of the agents intermediated by the bank.
In the case the bank is to be resolved, agents debts are to be restituted and reconnected in a new way, without the bank as an intermediator in the restituated debt contracts.

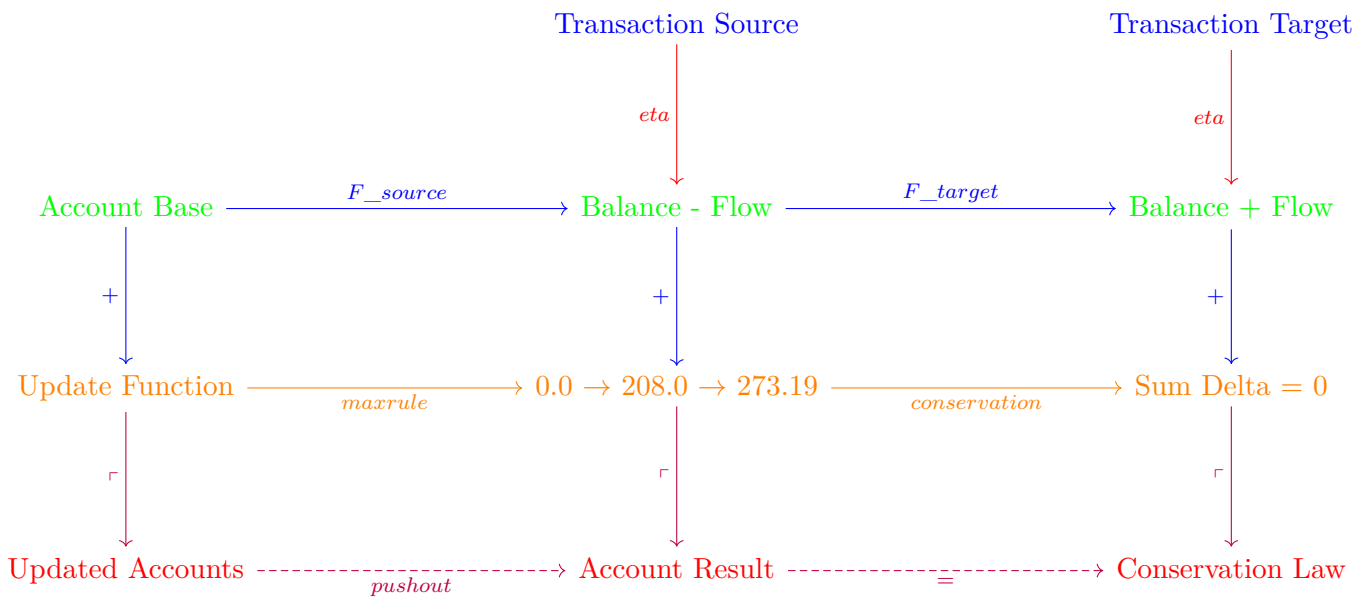
\begin{figure}[htbp]
\centering
\vspace{-1cm}
\begin{tikzcd}[column sep=3.5cm, row sep=1.8cm]\\
  & \textcolor{blue}{\text{Transaction Source}} 
    \arrow[d, "eta"', red] 
    & \textcolor{blue}{\text{Transaction Target}} 
      \arrow[d, "eta"', red] \\
  \textcolor{green}{\text{Account Base}} 
    \arrow[r, "F\_source", blue] 
    \arrow[d, "+"', blue] 
    & \textcolor{green}{\text{Balance - Flow}} 
      \arrow[r, "F\_target", blue] 
      \arrow[d, "+"', blue] 
      & \textcolor{green}{\text{Balance + Flow}} 
        \arrow[d, "+"', blue] \\
  \textcolor{orange}{\text{Update Function}} 
    \arrow[r, "max rule"', orange] 
    \arrow[d, "\ulcorner"', purple] 
    & \textcolor{orange}{\text{0.0 $\to$ 208.0 $\to$ 273.19}} 
      \arrow[r, "conservation"', orange] 
      \arrow[d, "\ulcorner"', purple] 
      & \textcolor{orange}{\text{Sum Delta = 0}} 
        \arrow[d, "\ulcorner"', purple] \\
  \textcolor{red}{\text{Updated Accounts}} 
    \arrow[r, "pushout"', dashed, purple] 
    & \textcolor{red}{\text{Account Result}} 
      \arrow[r, "="', dashed, purple] 
      & \textcolor{red}{\text{Conservation Law}}
\end{tikzcd}
\vspace{0.5cm}
\caption{Commutative Diagram: Account Update Natural Transformation}
\label{W}
\end{figure}

In Figure~\ref{W}, we see how money flows through the economy as a consistent compositionality of processes of bookings on accounts. 
When a transaction occurs, like when Labor provides services to a Company, two things happen simultaneously:
The real flow moves from source to target - Labor's services flow to the Company. At the same time, the nominal (money) flow moves in the opposite direction - 
from Company's bank account to Labor's bank account. The commutative diagram captures this dual movement through its categorical structure.
The top layer shows the transaction participants - a source account (like Labor's bank account) and a target account (like Company's bank account). 
The middle layer implements the core economic principle that values cannot be created or destroyed - they can only move between accounts. 
This is enforced by the update function that ensures when money leaves one account it must appear in another.

The bottom layer shows the final updated account balances and verifies that the fundamental law of accounting is preserved: the sum of all changes must equal zero. 
The categorical machinery (through morphisms like $\eta$ and $F_{source}, F_{target}$) ensures this rule to be mathematically guaranteed.
For example, when Labor's bank account grows from 0.0 to 208.0 to 273.19, we know with mathematical certainty that this money representing some value came from somewhere else in the system - 
it wasn't created from nothing. 
The pushout structure forces all money flows to be properly balanced, making the double-entry accounting principle a mathematical necessity.

This categorical view shows economic transactions as parallel synchronized flows of real goods and money that must always balance, 
governed by the mathematical laws of category theory rather than just accounting conventions.
The legal modalities that apear in these contracts can be lifted to the institutional level as a deontic logic as in the polycentric governance and management structures in~\cite{Ostrom2005Book}.
A categorical formulation of modal logic would add to the abstract syntax of the compiler and is a well researched path in programming language design.
A DSL for the modalities of legal contracts would be operating on the memory structures that collects and represents in our categorical setup the legal contracts.

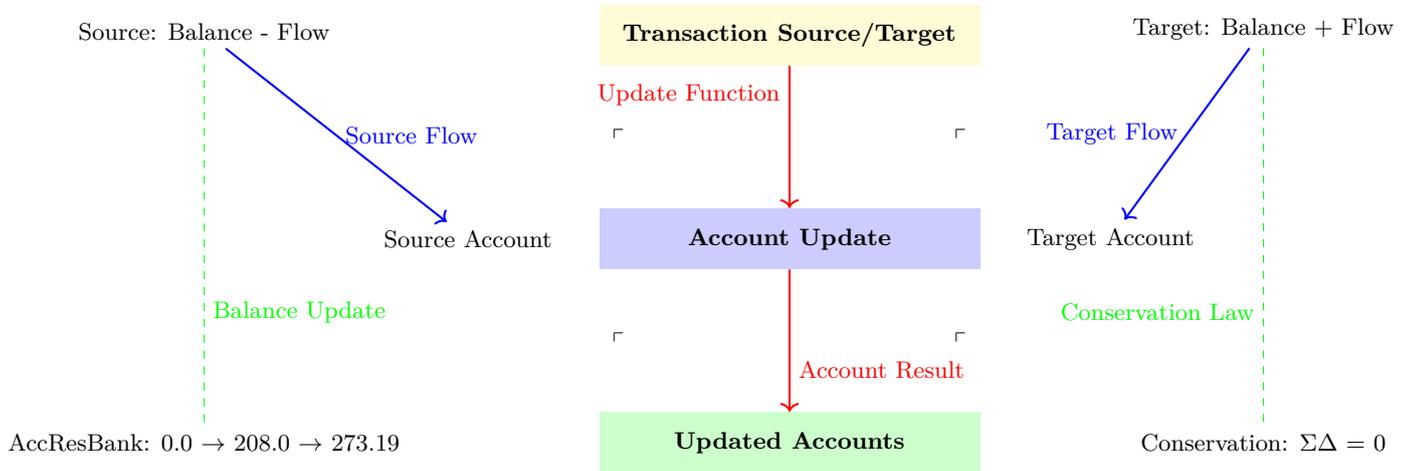
\begin{figure}[htbp]
\centering
{\footnotesize
\begin{tikzpicture}[scale=0.9]
\node[rectangle, fill=yellow!20, minimum width=5cm, minimum height=0.8cm] (transaction) at (0, 6) {\textbf{Transaction Source/Target}};
\node[rectangle, fill=blue!20, minimum width=5cm, minimum height=0.8cm] (update) at (0, 3) {\textbf{Account Update}};
\node[left=0.5cm of update] (sourceupdate) {Source Account};
\node[right=0.5cm of update] (targetupdate) {Target Account};
\node[rectangle, fill=green!20, minimum width=5cm, minimum height=0.8cm] (updated) at (0, 0) {\textbf{Updated Accounts}};
\node[above=0.5cm, left=2.5cm of updated] (balances) {AccResBank: 0.0 $\to$ 208.0 $\to$ 273.19};
\node[above=0.5cm, right=2.0cm of updated] (conservation) {Conservation: $\Sigma \Delta$ = 0};
\node[above=5.0cm of balances] (source) {Source: Balance - Flow};
\node[above=5.0cm of conservation] (target) {Target: Balance + Flow};
\node at (-2.5, 4.5) {$\ulcorner$};
\node at (2.5, 4.5) {$\ulcorner$};
\node at (-2.5, 1.5) {$\ulcorner$};
\node at (2.5, 1.5) {$\ulcorner$};
\draw[->, thick, red] (transaction) -- (update) node[pos=0.2,left] {Update Function};
\draw[->, thick, red] (update) -- (updated) node[pos=0.7,right] {Account Result};
\draw[->, thick, blue] (source) -- (sourceupdate) node[midway,right] {Source Flow};
\draw[->, thick, blue] (target) -- (targetupdate) node[midway,left] {Target Flow};
\draw[dashed, green] (source) -- (balances) node[pos=0.7,right] {Balance Update};
\draw[dashed, green] (target) -- (conservation) node[pos=0.7,left] {Conservation Law};
\node at (0,-1.0) {};
\end{tikzpicture}
}
\vspace{-1cm}
\caption{Commutative Diagram: Account Update Pushout Flow}
\label{X}
\end{figure}

This diagram~\ref{X} tells the economics of how money moves through the economy as the interaction of three levels: 
At the top level (yellow), we have two, a source and a target account, like when a company pays a worker's salary. 
The middle level (blue) shows the actual transaction happening - money leaves the company's account (source) and enters the worker's account (target). 
The bottom level (green) reveals the final result - we can track how the accounts changed (like the bank balance growing from 0 to 208 to 273.19), 
while making sure no money appears or disappears ($\sum \delta$ = 0) by the conservation law. 
The mathematical machinery of pushouts (shown by the corner marks $\ulcorner$) ensures this
as a mathematical guarantee of the program that money or values, i.e. the unit of account is conserved, like energy in physics, we enforce a conservation law. 
The arrows show how everything connects: 
yellow arrows track the transaction's progress, 
blue arrows show the money flows, 
and green dashed lines connect the initial balances to their final states. 
The categorical structure transforms banking from a set of rules into mathematically engineered consistent programs implementing economic universals.

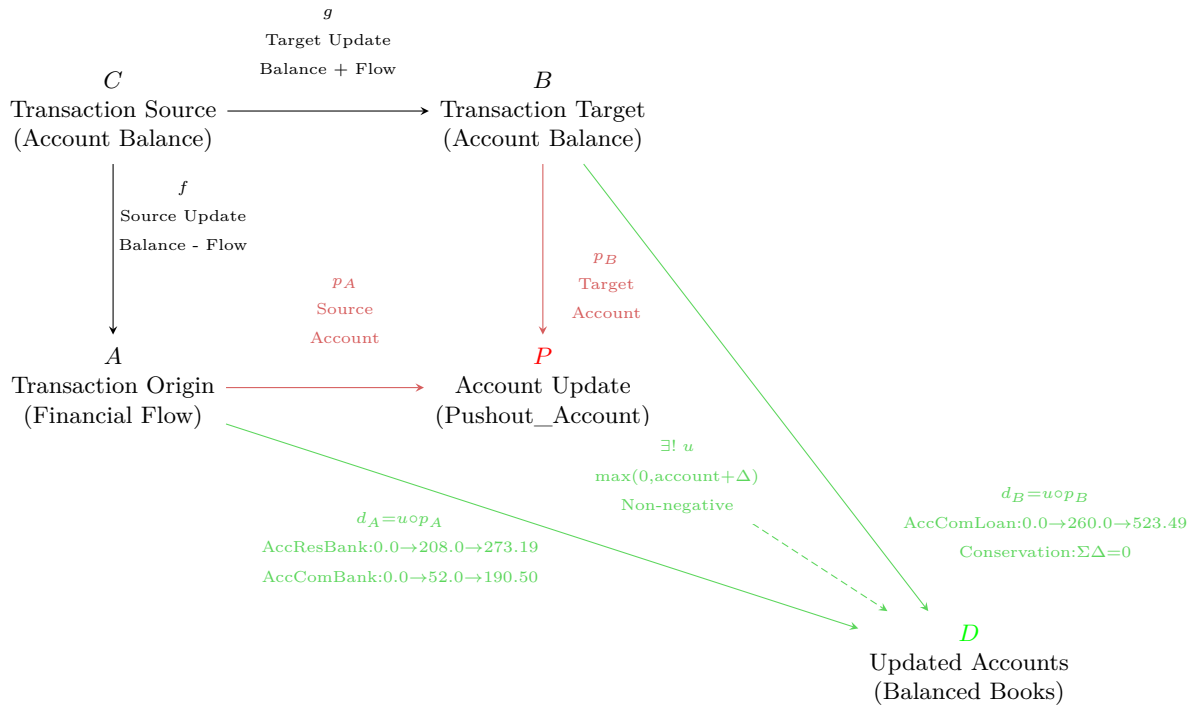
\begin{figure}[htbp]
    \centering
    {\footnotesize
    \begin{tikzcd}[column sep=8em, row sep=7em, arrows={-stealth}] 
    {\begin{matrix} C \\ \text{Transaction Source} \\ \text{(Account Balance)} \end{matrix}}
    & {\begin{matrix} B \\ \text{Transaction Target} \\ \text{(Account Balance)} \end{matrix}} & \\
    {\begin{matrix} A \\ \text{Transaction Origin} \\ \text{(Financial Flow)} \end{matrix}} 
    & {\begin{matrix} \red{P} \\ \text{Account Update} \\ \text{(Pushout\_Account)} \end{matrix}} & \\
    & & {\begin{matrix} \green{D} \\ \text{Updated Accounts} \\ \text{(Balanced Books)} \end{matrix}} \\
    \arrow["{\begin{matrix} f \\ \text{Source Update} \\ \text{Balance - Flow} \end{matrix}}"{right}, pos=0.3, from=1-1, to=2-1]
    \arrow["{\begin{matrix} g \\ \text{Target Update} \\ \text{Balance + Flow} \end{matrix}}"{above=0.3cm}, pos=0.5, from=1-1, to=1-2]
    \arrow["{\begin{matrix} p_A \\ \text{Source} \\ \text{Account} \end{matrix}}"{above=0.4cm}, pos=0.6, color={rgb,255:red,214;green,92;blue,92}, from=2-1, to=2-2]
    \arrow["{\begin{matrix} p_B \\ \text{Target} \\ \text{Account} \end{matrix}}"{right=0.3cm}, pos=0.7, color={rgb,255:red,214;green,92;blue,92}, from=1-2, to=2-2]
    \arrow["{\begin{matrix} d_A = u \circ p_A \\ \text{AccResBank}: 0.0 \to 208.0 \to 273.19 \\ \text{AccComBank}: 0.0 \to 52.0 \to 190.50 \end{matrix}}"{left=0.8cm}, pos=0.6, color={rgb,255:red,92;green,214;blue,92}, from=2-1, to=3-3]
    \arrow["{\begin{matrix} d_B = u \circ p_B \\ \text{AccComLoan}: 0.0 \to 260.0 \to 523.49 \\ \text{Conservation:} \Sigma \Delta = 0 \end{matrix}}"{right=0.5cm}, pos=0.8, color={rgb,255:red,92;green,214;blue,92}, from=1-2, to=3-3]
    \arrow["{\begin{matrix} \exists ! \;u \\ \max(0, \text{account} +  \Delta) \\ \text{Non-negative} \end{matrix}}"{description}, pos=0.2, color={rgb,255:red,92;green,214;blue,92}, dashed, from=2-2, to=3-3]
    \end{tikzcd}
    }
    \vspace{-2.5cm}
    \caption{Account Update Pushout: The account update pushout implements the 8 bookings through proper double-entry accounting, ensuring conservation of the unit of account ($\sum \delta$ = 0) and non-negative account balances.}
    \label{Y}        
\end{figure}

In Figure~\ref{Y} we see the account update pushout 
that implements the 8 macroeconomic bookings through proper double-entry accounting.
The diagram shows a pushout as a universal construction that combines information from multiple sources. 
Objects are 
C (Transaction Source): The account that money flows from (e.g., Lab's bank account).
B (Transaction Target): The account that money flows to (e.g., Com's bank account).
A (Transaction Origin): The original financial flow/transaction.
P (Account Update): The red pushout object that processes the update.
D (Updated Accounts): The green final state with balanced books.
Morphisms are
$f$: Source Update: subtracts the flow from the source account (Balance - Flow).
$g$: Target Update: adds the flow to the target account (Balance + Flow).
$p_A$, $p_B$: Project the updates to the pushout object.
$u$: The unique universal morphism ensuring mathematical consistency.
The pushout structure provides three critical Economic Guarantees:
Conservation: $\sum \delta$ = 0 (total money in the system remains constant),
Non-negativity: max(0, account + $ \Delta$) (accounts cannot go below zero),
Double-entry consistency: Every debit has a corresponding credit.
The Concrete Example shows actual values:
$Res^{Bank}_{\mathcal{N}}$: 0.0 $\to$ 208.0 $\to$ 273.19 (Resource owner's bank account grows),
$Com^{Bank}_{\mathcal{N}}$: 0.0 $\to$ 52.0 $\to$ 190.50 (Company's bank account grows),
$Com^{Loan}_{\mathcal{N}}$: 0.0 $\to$ 260.0 $\to$ 523.49 (Company's loan liability grows).

The categorical approach ensures that all 8 bookings respect these mathematical constraints. 
The pushout's universal property guarantees that there exists a unique way (morphism $u$) to update accounts 
that satisfies all constraints simultaneously, making errors impossible by construction rather than requiring manual checking.
This is the mathematical foundation that makes the categorical formulation more expressive, robust and compositionally extendable and maintainable than the recursive computational formalisation only.
    
\textbf{Economics: Double-Entry Accounting and Money Conservation}
The account update pushout embodies the fundamental principle of double-entry accounting through categorical colimits. The account update function $\max(0, a + \Delta)$ 
becomes a typed morphism $\text{Pushout}_{\text{Account}}: \mathbb{R}_{\geq 0} \times \mathbb{R} \rightarrow \mathbb{R}_{\geq 0}$ that transforms transaction flows 
into account balance updates while preserving the essential conservation law $\sum_i \delta_i = 0$ that ensures units of account are neither created nor destroyed.

Economically, this captures the fundamental accounting identity that underlies all monetary transactions: every debit has a corresponding credit, 
ensuring that economic value flows are conserved across the entire system. The pushout structure guarantees that individual account updates (AccResBank: 0.0 $\to$ 208.0 $\to$ 273.19) 
aggregate coherently into system-wide conservation, maintaining macroeconomic consistency as a macroeconomic foundation of microeconomic transactions.

The three-layer architecture of diagram~\ref{X} reveals the categorical nature of accounting theory: the yellow layer represents transaction sources and targets, 
the blue layer embodies the account update mechanism, and the green layer shows the resulting updated account balances. The pushout corners mark where individual 
transaction morphisms compose into the universal account update morphism. This categorical approach to accounting theory provides mathematical foundations 
for understanding monetary systems as categorical structures that preserve conservation laws, connecting accounting theory with category theory and 
enabling formal analysis of monetary consistency in complex economic systems.

% ------------------------------------------------------------------------------------------------------------------------------------------------------------
%\clearpage
\subsection{Categorical Evolution}\label{sec:categoricalstateevolution}

The categorical state transition integrates all universal constructions.
This is by stating that the model closes or that we have an endofunctor that is a monad.
We can trace the mathematical structure of the theory through the state transition functor.
Side effects as they are called in computer science are properly internalised or endogeneized as economists say.
The endofunctor implements the iteration of the categorical simulation
in endo loops of the universal pullbacks of verficication and pushouts of computation.
The result is a mathematically verified program and evolution of the system.

The monad is what is missing in the categorically typed recursive program in the Appendix~\ref{app:code_appendix_categorically_typed}.
The natural transformation helper functions \texttt{add\_copmponent!} and \texttt{add\_transformation!} 
are implemented as a function definition but are not used for running and implementing the simulation.
The categorical program is finally a program where evolution is a natural transformation.

\begin{definition}[Categorical Evolution]
The categorical evolution \textbf{endofunctor} $T: \mathcal{C}_t \rightarrow \mathcal{C}_{t+1}$ preserves all categorical structures:

\begin{align}
    T(\text{period}) &= Pullback_{Validation} \circ Pushout_{Computation} \circ Functor_{Transformation}(\text{period})
\end{align}

\begin{align}
    T&:& \mathcal{C}_{\text{Economy}} &\rightarrow \mathcal{C}_{\text{Economy}}\\
    T&:& (\mathcal{C}_{acc} \times \mathcal{C}_{flow} \times \mathcal{C}_{par}) &\rightarrow (\mathcal{C}_{acc} \times \mathcal{C}_{flow} \times \mathcal{C}_{par})
    && T(A, F, P) = (A', F', P')
    \end{align}
    where 
    \begin{align}
    A' &= \text{Pushout}_{acc}(\text{Pullback}_{\text{Validation}}(A, F, P))\nonumber\\
    F' &= \text{Functor}_{flows}(A', P)\nonumber\\
    P' &= \text{NatTrans}_{\text{Evolution}}(P)\nonumber
\end{align}
    
The commutative diagrams shows the categorical state transition as the temporal evolution of the system.

\begin{center}
\begin{tikzcd}[
    row sep=5.5em, 
    column sep=7em
    ]\\
    \mathcal{C}_{acc} \times \mathcal{C}_{flow} \times \mathcal{C}_{par} 
    \arrow[r, "T" description] 
    \arrow[d, "\text{Pullback}_{\text{Val}}" left] 
    & \mathcal{C}_{acc} \times \mathcal{C}_{flow} \times \mathcal{C}_{par} \\
    \mathcal{C}_{\text{Valid}} 
    \arrow[r, "\text{Pushout}_{\text{Comp}}" below] 
    \arrow[ur, "\text{Functor}_{\text{Trans}}" description, dashed] 
    & \mathcal{C}_{\text{Updated}}
    \end{tikzcd}
    \begin{tikzcd}[row sep=4.5em, column sep=6em]
    \mathcal{C}_t \arrow[r, "T" description] \arrow[d, "\text{decompose}" left] & \mathcal{C}_{t+1} \arrow[d, "\text{decompose}" right] \\
    (A_t, F_t, P_t) \arrow[r, "T_{typed}" description] \arrow[d, "\text{Pullback}" left] & (A_{t+1}, F_{t+1}, P_{t+1}) \\
    \mathcal{C}_{\text{Valid}} \arrow[r, "\text{Pushout}" below] \arrow[ur, "\text{Functor}" description, dashed] & \mathcal{C}_{\text{Economy}}
\end{tikzcd}
\end{center}

The composition of operations of the endofunctor of state transition is shown in the following string diagram:

\begin{center}
    \begin{tikzpicture}[scale=0.8]
    \draw[thick] (0,3) -- (1,3) node[above, pos=0] {$A_t$};
    \draw[thick] (0,2) -- (1,2) node[above, pos=0] {$F_t$};
    \draw[thick] (0,1) -- (1,1) node[above, pos=0] {$P_t$};
    \draw[rounded corners] (1,0.5) rectangle (4,3.5);
    \node at (2.5,2.3) {\scriptsize Pullback};
    \node at (2.5,1.7) {\scriptsize Validation};
    \draw[thick] (4,2.5) -- (5,2.5);
    \draw[thick] (4,2) -- (5,2);
    \draw[thick] (4,1.5) -- (5,1.5);
    \draw[rounded corners] (5,0.5) rectangle (8,3.5);
    \node at (6.5,2.3) {\scriptsize Pushout};
    \node at (6.5,1.7) {\scriptsize Computation};
    \draw[rounded corners] (9,0.5) rectangle (12,3.5);
    \node at (10.5,2.3) {\scriptsize Functor};
    \node at (10.5,1.7) {\scriptsize Transform};
    \draw[thick] (8,2.5) -- (9,2.5);
    \draw[thick] (8,2) -- (9,2);
    \draw[thick] (8,1.5) -- (9,1.5);
    \draw[thick] (12,3) -- (13,3) node[above, pos=1] {$A_{t+1}$};
    \draw[thick] (12,2) -- (13,2) node[above, pos=1] {$F_{t+1}$};
    \draw[thick] (12,1) -- (13,1) node[above, pos=1] {$P_{t+1}$};
    \end{tikzpicture}
\end{center}
    
\hfill$\blacksquare$
\end{definition}

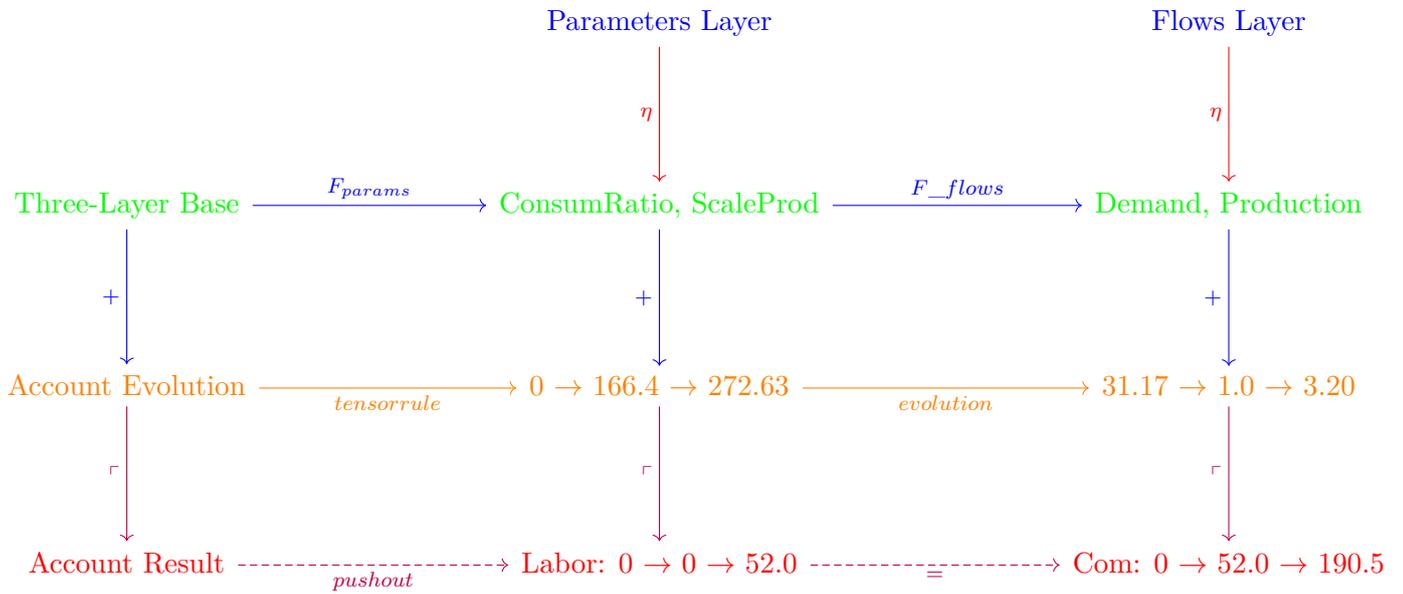
\begin{figure}[htbp]
\centering
\vspace{-1cm}
\begin{tikzcd}[column sep=3.0cm, row sep=1.8cm]\\
  & \textcolor{blue}{\text{Parameters Layer}} 
      \arrow[d, "\eta"', red] 
  & \textcolor{blue}{\text{Flows Layer}} 
      \arrow[d, "\eta"', red] \\
  \textcolor{green}{\text{Three-Layer Base}} 
      \arrow[r, "F_{params}", blue] 
      \arrow[d, "+"', blue] 
  & \textcolor{green}{\text{ConsumRatio, ScaleProd}} 
      \arrow[r, "F\_flows", blue] 
      \arrow[d, "+"', blue] 
  & \textcolor{green}{\text{Demand, Production}} 
      \arrow[d, "+"', blue] \\
     \textcolor{orange}{\text{Account Evolution}} 
       \arrow[r, "tensor rule"', orange] 
       \arrow[d, "\ulcorner"', purple] 
   & \textcolor{orange}{\text{0 $\to$ 166.4 $\to$ 272.63}} 
       \arrow[r, "evolution"', orange] 
       \arrow[d, "\ulcorner"', purple] 
   & \textcolor{orange}{\text{31.17 $\to$ 1.0 $\to$ 3.20}} 
       \arrow[d, "\ulcorner"', purple] \\
  \textcolor{red}{\text{Account Result}} 
      \arrow[r, "pushout"', dashed, purple] 
  & \textcolor{red}{\text{Labor: 0 $\to$ 0 $\to$ 52.0}} 
      \arrow[r, "="', dashed, purple] 
  & \textcolor{red}{\text{Com: 0 $\to$ 52.0 $\to$ 190.5}}
\end{tikzcd}
\vspace{0.5cm}
\caption{Commutative Diagram: Three-Layer Economic Evolution Natural Transformation}
\label{Z}
\end{figure}

The complete economic evolution as a commutative diagram~\ref{Z} is showing the parallel operations as needed for typing the eight bookings of the macroeconomic accounting systems.
The diagram illustrates the categorical structure of economic evolution through three interconnected layers. 
At the highest level, the Parameters Layer contains the fundamental economic ratios that govern behavior - consumption ratios and scale production factors. 
These parameters are transformed through natural transformations ($\eta$) to influence the Flows Layer, 
where actual economic activities like demand and production materialize. 
The flows represent the economic transactions, showing how demand grows from 0 to 166.4 to 272.63 units 
while production fluctuates from 31.17 to 1.0 to 3.20. 
Finally, these flows manifest in the concrete Account Layer through functorial transformations, 
tracking how Labor accounts grow from 0 to 52.0 units and Company accounts expand from 0 to 190.50 units. 
The categorical machinery of functors ($F_{\text{params}}$, $F_{\text{flows}}$) and natural transformations ensures consistency between layers, 
while tensor products and pushouts capture the parallel evolution and aggregation of economic quantities. 
This categorification reveals the mathematical structures underlying economic transformations - from abstract parameters to concrete account balances - 
while preserving the essential relationships between different aspects of the economy.

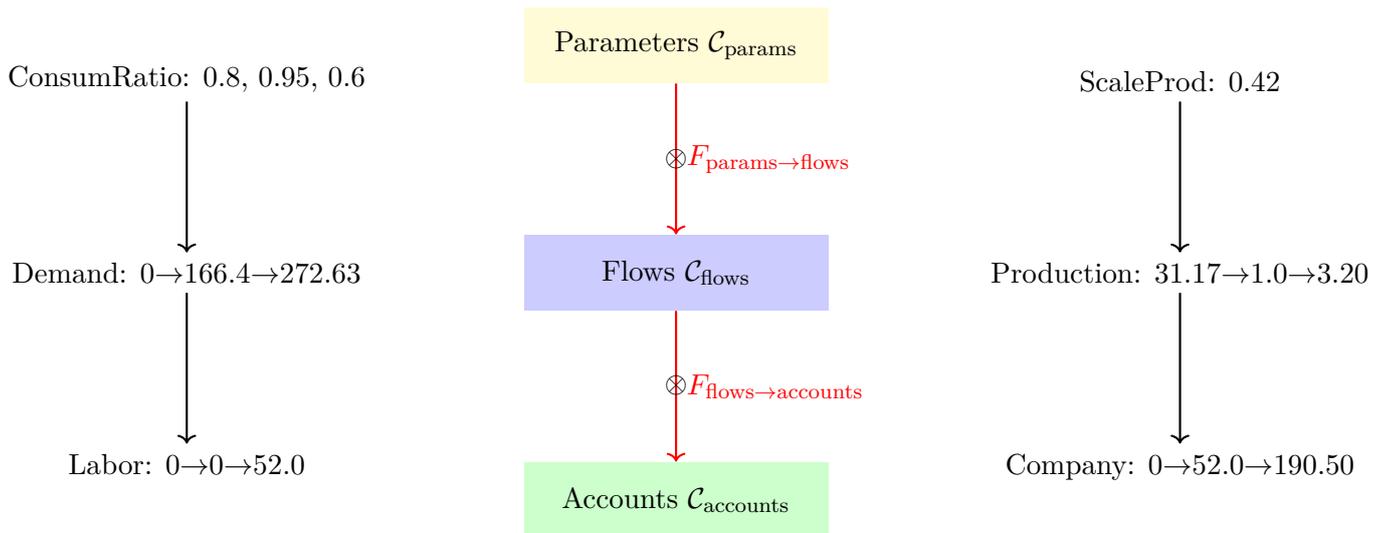
\begin{figure}[htbp]
\centering
\begin{tikzpicture}[node distance=2cm]
\node[rectangle, fill=yellow!20, minimum width=4cm, minimum height=1cm] (params) {Parameters $\mathcal{C}_{\text{params}}$};
\node[rectangle, fill=blue!20, minimum width=4cm, minimum height=1cm, below=of params] (flows) {Flows $\mathcal{C}_{\text{flows}}$};
\node[left=of flows] (demand) {Demand: 0$\to$166.4$\to$272.63};
\node[right=of flows] (prod) {Production: 31.17$\to$1.0$\to$3.20};
\node[rectangle, fill=green!20, minimum width=4cm, minimum height=1cm, below=of flows] (accounts) {Accounts $\mathcal{C}_{\text{accounts}}$};
\node[below=of demand] (lab) {Labor: 0$\to$0$\to$52.0};
\node[below=of prod] (com) {Company: 0$\to$52.0$\to$190.50};
\node[above=of demand] (cr) {ConsumRatio: 0.8, 0.95, 0.6};
\node[above=of prod] (sp) {ScaleProd: 0.42};
\draw[->, thick, red] (params) -- (flows) node[midway,right] {$F_{\text{params$\to$flows}}$};
\draw[->, thick, red] (flows) -- (accounts) node[midway,right] {$F_{\text{flows$\to$accounts}}$};
\draw[->, thick] (cr) -- (demand);
\draw[->, thick] (sp) -- (prod);
\draw[->, thick] (demand) -- (lab);
\draw[->, thick] (prod) -- (com);
\node at (0,-1.5) {$\otimes$}; % Between params and flows
\node at (0,-4.5) {$\otimes$}; % Between flows and accounts
\end{tikzpicture}
\caption{String Diagram: Parameters $\otimes$ Flows $\otimes$ Accounts Evolution}
\label{1}
\end{figure}

The diagram \ref{1} illustrates the three-layer categorical structure through a tensor product perspective. 
The Parameters layer ($\mathcal{C}_{\text{params}}$) contains the fundamental economic parameters - 
ConsumRatio evolving through values 0.8, 0.95, 0.6 and ScaleProd fixed at 0.42. 
These parameters are transformed through the functor $F_{\text{params} \to \text{flows}}$ into the Flows layer ($\mathcal{C}_{\text{flows}}$), 
where economic flows materialize as Demand (growing from 0 to 166.4 to 272.63) and Production (changing from 31.17 to 1.0 to 3.20). 
Finally, through $F_{\text{flows} \to \text{accouts}}$, these flows manifest in the Accounts layer ($\mathcal{C}_{\text{accouts}}$) as concrete account balances - 
Labor accounts evolving from 0 to 0 to 52.0 and Company accounts from 0 to 52.0 to 190.50. 
The tensor products ($\otimes$) indicate parallel operations occurring simultaneously at each layer, capturing the concurrent nature of economic transformations. 
This categorical structure ensures that changes in parameters propagate consistently through flows to account balances while maintaining parallel processing capabilities.

\begin{figure}[htbp]
    \centering
    \begin{tikzcd}[column sep=8em, row sep=4em]
    \mathcal{C}_0 \arrow[r, "T", blue] \arrow[d, "F_{acc}", green] & 
    \mathcal{C}_1 \arrow[r, "T", blue] \arrow[d, "F_{acc}", green] & 
    \mathcal{C}_2 \arrow[d, "F_{acc}", green] \\
    \mathcal{F}_0 \arrow[r, "T_F", orange] \arrow[ur, "\eta_0", purple, dashed] & 
    \mathcal{F}_1 \arrow[r, "T_F", orange] \arrow[ur, "\eta_1", purple, dashed] & 
    \mathcal{F}_2 \\
    \end{tikzcd}
    \begin{align}
    &\text{Period Evolution Values:} \nonumber\\
    &\text{Company Bank: } 0.0 \to 52.0 \to 190.50 \nonumber\\
    &\text{Labor Bank: } 0.0 \to 0.0 \to 52.0 \nonumber\\
    &\text{Resource Bank: } 0.0 \to 208.0 \to 273.19 \nonumber\\
    &\text{Investment: } 260.0 \to 289.49 \to 275.20 \nonumber\\
    &\text{Demand: } 0.0 \to 166.4 \to 272.63 \nonumber
    \end{align}
    \caption{Commutative Diagram: Period Evolution Natural Transformation}
    \label{2}
\end{figure}
    
The diagram~\ref{2} tells the economic story through categorical structures. 
The evolution occurs across three time periods, represented by categories $\mathcal{C}_0$, $\mathcal{C}_1$, and $\mathcal{C}_2$ in the top row. 
These categories capture the complete economic state at each moment. Below them, categories $\mathcal{F}_0$, $\mathcal{F}_1$, and $\mathcal{F}_2$ 
represent the actual flows of money and goods between economic agents. 
The blue arrows T show how the economy moves forward in time, while orange arrows $T_F$ track how flows evolve. 
The green vertical arrows $F_{acc}$ are functors that translate flows into account balance changes. 
Most importantly, the purple dashed arrows $\eta$ are natural transformations that ensure consistency 
between different ways of viewing the same economic evolution - whether through direct state changes or through accumulated flows. 
We can see this evolution quantitatively in the account values: 
The Company Bank account grows substantially from 0 to 190.50 units as business activity increases. 
The Labor Bank shows an interesting pattern - staying at 0 for two periods before jumping to 52 units, reflecting delayed wage payments. 
Meanwhile, Resource Bank builds up reserves from 0 to 273.19 units, and Investment fluctuates around 260-290 units 
while Demand steadily grows from 0 to 272.63 units. 
    
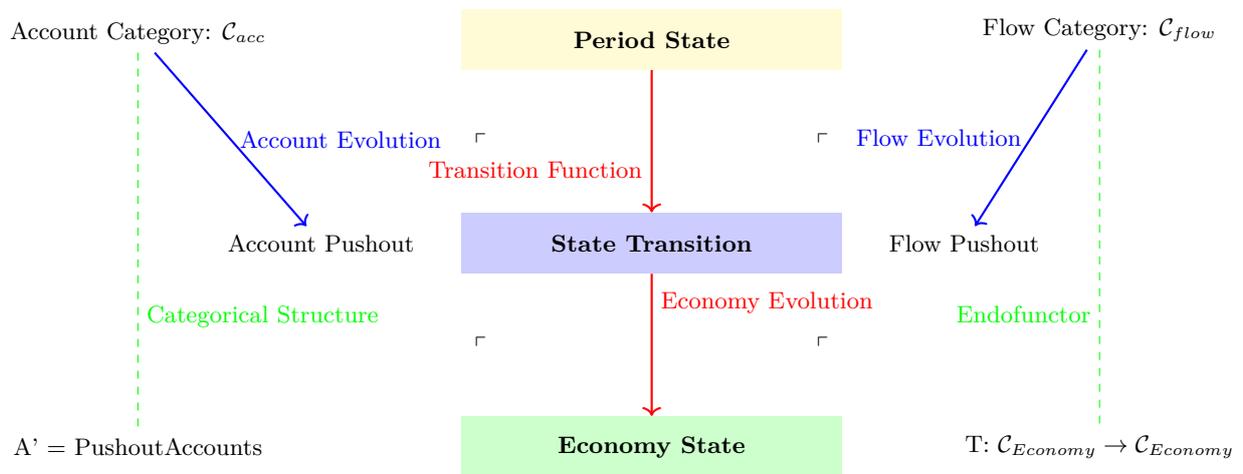
\begin{figure}[htbp]
    \centering
    {\footnotesize
    \begin{tikzpicture}[scale=0.9]
    \node[rectangle, fill=yellow!20, minimum width=5cm, minimum height=0.8cm] (period) at (0, 6) {\textbf{Period State}};
    \node[rectangle, fill=blue!20, minimum width=5cm, minimum height=0.8cm] (transition) at (0, 3) {\textbf{State Transition}};
    \node[left=0.5cm of transition] (accountpush) {Account Pushout};
    \node[right=0.5cm of transition] (flowpush) {Flow Pushout};
    \node[rectangle, fill=green!20, minimum width=5cm, minimum height=0.8cm] (economy) at (0, 0) {\textbf{Economy State}};
    \node[above=0.1cm, left=2.5cm of economy] (newaccounts) {A' = PushoutAccounts};
    \node[above=0.1cm, right=1.5cm of economy] (endofunctor) {T: $\mathcal{C}_{Economy} \rightarrow \mathcal{C}_{Economy}$};
    \node[above=5.0cm of newaccounts] (accounts) {Account Category: $\mathcal{C}_{acc}$};
    \node[above=5.0cm of endofunctor] (flows) {Flow Category: $\mathcal{C}_{flow}$};
    \node at (-2.5, 4.5) {$\ulcorner$};
    \node at (2.5, 4.5) {$\ulcorner$};
    \node at (-2.5, 1.5) {$\ulcorner$};
    \node at (2.5, 1.5) {$\ulcorner$};
    \draw[->, thick, red] (period) -- (transition) node[pos=0.7,left] {Transition Function};
    \draw[->, thick, red] (transition) -- (economy) node[pos=0.2,right] {Economy Evolution};
    \draw[->, thick, blue] (accounts) -- (accountpush) node[midway,right] {Account Evolution};
    \draw[->, thick, blue] (flows) -- (flowpush) node[midway,left] {Flow Evolution};
    \draw[dashed, green] (accounts) -- (newaccounts) node[pos=0.7,right] {Categorical Structure};
    \draw[dashed, green] (flows) -- (endofunctor) node[pos=0.7,left] {Endofunctor};
    \node at (0,-1.0) {};    
    \end{tikzpicture}
    }
    \vspace{-1cm}
    \caption{Commutative Diagram: Categorical State Transition Pushout Flow}
    \label{3}        
\end{figure}

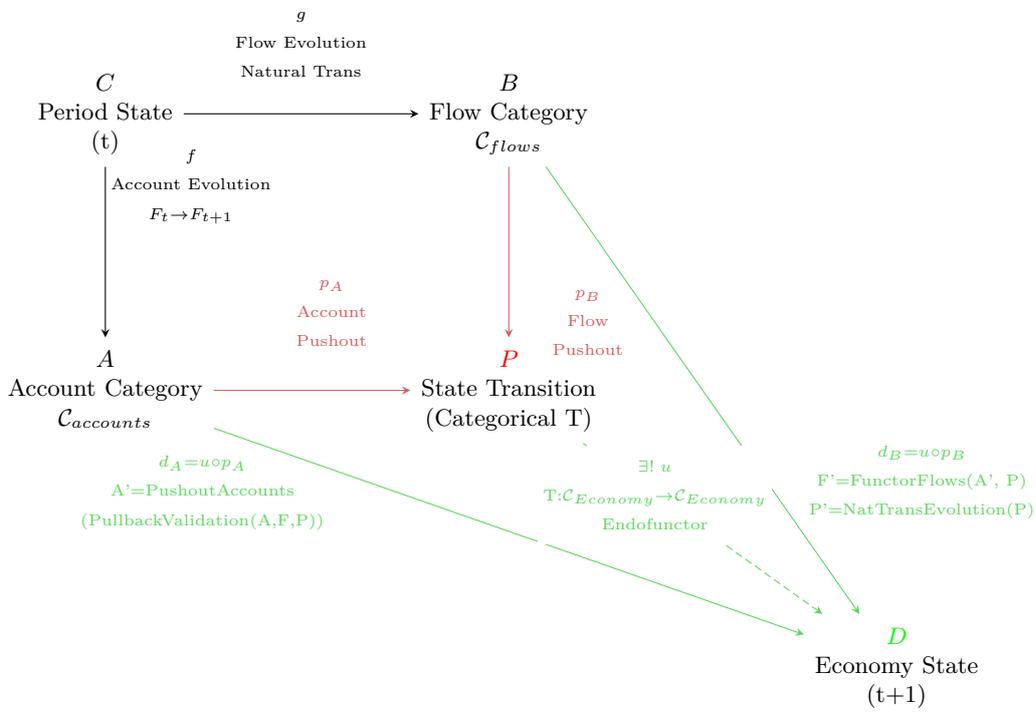
\begin{figure}[htbp]
    \centering
    {\footnotesize
    \begin{tikzcd}[column sep=8em, row sep=7em, arrows={-stealth}]
        {\begin{matrix} C \\ \text{Period State} \\ \text{(t)} \end{matrix}}
        & {\begin{matrix} B \\ \text{Flow Category} \\ \mathcal{C}_{flows} \end{matrix}} \\
        {\begin{matrix} A \\ \text{Account Category} \\ \mathcal{C}_{accounts} \end{matrix}} 
        & {\begin{matrix} \red{P} \\ \text{State Transition} \\ \text{(Categorical T)} \end{matrix}} & \\
        && {\begin{matrix} \green{D} \\ \text{Economy State} \\ \text{(t+1)} \end{matrix}} & \\
        \arrow["{\begin{matrix} f \\ \text{Account Evolution} \\ F_t \to F_{t+1} \end{matrix}}"{right}, pos=0.1, from=1-1, to=2-1]
        \arrow["{\begin{matrix} g \\ \text{Flow Evolution} \\ \text{Natural Trans} \end{matrix}}"{above=0.3cm}, pos=0.5, from=1-1, to=1-2]
        \arrow["{\begin{matrix} p_A \\ \text{Account} \\ \text{Pushout} \end{matrix}}"{above=0.4cm}, pos=0.6, color={rgb,255:red,214;green,92;blue,92}, from=2-1, to=2-2]
        \arrow["{\begin{matrix} p_B \\ \text{Flow} \\ \text{Pushout} \end{matrix}}"{right=0.5cm}, pos=0.9, color={rgb,255:red,214;green,92;blue,92}, from=1-2, to=2-2]
        \arrow["{\begin{matrix} d_A = u \circ p_A \\ \text{A'} = \text{PushoutAccounts} \\ \text{(PullbackValidation(A,F,P))} \end{matrix}}"{left=0.8cm}, pos=0.3, color={rgb,255:red,92;green,214;blue,92}, from=2-1, to=3-3]
        \arrow["{\begin{matrix} d_B = u \circ p_B \\ \text{F'} = \text{FunctorFlows(A', P)} \\ \text{P'} = \text{NatTransEvolution(P)} \end{matrix}}"{right=0.5cm}, pos=0.7, color={rgb,255:red,92;green,214;blue,92}, from=1-2, to=3-3]
        \arrow["{\begin{matrix} \exists ! \;u \\ \text{T}: \mathcal{C}_{Economy} \to \mathcal{C}_{Economy} \\ \text{Endofunctor} \end{matrix}}"{description}, pos=0.3, color={rgb,255:red,92;green,214;blue,92}, dashed, from=2-2, to=3-3]
    \end{tikzcd}
    }
    \vspace{-2.5cm}
    \caption{Categorical State Transition Pushout}
    \label{4}
\end{figure}

The diagram in Figure \ref{4} describes the eolution in time by categorical constructions.
At any moment, the economy has two main aspects: the accounts that show what everyone owns (represented in category $\mathcal{C}_{accounts}$), 
and the flows that show how money and goods move between agents (captured in $\mathcal{C}_{flows}$). 
As time passes from period $t$ to $t+1$, both accounts and flows change - wages get paid, goods are bought and sold, and investments are made. 
These changes don't happen independently - when money flows from one account to another, both accounts must update consistently. 
Pushouts verify that all changes fit together properly. 
The final result is a single mathematical object - an endofunctor T - 
that takes the entire economic state and transforms it into the next period's state while preserving all the important relationships between accounts and flows. 
This mathematical framework helps us understand how individual economic actions combine to create the overall evolution of the economy as a macroeconomic foundation of microeconomic transactions.

\textbf{Economics: Complete Economic Evolution and Systemic Integration:}
The categorical state transition pushout represents the ultimate synthesis of all economic mechanisms through categorical colimits. 
The endofunctor 
\begin{equation}
    T: \mathcal{C}_{Economy} \rightarrow \mathcal{C}_{Economy}
\end{equation}
integrates pullback validation, pushout computation, 
and functor transformation into a single categorical morphism that preserves all economic structures while enabling systemic evolution.

Economically, this captures the complete economic evolution process where individual categorical structures 
(Account Category: $\mathcal{C}_{acc}$, Flow Category: $\mathcal{C}_{flow}$) 
compose through state transitions to generate coherent economic development. The pushout ensures that economic evolution respects both microeconomic level constraints 
and macroeconomic consistency, creating new economic states (A' = PushoutAccounts) that emerge naturally from previous states while maintaining categorical properties.
The three-layer architecture of diagram~\ref{3} reveals the categorical nature of complete economic systems: the yellow layer 
represents current period's economic state, the blue layer embodies the state transition mechanism, and the green layer shows the evolved economy state. 
The pushout corners mark where individual categorical structures compose into the universal economic evolution morphism. 
This categorical approach to economic system theory provides mathematical foundations for understanding economic development 
as a categorical process, connecting all economic theory through categorical mathematics and enabling formal analysis of economic systems 
as evolving categorical structures that preserve essential economic relationships while adapting to changing conditions and maintaining systemic coherence.
The complete set of economic variables across periods 0, 1, 2 is in Table~\ref{5}.

\begin{table}[htbp]
\centering
\small
\begin{tabular}{|l|r|r|r||l|r|r|r|}
\hline
\textbf{Variable} & \textbf{$t=0$} & \textbf{$t=1$} & \textbf{$t=2$} & \textbf{Variable} & \textbf{$t=0$} & \textbf{$t=1$} & \textbf{$t=2$} \\
\hline
ConsumRes & 0.0 & 166.4 & 218.55 & AccLabBank & 0.0 & 0.0 & 52.0 \\
ConsumLab & 0.0 & 0.0 & 49.4 & AccLabLab & 0.0 & 100.0 & 195.67 \\
ConsumCap & 0.0 & 0.0 & 4.68 & AccLabGood & 0.0 & 0.0 & 0.0 \\
Demand & 0.0 & 166.4 & 272.63 & AccResBank & 0.0 & 208.0 & 273.19 \\
DemandPlan & 0.0 & 117.0 & 247.27 & AccResRes & 0.0 & 91.68 & 182.42 \\
DemandSurplus & 0.0 & 49.4 & 25.36 & AccResGood & 0.0 & 0.0 & 1.17 \\
GoodProduction & 31.17 & 1.0 & 3.20 & AccCapBank & 0.0 & 0.0 & 7.8 \\
GoodPrice & 30.0 & 141.7 & 89.94 & AccCapDiv & 0.0 & 7.8 & 41.57 \\
Investment & 260.0 & 289.49 & 275.20 & AccCapGood & 0.0 & 91.68 & 182.42 \\
InvestmentRes & 208.0 & 231.59 & 220.16 & AccComBank & 0.0 & 52.0 & 190.50 \\
InvestmentLab & 52.0 & 57.90 & 55.04 & AccComLoan & 0.0 & 260.0 & 523.49 \\
Repayment & 26.0 & 28.95 & 27.52 & AccComDiv & 0.0 & 7.8 & 41.57 \\
WagesPayment & 0.0 & 52.0 & 109.90 & AccComRes & 20.0 & 8.32 & 9.26 \\
RepaysPayment & 0.0 & 26.0 & 54.95 & AccComLab & 110.0 & 0.0 & 4.33 \\
Diff & 52.0 & 138.50 & 121.25 & AccComGood & 0.0 & 31.17 & 30.99 \\
DividendDecision & 7.8 & 41.57 & 94.39 & AccBankComLoan & 0.0 & 260.0 & 523.49 \\
DividendPayment & 0.0 & 7.8 & 41.57 & AccBankComBank & 0.0 & 52.0 & 190.50 \\
& & & & AccBankLabBank & 0.0 & 0.0 & 52.0 \\
& & & & AccBankResBank & 0.0 & 208.0 & 273.19 \\
& & & & AccBankCapBank & 0.0 & 0.0 & 7.8 \\
\hline
\end{tabular}
\caption{Complete Economic Variable Evolution for Periods 0, 1, 2}
\label{5}
\end{table}

% ------------------------------------------------------------------------------------------------------------------------------------------------------------
\section{Conclusion}

The main economic result of this paper is the construction of a stable monetary accounting system.
Money primarily serves as a means to measure and repay debts and liabilities by transferring ownership of physical tokens 
like paper bills and coins, or references to them such as checking account balances representing defined units of account.
These monetary instruments are necessary not only for loan contracts but also for any exchange contracts where goods are traded for money, tit for tat,
with the debt being denominated in terms of both the good and the money transferred to fulfill the contract.

Our MoMaT demonstrates that monetary systems are essential for economies organised by a division of labour.
The hierarchical nature of these systems provides a risk-sharing technology: risk originates at the microeconomic company level and is ultimately socialised 
at the central bank level through the banking sector. 
However, a critical missing piece has been an accounting consistency technology at the macroeconomic level that can integrate consistent microeconomic double-entry accounting systems. 
This paper's core contribution is accordingly providing the mathematical tools for these macroeconomic systems.

The ultimate goal of sharing in economies and societies concerns GDP distribution, beginning with the organisation of labour sharing in modern societies. 
Effective economic policy and distributional theory require a deep understanding of the underlying processes.
This has been the primary aim of our paper, which we hope to have achieved to some extent. 
We believe this work represents important progress toward enabling the economics profession to better address these fundamental societal challenges.

A more abstract perspective on monetary systems in modern economies reveals that hierarchical levels form an institutional framework designed to synchronise payment flows. 
This synchronisation is crucial because input providers, who are risk-averse, must be paid before companies receive payments from consumers of their products.
This timing challenge parallels the modern biological understanding of cognitive systems described in~\cite{EV2007}, 
where hierarchical synchronisation of time scales occurs across different organisational levels. 
For instance, top-level corporate management must operate with much longer planning horizons compared to assembly line workers or middle management.
Our categorical approach to economic modelling, both at the foundational accounting level and through strategic game theory in open games in~\cite{GHWZ2018}, 
is particularly well-suited for analysing decentralised control and management of private and common pool resources, 
as explored in polycentric management by Elinor~\cite{Ostrom2005Book}.
We demonstrate how category theory provides rigorous mathematical tools for economic modelling and simulation programming. 
Our categorical formulation of macroeconomic accounting shows that complex economic phenomena - from local individual decisions to global macroeconomic systemic outcomes - 
can be effectively understood through categories, functors, and natural transformations in universal constructions.

% Theoretical Contributions
We have shown in our MoMaT that eight fundamental macroeconomic bookings naturally 
arise as typed morphisms in product categories $\mathcal{C}_{\text{Real}} \times \mathcal{C}_{\text{Nominal}}$, 
revealing the deep categorical structure underlying double-entry bookkeeping. 
The contravariant relationship between real and nominal flows emerges not as an accounting convention, but as a mathematical necessity encoded in the categorical structure itself. 
String diagrams provide visualisations of economic evolution, offering economists a new language for understanding temporal dynamics through functors and natural transformations.

% Practical Advantages
The categorical approach delivers concrete computational benefits: pullbacks ensure constraint validation before transactions occur, 
pushouts guarantee consistency in state transitions, and natural transformations preserve economic relationships across time periods. 
Unlike the recursive formulation, our categorical approach provides type safety for all economic variables and universal property guarantees for all calculations.

% New Economic Insights
The categorical perspective reveals previously hidden economic structures. Investment 
decisions emerge as universal constructions with functors transforming demand surplus into capital allocation, while market equilibration 
appears as pullback constructions ensuring supply-demand consistency. The three-layer categorical structure 
(Parameters $\to$ Flows $\to$ Accounts) shows how abstract economic behavioural specifications materialize into concrete account balances 
through functorial transformations, providing a mathematical foundation for understanding the propagation and communication of 
economic decisions through hierarchical layers.

% Broader Significance
This work establishes category theory as more than a mathematical abstraction in economics. 
It becomes a practical tool for economic analysis and policy design, gives economic theories a formal semantics 
expressable in a domain specific language (DSL) for economics as calculations and information processing of double-entry bookkeeping programs.
Our approach suggests that many classical economic problems, from stability analysis to welfare optimisation, might benefit from categorifications 
which amounts to a typing discipline for units of accounts. 
The framework opens new research directions in computational economics, monetary theory, and macroeconomic modelling.

% Limitations and Future Work
While our theory captures key macroeconomic dynamics, it currently focuses on a simplified five-sector-agent economy with a bank as the banking sector which is usually about risk management.
However, extensions with central banks are now compositional and for consistency and decision structures it is clear how to integrate them into the categorical framework.
Extensions to incorporate heterogeneous agents, complex financial instruments, and international trade represent important avenues for future research. 
Additionally, the relationship between our categorical constructions and established economic theories requires deeper investigations
by using the language of categories to formulate them and check for consistencies and inconsistencies of different points of view.

% Categories, functors and natural transformations
Categories, functors and natural transformations, as shown by Lawvere, categorify the concepts of theory and models with homomorphic consistency in Tarski's logic,
ready to use for the epistemology of econom(etr)ics.
Categorical universals generalise maximisation in economics and econometrics which is useful to unite operational and institutional dynamics.
A monad is an endofunctor as well and formalises side effects in programs which is useful to formally capture institutional dynamics versus operational dynamics 
within the institutional setup.
Category theory's anecdotical {\it abstract nonsense} does make sense when syntax and semantics are designed for domain-specific programming languages (DSL) and their compilers. 
In economic systems syntax and semantics have to be separated to type the observing and observed agents in the structural econometrics that emerged after the Lucas Critique.
Custom units of account and their creation are definable by functors and natural transformations.
The national accounting system is extendable compositionally to also include central banks and governments in institutional hierarchies.
New agents can be added with accounts, hierarchies, bookings and consistency conditions to design, program and control monetary institutional contractual evolutions.
An earlier categorically programmed double-entry system can be composed with the macroeconomic accounting of this paper by suitably specified type theories 
to industrially scalable programs for managing many agents' accounts, decisions, hierarchies of strategic game theoretic, econometric or multi-agent systems 
as digital twins of holdings, platforms, banks, crypto currencies, stable coins, central banks 
or at the top level of monetary economics as exchange rate regime dynamics booked in the bank of central banks.

Leibniz might have probably said {\it calculemus}, which is Latin for {\it let us calculate}.
When he used this term, he was not only referring to mathematical calculations in the modern sense.
He was expressing a much broader vision for resolving all forms of human disputes.
In our case major economic disputes are about the distribution of the GDP, ever since Marx's {\it struggle of the classes}.
Leibniz wished that when people had a disagreement on philosophical, legal, political, or scientific questions 
they could sit down, take out their pens, and say to each other, {\it Calculemus!} and 
then work through the problem logically to see who was right.
In essence, {\it calculemus} called for:
\begin{itemize}
\item Objectivity: It was his belief that logic and reason, if applied systematically, could transcend subjective opinions and emotional arguments. 
\item Automation: It was a precursor to modern ideas of computation and artificial intelligence, where complex reasoning could be automated through formal, rule-based processes.
\item Notation: It was a testament to his deep conviction that the right notation (like the one he developed for calculus) could make complex ideas clear and tractable.
\end{itemize}
In this sense, we try to lay out an approach to {\it calculemus} for economics.
Who owes what of the common product, the GDP, to whom in the production process of an economy organised by a division of labour and hierarchical risk sharing.

% ------------------------------------------------------------------------------------------------------------------------------------------------------------
%\clearpage
\appendix
\section{Appendix}
\subsection{Category Theory}\label{app:category_theory}
% ------------------------------------------------------------------------------------------------------------------------------------------------------------
We use in this paper the three fundamental categorical data structures of category, functor and natural transformation.
Categories contain objects and morphisms, like a graph with vertices and edges, while the morphisms in a category are closed under composition.
This means that a structure to qualify as a category, for all morphisms which are composable, the composed morphism must be part of the category as well.
In short, categories are structures where composable morphisms are composed.
It is the mathematics of composable structures and thus perfect for the study of complex systems, like mathematics itself.
Examples are formed by the category of Sets as objects with morphisms being functions - linear or continuous or relations - between them.
Other categories contain topological spaces, dynamical systems, strategic games and other subjects that can be made into a category 
with suitable definitions of what morphisms and their compositions are or should be for some modelling purpose.

Category theory allows to switch between an original category with morphisms between objects to a meta category 
with the original morphisms being objects and the morphisms are then morphisms between the morphisms of the original category.
Hence, we can switch from an {\it object oriented} view to a {\it processes oriented} view of the structures being analysed.
Categories are for consistently bookkeeping morphisms and meta morphisms in mathematics.
We have made them available for macroeconomic bookkeeping as well.
Another view is that category theory is the grammar of mathematics and accordingly in our macroeconomic accounting system categories seem to be useful as the grammar of economies and economics.

We have shown that exchanges of goods and money are naturally expressed in categories.
The resulting morphisms between morphisms bind opposing processes in value transfers in economies subject to the legal principles of {\it tit for tat} or {\it quid pro quo},
for example underlying the restitution claims in paragraph §812 of the BGB (German Civil Code), see more details in~\cite{MW2025}.

Category theory provides an extension of set based mathematics where modelling takes place in terms of custom categories.
Functors can be used to map them for example into the standard category of sets for a usual mathematical representation.
But not only algebra is inherently categorical but also geometry and logic, see~\cite{Dagstuhl2014} for a survey.
The economic models are typed, decomposed and recomposed in a compositional and modular way and equipped with a programming language for economics.
They help in understanding, programming, simulating or learning parameters and patterns from data.
Modern software architectures and compiler technology can not be thought without category theory nowadays and 
accordingly our paper is a step to unite computer science and economics for designing and implementing the informational structures of digital twins of complex real world systems.

A topos, a top level categorical construction, has an internal language, which is a formal language that is interpreted within the topos itself. 
This is often an intuitionistic higher-order logic. 
From a computer scientific view category theory is useful in designing different semantics of functional and domain specific (DSL) programming languages, see~\cite{Dagstuhl2014}.
This concept is a powerful idea that allows one to treat a topos as a universe where one can do mathematics and define new objects and relationships. 
The {\it internal} nature means that the language's syntax and semantics are completely defined by the categorical structure of the topos, rather than being imposed from the outside.
In this sense this paper is ultimately aiming to the define the topos of economics.
The invariance based approach to complex systems mimicks the {\it Erlangen program} of Felix Klein to study geometries by their invariants.
Similarly the {\it Emmy Noether program} is to understand symmetries and invariants of physical systems by understanding what is conserved during transformations.
In economics values are conserved during bookings.
Accordingly, a value based money theory, as is often attempted in order to understand money, is from our point of view doomed to fail.

Using categories as states with functors as generalised functions and natural transformations as consistency conditions and equations 
over composed functions and universals as adjunctions for expressing optimality is a rather general and powerful approach to modelling.
The power comes from using categories for keeping consistency in local to global structures and evolutionary transitions over time.
This is where category theory shines and which is the core of using them in macroeconomic accounting as an example of a complex system with a whole and local parts.
It is a language for describing and reasoning about complex systems and their interactions.

Functors map objects and morphisms of one category to another category preserving the identities and the compositional structure.
Natural transformations are maps between functors, also preserving the identities and the compositional structure.
We use cateogry theory as the mathematics of function composition.
The usual elements of sets $x\in X$ can be represented as a function from a singleton set like $\{*\}$ to $X$ picking the element $x\in X$.
Function application $f(x)$ then becomes function composition $f\circ x:\{*\}\rightarrow X$,
making the usual set and element based mathematics into a special case of a categorical morphism based mathematics.
Categories can also be conceived as the module system of mathematics being able to represent algebraic, geometric and logical structures and how they morph into each other.

For economics another aspect seems to be most powerfull - namely that category theory provides a calculus for morphisms 
where systems of equations are solved for functions, processes, morphisms or relations and not only numbers.
This high-levelness provides much expressivity for economic concepts which are often more appropriately described by processes and not by objects, 
much in line with the ideas usually attributed to Heraklit - the world is a flux of change.

Pullbacks and pushouts are the tools we use for verifying consistency conditions and composing calculations, respectively.
They are examples of limits and (dual) colimits which are universal constructions.
The universality can represent uniqueness, optimality, best, biggest or similar concepts.
For economics they massively generalise economic principles of optimality, see~\cite{Dagstuhl2015}.
The categorical machinery comes with formal and visual 2-dimensional mathematical tools like string and commutative diagrams, see~\cite{Selinger2009a}
and~\cite{Selinger2011} where they are used in a categorical reformulation of physics.
The results are now used in compilers for quantum computers, for example at the company Quantinuum,
based on Bob Coecke's and Samson Abramsky's original work in~\cite{AC2004} or its results later in~\cite{Coecke2017}.
This work has been a direct inspiration for developing open games for economic applications in \cite{GHWZ2018} and this paper's tools.

Diagrams can be used to visually represent systems of equations of functions.
String diagrams are in a way the process oriented visualisations of functions while commutative diagrams are visual representations of equations.
Commuting diagrams show paths of sequentially composed morphisms as arrows.
Their commuting property is that different paths and by that composed functions must be equal.
String diagrams can do the same but turn morphisms into the object of interest 
by boxing them while the domain and codomains (types) are taken to be the wires.
Usual pictures of functions $\boxed{A}\stackrel{f}{\longrightarrow} \boxed{B}$ are an object oriented view 
while string diagrams $\stackrel{A}{\longrightarrow} \boxed{f} \stackrel{B}{\longrightarrow}$ focus on the morphisms as a process oriented view.
Function composition in formulas is denoted by a circle in $f \circ \red{p_A}$ and reads as {\it apply $f$ after $\red{p_A}$}.
Equality on paths is then written as $f \circ \red{p_A} = g \circ \red{p_B}$ which is what commutative diagrams represent in a visual way.
For some people equations are more intutitive while others prefer visual representations in diagrams.

We now define formulas, commuting and string diagrams of categories, functors, natural transformations, pullbacks and pushouts in their abstract form 
which we apply to macroeconomics in the main text.

%-------------------------------------------------------------------------------------------------------------------------------------------------------------------------
\subsubsection{Category, Functor, Natural Transformation}
A category contains objects and morphisms being functions or relations between them.
A usual set is a category with objects without any morphisms between them.
By that we can add relations and morphisms between objects and take these structures into account which allows us to model emergent efffects of geometric and macroeconomic wholenesses.
These concepts are modelable in a very elegant way in categories, functors and universals as optimal decisions in spaces where morphisms and relations between objects are to be composed.
The important point of category theory is that properties of objects are not defined by their internal structure which one needs to look up within the set.
Instead mathematical properties are defined by their behaviour at their interface when being used by their context in operations.
Mathematics becomes by that a kind of social science of relational behaviours at interfaces of systems.
These features make categories into attractive representations of states, relations or structures in a social science like economics.

\begin{definition}[Category]\label{def:category}
A category $\mathcal{A}$ in the category of categories $\mathcal{A}\in\mathcal{C}$ has the following structure.
\begin{itemize}
    \item Objects Ob: objects of $\mathcal{A}$ have a unique identity morphism, $\forall A\in\text{Ob}(\mathcal{A}), \exists !\;\text{id}_A\in\text{Hom}(\mathcal{A}) :: A \to A$
    \item Morphisms Hom: morphisms of $\mathcal{A}$ from objects to objects, $\forall f\in\text{Hom}(\mathcal{A}):: A_s \to A_t$, $A_s, A_t\in\text{Ob}(\mathcal{A})$
    \item Composition Hom: that are composable, $\forall f_1, f_2, f_3, \circ::\text{Hom}(A_2,A_3) \times \text{Hom}(A_1,A_2) \to \text{Hom}(A_1,A_3)$
    \item Identity of $\circ$: can be composed obeying to identity law, $f \circ \text{id}_{A_1} = f = \text{id}_{A_2} \circ f$
    \item Associativity of $\circ$: and associativity law, $(f_3\circ f_2) \circ f_1 = f_3\circ f_2\circ f_1 = f_3 \circ (f_2 \circ f_1)$.
        \item Diagram of category with identities and associative compositions.\\
\begin{tikzcd}[column sep=3em, row sep=3em]
    A_1 \arrow[loop below, "\text{id}_{A_1}"{right=0.2cm}]
        \arrow[r, "f_1"] 
        \arrow[rr, bend left=80, looseness=1.1, "f_2 \circ f_1"'{above}] 
        \arrow[rrr, bend left=80, looseness=1.2, "(f_3\circ f_2) \circ f_1 = f_3 \circ (f_2 \circ f_1)=f_3\circ f_2\circ f_1"{above}] & 
    A_2 \arrow[loop below, "\text{id}_{A_2}"{right=0.2cm}]
        \arrow[r, "f_2"] 
        \arrow[rr, bend left=80, looseness=1.1, "f_3 \circ f_2"'{above}] & 
    A_3 \arrow[loop below, "\text{id}_{A_3}"{right=0.2cm}]
        \arrow[r, "f_3"] &
    A_4 \arrow[loop below, "\text{id}_{A_4}"{right=0.2cm}] \\
\end{tikzcd}
        \item Commutative diagram of associativity and identity law of composition.\\
\begin{tikzcd}[column sep=3em, row sep=3em]
    A_1 \arrow[r, "f_1"] 
        \arrow[rr, bend left=80, looseness=1.1, "f_2 \circ f_1"'{above}] 
        \arrow[rrr, bend left=80, looseness=1.2, "f_3 \circ (f_2 \circ f_1)"{above}]
        \arrow[d, "\text{id}_{A_1}"{right=0.2cm}]  & 
    A_2 \arrow[r, "f_2"] \arrow[d, "\text{id}_{A_2}"{right=0.2cm}] & 
    A_3 \arrow[r, "f_3"] \arrow[d, "\text{id}_{A_3}"{right=0.2cm}] &
    A_4 \arrow[d, "\text{id}_{A_4}"{right=0.2cm}] \\
    A_1 \arrow[r, "f_1"']  
        \arrow[rrr, bend right=80, looseness=1.2, "(f_3 \circ f_2)\circ f_1"'{below}] & 
    A_2 \arrow[r, "f_2"'] 
        \arrow[rr, bend right=80, looseness=1.1, "f_3 \circ f_2"'{below}] & 
    A_3 \arrow[r, "f_3"'] &
    A_4
\end{tikzcd}
        \item String diagram of preservation of identity law for identity morphisms.\\
            % define the string diagram style
\tikzset{
    stringdiagram/.style={
        baseline=(current bounding box.center),
        every node/.style={draw=none},
        every path/.style={thick}
        }
    }
            \begin{tikzpicture}[stringdiagram]
                % Left diagram showing identity morphisms
                \begin{scope}[shift={(-4,0)}]        
                    % Left diagram showing f ∘ id_A1
                    \node at (4,2.25) {$A_1$};
                    \node[draw, rectangle, minimum width=1cm, minimum height=0.8cm] (id1) at (3.5,1.5) {$id_{A_1}$};
                    \node at (4,0.5) {$A_1$};
                    \node[draw, rectangle, minimum width=1cm, minimum height=0.8cm] (f1) at (3.5,-0.5) {$f$};
                    \node at (4,-1.25) {$A_2$};
                    \draw (3.5,-1.5) -- (f1);
                    \draw (f1) -- (3.5,0.5);
                    \draw (3.5,0.5) -- (id1);
                    \draw (id1) -- (3.5,2.5);
                    
                    % Equals sign
                    \node at (5.5,0.5) {$=$};
                    
                    % Middle diagram showing just f
                    \node at (7.5,2.25) {$A_1$};
                    \node[draw, rectangle, minimum width=1cm, minimum height=0.8cm] (f2) at (7,0.5) {$f$};
                    \node at (7.5,-1.25) {$A_2$};
                    \draw (7,-1.5) -- (f2);
                    \draw (f2) -- (7,2.5);
                    
                    % Equals sign
                    \node at (9,0.5) {$=$};
                    
                    % Right diagram showing id_A2 ∘ f
                    \node at (11,2.25) {$A_1$};
                    \node[draw, rectangle, minimum width=1cm, minimum height=0.8cm] (f3) at (10.5,1.5) {$f$};
                    \node at (11,0.5) {$A_2$};
                    \node[draw, rectangle, minimum width=1cm, minimum height=0.8cm] (id2) at (10.5,-0.5) {$id_{A_2}$};
                    \node at (11,-1.25) {$A_2$};
                    \draw (10.5,-1.5) -- (id2);
                    \draw (id2) -- (10.5,0.5);
                    \draw (10.5,0.5) -- (f3);
                    \draw (f3) -- (10.5,2.5);

                    \node at (13,0.5) {where};
                    % Right diagram showing id as a wire
                    \node[draw, rectangle, minimum width=1cm, minimum height=0.8cm] (id3) at (15,0.5) {$id_A$};
                    \node at (15.5,2.25) {$A$};
                    \node at (15.5,-1.25) {$A$};
                    \draw (15,-1.5) -- (id3);
                    \draw (id3) -- (15,2.5);
                    \node at (17,0.5) {$=$};
                    \node at (19,2.25) {$A$};
                    \node at (19,-1.25) {$A$};
                    \draw (18.5,-1.5) -- (18.5,2.5);
                \end{scope}
            \end{tikzpicture}
        \item String diagram of preservation of associativity law for composition of morphisms.\\
            \begin{tikzpicture}[stringdiagram]
                % Right diagram showing composition
                \begin{scope}[shift={(0,0)}]        
                    % Draw boxes (foreground)
                    \node at (0.75,3.75) {$A_1$};
                    \node[draw, rectangle] (f4) at (0,3) {$f_1$};
                    \node[draw, rectangle] (f5) at (0,2) {$f_3 \circ f_2$};
                    \node at (0.75,0.25) {$A_4$};
                    \node at (2,2) {$=$};
                    \node at (5.0,3.75) {$A_1$};
                    \node[draw, rectangle] (f1) at (4,3) {$f_1$};
                    \node[draw, rectangle] (f2) at (4,2) {$f_2$};
                    \node[draw, rectangle] (f3) at (4,1) {$f_3$};
                    \node at (5.0,2.5) {$A_2$};
                    \node at (5.0,1.5) {$A_3$};
                    \node at (5.0,0.25) {$A_4$};
                    \node at (6,2) {$=$};
                    \node at (9.0,3.75) {$A_1$};
                    \node[draw, rectangle] (f7) at (8,2) {$f_2 \circ f_1$};
                    \node[draw, rectangle] (f8) at (8,1) {$f_3$};
                    \node at (9.0,1.5) {$A_3$};
                    \node at (9.0,0.25) {$A_4$};
                    % Draw visible wire segments
                    \draw (0,0) -- (f5); % left wire
                    \draw (f5) -- (f4);
                    \draw (f4) -- (0,4);
                    \draw (4,0) -- (f3); % middle wire
                    \draw (f3) -- (f2);
                    \draw (f2) -- (f1);
                    \draw (f1) -- (4,4);
                    \draw (8,0) -- (f8); % right wire
                    \draw (f8) -- (f7);
                    \draw (f7) -- (8,4);
                \end{scope}
            \end{tikzpicture}
        \end{itemize}
    \end{definition}

Functors are views on categories.
They allow us to focus on parts, 
build the blocks of composites with natural transformations between functors making sure that these composites are consistent in a definable way.
We use functors to lift all sectoral decisions of the economy from being a function to being a functor which also take care of the 
domain and codomain maps for managing units of accounts and units in economic systems.
Domains and codomains also change when objects are created and deleted, formalisable by initial and final objects.
By that any parts of the economy can be composed and made consistent by suitable natural transformations to the functors.

    \begin{definition}[Functor]
A functor $F: \mathcal{A} \to \mathcal{B}$ between categories $\mathcal{A}$ and $\mathcal{B}$ consists of:
\begin{itemize}
    \item An object mapping: $F_o: \text{Ob}(\mathcal{A}) \to \text{Ob}(\mathcal{B})$
    \item A morphism mapping: $F_m: \text{Hom}_{\mathcal{A}}(A_1,A_2) \to \text{Hom}_{\mathcal{B}}(F_o(A_1),F_o(A_2))$
    \item which preserve identity:
            $F_m(id^{\red{\mathcal{A}}}_A) = id_{F_o(A)}^{\red{\mathcal{B}}}$, $\forall A$
    \item  and composition: 
            $F_m(g\circ_{{\red{\mathcal{A}}}} f)=F_m(g)\circ_{\red{\mathcal{B}}} F_m(f)$, $\forall A, f, g $\\
        \item Commutative diagram of identity and composition preservation:\\
\begin{tikzcd}[column sep=2em, row sep=2em]
A_1 \arrow[r, "f"] 
    \arrow[d, "F_o"'] 
    \arrow[rr, "g \circ_{\red{\mathcal{A}}} f", bend left=70] 
    \arrow[loop above, "\text{id}_{A_1}^{\red{\mathcal{A}}}"] 
    & 
A_2 \arrow[r, "g"] 
    \arrow[d, "F_o"] 
    \arrow[loop above, "\text{id}_{A_2}^{\red{\mathcal{A}}}"] 
    & 
A_3 \arrow[d, "F_o"] 
    \arrow[loop above, "\text{id}_{A_3}^{\red{\mathcal{A}}}"] 
    \\
F_o(B_1) 
    \arrow[r, "F_m(f)"'] 
    \arrow[rr, "F_m(g) \circ_{\red{\mathcal{B}}} F_m(f)"', bend right=70] 
    \arrow[loop below, "\text{id}_{F_o(A_1)}^{\red{\mathcal{B}}}"] 
    & 
F_o(B_2) 
    \arrow[r, "F_m(g)"'] 
    \arrow[loop below, "\text{id}_{F_o(A_2)}^{\red{\mathcal{B}}}"] 
    &
F_o(B_3) 
    \arrow[loop below, "\text{id}_{F_o(A_3)}^{\red{\mathcal{B}}}"] 
   \\
\end{tikzcd}
        \item String diagram of identity preservation:\\
            \tikzset{
                stringdiagram/.style={
                    baseline=(current bounding box.center),
                    every node/.style={draw=none},
                    every path/.style={thick}
                    }
                }
            \begin{tikzpicture}[stringdiagram]
                \begin{scope}[shift={(-2,0)}]
                    % left right wires, wire caption domain - functions - wire caption codomain, top - down: A_1 - f - A2 , = , B_1 - F(f) - B_2
                    \node at (1.0,1.75) {$A_1$};
                    \node[draw, rectangle] (f1) at (0,1) {$id_{A_1}$};
                    \node at (1.0,0.25) {$A_1$};
                    \node at (2,1) {$=$};
                    \node at (5.5,2.0) {$F_o(A_1)$};
                    \node[draw, rectangle] (Ff) at (4,1) {$id_{F_o(A_1)}$};
                    \node at (5.5,0.25) {$F_o(A_1)$};
                    \draw (0,0) -- (f1);
                    \draw (f1) -- (0,2);
                    \draw (4,0) -- (Ff);
                    \draw (Ff) -- (4,2);
                \end{scope}
            \end{tikzpicture}
        \item String diagram of composition preservation:\\
            \begin{tikzpicture}[stringdiagram]
                % Right diagram showing composition
                \begin{scope}[shift={(4,0)}]        
                    % Draw boxes (foreground)
                    \node at (0.5,2.75) {$A_f$};
                    \node[draw, rectangle] (f1) at (0,2) {$f$};
                    \node[draw, rectangle] (f2) at (0,1) {$g$};
                    \node at (0.5,0.25) {$A_g$};
                    \node at (2,1) {$=$};
                    \node at (5.5,2.75) {$A_f$};
                    \node[draw, rectangle] (f3) at (4,1) {$F_m(f\circ_{\red{\mathcal{A}}} g)$};
                    \node at (5.5,0.25) {$A_g$};
                    \node at (6,1) {$=$};
                    \node at (9.75,2.75) {$A_f$};
                    \node[draw, rectangle] (f4) at (8,1) {$F_m(f)\circ_{\red{\mathcal{B}}} F(g)$};
                    \node at (9.75,0.25) {$A_g$};
                    % Draw visible wire segments
                    \draw (0,0) -- (f2); % left wire
                    \draw (f2) -- (f1);
                    \draw (f1) -- (0,3);
                    \draw (4,0) -- (f3); % middle wire
                    \draw (f3) -- (4,3);
                    \draw (8,0) -- (f4); % right wire
                    \draw (f4) -- (8,3);
                \end{scope}
            \end{tikzpicture}
    \end{itemize}
    \end{definition}

A natural transformation is a morphism between functors.
It preserves identities and composition meaning they keep the systems consistent.
We can choose in what way they are to be kept consistent.
In economics usually some optimality or decision is formulated as a maximisation of some object or flow by some agent.
In category theory these are then the universals which formulate in what sense the choise is best, maximal or what ever way optimality is useful or optimal to be defined.
\begin{definition}[Natural Transformation]
A natural transformation $\eta: F \Rightarrow G$ between functors $F,G: \mathcal{A} \to \mathcal{B}$ consists of:
\begin{itemize}
    \item For each object a morphism: $\eta_A: F(A) \to G(A)$ for each object $A \in \mathcal{A}$, 
    \item which preserves identities $\eta_{A_1} \circ_{\red{\mathcal{B}}} F(id_{A_1}^{\red{\mathcal{A}}}) = \eta_{A_1} = G(id_{A_1}^{\red{\mathcal{A}}}) \circ_{\red{\mathcal{B}}} \eta_{A_1}$
    \item and composition $\eta_{A_1} \circ_{\red{\mathcal{B}}} F(g\circ_{\red{\mathcal{A}}} f) = \eta_{A_1} \circ_{\red{\mathcal{B}}} F(g) \circ_{\red{\mathcal{B}}} F(f)$
        \item Commutative diagram of identity and composition preservation:\\
\begin{tikzcd}[column sep=2em, row sep=2em]
    &&& 
    G(B_1) \arrow[loop above, "\text{id}_{G(A_1)}^{\red{\mathcal{B}}}"] 
           \arrow[rr, "F(g) \circ_{\red{\mathcal{B}}} F(f)"{above}, bend left=70] 
           \arrow[r,"G(f)"{above}] &
    G(B_2) \arrow[loop above, "\text{id}_{G(A_2)}^{\red{\mathcal{B}}}"]  
           \arrow[r,"G(g)"{above}] & 
    G(B_3) \arrow[loop above, "\text{id}_{G(A_3)}^{\red{\mathcal{B}}}"]  \\ 
    A_1 \arrow[r, "f"{below=0.0cm}] 
        \arrow[rrru, "G"] 
        \arrow[rrrd, "F"']  
        \arrow[loop above, "\text{id}_{A_1}^{\red{\mathcal{A}}}"]  &
    A_2 \arrow[r, "g"{below=0.0cm}] 
        \arrow[rrru, "G"] 
        \arrow[rrrd, "F"'] 
        \arrow[loop above, "\text{id}_{A_2}^{\red{\mathcal{A}}}"] &
    A_3 \arrow[rrru, "G"] 
        \arrow[rrrd, "F"'] 
        \arrow[loop above, "\text{id}_{A_3}^{\red{\mathcal{A}}}"]         \\
    &&& 
    F(B_1) \arrow[r, "F(f)"'] 
           \arrow[loop below, "\text{id}_{F(A_1)}^{\red{\mathcal{B}}}"] 
           \arrow[rr, "F(g) \circ_{\red{\mathcal{B}}} F(f)"', bend right=70] 
           \arrow[uu,  "\eta_{A_1}"{right}] &
    F(B_2) \arrow[r, "F(g)"'] 
           \arrow[loop below, "\text{id}_{F(A_2)}^{\red{\mathcal{B}}}"] 
           \arrow[uu,  "\eta_{A_2}"{right}] & 
    F(B_3) \arrow[loop below, "\text{id}_{F(A_3)}^{\red{\mathcal{B}}}"]
           \arrow[uu,  "\eta_{A_3}"{right}] \\
\end{tikzcd}
        \item String diagram of identity preservation:\\
            \tikzset{
                stringdiagram/.style={
                    baseline=(current bounding box.center),
                    every node/.style={draw=none},
                    every path/.style={thick}
                    }
                }
            \begin{tikzpicture}[stringdiagram]
                \begin{scope}[shift={(-2,0)}]
                    % Left diagram
                    \node at (1.0,2.75) {$F(A_1)$};
                    \node[draw, rectangle] (eta1) at (0,2) {$\eta_{A_1}$};
                    \node[draw, rectangle] (Fid) at (0,1) {$F(id_{A_1})$};
                    \node at (1.0,0.25) {$F(A_1)$};
                    
                    % Middle diagram
                    \node at (2,1.5) {$=$};
                    
                    % Center diagram 
                    \node at (5.0,2.75) {$F(A_1)$};
                    \node[draw, rectangle] (eta2) at (4,1.5) {$\eta_{A_1}$};
                    \node at (5.0,0.25) {$G(A_1)$};
                    
                    % Right equals
                    \node at (6,1.5) {$=$};
                    
                    % Right diagram
                    \node at (9.0,2.75) {$F(A_1)$}; 
                    \node[draw, rectangle] (eta3) at (8,2) {$\eta_{A_1}$};
                    \node[draw, rectangle] (Gid) at (8,1) {$G(id_{A_1})$};
                    \node at (9.0,0.25) {$G(A_1)$};

                    % Draw wires
                    \draw (0,0) -- (Fid);
                    \draw (Fid) -- (eta1);
                    \draw (eta1) -- (0,3);
                    
                    \draw (4,0) -- (eta2);
                    \draw (eta2) -- (4,3);
                    
                    \draw (8,0) -- (Gid);
                    \draw (Gid) -- (eta3);
                    \draw (eta3) -- (8,3);
                \end{scope}
            \end{tikzpicture}
        \item String diagram of composition preservation:\\
            \begin{tikzpicture}[stringdiagram]
                % Right diagram showing composition
                \begin{scope}[shift={(4,0)}]        
                    % Draw boxes (foreground)
                    \node at (1.0,3.75) {$A_1$};
                    \node[draw, rectangle] (f3) at (0.5,2) {$F_m(f\circ_{\red{\mathcal{A}}} g)$};
                    \node at (1.0,0.25) {$A_3$};
                    \node at (2.0,2) {$=$};
                    \node at (5.0,3.75) {$F_o(A_1)$};
                    \node[draw, rectangle] (f4) at (4.0,2) {$F_m(f)\circ_{\red{\mathcal{B}}} F_m(g)$};
                    \node at (5.0,0.25) {$F_o(A_3)$};
                    \node at (6.5,2) {$=$};
                    \node at (8.0,0.25) {$A_3$};
                    \node[draw, rectangle] (f1) at (7.5,3) {$f$};
                    \node at (8.0,2) {$A_2$};
                    \node[draw, rectangle] (f2) at (7.5,1) {$g$};
                    \node at (8.0,3.75) {$A_1$};
                    \node at (9.0,2) {$=$};
                    \node at (12.0,3.75) {$G_o(A_1)$};
                    \node[draw, rectangle] (f5) at (11.0,2) {$G_m(f\circ_{\red{\mathcal{A}}} g)$};
                    \node at (12.0,0.25) {$G_o(A_3)$};
                    \node at (12.75,2) {$=$};
                    \node at (15.8,3.75) {$G_o(A_1)$};
                    \node[draw, rectangle] (f6) at (15.0,2) {$G_m(f)\circ_{\red{\mathcal{B}}} G_m(g)$};
                    \node at (15.8,0.25) {$G_o(A_3)$};
                    % Draw visible wire segments
                    \draw (0.5,0) -- (f3); % left wire
                    \draw (f3) -- (0.5,4);
                    \draw (4,0) -- (f4); % next wire
                    \draw (f4) -- (4,4);
                    \draw (7.5,0) -- (f2); % next wire
                    \draw (f2) -- (f1);
                    \draw (f1) -- (7.5,4);
                    \draw (11,0) -- (f5); % next wire
                    \draw (f5) -- (11,4);
                    \draw (15,0) -- (f6); % right wire
                    \draw (f6) -- (15,4);
                \end{scope}
            \end{tikzpicture}
        \end{itemize}
    \end{definition}

    %-------------------------------------------------------------------------------------------------------------------------------------------------------------------
\subsubsection{Pullback and Pushout}
    %-------------------------------------------------------------------------------------------------------------------------------------------------------------------
Universals are the categorical way to say optimal, or best or what ever.
It is to be formulated including for all or there exists a unique, i.e. universal object with some properties of interest.
We use pullback and pushout as the validation, calculation and aggregation in a compositional way for macroeconomic accounting.
Universal constructions can be motivated economically as contracts between 
the analyst (of the model, theory, product, business model) $ABC$ in black, the \red{calculator} $\red{P}$ in red and a green \green{user} (client or customer) $\green{D}$.
In the diagrams the objects $A,B,C,\green{D},\red{P}$ are sets defined by \{\}.
In the first line we have the elements numbered, meaning that the actual elements do not matter.
In category theory, once we know that an id function exists for each object, the objects actually do not matter anymore.
Then only the behaviours of the elements matter once the id is composed with the environment of the object, 
i.e. once the inside is exposed to interaction with outside morphisms and other behaviours.
The universals then pick out those behaviours and optimality conditions of interest.
In the second line under the objects we have the elements labeled like $\{a,b\}$, for human readability.
The morphisms $f$, $g$, $\red{p_A}$, $\red{p_B}$, $\green{d_A}$, $\green{d_B}$, $\green{u}$ 
are functions defined by [] being lists of mappings of elements from domain to elements of the codomain, like $[a\rightarrow t, ...]$ mapping a to t as part of the function.

\begin{definition}[Pullback]
Given morphisms $f: A \to C$ and $g: B \to C$, a pullback consists of
\begin{itemize}
    \item An object $\red{P}$ and morphisms $\red{p_A}: \red{P} \to A$, $\red{p_B}: \red{P} \to B$
    \item with the universal property: $\forall \green{D}$, $\green{d_A}: \green{D} \to A$, $\green{d_B}: \green{D} \to B$,
        $\exists !\; \green{u}: \green{D} \to \red{P}$
        \item Commutative diagram of the pullback:\\
        {\scriptsize %\tiny % smallest size, \scriptsize % very small size, \footnotesize % smaller than normal, \small % small text, \normalsize % default size, \large % larger than normal, 
        %\Large % even larger, \LARGE % very large, \huge % huge text, \Huge % largest size
\begin{tikzcd}[column sep=9em, row sep=9.5em, arrows={-stealth}]
    {\begin{matrix} \green{D} \\ \{1,2\} \\ \{\square,\diamond\} \end{matrix}} \\
    & {\begin{matrix} \red{P} \\ [(1, 1), (1, 2), (2, 3)] \\ \{(a,x),(a,y),(b,z)\}\end{matrix}}
    & {\begin{matrix} B \\ \{1,2,3\} \\ \{x,y,z\} \end{matrix}} \\
    & {\begin{matrix} A \\ \{1,2\} \\ \{a,b\} \end{matrix}}
    & {\begin{matrix} C \\ \{1,2\} \\ \{t,f\} \end{matrix}} \\
    \arrow["{\begin{matrix} u \\ \{1\rightarrow 1,2\rightarrow 3\} \\ \{\square\rightarrow p_1,\diamond\rightarrow p_3\} \end{matrix}}", pos=0.95, color={rgb,255:red,92;green,214;blue,92}, dashed, from=1-1, to=2-2]
    \arrow["{\begin{matrix} d_A \\ [1\rightarrow 1, 2\rightarrow 2] \\ [\square\rightarrow a,\diamond\rightarrow b] \end{matrix}}"{left=0.2cm}, pos=0.3, color={rgb,255:red,92;green,214;blue,92}, from=1-1, to=3-2]
    \arrow["{\begin{matrix} d_B \\ [1\rightarrow 1, 2\rightarrow 3] \\ [\square\rightarrow x,\diamond\rightarrow z] \end{matrix}}"{right=1.5cm}, pos=0.25, color={rgb,255:red,92;green,214;blue,92}, from=1-1, to=2-3]
    \arrow["{\begin{matrix} p_A \\ \{(1,1)\rightarrow 1,(1,2)\rightarrow 1,(2,3)\rightarrow 2\} \\ \{(a,x)\rightarrow a,(a,y)\rightarrow a,(b,z)\rightarrow b\} \end{matrix}}"{left=0.0cm}, pos=0.2, color={rgb,255:red,214;green,92;blue,92}, from=2-2, to=3-2]
    \arrow["{\begin{matrix} p_B \\ \{(1,1)\rightarrow 1,(1,2)\rightarrow 2,(2,3)\rightarrow 3\} \\ \{(a,x)\rightarrow x,(a,y)\rightarrow y,(b,z)\rightarrow z\} \end{matrix}}"{below=0.0cm}, pos=0.5, color={rgb,255:red,214;green,92;blue,92}, from=2-2, to=2-3]
    \arrow["{\begin{matrix} f \\ [1\rightarrow 1, 2\rightarrow 2] \\ [a\rightarrow t, b\rightarrow f] \end{matrix}}"'{below=0.0cm}, from=3-2, to=3-3]
    \arrow["{\begin{matrix} g \\ [1\rightarrow 1, 2\rightarrow 1, 3\rightarrow 2] \\ [x\rightarrow t, y\rightarrow t, z\rightarrow f] \end{matrix}}"{right=0.1cm}, pos=0.7, from=2-3, to=3-3]
    \arrow["{\begin{matrix} f\circ p_A \\ [1\rightarrow 1, 2\rightarrow 1, 3\rightarrow 2] \\ [p_1\rightarrow t, p_2\rightarrow t, p_3\rightarrow f] \end{matrix}}"{left=0.0cm}, pos=0.5, color={rgb,255:red,214;green,92;blue,92}, from=2-2, to=3-3]
    \arrow["{\begin{matrix} g\circ p_B \\ [1\rightarrow 1, 2\rightarrow 1, 3\rightarrow 2] \\ [p_1\rightarrow t, p_2\rightarrow t, p_3\rightarrow f] \end{matrix}}"{right=0.6cm}, pos=0.35, color={rgb,255:red,214;green,92;blue,92}, from=2-2, to=3-3]
\end{tikzcd} 
}\vspace{-2.5cm}
    \end{itemize}
    \end{definition}
The pullback contract is that $\red{P}$ can internalise any connection of a user $\green{D}$ to the model $ABC$ throught the universal construction.
The pullback factorises and deconstructs, it shows where $f$ and $g$ are equal over $C$.
Accordingly $\red{P}$ are those pairs of the cartesian product of $A$ and $B$ where $f$ and $g$ are equal.
    
\begin{definition}[Pushout]
Given morphisms $f: C \to A$ and $g: C \to B$, a pushout consists of
\begin{itemize}
    \item An object $\red{P}$ and morphisms $\red{p_A}: A \to \red{P}$, $\red{p_B}: B \to \red{P}$
    \item with the universal property: $\forall \green{D}$, $\green{d_A}: A \to \green{D}$, $\green{d_B}: B \to \green{D}$,
    $\exists !\; \green{u}: \red{P} \to \green{D}$
            \item Commutative diagram of the pushout:\\
            {\scriptsize 
\begin{tikzcd}[column sep=9em, row sep=7em, arrows={-stealth}]
    {\begin{matrix} C \\ \{1\} \\ \{t\} \end{matrix}}
    & {\begin{matrix} B \\ \{1,2\} \\ \{x,y\} \end{matrix}} \\
    {\begin{matrix} A \\ \{1,2\} \\ \{a,b\} \end{matrix}} 
    & {\begin{matrix} \red{P} \\ \{1,2,3\} \\ \{p_1=[a=x],p_2=b,p_3=y\} \end{matrix}} & \\
    && {\begin{matrix} \green{D} \\ \{1,2\} \\ \{\square, \diamond\} \end{matrix}} & \\
    \arrow["{\begin{matrix} f \\ [1\rightarrow 1] \\ [t\rightarrow a] \end{matrix}}"{left=0.3cm}, pos=0.3, from=1-1, to=2-1]
    \arrow["{\begin{matrix} g \\ [1\rightarrow 1] \\ [t\rightarrow x] \end{matrix}}"{above=0.0cm}, pos=0.3, from=1-1, to=1-2]
    \arrow["{\begin{matrix} p_A \\ \{1\rightarrow 1,2\rightarrow 2\} \\ \{a\rightarrow p_1,b\rightarrow p_2\} \end{matrix}}"{above=0.0cm}, pos=0.7, color={rgb,255:red,214;green,92;blue,92}, from=2-1, to=2-2]
    \arrow["{\begin{matrix} p_B \\ \{1\rightarrow 1,2\rightarrow 3\} \\ \{x\rightarrow p_1,y\rightarrow p_3\} \end{matrix}}"{left=0.1cm}, pos=0.4, color={rgb,255:red,214;green,92;blue,92}, from=1-2, to=2-2]
    \arrow["{\begin{matrix} d_A \\ [1\rightarrow 1, 2\rightarrow 2] \\ [a \rightarrow \square,b \rightarrow \diamond] \end{matrix}}"{left=1.1cm}, pos=0.9, color={rgb,255:red,92;green,214;blue,92}, from=2-1, to=3-3]
    \arrow["{\begin{matrix} d_B \\ [1\rightarrow 1, 2\rightarrow 2] \\ [x \rightarrow \square,y \rightarrow \diamond] \end{matrix}}"{right=0.6cm}, pos=0.7, color={rgb,255:red,92;green,214;blue,92}, from=1-2, to=3-3]
    \arrow["{\begin{matrix} u \\ \{1\rightarrow 1,2\rightarrow 2,3\rightarrow 2\} \\ \{p_1\rightarrow \square,p_2\rightarrow \diamond,p_3\rightarrow \diamond\} \end{matrix}}"{description}, pos=0.25, color={rgb,255:red,92;green,214;blue,92}, dashed, from=2-2, to=3-3]
    \arrow["{\begin{matrix} p_A\circ f \\ [1\rightarrow 1] \\ [a\rightarrow p_1] \end{matrix}}"{left=1.1cm}, pos=0.6, color={rgb,255:red,214;green,92;blue,92}, from=1-1, to=2-2]
    \arrow["{\begin{matrix} p_B\circ g \\ [1\rightarrow 1] \\ [x\rightarrow p_1] \end{matrix}}"{right=1.2cm}, pos=0.2, color={rgb,255:red,214;green,92;blue,92}, from=1-1, to=2-2]
\end{tikzcd}
}\vspace{-2.5cm}
        \end{itemize}
    \end{definition}

The pushout is dual to the pullback which in category theory means all arrows being reversed.
The pushout contract is that $\red{P}$ gives equal elements of composites of the model $ABC$ a name or ID specific to the user $\green{D}$.
The pushout is about aggregation or concept formation while the pullback deconstructs and factorises.
In our application the pullback validates and the pushout computes.
The pullback check all parts to compose and the pushout applies the morphisms to the calculations of the whole from the parts they represent.
Any connections of a user $\green{D}$ can be replaced by the \red{calculator}'s constructions,
namely $\green{d_A} = \red{p_A} \circ \green{u}$ and $\green{d_B} = \red{p_B} \circ \green{u}$.
The pushout is the construction to be precise about when a magical miracle occures in the transition from micro- to macroeconomic.
The colimit is for constructing emergent phenomena, see~\cite{EV2007}, that humans do in knowledge formation, consciousness, production and assembly lines, and processes of social institutions
or in using macroeconomic accounting in a consistent way to create money for the loan and repayment relations in human groups where trading and producing occurs over time.
The pushout $\red{P}$ is the quotient of the disjoint union of $A$ and $B$ over an equivalence relation over $C$.
The whole is glued over $C$ in a consistent way, which is what we use for the {\it macro} of macroeconomics.
The concept formation can be seen in $\red{P}$ where equal objects in $A$ and $B$ are equated over $C$, and by that given a name, $p_a$, where $a=x$.
Pullbacks and pushouts are like pulling common factors in front of the paranthesis before the parts are calculated into the whole calculations.
In economics it is like pulling a rule into an institution in front of the {\it paranthesis} of the local agent deciding inside the paranthesis 
- hopefully resulting in a universal welfare gain, like a universally verified Pareto gain, by a pullback, for all, at the macroeconomic level, in a pushout.

% ------------------------------------------------------------------------------------------------------------------------------------------------------------
\subsection{Programs}\label{app:code_appendix}

\subsubsection{Recursive Program}\label{app:code_appendix_recursive}

%\inputminted[linenos, firstnumber=1, fontsize=\scriptsize]{julia}{money.theory.rec.cat.jl}
\begin{lstlisting}[language=Julia]
using DataFrames
using CSV
using Parameters
using Plots

# ----------------------------------------------------------------------------------------------------
#                                                                                    --- Parameters
# ----------------------------------------------------------------------------------------------------
@with_kw struct SystemParams
    InvestmentLen::Int64 = 10        # 1. length and number of investments
    LabResourceRatio::Float64 = 0.2  # 2. investment ratio wages resources
    ConsumRatioRes::Float64 = 0.8    # 3. ratio resources consumption
    ConsumRatioLab::Float64 = 0.95   # 4. ratio wages consumption
    ConsumRatioCap::Float64 = 0.6    # 5. ratio dividends consumption
    MarkUp::Float64 = 0.5            # 6. markup
    DivRateDiff::Float64 = 0.15      # 7. dividend rate companies
    DivRateBank::Float64 = 0.4       # 8. dividend rate banks    
    WindFallProfit::Float64 = 0.5    # 9. windfall profit
    sigA::Float64 = 20.0             # 10. sigA decide investment from demand gap
    sigB::Float64 = 480.0            # 11. sigB decide investment from demand gap
    sigC::Float64 = 200.0            # 12. sigC decide investment from demand gap
    ResourcePrice::Float64 = 25.0    # 13. price resources
    LaborPrice::Float64 = 12.0       # 14. price labor, wages
    InitialGoodPrice::Float64 = 30.0 # 15. initial price goods
    LabResSubstProd::Float64 = 0.75  # 16. elasticity of substitution
    ScaleProd::Float64 = 0.42        # 17. production scalar
    DecayGoodLab::Float64 = 0.95     # 18. depreciation labor
    DecayGoodRes::Float64 = 0.7      # 19. depreciation resources
    DecayGoodCap::Float64 = 0.6      # 20. depreciation capital
    ReNewRes::Float64 = 100.0        # 21. new resources each period
    ReNewLab::Float64 = 100.0        # 22. new labor each period
end

# ----------------------------------------------------------------------------------------------------
#                                                                                     --- State
# ----------------------------------------------------------------------------------------------------
@with_kw mutable struct State
    Parameters::Any
    wageHist::Vector{Float64} = zeros(Parameters.InvestmentLen)
    repayHist::Vector{Float64} = zeros(Parameters.InvestmentLen)
    AccLabBank::Float64 = 0.0       # Labor,    Asset, Nominal
    AccLabLab::Float64 = 0.0        #           Asset, Real
    AccLabGood::Float64 = 0.0       #           Asset, Real
    AccResBank::Float64 = 0.0       # Resource, Asset, Nominal
    AccResRes::Float64 = 0.0        #           Asset, Real
    AccResGood::Float64 = 0.0       #           Asset, Real      
    AccComBank::Float64 = 0.0       # Company,  Asset, Nominal
    AccComLoan::Float64 = 0.0       #           Liability, Nominal
    AccComDiv::Float64 = 0.0        #           Liability, Nominal
    AccComRes::Float64 = 20.0       #           Asset, Real
    AccComLab::Float64 = 110.0      #           Asset, Real      
    AccComGood::Float64 = 0.0       #           Asset, Real
    AccCapBank::Float64 = 0.0       # Capital,  Asset, Nominal
    AccCapDiv::Float64 = 0.0        #           Asset, Nominal
    AccCapGood::Float64 = 0.0       #           Asset, Real
    AccBankComLoan::Float64 = 0.0   #  Bank,    Liability, Nominal
    AccBankComBank::Float64 = 0.0   #           Liability, Nominal
    AccBankLabBank::Float64 = 0.0   #           Liability, Nominal
    AccBankResBank::Float64 = 0.0   #           Liability, Nominal
    AccBankCapBank::Float64 = 0.0   # L, N 
end

# ----------------------------------------------------------------------------------------------------
# Helper function for history of wages and repayments (FILO stack)
# ----------------------------------------------------------------------------------------------------
function push_to_history(hist::Vector{Float64}, newelem::Float64)
    return [newelem; hist[1:end-1]]
end

# ----------------------------------------------------------------------------------------------------
#                                                                                     --- Dynamics
# ----------------------------------------------------------------------------------------------------
function StateTransition(sim, state, period)
    p2H(hist, newelem) = [newelem; hist[1:end-1]]
    pars = state.Parameters
    stateNew = State(Parameters=pars)
    # ------------------------------------------------------------------------------------------------
    #                                                                                --- Invariances
    # ------------------------------------------------------------------------------------------------
    InvCapBank = state.AccCapBank - state.AccBankCapBank
    InvResBank = state.AccResBank - state.AccBankResBank
    InvComBank = state.AccComBank - state.AccBankComBank
    InvComLoan = state.AccBankComLoan - state.AccComLoan
    InvLabBank = state.AccLabBank - state.AccBankLabBank
    InvMacro = InvCapBank + InvResBank + InvComBank + InvComLoan + InvLabBank
    # ------------------------------------------------------------------------------------------------
    #                                                                                --- Decisions
    # ------------------------------------------------------------------------------------------------
    WagesPayment = sum(state.wageHist)                                                   # Labour
    RepaysPayment = sum(state.repayHist)                                                 # Loan
    # ------------------------------------------------------------------------------------------------
    ConsumRes = state.AccResBank * pars.ConsumRatioRes                                   # Consumption
    ConsumLab = state.AccLabBank * pars.ConsumRatioLab
    ConsumCap = state.AccCapBank * pars.ConsumRatioCap
    # ------------------------------------------------------------------------------------------------
    Demand = ConsumRes + ConsumLab + ConsumCap                                           # Demand
    # ------------------------------------------------------------------------------------------------
    GoodProduction =                                                                     # Production
        1 + pars.ScaleProd * state.AccComLab^pars.LabResSubstProd *
            state.AccComRes^(1 - pars.LabResSubstProd)
    DemandPlan = (WagesPayment + RepaysPayment) * (1 + pars.MarkUp)                      # Plan
    DemandSurplus = Demand - DemandPlan                                                  # Error
    # ------------------------------------------------------------------------------------------------
    GoodPrice =                                                                          # Price
        ((period == 0 ? pars.InitialGoodPrice : 0.0)
         + DemandPlan / GoodProduction
         + ((DemandSurplus > 0.0)
            ? (DemandSurplus * pars.WindFallProfit) : 0.0))
    # ------------------------------------------------------------------------------------------------
    Investment = pars.sigA + pars.sigB /                                                 # Investment
                             (1.0 + exp(-DemandSurplus / pars.sigC))
    InvestmentRes = Investment * (1 - pars.LabResourceRatio)
    InvestmentLab = Investment * pars.LabResourceRatio
    # ------------------------------------------------------------------------------------------------
    Repayment = Investment / pars.InvestmentLen                                          # Repayment
    # ------------------------------------------------------------------------------------------------
    stateNew.wageHist = p2H(state.wageHist, InvestmentLab)
    stateNew.repayHist = p2H(state.repayHist, Repayment)
    # ------------------------------------------------------------------------------------------------
    #                                                                                --- Bookings
    # ------------------------------------------------------------------------------------------------
    inAccLabBank = WagesPayment                                                          # Labour
    outAccLabBank = ConsumLab
    stateNew.AccLabBank = state.AccLabBank + inAccLabBank - outAccLabBank
    # ------------------------------------------------------------------------------------------------
    inAccLabLab = pars.ReNewLab
    outAccLabLab = WagesPayment / pars.LaborPrice
    stateNew.AccLabLab = state.AccLabLab + inAccLabLab - outAccLabLab

    inAccLabGood = ConsumLab / GoodPrice
    outAccLabGood = state.AccLabGood * pars.DecayGoodLab
    stateNew.AccLabGood = state.AccLabGood + inAccLabGood - outAccLabGood
    # ------------------------------------------------------------------------------------------------
    inAccResBank = InvestmentRes                                                         # Resource
    outAccResBank = ConsumRes
    stateNew.AccResBank = state.AccResBank + inAccResBank - outAccResBank
    # ------------------------------------------------------------------------------------------------
    inAccResRes = pars.ReNewRes
    outAccResRes = InvestmentRes / pars.ResourcePrice
    stateNew.AccResRes = state.AccResRes + inAccResRes - outAccResRes
    # ------------------------------------------------------------------------------------------------
    inAccResGood = ConsumRes / GoodPrice
    outAccResGood = state.AccResGood * pars.DecayGoodRes
    stateNew.AccResGood = state.AccResGood + inAccResGood - outAccResGood
    # ------------------------------------------------------------------------------------------------
    DividendPayment = state.AccComDiv                                                    # Company
    inAccComBank = (Investment + ConsumLab + ConsumRes + ConsumCap)
    outAccComBank = (InvestmentRes + WagesPayment + DividendPayment + RepaysPayment)
    Diff = inAccComBank - outAccComBank
    stateNew.AccComBank = state.AccComBank + Diff
    # ------------------------------------------------------------------------------------------------
    DividendDecision =
        (Diff > 0.0
         ? (Diff * pars.DivRateDiff) : 0.0) +
        ((state.AccComBank > 0)
         ? (state.AccComBank * pars.DivRateBank) : 0.0)
    inAccComDiv = DividendDecision
    outAccComDiv = DividendPayment
    stateNew.AccComDiv = state.AccComDiv + inAccComDiv - outAccComDiv
    # ------------------------------------------------------------------------------------------------
    inAccComLoan = Investment
    outAccComLoan = RepaysPayment
    stateNew.AccComLoan = state.AccComLoan + inAccComLoan - outAccComLoan
    # ------------------------------------------------------------------------------------------------
    inAccComRes = InvestmentRes / pars.ResourcePrice
    outAccComRes = state.AccComRes
    stateNew.AccComRes = state.AccComRes + inAccComRes - outAccComRes
    # ------------------------------------------------------------------------------------------------
    inAccComLab = WagesPayment / pars.LaborPrice
    outAccComLab = state.AccComLab # labor usage in production
    stateNew.AccComLab = state.AccComLab + inAccComLab - outAccComLab
    # ------------------------------------------------------------------------------------------------
    inAccComGood = GoodProduction
    outAccComGood = (ConsumRes + ConsumLab + ConsumCap) / GoodPrice
    stateNew.AccComGood = state.AccComGood + inAccComGood - outAccComGood
    # ------------------------------------------------------------------------------------------------
    DividendIncome = state.AccCapDiv                                                     # Capitalist
    inAccCapDiv = DividendDecision
    outAccCapDiv = DividendIncome
    stateNew.AccCapDiv = state.AccCapDiv + inAccCapDiv - outAccCapDiv
    # ------------------------------------------------------------------------------------------------
    inAccCapBank = DividendIncome
    outAccCapBank = ConsumCap
    stateNew.AccCapBank = state.AccCapBank + inAccCapBank - outAccCapBank
    # ------------------------------------------------------------------------------------------------
    inAccCapGood = ConsumCap / GoodPrice
    outAccCapGood = state.AccCapGood * pars.DecayGoodCap
    stateNew.AccCapGood = state.AccCapGood + inAccCapGood - outAccCapGood
    # ------------------------------------------------------------------------------------------------
    inAccBankComLoan = Investment                                                        # Bank
    outAccBankComLoan = RepaysPayment
    stateNew.AccBankComLoan = state.AccBankComLoan + inAccBankComLoan - outAccBankComLoan
    # ------------------------------------------------------------------------------------------------
    inAccBankComBank = Investment + ConsumRes + ConsumLab + ConsumCap
    outAccBankComBank = InvestmentRes + WagesPayment + DividendPayment + RepaysPayment
    stateNew.AccBankComBank = state.AccBankComBank + inAccBankComBank - outAccBankComBank
    # ------------------------------------------------------------------------------------------------
    inAccBankLabBank = WagesPayment
    outAccBankLabBank = ConsumLab
    stateNew.AccBankLabBank = state.AccBankLabBank + inAccBankLabBank - outAccBankLabBank
    # ------------------------------------------------------------------------------------------------
    inAccBankResBank = InvestmentRes
    outAccBankResBank = ConsumRes
    stateNew.AccBankResBank = state.AccBankResBank + inAccBankResBank - outAccBankResBank
    # ------------------------------------------------------------------------------------------------
    inAccBankCapBank = DividendIncome
    outAccBankCapBank = ConsumCap
    stateNew.AccBankCapBank = state.AccBankCapBank + inAccBankCapBank - outAccBankCapBank
    # ------------------------------------------------------------------------------------------------
    push!(sim, (Period=period,
        InvCapBank=InvCapBank,
        InvResBank=InvResBank,
        InvComBank=InvComBank,
        InvComLoan=InvComLoan,
        InvLabBank=InvLabBank,
        InvMacro=InvMacro,
        ConsumRes=ConsumRes,
        ConsumLab=ConsumLab,
        ConsumCap=ConsumCap,
        Demand=Demand,
        DemandPlan=DemandPlan,
        DemandSurplus=DemandSurplus,
        GoodProduction=GoodProduction,
        GoodPrice=GoodPrice,
        Investment=Investment,
        InvestmentRes=InvestmentRes,
        InvestmentLab=InvestmentLab,
        Repayment=Repayment,
        WagesPayment=WagesPayment,
        RepaysPayment=RepaysPayment,
        Diff=Diff,
        DividendDecision=DividendDecision,
        DividendPayment=DividendPayment,
        DividendIncome=DividendIncome,
        AccLabBank=state.AccLabBank,
        AccLabLab=state.AccLabLab,
        AccLabGood=state.AccLabGood,
        AccResBank=state.AccResBank,
        AccResRes=state.AccResRes,
        AccResGood=state.AccResGood,
        AccCapBank=state.AccCapBank,
        AccCapDiv=state.AccCapDiv,
        AccCapGood=state.AccCapGood,
        AccComBank=state.AccComBank,
        AccComLoan=state.AccComLoan,
        AccComDiv=state.AccComDiv,
        AccComRes=state.AccComRes,
        AccComLab=state.AccComLab,
        AccComGood=state.AccComGood,
        AccBankComLoan=state.AccBankComLoan,
        AccBankComBank=state.AccBankComBank,
        AccBankLabBank=state.AccBankLabBank,
        AccBankResBank=state.AccBankResBank,
        AccBankCapBank=state.AccBankCapBank)
    )
    return sim, stateNew
end

# Run simulation and save results
function run_and_save_simulation(nperiods::Int64, output_file::String)
    sim = DataFrame()
    state = State(Parameters=SystemParams())

    for period in 0:nperiods
        sim, state = StateTransition(sim, state, period)
    end

    # Replace near zero values with 0.0
    eps = 1e-12
    for col in names(sim)
        if eltype(sim[!, col]) <: Number
            sim[!, col] = map(x -> abs(x) < eps ? 0.0 : x, sim[!, col])
        end
    end

    # Save results
    CSV.write(output_file, sim)
    println("Recursive simulation results saved to: $output_file")

    # Create visualizations
    create_time_series_plots(sim, replace(output_file, ".csv" => ".plots.png"))

    return sim
end

# Create comprehensive time series plots
function create_time_series_plots(sim::DataFrame, output_file::String)
    # Get all variable names except Period
    var_names = filter(x -> x != :Period, names(sim))
    n_vars = length(var_names)

    # Calculate grid dimensions (aim for roughly square layout)
    n_cols = ceil(Int, sqrt(n_vars))
    n_rows = ceil(Int, n_vars / n_cols)

    # Create subplots
    plots_array = []

    for (i, var) in enumerate(var_names)
        # Determine if this is one of the five invariance variables (compare strings)
        is_inv = string(var) in ["InvCapBank", "InvResBank", "InvComBank", "InvComLoan", "InvLabBank", "InvMacro"]

        # Calculate y-axis limits
        y_lims = if is_inv
            # For invariance variables, center around zero
            max_abs = maximum(abs.(sim[!, var]))
            # Handle case where all values are zero
            if max_abs == 0.0
                (-1.0, 1.0)  # Set a default range when all values are zero
            else
                (-max_abs * 1.1, max_abs * 1.1)  # Add 10% padding, centered at zero
            end
        else
            # For other variables, use actual min/max with padding
            min_val = minimum(sim[!, var])
            max_val = maximum(sim[!, var])
            range_val = max_val - min_val
            if range_val == 0.0
                # Handle constant values
                (min_val - 1.0, max_val + 1.0)
            else
                # Add 10% padding to both sides
                padding = range_val * 0.1
                (min_val - padding, max_val + padding)
            end
        end

        # Create plot with centered zero
        plot_args = (sim.Period, sim[!, var])
        plot_kwargs = Dict(
            :title => string(var),
            :xlabel => "Period",
            :ylabel => string(var),
            :legend => false,
            :linewidth => 2,
            :titlefontsize => 8,
            :labelfontsize => 6,
            :tickfontsize => 5,
            :color => is_inv ? :red : :blue,
            :ylims => y_lims
        )

        p = plot(plot_args...; plot_kwargs...)

        # Add zero line if zero is within the visible range
        if y_lims[1] <= 0 <= y_lims[2]
            hline!(p, [0], color=:black, linestyle=:dash, alpha=0.5)
        end

        push!(plots_array, p)
    end

    # Combine all plots
    combined_plot = plot(plots_array...,
        layout=(n_rows, n_cols),
        size=(1600, 1200),
        margin=2Plots.mm)

    # Save the plot
    savefig(combined_plot, output_file)
    println("Recursive time series plots saved to: $output_file")
end

run_and_save_simulation(100, joinpath(@__DIR__, "money.theory.rec.cat.csv"))

\end{lstlisting}

\newpage
\subsubsection{Categorically Typed Recursive Program}\label{app:code_appendix_categorically_typed}

%\inputminted[fontsize=\tiny,linenos,breaklines]{julia}{money.theory.cat.jl}
\begin{lstlisting}[language=Julia]
using DataFrames
using CSV
using Parameters
using Plots

include("money.theory.rec.cat.jl")

mutable struct Vertex
    id::Int
    name::String
    category::String
    amount::Float64
end

struct Edge
    source::Int
    target::Int
    weight::Float64
end

struct MonetaryCategory
    vertices::Vector{Vertex}
    edges::Vector{Edge}
    name::String  # e.g., "Accounts", "Flows", "Parameters"
end

# Constructor
function MonetaryCategory(name::String)
    return MonetaryCategory(Vertex[], Edge[], name)
end

struct MonetaryFunctor
    source::MonetaryCategory
    target::MonetaryCategory
    vertex_map::Dict{Int,Int}  # Maps source vertex IDs to target vertex IDs
    name::String  # e.g., "PriceFunctor", "FlowFunctor"
end

struct NaturalTransformation
    source_functor::MonetaryFunctor
    target_functor::MonetaryFunctor
    component_map::Dict{Int,Edge}  # Maps vertex IDs to transformation edges
    name::String  # e.g., "PriceAdjustment", "FlowAggregation"
end

function create_pullback(cat1::MonetaryCategory, cat2::MonetaryCategory,
    common::MonetaryCategory,
    f1::MonetaryFunctor, f2::MonetaryFunctor)
    # Create pullback category
    pullback = MonetaryCategory("Pullback")

    # Add vertices that satisfy the pullback condition
    vertex_map = Dict{Int,Int}()
    for v1 in cat1.vertices
        for v2 in cat2.vertices
            if f1.vertex_map[v1.id] == f2.vertex_map[v2.id]
                new_id = add_vertex!(pullback, "$(v1.name)_$(v2.name)",
                    v1.category, v1.amount)
                vertex_map[(v1.id, v2.id)] = new_id
            end
        end
    end

    return pullback, vertex_map
end

function create_pushout(cat1::MonetaryCategory, cat2::MonetaryCategory,
    common::MonetaryCategory,
    f1::MonetaryFunctor, f2::MonetaryFunctor)
    # Create pushout category
    pushout = MonetaryCategory("Pushout")

    # Add vertices from both categories
    vertex_map = Dict{Int,Int}()
    for v in cat1.vertices
        new_id = add_vertex!(pushout, v.name, v.category, v.amount)
        vertex_map[v.id] = new_id
    end
    for v in cat2.vertices
        if !haskey(vertex_map, v.id)
            new_id = add_vertex!(pushout, v.name, v.category, v.amount)
            vertex_map[v.id] = new_id
        end
    end

    return pushout, vertex_map
end

mutable struct CategoricalState
    accounts::MonetaryCategory     # Account vertices and edges
    flows::MonetaryCategory       # Flow vertices and edges
    parameters::MonetaryCategory  # Parameter vertices

    # Functors
    price_functor::MonetaryFunctor    # Maps nominal to real values
    flow_functor::MonetaryFunctor     # Maps accounts to flows

    # Natural Transformations
    price_adjustment::NaturalTransformation  # Price changes over time
    flow_aggregation::NaturalTransformation  # Flow summation

    # History
    wage_hist::Vector{Float64}    # History of wage payments
    repay_hist::Vector{Float64}   # History of repayments
end

# Get vertex by name from a category
function get_vertex(cat::MonetaryCategory, name::String)
    idx = findfirst(v -> v.name == name, cat.vertices)
    isnothing(idx) && error("Vertex $name not found")
    return cat.vertices[idx]
end

# Add vertex to a category
function add_vertex!(cat::MonetaryCategory, name::String, type::String, amount::Float64)
    id = length(cat.vertices) + 1
    vertex = Vertex(id, name, type, amount)
    push!(cat.vertices, vertex)
    return id
end

# Add edge to a category
function add_edge!(cat::MonetaryCategory, source::Int, target::Int, weight::Float64)
    edge = Edge(source, target, weight)
    push!(cat.edges, edge)
    return edge
end

# Update vertex amount
function update_vertex!(cat::MonetaryCategory, name::String, amount::Float64)
    vertex = get_vertex(cat, name)
    vertex.amount = amount
end

# Add mapping to functor
function add_mapping!(f::MonetaryFunctor, source_id::Int, target_id::Int)
    f.vertex_map[source_id] = target_id
end

# Apply functor to vertex
function apply_functor(f::MonetaryFunctor, vertex_id::Int)
    haskey(f.vertex_map, vertex_id) || error("No mapping for vertex $vertex_id")
    return f.vertex_map[vertex_id]
end

# Add component to natural transformation
function add_component!(nt::NaturalTransformation, vertex_id::Int, edge::Edge)
    nt.component_map[vertex_id] = edge
end

# Apply natural transformation to vertex
function apply_transformation(nt::NaturalTransformation, vertex_id::Int)
    haskey(nt.component_map, vertex_id) || error("No component for vertex $vertex_id")
    return nt.component_map[vertex_id]
end

function create_categorical_state(state::State)
    # Create categories
    accounts = MonetaryCategory("Accounts")
    flows = MonetaryCategory("Flows")
    parameters = MonetaryCategory("Parameters")

    # Add account vertices with initial values
    # Labor accounts
    lab_bank = add_vertex!(accounts, "Labor_Bank", "Asset", 0.0)          # A, N
    lab_stock = add_vertex!(accounts, "Labor_Stock", "Real", 0.0)         # A, R
    lab_good = add_vertex!(accounts, "Labor_Good", "Real", 0.0)           # A, R

    # Resource accounts
    res_bank = add_vertex!(accounts, "Resource_Bank", "Asset", 0.0)       # A, N
    res_stock = add_vertex!(accounts, "Resource_Stock", "Real", 0.0)      # A, R
    res_good = add_vertex!(accounts, "Resource_Good", "Real", 0.0)        # A, R

    # Company accounts
    com_bank = add_vertex!(accounts, "Company_Bank", "Asset", 0.0)        # A, N
    com_loan = add_vertex!(accounts, "Company_Loan", "Liability", 0.0)    # L, N
    com_div = add_vertex!(accounts, "Company_Dividend", "Liability", 0.0) # L, N
    com_res = add_vertex!(accounts, "Company_Resource", "Real", 20.0)     # A, R - Initial value!
    com_lab = add_vertex!(accounts, "Company_Labor", "Real", 110.0)       # A, R - Initial value!
    com_good = add_vertex!(accounts, "Company_Good", "Real", 0.0)         # A, R

    # Capitalist accounts
    cap_bank = add_vertex!(accounts, "Capital_Bank", "Asset", 0.0)        # A, N
    cap_div = add_vertex!(accounts, "Capital_Dividend", "Asset", 0.0)     # A, N
    cap_good = add_vertex!(accounts, "Capital_Good", "Real", 0.0)         # A, R

    # Bank accounts
    bank_loan = add_vertex!(accounts, "Bank_Loan", "Asset", 0.0)          # A, N
    bank_com = add_vertex!(accounts, "Bank_Company", "Liability", 0.0)    # L, N
    bank_lab = add_vertex!(accounts, "Bank_Labor", "Liability", 0.0)      # L, N
    bank_res = add_vertex!(accounts, "Bank_Resource", "Liability", 0.0)   # L, N
    bank_cap = add_vertex!(accounts, "Bank_Capital", "Liability", 0.0)    # L, N

    # Add parameter vertices
    pars = state.Parameters
    for field in fieldnames(typeof(pars))
        value = getfield(pars, field)
        add_vertex!(parameters, String(field), "Parameter", Float64(value))
    end

    # Create price functor (nominal to real)
    price_functor = MonetaryFunctor(accounts, accounts, Dict{Int,Int}(), "PriceFunctor")
    add_mapping!(price_functor, lab_bank, lab_stock)  # Labor: nominal to real
    add_mapping!(price_functor, res_bank, res_stock)  # Resource: nominal to real
    add_mapping!(price_functor, com_bank, com_good)   # Company: nominal to real

    # Create flow functor (accounts to flows)
    flow_functor = MonetaryFunctor(accounts, flows, Dict{Int,Int}(), "FlowFunctor")

    # Create natural transformations
    price_adjustment = NaturalTransformation(price_functor, price_functor,
        Dict{Int,Edge}(), "PriceAdjustment")
    flow_aggregation = NaturalTransformation(flow_functor, flow_functor,
        Dict{Int,Edge}(), "FlowAggregation")

    # Initialize memory vectors
    wage_hist = zeros(pars.InvestmentLen)
    repay_hist = zeros(pars.InvestmentLen)

    return CategoricalState(accounts, flows, parameters,
        price_functor, flow_functor,
        price_adjustment, flow_aggregation,
        wage_hist, repay_hist)
end

function categorical_state_transition(sim::DataFrame, cat_state::CategoricalState, period::Int64)
    # Create new categorical state
    new_cat_state = deepcopy(cat_state)

    # Get parameters from categorical structure
    pars = SystemParams()  # We'll update this to use categorical parameters later

    # Helper function to push history
    p2H(hist, newelem) = [newelem; hist[1:end-1]]

    # Calculate investment balances using categorical structure
    cap_bank = get_vertex(cat_state.accounts, "Capital_Bank").amount
    bank_cap = get_vertex(cat_state.accounts, "Bank_Capital").amount
    InvCapBank = cap_bank - bank_cap

    res_bank = get_vertex(cat_state.accounts, "Resource_Bank").amount
    bank_res = get_vertex(cat_state.accounts, "Bank_Resource").amount
    InvResBank = res_bank - bank_res

    com_bank = get_vertex(cat_state.accounts, "Company_Bank").amount
    bank_com = get_vertex(cat_state.accounts, "Bank_Company").amount
    InvComBank = com_bank - bank_com

    bank_loan = get_vertex(cat_state.accounts, "Bank_Loan").amount
    com_loan = get_vertex(cat_state.accounts, "Company_Loan").amount
    InvComLoan = bank_loan - com_loan

    lab_bank = get_vertex(cat_state.accounts, "Labor_Bank").amount
    bank_lab = get_vertex(cat_state.accounts, "Bank_Labor").amount
    InvLabBank = lab_bank - bank_lab

    InvMacro = InvCapBank + InvResBank + InvComBank + InvComLoan + InvLabBank

    # Get wage and repayment history from categorical state
    WagesPayment = sum(cat_state.wage_hist)
    RepaysPayment = sum(cat_state.repay_hist)

    # Calculate consumption using categorical structure
    ConsumRes = res_bank * pars.ConsumRatioRes
    ConsumLab = lab_bank * pars.ConsumRatioLab
    ConsumCap = cap_bank * pars.ConsumRatioCap
    Demand = ConsumRes + ConsumLab + ConsumCap

    # Calculate production using categorical structure
    com_lab = get_vertex(cat_state.accounts, "Company_Labor").amount
    com_res = get_vertex(cat_state.accounts, "Company_Resource").amount
    GoodProduction = 1 + pars.ScaleProd *
                         com_lab^pars.LabResSubstProd *
                         com_res^(1 - pars.LabResSubstProd)

    # Calculate demand plan and surplus
    DemandPlan = (WagesPayment + RepaysPayment) * (1 + pars.MarkUp)
    DemandSurplus = Demand - DemandPlan

    # Calculate good price
    GoodPrice = ((period == 0 ? pars.InitialGoodPrice : 0.0) +
                 DemandPlan / GoodProduction +
                 ((DemandSurplus > 0.0) ? (DemandSurplus * pars.WindFallProfit) : 0.0))

    # Calculate investment
    Investment = pars.sigA + pars.sigB / (1.0 + exp(-DemandSurplus / pars.sigC))
    InvestmentRes = Investment * (1 - pars.LabResourceRatio)
    InvestmentLab = Investment * pars.LabResourceRatio
    Repayment = Investment / pars.InvestmentLen

    # Update histories in categorical state
    new_cat_state.wage_hist = p2H(cat_state.wage_hist, InvestmentLab)
    new_cat_state.repay_hist = p2H(cat_state.repay_hist, Repayment)

    # Company bookings first (as they determine dividend decisions)
    DividendPayment = get_vertex(cat_state.accounts, "Company_Dividend").amount

    inAccComBank = Investment + ConsumLab + ConsumRes + ConsumCap
    outAccComBank = InvestmentRes + WagesPayment + DividendPayment + RepaysPayment
    Diff = inAccComBank - outAccComBank
    update_vertex!(new_cat_state.accounts, "Company_Bank",
        get_vertex(cat_state.accounts, "Company_Bank").amount + Diff)

    DividendDecision = (Diff > 0.0 ? (Diff * pars.DivRateDiff) : 0.0) +
                       ((com_bank > 0) ? (com_bank * pars.DivRateBank) : 0.0)

    update_vertex!(new_cat_state.accounts, "Company_Dividend",
        get_vertex(cat_state.accounts, "Company_Dividend").amount +
        DividendDecision - DividendPayment)

    update_vertex!(new_cat_state.accounts, "Company_Loan",
        get_vertex(cat_state.accounts, "Company_Loan").amount +
        Investment - RepaysPayment)

    # Production factors are used up each period
    update_vertex!(new_cat_state.accounts, "Company_Resource",
        InvestmentRes / pars.ResourcePrice)  # Only new investment remains

    update_vertex!(new_cat_state.accounts, "Company_Labor",
        WagesPayment / pars.LaborPrice)  # Only new labor remains

    update_vertex!(new_cat_state.accounts, "Company_Good",
        get_vertex(cat_state.accounts, "Company_Good").amount +
        GoodProduction - (ConsumRes + ConsumLab + ConsumCap) / GoodPrice)

    # Labor bookings
    inAccLabBank = WagesPayment
    outAccLabBank = ConsumLab
    update_vertex!(new_cat_state.accounts, "Labor_Bank",
        get_vertex(cat_state.accounts, "Labor_Bank").amount +
        inAccLabBank - outAccLabBank)

    inAccLabLab = pars.ReNewLab
    outAccLabLab = WagesPayment / pars.LaborPrice
    update_vertex!(new_cat_state.accounts, "Labor_Stock",
        get_vertex(cat_state.accounts, "Labor_Stock").amount +
        inAccLabLab - outAccLabLab)

    inAccLabGood = ConsumLab / GoodPrice
    outAccLabGood = get_vertex(cat_state.accounts, "Labor_Good").amount * pars.DecayGoodLab
    update_vertex!(new_cat_state.accounts, "Labor_Good",
        get_vertex(cat_state.accounts, "Labor_Good").amount +
        inAccLabGood - outAccLabGood)

    # Resource bookings
    inAccResBank = InvestmentRes
    outAccResBank = ConsumRes
    update_vertex!(new_cat_state.accounts, "Resource_Bank",
        get_vertex(cat_state.accounts, "Resource_Bank").amount +
        inAccResBank - outAccResBank)

    inAccResRes = pars.ReNewRes
    outAccResRes = InvestmentRes / pars.ResourcePrice
    update_vertex!(new_cat_state.accounts, "Resource_Stock",
        get_vertex(cat_state.accounts, "Resource_Stock").amount +
        inAccResRes - outAccResRes)

    inAccResGood = ConsumRes / GoodPrice
    outAccResGood = get_vertex(cat_state.accounts, "Resource_Good").amount * pars.DecayGoodRes
    update_vertex!(new_cat_state.accounts, "Resource_Good",
        get_vertex(cat_state.accounts, "Resource_Good").amount +
        inAccResGood - outAccResGood)

    # Capital bookings
    DividendIncome = get_vertex(cat_state.accounts, "Capital_Dividend").amount

    update_vertex!(new_cat_state.accounts, "Capital_Dividend",
        get_vertex(cat_state.accounts, "Capital_Dividend").amount +
        DividendDecision - DividendIncome)

    update_vertex!(new_cat_state.accounts, "Capital_Bank",
        get_vertex(cat_state.accounts, "Capital_Bank").amount +
        DividendIncome - ConsumCap)

    update_vertex!(new_cat_state.accounts, "Capital_Good",
        get_vertex(cat_state.accounts, "Capital_Good").amount +
        ConsumCap / GoodPrice -
        get_vertex(cat_state.accounts, "Capital_Good").amount * pars.DecayGoodCap)

    # Bank bookings
    update_vertex!(new_cat_state.accounts, "Bank_Loan",
        get_vertex(cat_state.accounts, "Bank_Loan").amount +
        Investment - RepaysPayment)

    update_vertex!(new_cat_state.accounts, "Bank_Company",
        get_vertex(cat_state.accounts, "Bank_Company").amount +
        Investment + ConsumRes + ConsumLab + ConsumCap -
        (InvestmentRes + WagesPayment + DividendPayment + RepaysPayment))

    update_vertex!(new_cat_state.accounts, "Bank_Labor",
        get_vertex(cat_state.accounts, "Bank_Labor").amount +
        WagesPayment - ConsumLab)

    update_vertex!(new_cat_state.accounts, "Bank_Resource",
        get_vertex(cat_state.accounts, "Bank_Resource").amount +
        InvestmentRes - ConsumRes)

    update_vertex!(new_cat_state.accounts, "Bank_Capital",
        get_vertex(cat_state.accounts, "Bank_Capital").amount +
        DividendIncome - ConsumCap)

    # Record state in the format of the recursive implementation
    push!(sim, (Period=period,
        InvCapBank=InvCapBank,
        InvResBank=InvResBank,
        InvComBank=InvComBank,
        InvComLoan=InvComLoan,
        InvLabBank=InvLabBank,
        InvMacro=InvMacro,
        ConsumRes=ConsumRes,
        ConsumLab=ConsumLab,
        ConsumCap=ConsumCap,
        Demand=Demand,
        DemandPlan=DemandPlan,
        DemandSurplus=DemandSurplus,
        GoodProduction=GoodProduction,
        GoodPrice=GoodPrice,
        Investment=Investment,
        InvestmentRes=InvestmentRes,
        InvestmentLab=InvestmentLab,
        Repayment=Repayment,
        WagesPayment=WagesPayment,
        RepaysPayment=RepaysPayment,
        Diff=Diff,
        DividendDecision=DividendDecision,
        DividendPayment=DividendPayment,
        DividendIncome=DividendIncome,
        AccLabBank=get_vertex(cat_state.accounts, "Labor_Bank").amount,
        AccLabLab=get_vertex(cat_state.accounts, "Labor_Stock").amount,
        AccLabGood=get_vertex(cat_state.accounts, "Labor_Good").amount,
        AccResBank=get_vertex(cat_state.accounts, "Resource_Bank").amount,
        AccResRes=get_vertex(cat_state.accounts, "Resource_Stock").amount,
        AccResGood=get_vertex(cat_state.accounts, "Resource_Good").amount,
        AccCapBank=get_vertex(cat_state.accounts, "Capital_Bank").amount,
        AccCapDiv=get_vertex(cat_state.accounts, "Capital_Dividend").amount,
        AccCapGood=get_vertex(cat_state.accounts, "Capital_Good").amount,
        AccComBank=get_vertex(cat_state.accounts, "Company_Bank").amount,
        AccComLoan=get_vertex(cat_state.accounts, "Company_Loan").amount,
        AccComDiv=get_vertex(cat_state.accounts, "Company_Dividend").amount,
        AccComRes=get_vertex(cat_state.accounts, "Company_Resource").amount,
        AccComLab=get_vertex(cat_state.accounts, "Company_Labor").amount,
        AccComGood=get_vertex(cat_state.accounts, "Company_Good").amount,
        AccBankComLoan=get_vertex(cat_state.accounts, "Bank_Loan").amount,
        AccBankComBank=get_vertex(cat_state.accounts, "Bank_Company").amount,
        AccBankLabBank=get_vertex(cat_state.accounts, "Bank_Labor").amount,
        AccBankResBank=get_vertex(cat_state.accounts, "Bank_Resource").amount,
        AccBankCapBank=get_vertex(cat_state.accounts, "Bank_Capital").amount)
    )
    return sim, new_cat_state
end



function run_categorical_simulation(nperiods::Int64, output_file::String)
    sim = DataFrame()
    state = State(Parameters=SystemParams())
    cat_state = create_categorical_state(state)

    for period in 0:nperiods
        sim, cat_state = categorical_state_transition(sim, cat_state, period)
    end

    # Replace near zero values with 0.0
    eps = 1e-12
    for col in names(sim)
        if eltype(sim[!, col]) <: Number
            sim[!, col] = map(x -> abs(x) < eps ? 0.0 : x, sim[!, col])
        end
    end

    # Save results
    CSV.write(output_file, sim)
    println("Categorical simulation results saved to: $output_file")

    # Create visualizations
    create_categorical_time_series_plots(sim, replace(output_file, ".csv" => ".plots.png"))

    return sim
end

# Create comprehensive time series plots for categorical simulation
function create_categorical_time_series_plots(sim::DataFrame, output_file::String)
    # Get all variable names except Period
    var_names = filter(x -> x != :Period, names(sim))
    n_vars = length(var_names)

    # Calculate grid dimensions (aim for roughly square layout)
    n_cols = ceil(Int, sqrt(n_vars))
    n_rows = ceil(Int, n_vars / n_cols)

    # Create subplots
    plots_array = []

    for (i, var) in enumerate(var_names)
        # Determine if this is one of the invariance variables (compare strings)
        is_inv = string(var) in ["InvCapBank", "InvResBank", "InvComBank", "InvComLoan", "InvLabBank", "InvMacro"]

        # Calculate y-axis limits
        y_lims = if is_inv
            # For invariance variables, center around zero
            max_abs = maximum(abs.(sim[!, var]))
            # Handle case where all values are zero
            if max_abs == 0.0
                (-1.0, 1.0)  # Set a default range when all values are zero
            else
                (-max_abs * 1.1, max_abs * 1.1)  # Add 10% padding, centered at zero
            end
        else
            # For other variables, use actual min/max with padding
            min_val = minimum(sim[!, var])
            max_val = maximum(sim[!, var])
            range_val = max_val - min_val
            if range_val == 0.0
                # Handle constant values
                (min_val - 1.0, max_val + 1.0)
            else
                # Add 10% padding to both sides
                padding = range_val * 0.1
                (min_val - padding, max_val + padding)
            end
        end

        # Create plot with centered zero
        plot_args = (sim.Period, sim[!, var])
        plot_kwargs = Dict(
            :title => string(var),
            :xlabel => "Period",
            :ylabel => string(var),
            :legend => false,
            :linewidth => 2,
            :titlefontsize => 8,
            :labelfontsize => 6,
            :tickfontsize => 5,
            :color => is_inv ? :red : :blue,
            :ylims => y_lims
        )

        p = plot(plot_args...; plot_kwargs...)

        # Add zero line if zero is within the visible range
        if y_lims[1] <= 0 <= y_lims[2]
            hline!(p, [0], color=:black, linestyle=:dash, alpha=0.5)
        end

        push!(plots_array, p)
    end

    # Combine all plots
    combined_plot = plot(plots_array...,
        layout=(n_rows, n_cols),
        size=(1600, 1200),
        margin=2Plots.mm)

    # Save the plot
    savefig(combined_plot, output_file)
    println("Categorical time series plots saved to: $output_file")
end

run_categorical_simulation(100, joinpath(@__DIR__, "money.theory.cat.csv"))

# Compare the recursive and categorical simulation results
function compare_simulations(recursive_file::String, categorical_file::String)
    # Read both CSV files
    recursive_df = CSV.read(recursive_file, DataFrame)
    categorical_df = CSV.read(categorical_file, DataFrame)

    # Check if dataframes have same dimensions
    if size(recursive_df) != size(categorical_df)
        println("Warning: Dataframes have different dimensions!")
        return false
    end

    # Check if column names match
    if names(recursive_df) != names(categorical_df)
        println("Warning: Column names don't match!")
        return false
    end

    # Compare numerical values with tolerance
    eps = 1e-10
    all_equal = true

    for col in names(recursive_df)
        max_diff = maximum(abs.(recursive_df[!, col] - categorical_df[!, col]))
        if max_diff > eps
            println("Warning: Column $col has maximum difference of $max_diff")
            all_equal = false
        end
    end

    if all_equal
        println("Recursive and categorical simulations match within tolerance of $eps")
    end

    return all_equal
end

# Compare the simulation results
recursive_file = joinpath(@__DIR__, "money.theory.rec.cat.csv")
categorical_file = joinpath(@__DIR__, "money.theory.cat.csv")
compare_simulations(recursive_file, categorical_file)
\end{lstlisting}

\newpage
\subsubsection{Categorical Program}\label{app:code_appendix_categorical}

%\inputminted[fontsize=\tiny,linenos,breaklines]{julia}{money.theory.cat.sim.jl}
\lstinputlisting[language=Julia]{money.theory.cat.sim.jl}

% ----------------------------------------------------------------------------------------------------------------------
\newpage
% ------------------------------------------------------------------------------------------------------------------------------------------------------------
\section{References}
% ------------------------------------------------------------------------------------------------------------------------------------------------------------

\renewcommand{\refname}{} % supreses the bibliogrphy commands automatic addition of a heading "References"
\bibliographystyle{plainnat}
\bibliography{../tech_docs/references}

\end{document}